\documentclass[a4paper,12pt,twoside]{book}
\usepackage[english]{babel}
\usepackage[utf8]{inputenc}
\usepackage[T1]{fontenc}
\usepackage{times}
\usepackage{amssymb,amsmath,amsfonts,amsthm}
\usepackage{hyperref}
\usepackage{url}
\usepackage{graphicx}
\usepackage{epsfig}
\usepackage{textcomp}
\usepackage{listings}
\def\be{\begin{equation}}
\def\ee{\end{equation}}
\def\bea{\begin{eqnarray}}
\def\eea{\end{eqnarray}}
\def\vec#1{\mbox{\boldmath$#1$}}
\def\hsp5{\hspace{5mm}}
\def\lb{\label}
\def\bi{\bibitem}
\def\ct{\cite}

\newcommand{\R}{\textsf{R}}

\def\case#1/#2{\textstyle\frac{#1}{#2}}
\newcommand{\leftout}[1]{}
\begin{document}

\vspace*{1cm}
\begin{center}
{\Huge\sc
                    Foundations of \\
                    Descriptive and Inferential \\
                    Statistics
\par}
\par\vfill\vfill\vfill
              Lecture notes for a quantitative–methodological
							module at the Bachelor degree (B.Sc.) level
\par\vfill
\par\vfill
 
\par\vfill\vfill\vfill\vfill

                    {\large\sc Henk van Elst}

\par\vfill

                 August 30, 2019
\par\vfill\vfill

                 parcIT GmbH \\
								 Erftstra\ss e 15 \\
								 50672 K\"{o}ln \\
                 Germany
\par\vfill
                 E--Mail: \texttt{Henk.van.Elst@parcIT.de}
\par\vfill\vfill
                 E--Print: 
\href{http://arxiv.org/abs/1302.2525}{arXiv:1302.2525v4 [stat.AP]}
\par\vfill\vfill
								 \copyright\ 2008--2019 Henk van Elst 
\end{center}
\vspace*{1cm}
\sloppy
      \renewcommand{\thepage}{}                 
\addcontentsline{toc}{chapter}{Abstract}
\chapter*{}
\vspace{-8ex}
\section*{Abstract}
{\small These lecture notes were written with the aim to provide 
an accessible though technically solid introduction to the logic 
of systematical analyses of statistical data to both undergraduate 
and postgraduate students, in particular in the Social Sciences, 
Economics, and the Financial Services. They may also serve as a
general reference for the application of quantitative--empirical
research methods. In an attempt to encourage the adoption of an 
interdisciplinary perspective on quantitative problems arising in
practice, the notes cover the four broad topics (i)~descriptive
statistical processing of raw data, (ii)~elementary probability
theory, (iii)~the operationalisation of one-dimensional latent
statistical variables according to Likert's widely used scaling
approach, and (iv)~null hypothesis significance testing within the
frequentist approach to probability theory concerning 
(a)~distributional differences of variables between subgroups of a 
target population, and (b)~statistical associations between two 
variables. The relevance of effect sizes for making inferences is
emphasised. These lecture notes are fully hyperlinked, thus
providing a direct route to original scientific papers as well as
to interesting biographical information. They also list many
commands for running statistical functions and data analysis
routines in the software packages \R{}, SPSS, EXCEL and
OpenOffice. The immediate involvement in actual data analysis
practices is strongly recommended.}

\vspace{10mm}
\noindent
\underline{Cite as:} 
\href{http://arxiv.org/abs/1302.2525}{arXiv:1302.2525v4 [stat.AP]}
\vfill

\medskip
\noindent
These lecture notes were typeset in \LaTeXe.

      \newpage \thispagestyle{empty}
\tableofcontents
      \newpage \thispagestyle{empty}
      \cleardoublepage \pagenumbering{arabic}
\addcontentsline{toc}{chapter}{Introductory remarks}
\chapter*{Introductory remarks}
\textbf{Statistical methods of data analysis} form the cornerstone
of quantitative--empirical research in the \textbf{Social
Sciences}, \textbf{Humanities}, and \textbf{Economics}.
Historically, the bulk of knowledge available in
\textbf{Statistics} emerged in the context of the analysis of
(nowadays large) data sets from observational and experimental
measurements in the \textbf{ Natural Sciences}. The purpose of the
present lecture notes is to provide its readers with a solid and
thorough, though accessible introduction to the basic concepts of
\textbf{Descriptive and Inferential Statistics}. When discussing
methods relating to the latter subject, we will here take the
perspective of the \textbf{frequentist approach} to
\textbf{Probability Theory}. (See Ref.~\ct{hve2018} for a
methodologically different approach.)

\medskip
\noindent
The concepts to be introduced and the topics to be covered 
have been selected in order to make available a fairly 
self-contained basic statistical tool kit for thorough analysis 
at the \textbf{univariate} and \textbf{bivariate} levels of
complexity of data, gained by means of opinion polls, surveys or
observation.
%

\medskip
\noindent
In the \textbf{Social Sciences}, \textbf{Humanities}, and
\textbf{Economics} there are two broad families of empirical
research tools available for studying behavioural features of and
mutual interactions between human individuals on the one-hand side,
and the social systems and organisations that these form on the
other. \textbf{Qualitative--empirical methods} focus their view 
on the individual with the aim to account for her/his/its 
particular characteristic features, thus probing the ``small 
scale-structure'' of a social system, while
\textbf{quantitative--empirical methods} strive to recognise
patterns and regularities that pertain to a large number of
individuals and so hope to gain insight on the ``large-scale
structure'' of a social system.

\medskip
\noindent
Both approaches are strongly committed to pursuing the principles 
of the \textbf{scientific method}. These entail the systematic
observation and measurement of phenomena of interest on the basis 
of well-defined statistical variables, the structured analysis of 
data so generated, the attempt to provide compelling theoretical 
explanations for effects for which there exists conclusive 
evidence in the data, the derivation from the data of predictions 
which can be tested empirically, and the publication of all 
relevant data and the analytical and interpretational tools 
developed and used, so that the pivotal \textbf{replicability} of a 
researcher's findings and associated conclusions is ensured. By 
complying with these principles, the body of scientific knowledge 
available in any field of research and its practical applications
undergoes a continuing process of updating and expansion.

\medskip
\noindent
Having thoroughly worked through these lecture notes, a reader 
should have obtained a good understanding of the use and 
efficiency of descriptive and frequentist inferential statistical
methods for handling quantitative issues, as they often arise in a
manager's everyday business life. Likewise, a reader should feel
well-prepared for a smooth entry into any Master degree programme
in the \textbf{Social Sciences} or \textbf{Economics} which puts
emphasis on quantitative--empirical methods.

\medskip
\noindent
Following a standard pedagogical concept, these lecture notes are 
split into three main parts: Part~I, comprising
Chapters~\ref{ch1} to \ref{ch5}, covers the basic considerations 
of \textbf{Descriptive Statistics}; Part~II, which consists 
of Chapters~\ref{ch6} to \ref{ch8}, introduces the foundations of 
\textbf{Probability Theory}. Finally, the material of Part~III, 
provided in Chapters~\ref{ch9} to \ref{ch13}, first reviews a
widespread method for operationalising latent statistical 
variables, and then introduces a number of standard uni- and 
bivariate analytical tools of \textbf{Inferential Statistics}
within the \textbf{frequentist framework} that prove valuable in 
applications. As such, the contents of Part~III are the most 
important ones for quantitative--empirical research work. Useful 
mathematical tools and further material have been gathered in 
appendices.

\medskip
\noindent
Recommended introductory textbooks, which may be used for study in 
parallel to these lecture notes, are Levin \textit{et al} 
(2010)~\ct{levetal2009}, Hatzinger and Nagel 
(2013)~\ct{hatnag2013}, Weinberg and Abramowitz
(2008)~\ct{weiabr2008}, Wewel (2014)~\ct{wew2014}, Toutenburg 
(2005) \ct{tou2005}, or Duller (2007)~\ct{dul2007}.

\medskip
\noindent
There are \textit{not} included in these lecture notes any explicit 
exercises on the topics to be discussed. These are 
reserved for lectures given throughout term time.

\medskip
\noindent
The present lecture notes are designed to be dynamical in 
character. On the one-hand side, this means that they will be 
updated on a regular basis. On the other, that the *.pdf version 
of the notes contains interactive features such as fully 
hyperlinked references to original publications at the websites 
\href{https://doi.org}{\texttt{doi.org}} and 
\href{http://www.jstor.org}{\texttt{jstor.org}}, as well as many 
active links to biographical information on scientists that have 
been influential in the historical development of
\textbf{Probability Theory} and \textbf{Statistics}, hosted by the
websites \href{http://www-history.mcs.st-and.ac.uk/}{The MacTutor
History of Mathematics archive
({\tt www-history.mcs.st-and.ac.uk})} and
\href{http://en.wikipedia.org/wiki/Main_Page}{\texttt{en.wikipedia.org}}.

\medskip
\noindent
Throughout these lecture notes references have been provided to
respective descriptive and inferential statistical functions and
routines that are available in the excellent and widespread
statistical software package \R{}, on a standard graphic display
calculator (GDC), and in the statistical software packages EXCEL,
OpenOffice and SPSS (Statistical Program for the Social Sciences).
\R{}~and its exhaustive documentation are distributed by the
R~Core Team (2019)~\ct{rct2019} via the website
\href{http://cran.r-project.org}{\texttt{cran.r-project.org}}.
\R{}, too, has been employed for generating all the figures
contained in these lecture notes. Useful and easily accessible
textbooks on the application of \R{} for statistical data analysis
are, e.g., Dalgaard (2008)~\ct{dal2008}, or Hatzinger \textit{et
al} (2014)~\ct{hatetal2014}. Further helpful information and 
assistance is available from the website 
\href{http://www.r-tutor.com/}{\texttt{www.r-tutor.com}}. For
active statistical data analysis with \R{}, we strongly recommend
the use of the convenient custom-made work environment \R{}~Studio,
provided free of charge at \href{http://www.rstudio.com}{\texttt{www.rstudio.com}}. Another user friendly statistical software
package is GNU PSPP. This is available as shareware from 
\href{http://www.gnu.org/software/pspp/}{\texttt{www.gnu.org/software/pspp/}}.

\medskip
\noindent
A few examples from the inbuilt \R{} data sets package have
been related to in these lecture notes in the context of the
visualisation of distributional features of statistical data.
Further information on these data sets can be obtained by typing
\texttt{library(help = "datasets")} at the \R{} prompt.

\medskip
\noindent
Lastly, we hope the reader will discover something useful or/and
enjoyable for her/him-self when working through these lecture
notes. Constructive criticism is always welcome.

\vfill
\medskip
\noindent
\textit{Acknowledgments:} I am grateful to Kai Holschuh, Eva 
Kunz and Diane Wilcox for valuable comments on an earlier draft of 
these lecture notes, to Isabel Passin for being a critical sparing
partner in evaluating pedagogical considerations concerning
cocreated accompanying lectures, and to Michael R\"{u}ger for
compiling an initial list of online survey tools for the Social
Sciences.

\chapter[Statistical variables]{\href{https://www.youtube.com/watch?v=DXkHcaiRcd4}{Statistical variables}}
\lb{ch1}
A central task of an empirical scientific discipline is the
\textbf{observation} or \textbf{measurement} of a  finite set of
characteristic \textbf{variable features} of a given
\textbf{system of objects} chosen for study. The hope is to be able
to recognise in a sea of data, typically guided by
\textbf{randomness}, meaningful patterns and regularities that
provide evidence for possible \textbf{associations}, or, stronger
still, \textbf{causal relationships} between these variable
features. Based on a combination of \textbf{inductive} and
\textbf{deductive methods of data analysis}, one aims at gaining
insights of a qualitative and/or quantitative nature into the
intricate and often complex interdependencies of such variable
features for the purpose of (i)~obtaining explanations for
phenomena that have been observed, and (ii)~making predictions
which, subsequently, can be tested. 
The acceptance of the validity of a particular empirical 
scientific framework generally increases with the number of 
successful \textbf{replications} of its predictions.\footnote{A 
particularly sceptical view on the ability of making reliable
predictions in certain empirical scientific disciplines is voiced 
in Taleb (2007)~\ct[pp~135--211]{tal2007}.} It is the interplay of 
observation, experimentation and theoretical modelling, 
systematically coupled to one another by a number of 
feedback loops, which gives rise to progress in learning and 
understanding in all empirical scientific activities. This 
procedure, which focuses on replicable \textbf{facts}, is
referred to as the \textbf{scientific method}.

\medskip
\noindent
More specifically, the general intention of empirical 
scientific activities is to modify or strengthen the
\textbf{theoretical foundations} of an empirical scientific
discipline by means of observational and/or experimental
\textbf{testing} of sets of \textbf{hypotheses}; see
Ch.~\ref{ch11}. This is generally 
achieved by employing the quantitative--empirical techniques that 
have been developed in \textbf{Statistics}, in particular in the 
course of the $20^\mathrm{th}$ Century. At the heart of these 
techniques is the concept of a \textbf{statistical variable} $X$ as 
an entity which represents a single common aspect of the system of 
objects selected for analysis --- the \textbf{target population} 
$\boldsymbol{\Omega}$ of a \textbf{statistical investigation}. In
the ideal case, a variable entertains a one-to-one correspondence
with an \textbf{observable}, and thus is directly amenable to
\textbf{measurement}. In the \textbf{Social Sciences},
\textbf{Humanities}, and \textbf{Economics}, however, one needs to
carefully distinguish between \textbf{manifest variables}
corresponding to observables on the one-hand side, and
\textbf{latent variables} representing in general unobservable
``social constructs'' on the other. It is this latter kind of
variables which is commonplace in the fields mentioned. Hence, it
becomes an unavoidable task to thoroughly address the issue of a
reliable, valid and objective \textbf{operationalisation} of any
given latent variable one has identified as providing essential
information on the objects under investigation. A standard approach
to dealing with the important matter of rendering latent variables
measurable is reviewed in Ch.~\ref{ch9}.

\medskip
\noindent
In \textbf{Statistics}, it has proven useful to classify variables
on the basis of their intrinsic information content into one of
three hierachically ordered categories, referred to as the
\textbf{scale levels of measurement}; cf. Stevens
(1946)~\ct{ste1946}. We provide the definition of these scale
levels next.

\section[Scale levels of measurement]{Scale levels of measurement}
\lb{merkskal}
\underline{\textbf{Def.:}} Let $X$ be a one-dimensional
\textbf{statistical variable} with $k \in \mathbb{N}$ (countably
many) resp.\ $k \in \mathbb{R}$ (uncountably many) possible
\textbf{values}, \textbf{attributes}, or \textbf{categories}
$a_{j}$ ($j=1,\ldots,k$). Statistical variables
are classified as belonging into one of three hierachically ordered 
\textbf{scale levels of measurement}. This is done on the basis of 
three criteria for distinguishing information contained in the 
values of actual \textbf{data} for these variables. One thus
defines:

\begin{itemize}

\item \textbf{Metrically scaled variables} $X$ \hfill
\textbf{(quantitative/numerical)}\\
Possible values can be distinguished by
\begin{itemize}
\item[(i)] their \textit{names}, $a_{i} \neq a_{j}$,
\item[(ii)] they allow for a \textit{natural rank order}, $a_{i} < 
a_{j}$, and
\item[(iii)] \textit{distances} between them, $a_{i}-a_{j}$,
are uniquely determined.
\end{itemize}
	\begin{itemize}
	\item \textbf{Ratio scale}: $X$ has an \textit{absolute zero
	point} and otherwise only non-negative values;	analysis of both 
	differences	$a_{i}-a_{j}$ and ratios $a_{i}/a_{j}$ is meaningful.
	
	\underline{Examples:} body height, monthly net income, \ldots.
	
	\item \textbf{Interval scale}: $X$ has no \textit{absolute
	zero point}; only differences $a_{i}-a_{j}$
	are meaningful.
	
	\underline{Examples:} year of birth, temperature in centigrades,
	Likert scales (cf. Ch.~\ref{ch9}), \ldots.
	\end{itemize}
	
Note that the values obtained for a metrically scaled variable 
(e.g. in a survey) always constitute definite numerical 
multiples of a specific \textbf{unit of measurement}.

\item \textbf{Ordinally scaled variables} $X$ \hfill
\textbf{(qualitative/categorical)}\\
Possible values, attributes, or categories can be distinguished by
\begin{itemize}
\item[(i)] their \textit{names}, $a_{i} \neq a_{j}$, and
\item[(ii)] they allow for a \textit{natural rank order}, $a_{i} 
< a_{j}$.
\end{itemize}
\underline{Examples:} Likert item rating scales (cf. 
Ch.~\ref{ch9}), grading of commodities, \ldots.

\item \textbf{Nominally scaled variables} $X$ \hfill
\textbf{(qualitative/categorical)}\\
Possible values, attributes, or categories can be distinguished 
only by
\begin{itemize}
\item[(i)] their \textit{names}, $a_{i} \neq a_{j}$.
\end{itemize}
\underline{Examples:} first name, location of birth, \ldots.

\end{itemize}

\noindent
\underline{\textbf{Remark:}} As we will see later in Ch.~\ref{ch12}
and~\ref{ch13}, the applicability of specific methods of
\textbf{statistical data analysis} crucially depends on the
\textbf{scale level of measurement} of the variables involved in
the respective procedures. Metrically scaled data offers the
largest variety of powerful methods for this purpose!

\section[Raw data sets and data matrices]{Raw data sets and data
matrices}
\lb{urlist}
To set the stage for subsequent considerations, we here introduce
some formal representations of entities which assume central roles
in statistical data analyses.

\medskip
\noindent
Let $\boldsymbol{\Omega}$ denote the \textbf{target population} of 
study objects of interest (e.g., human individuals forming a 
particular social system) relating to some \textbf{statistical 
investigation}. This set~$\boldsymbol{\Omega}$ shall comprise a 
total of $N \in \mathbb{N}$ \textbf{statistical units}, i.e., its 
size be $|\boldsymbol{\Omega}|=N$.

\medskip
\noindent
Suppose one intends to determine the \textbf{frequency
distributional properties} in $\boldsymbol{\Omega}$ of a portfolio
of $m \in \mathbb{N}$ \textbf{statistical variables} $X$, $Y$,
\ldots, and $Z$, with \textbf{spectra of values} $a_{1}, a_{2},
\ldots, a_{k}$, $b_{1}, b_{2}, \ldots, b_{l}$, \ldots, and $c_{1},
c_{2}, \ldots, c_{p}$, respectively ($k,l,p \in \mathbb{N}$). A
\textbf{survey} typically obtains from~$\boldsymbol{\Omega}$ a
\textbf{statistical sample} $\boldsymbol{S_{\Omega}}$ of size 
$|\boldsymbol{S_{\Omega}}|=n$ ($n \in \mathbb{N}$, $n < N$), 
unless one is given the rare opportunity to conduct a proper
\textbf{census} on $\boldsymbol{\Omega}$ (in which case $n=N$).
The \textbf{data} thus generated consists of \textbf{observed 
values}~$\{x_{i}\}_{i=1,\ldots,n}$, 
$\{y_{i}\}_{i=1,\ldots,n}$, \ldots, and $\{z_{i}\}_{i=1,\ldots,n}$.
It constitutes the \textbf{raw data set} $\{(x_{i}, y_{i}, \ldots, 
z_{i})\}_{i=1,\ldots,n}$ of a statistical investigation and may be 
conveniently assembled in the form of an $\boldsymbol{(n \times 
m)}$ \textbf{data matrix} $\boldsymbol{X}$ given by
\begin{center}
\begin{tabular}[h]{|c||c|c|c|c|}
\hline
 & & & & \\
\textbf{sampling} & \textbf{variable} & \textbf{variable} &
\ldots & \textbf{variable} \\
\textbf{unit} & $X$ & $Y$ & & $Z$ \\
 & & & & \\
\hline\hline
 & & & & \\
$1$ & $x_{1}=a_{5}$ & $y_{1}=b_{9}$ & \ldots & $z_{1}=c_{3}$ \\
 & & & & \\
\hline
 & & & & \\
$2$ & $x_{2}=a_{2}$ & $y_{2}=b_{12}$ & \ldots & $z_{2}=c_{8}$ \\
 & & & & \\
\hline
 & & & & \\
\vdots & \vdots & \vdots & \vdots & \vdots \\
 & & & & \\
\hline
 & & & & \\
$n$ & $x_{n}=a_{8}$ & $y_{n}=b_{9}$ & \ldots & $z_{n}=c_{15}$ \\
 & & & & \\
\hline
\end{tabular}
\end{center}
To systematically record the information obtained from measuring 
the values of a portfolio of statistical variables in a 
statistical sample $\boldsymbol{S_{\Omega}}$, in the 
$\boldsymbol{(n \times m)}$~\textbf{data matrix} 
$\boldsymbol{X}$ every one of the $n$~\textbf{sampling units} 
investigated is assigned a particular \textit{row}, while every one 
of the $m$~\textbf{statistical variables} measured is assigned a 
particular \textit{column}. In the following, $X_{ij}$ denotes the 
data entry in the $i$th row ($i=1,\ldots,n$) and the $j$th column 
($i=1,\ldots,m$) of $\boldsymbol{X}$. To clarify standard 
terminology used in \textbf{Statistics}, a \textbf{raw data set} is 
referred to as
\begin{itemize}
\item[(i)] \textbf{univariate}, when $m=1$,
\item[(ii)] \textbf{bivariate}, when $m=2$, and
\item[(iii)] \textbf{multivariate}, when $m \geq 3$.
\end{itemize}
According to Hair \textit{et al} (2010) 
\ct[pp~102, 175]{haietal2010}, a rough rule of thumb concerning an 
adequate \textbf{sample size} $|\boldsymbol{S_{\Omega}}|=n$ for 
\textbf{multivariate data analysis} is given 
by
\be
\lb{eq:samplesize}
n \geq 10m \ .
\ee
Considerations of \textbf{statistical power} of particular methods
of data analysis lead to more refined recommendations; cf. 
Sec.~\ref{sec:testgen}.

\medskip
\noindent
\textbf{``Big data''} scenarios apply when $n, m 
\gg 1$ (i.e., when $n$ is typically on the order of $10^{4}$, or
very much larger still, and $m$ is on the order of $10^{2}$, or
larger).

\medskip
\noindent
In general, an $(n \times m)$ data matrix $\boldsymbol{X}$ is the 
starting point for the application of a \textbf{statistical
software package} such as \R{}, SPSS, GNU PSPP, or other for the
purpose of systematic data analysis. When the sample comprises
exclusively \textbf{metrically scaled data}, the data matrix is
real-valued, i.e.,
\be
\lb{eq:metrdatamatrix}
\boldsymbol{X} \in \mathbb{R}^{n \times m} \ ;
\ee
cf. Ref.~\ct[Sec.~2.1]{hve2009}. Then the information contained in 
$\boldsymbol{X}$ uniquely positions a collection of $n$ sampling 
units according to $m$ quantitative characteristic variable 
features in (a subset of) an $m$-dimensional \textbf{Euclidian
space}~$\mathbb{R}^{m}$.

\medskip
\noindent
\underline{\R:} \texttt{datMat <- data.frame(x = c($x_{1}$,\ldots,$x_{n}$),
y = c($y_{1}$,\ldots,$y_{n}$), \ldots, \\
z = c($z_{1}$,\ldots,$z_{n}$))} 

\vspace{5mm}
\noindent
We next turn to describe phenomenologically the \textbf{univariate 
frequency distribution} of a single one-dimensional statistical 
variable $X$ in a specific statistical sample 
$\boldsymbol{S_{\Omega}}$ of size $n$,  drawn in the context of a 
survey from some target population of study objects 
$\boldsymbol{\Omega}$ of size $N$.


\chapter[Univariate frequency distributions]{\href{https://www.youtube.com/watch?v=CWzG2gphFNU}{Univariate frequency distributions}}
\lb{ch2}
The first task at hand in unravelling the intrinsic structure 
potentially residing in a given raw data set 
$\{x_{i}\}_{i=1,\ldots,n}$ for some statistical variable $X$ 
corresponds to Cinderella's task of separating the ``good peas'' 
from the ``bad peas,'' and collecting them in respective bowls (or 
bins). This is to say, the first question to be answered requires 
determination of the \textbf{frequency} with which a value (or 
attribute, or category)~$a_{j}$ in the spectrum of possible values 
of $X$ was observed in a statistical sample 
$\boldsymbol{S_{\Omega}}$ of size~$n$.

\section[Absolute and relative frequencies]{Absolute and relative
frequencies}
\lb{haeufig}
\underline{\textbf{Def.:}} Let $X$ be a nominally, ordinally or
metrically scaled one-dimensional \textbf{statistical variable},
with a spectrum of $k$ different \textbf{values} or
\textbf{attributes} $a_{j}$ resp.\ $k$ different
\textbf{categories} (or bins) $K_{j}$ ($j=1,\ldots,k$). If, for
$X$, we have a univariate \textbf{raw data set} comprising
$n$ \textbf{observed values}~$\{x_{i}\}_{i=1,\ldots,n}$, we define
by
\be
o_{j}:=
\begin{cases}
o_{n}(a_{j}) & = \text{number\ of}\ x_{i}\ \text{with}
\ x_{i}=a_{j} \\
 & \\
o_{n}(K_{j}) & = \text{number\ of}\ x_{i}\ \text{with}
\ x_{i} \in K_{j}
\end{cases}
\ee
($j=1,\ldots,k$) the \textbf{absolute (observed) frequency} of $a_{j}$
resp.\ $K_{j}$, and, upon division of the $o_{j}$ by the sample 
size $n$, we define by
\be
\lb{eq:relfreq}
h_{j}:=
\begin{cases}
{\displaystyle \frac{o_{n}(a_{j})}{n}} & \\
 & \\
{\displaystyle \frac{o_{n}(K_{j})}{n}} &
\end{cases}
\ee
($j=1,\ldots,k$) the \textbf{relative frequency} of $a_{j}$
resp.\ $K_{j}$. Note that for all $j=1,\ldots,k$, we have $0 \leq 
o_{j} \leq n$ with $\displaystyle\sum_{j=1}^{k}o_{j}=n$, and $0 
\leq h_{j} \leq 1$ with $\displaystyle\sum_{j=1}^{k}h_{j}=1$.

\medskip
\noindent
The $k$ value pairs $(a_{j},o_{j})_{j=1,\ldots,k}$ resp.\ 
$(K_{j},o_{j})_{j=1,\ldots,k}$ represent the univariate \textbf{
distribution of absolute frequencies}, the $k$ value pairs 
$(a_{j},h_{j})_{j=1,\ldots,k}$
resp.\ $(K_{j},h_{j})_{j=1,\ldots,k}$ represent the univariate 
\textbf{distribution of relative frequencies} of the $a_{j}$
resp.\ $K_{j}$ in $\boldsymbol{S_{\Omega}}$.

\medskip
\noindent
\underline{\R:} \texttt{table(\textit{variable})},
\texttt{prop.table(\textit{variable})} \\
\underline{EXCEL, OpenOffice:} \texttt{FREQUENCY} (dt.:
\texttt{H\"AUFIGKEIT})\\
\underline{SPSS:} Analyze $\rightarrow$ Descriptive Statistics
$\rightarrow$ Frequencies \ldots

\medskip
\noindent
Typical graphical representations of univariate \textbf{relative 
frequency distributions}, regularly employed in visualising 
results of \textbf{descriptive statistical data analyses}, are the
\begin{itemize}

\item \textbf{histogram} for \textit{metrically} scaled data;
cf. Fig.~\ref{fig:histogram},\footnote{The appearance of graphs
generated in \R{} can be prettified by employing the advanced
graphical package \texttt{ggplot2} by Wickham (2016)~\ct{wic2016}.}

\item \textbf{bar chart} for \textit{ordinally} scaled data;
cf. Fig.~\ref{fig:barchart},

\item \textbf{pie chart} for \textit{nominally} scaled data;
cf. Fig.~\ref{fig:piechart}.
\end{itemize}

\medskip
\noindent
\underline{\R:} \texttt{hist(\textit{variable}, freq = FALSE)}, \\
\texttt{barplot(table(\textit{variable}))},
\texttt{barplot(prop.table(table(\textit{variable})))}, \\
\texttt{pie(table(\textit{variable}))},
\texttt{pie(prop.table(table(\textit{variable})))}

\begin{figure}[!htb]
\begin{center}
\includegraphics[scale=0.8]{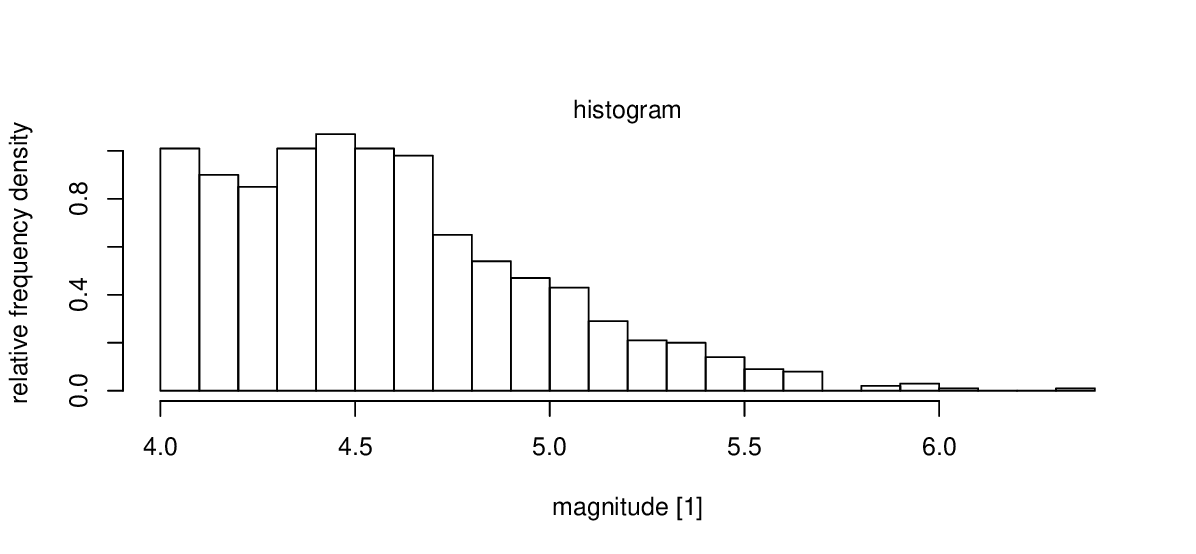}
\end{center}
\caption{Example of a histogram, representing the relative
frequency density for the variable ``magnitude'' in the
\R{} data set ``quakes.''
\newline
\underline{\R:} \newline
\texttt{data("quakes")} \newline
\texttt{?quakes} \newline
\texttt{hist( quakes\$mag , breaks = 20 , freq = FALSE )}}
\lb{fig:histogram}
\end{figure}
\begin{figure}[!htb]
\begin{center}
\includegraphics[scale=0.8]{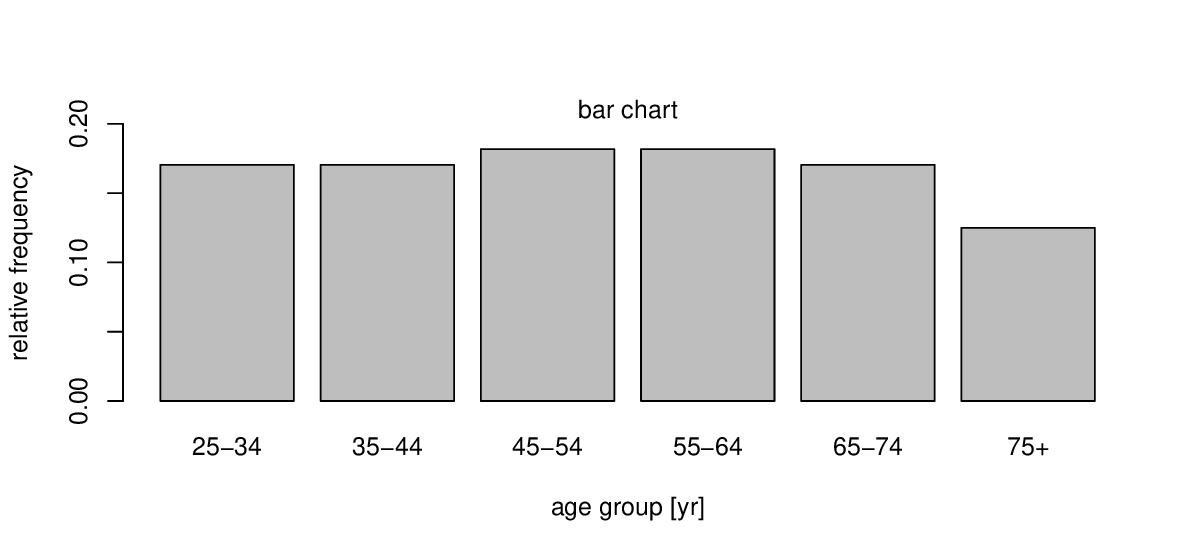}
\end{center}
\caption{Example of a bar chart, representing the relative
frequency distribution for the variable ``age group'' in the
\R{} data set ``esoph.'' \newline
\underline{\R:} \newline
\texttt{data("esoph")} \newline
\texttt{?esoph} \newline
\texttt{barplot( prop.table( table( esoph\$agegp ) ) )}}
\lb{fig:barchart}
\end{figure}
\begin{figure}[!htb]
\begin{flushleft}
\includegraphics[scale=0.8]{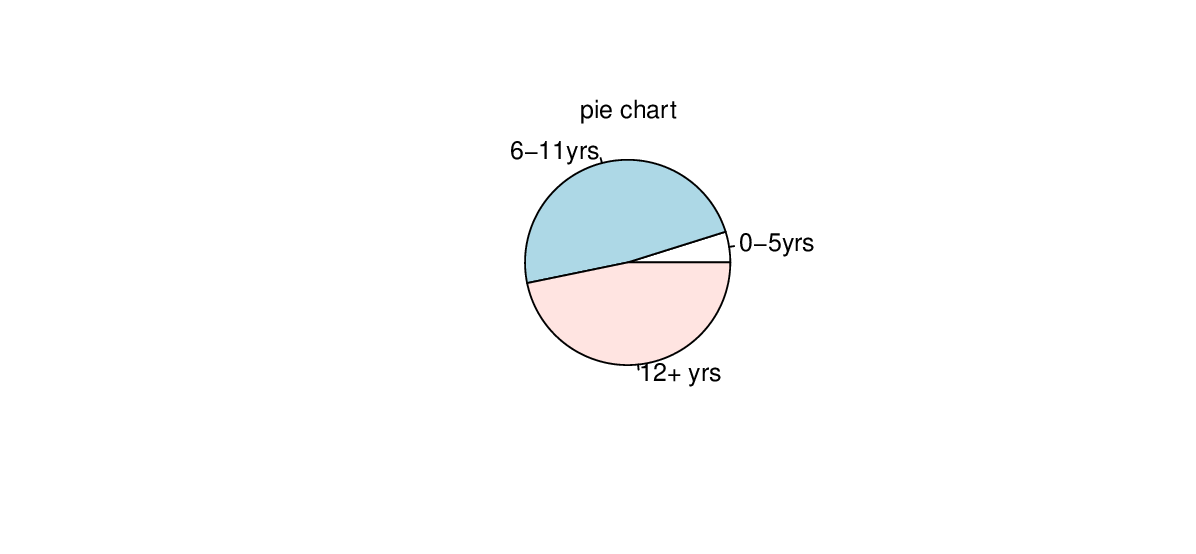}
\end{flushleft}
\caption{Example of a pie chart, representing the relative
frequency distribution for the variable ``education'' in the
\R{} data set ``infert.'' \newline
\underline{\R:} \newline
\texttt{data("infert")} \newline
\texttt{?infert} \newline
\texttt{pie( table( infert\$education ) )}}
\lb{fig:piechart}
\end{figure}

\medskip
\noindent
It is standard practice in \textbf{Statistics} to compile from the 
univariate relative frequency distribution 
$(a_{j},h_{j})_{j=1,\ldots,k}$
resp.\ $(K_{j},h_{j})_{j=1,\ldots,k}$ of data for some ordinally 
or metrically scaled one-dimensional statistical variable $X$ the 
associated empirical cumulative distribution function. Hereby it 
is necessary to distinguish the case of data for a variable with a 
discrete spectrum of values from the case of data for a variable 
with a continuous spectrum of values. We will discuss this issue 
next.

\section[Empirical cumulative distribution function (discrete
data)]{Empirical cumulative distribution function (discrete data)}
\lb{sec:empvert}
\underline{\textbf{Def.:}} Let $X$ be an ordinally or metrically
scaled one-dimensional statistical variable, the spectrum of values 
$a_{j}$ ($j=1,\ldots,k$) of which vary \textit{discretely}. Suppose 
given for $X$ a statistical sample $\boldsymbol{S_{\Omega}}$ of 
size $|\boldsymbol{S_{\Omega}}|=n$ comprising observed values 
$\{x_{i}\}_{i=1,\ldots,n}$, which we assume arranged in an 
ascending fashion according to the natural order $a_{1} < a_{2} < 
\ldots < a_{k}$. The corresponding univariate relative frequency 
distribution is $(a_{j},h_{j})_{j=1,\ldots,k}$. For all real 
numbers $x \in \mathbb{R}$, we then define by
\be
\lb{eq:empvert}
\fbox{$\displaystyle
F_{n}(x) :=
\begin{cases}
0 & \text{for}\ x < a_{1} \\
 & \\
{\displaystyle\sum_{i=1}^{j}h_{n}(a_{i})} & \text{for}
\ a_{j} \leq x < a_{j+1} 
\qquad (j=1,\ldots,k-1) \\
 & \\
1 & \text{for}\ x \geq a_{k}
\end{cases}
$}
\ee
the \textbf{empirical cumulative distribution function} for $X$.
The value of $F_{n}$ at $x \in \mathbb{R}$ represents the
cumulative relative frequencies of all $a_{j}$ which are less
or equal to $x$; cf. Fig.~\ref{fig:ecdf}. $F_{n}(x)$ has the
following properties:
\begin{itemize}

\item its domain is $D(F_{n})=\mathbb{R}$, and its range is 
$W(F_{n})=[0,1]$; hence, $F_{n}$ is bounded from above and from 
below,

\item it is continuous from the right and monotonously increasing,

\item it is constant on all half-open intervals $[a_{j},a_{j+1})$, 
but exhibits jump discontinuities of size $h_{n}(a_{j+1})$ at all 
$a_{j+1}$, and,

\item asymptotically, it behaves as ${\displaystyle \lim_{x\to 
-\infty}F_{n}(x)=0}$ and ${\displaystyle \lim_{x\to 
+\infty}F_{n}(x)=1}$.

\end{itemize}
%

\medskip
\noindent
\underline{\R:} \texttt{ecdf(\textit{variable})},
\texttt{plot(ecdf(\textit{variable}))}

\begin{figure}[!htb]
\begin{flushleft}
\includegraphics[scale=0.8]{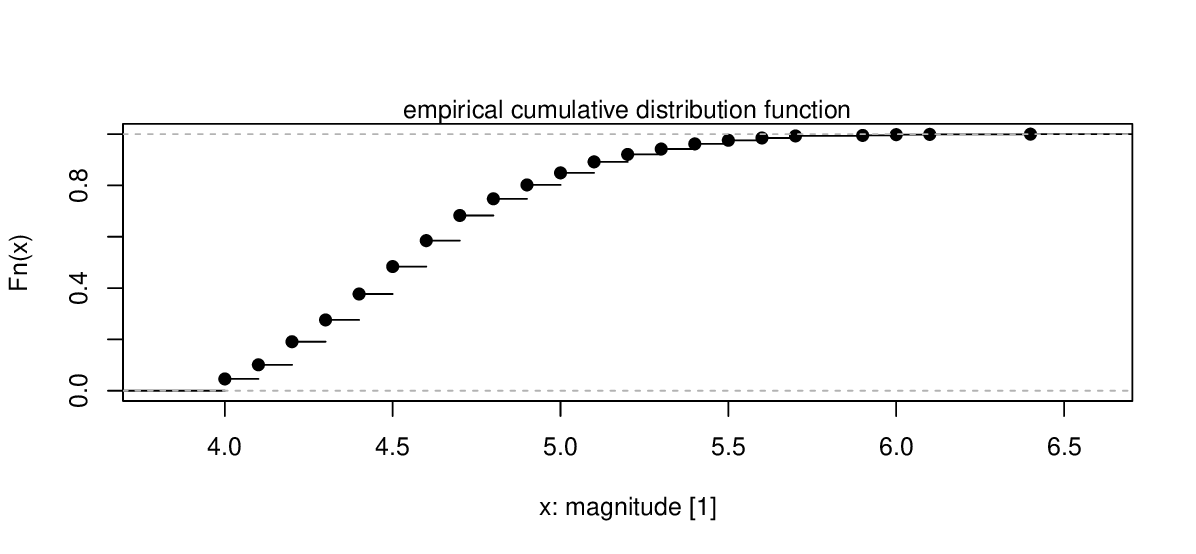}
\end{flushleft}
\caption{Example of an empirical cumulative distribution function,
here for the variable ``magnitude'' in the \R{} data set
``quakes.'' \newline
\underline{\R:} \newline
\texttt{data("quakes")} \newline
\texttt{?quakes} \newline
\texttt{plot( ecdf( quakes\$magnitude ) )}}
\lb{fig:ecdf}
\end{figure}

\medskip
\noindent
\textbf{Computational rules for} $\boldsymbol{F_{n}(x)}$

\begin{enumerate}
\item $h(x \leq d) = F_{n}(d)$
\item $h(x < d) = F_{n}(d) - h_{n}(d)$
\item $h(x \geq c) = 1 - F_{n}(c) + h_{n}(c)$
\item $h(x > c) =  1- F_{n}(c)$
\item $h(c \leq x \leq d) = F_{n}(d) - F_{n}(c) + h_{n}(c)$
\item $h(c < x \leq d) = F_{n}(d) - F_{n}(c)$
\item $h(c \leq x < d) = F_{n}(d) - F_{n}(c) - h_{n}(d)
+ h_{n}(c)$
\item $h(c < x < d) = F_{n}(d) - F_{n}(c) - h_{n}(d)$,
\end{enumerate}
wherein $c$ denotes an arbitrary \textbf{lower bound}, and $d$ 
denotes an arbitrary \textbf{upper bound}, on the argument $x$ of 
$F_{n}(x)$.

\section[Empirical cumulative distribution function (continuous
data)]{Empirical cumulative distribution function (continuous
data)}
\lb{empvertklass}
\underline{\textbf{Def.:}} Let $X$ be a metrically scaled
one-dimensional statistical variable, the spectrum of values of 
which vary \textit{continuously}, and let observed values 
$\{x_{i}\}_{i=1,\ldots,n}$ for $X$ from a statistical sample 
$\boldsymbol{S_{\Omega}}$ of size $|\boldsymbol{S_{\Omega}}|=n$ be 
binned into a finite set of $k$ (with $k \approx \sqrt{n}$) 
ascendingly ordered exclusive \textbf{class intervals} (or bins) 
$K_{j}$ ($j=1,\ldots,k$), of width $b_{j}$, and with lower 
boundary~$u_{j}$ and upper boundary~$o_{j}$. The univariate 
distribution of relative frequencies of the class intervals be 
$(K_{j},h_{j})_{j=1,\ldots,k}$. Then, for all real numbers $x \in 
\mathbb{R}$,
\be
\lb{klempvert}
\fbox{$\displaystyle
\tilde{F}_{n}(x) :=
\begin{cases}
0 & \text{for}\ x< u_{1} \\
 & \\
{\displaystyle \sum_{i=1}^{j-1}h_{i}
+ \frac{h_{j}}{b_{j}}(x-u_{j})}
& \text{for}\ x \in K_{j} \\
 & \\
1 & \text{for}\ x > o_{k}
\end{cases}
$}
\ee
defines the \textbf{empirical cumulative distribution function}
for $X$. $\tilde{F}_{n}(x)$ has the following properties:
\begin{itemize}

\item its domain is $D(\tilde{F}_{n})=\mathbb{R}$, and its
range is $W(\tilde{F}_{n})=[0,1]$; hence, $\tilde{F}_{n}$ is 
bounded from above and from below,

\item it is continuous and monotonously increasing, and,

\item asymptotically, it behaves as ${\displaystyle\lim_{x\to 
-\infty}\tilde{F}_{n}(x)=0}$ and ${\displaystyle\lim_{x\to 
+\infty}\tilde{F}_{n}(x)=1}$.
\end{itemize}

\medskip
\noindent
\underline{\R:} \texttt{ecdf(\textit{variable})},
\texttt{plot(ecdf(\textit{variable}))}

\medskip
\noindent
\textbf{Computational rules for} $\boldsymbol{\tilde{F}_{n}(x)}$
\begin{enumerate}

\item $h(x<d)=h(x\leq d) = \tilde{F}_{n}(d)$

\item $h(x>c)=h(x\geq c) = 1-\tilde{F}_{n}(c)$

\item $h(c<x<d) = h(c\leq x<d) = h(c<x\leq d)
= h(c\leq x\leq d) = \tilde{F}_{n}(d)-\tilde{F}_{n}(c)$,
\end{enumerate}
wherein $c$ denotes an arbitrary \textbf{lower bound}, and $d$ 
denotes an arbitrary \textbf{upper bound}, on the argument $x$ of
$\tilde{F}_{n}(x)$.

\vspace{5mm}
\noindent
Our next steps comprise the introduction of a set of 
scale-level-dependent standard \textbf{descriptive measures} which 
characterise specific properties of univariate and bivariate 
relative frequency distributions of statistical variables $X$ 
resp.\ $(X,Y)$.


\chapter[Measures for univariate distributions]{\href{https://www.youtube.com/watch?v=ssnLQRR4kus}{Descriptive measures for univariate frequency 
distributions}}
\lb{ch3}
There are four families of scale-level-dependent standard measures 
one employs in \textbf{Statistics} to describe characteristic 
properties of univariate relative frequency distributions. On a 
technical level, the determination of the values of these measures 
from available data does not go beyond application of the four 
fundamental arithmetical operations: addition, subtraction, 
multiplication and division. We will introduce these measures in 
turn. In the following we suppose given from a \textbf{survey} for 
some one-dimensional statistical variable $X$ either (i)~a
\textbf{raw data set}~$\{x_{i}\}_{i=1,\ldots,n}$ of $n$ 
measured values, or (ii)~a \textbf{relative frequency distribution}
$(a_{j},h_{j})_{j=1,\ldots,k}$ resp.\ 
$(K_{j},h_{j})_{j=1,\ldots,k}$.

\section[Measures of central tendency]{Measures of central
tendency}
\lb{sec:lage}
Let us begin with the \textbf{measures of central tendency} which 
intend to convey a notion of ``middle'' or ``centre'' of a 
univariate relative frequency distribution.

\subsection[Mode]{Mode}
The \textbf{mode} $x_\mathrm{mod}$ (nom, ord, metr) of the relative 
frequency distribution for any one-dimensional variable $X$ is that 
value $a_{j}$ in $X$'s spectrum which was observed with the 
highest relative frequency in a statistical sample 
$\boldsymbol{S_{\Omega}}$. Note that the mode does not necessarily 
take a unique value.

\medskip
\noindent
\underline{\textbf{Def.:}} $h_{n}(x_\mathrm{mod}) \geq h_{n}(a_{j})$
for all $j=1,\ldots,k$.

\medskip
\noindent
\underline{EXCEL, OpenOffice:} \texttt{MODE.SNGL} (dt.:
\texttt{MODUS.EINF}, \texttt{MODALWERT}) \\
\underline{SPSS:} Analyze $\rightarrow$ Descriptive Statistics
$\rightarrow$ Frequencies \ldots $\rightarrow$ Statistics
\ldots: Mode

\subsection[Median]{Median}
To determine the \textbf{median} $\tilde{x}_{0.5}$ (or $Q_{2}$)
(ord, metr) of the relative frequency distribution for an ordinally
or metrically scaled one-dimensional variable $X$, it is necessary
to first arrange the $n$ observed values $\{x_{i}\}_{i=1,\ldots,n}$ 
in their ascending natural \textbf{rank order}, i.e., $x_{(1)} \leq 
x_{(2)} \leq \ldots \leq x_{(n)}$.

\medskip
\noindent
\underline{\textbf{Def.:}} For the ascendingly ordered $n$
observed values $\{x_{i}\}_{i=1,\ldots,n}$, at most 50\% have a 
rank lower or equal to resp.\ are less or equal to the median 
value $\tilde{x}_{0.5}$, and at most 50\% have a rank higher or 
equal to resp.\ are greater or equal to the median value 
$\tilde{x}_{0.5}$.

%
\begin{itemize}

\item[(i)] Discrete data \hfill
$\fbox{$F_{n}(\tilde{x}_{0.5})\geq 0.5$}$
\be
\tilde{x}_{0.5}=\begin{cases}
x_{(\frac{n+1}{2})} & \text{if}\ n\ \text{is odd} \\
\frac{1}{2}[x_{(\frac{n}{2})}+x_{(\frac{n}{2}+1)}] & \text{if}
\ n\ \text{is even}
\end{cases} \ .
\ee

\item[(ii)] Binned data \hfill
$\fbox{$\tilde{F}_{n}(\tilde{x}_{0.5})=0.5$}$

The class interval $K_{i}$ contains the median value 
$\tilde{x}_{0.5}$, if
$\displaystyle\sum_{j=1}^{i-1}h_{j}<0.5$ and 
$\displaystyle\sum_{j=1}^{i}h_{j}\geq 0.5$.
Then
\be
\tilde{x}_{0.5}=u_{i}+\frac{b_{i}}{h_{i}}
\left(0.5-\sum_{j=1}^{i-1}h_{j}\right) \ .
\ee
Alternatively, the median of a statistical sample 
$\boldsymbol{S_{\Omega}}$ for a continuous variable $X$ with 
binned data $(K_{j},h_{j})_{j=1,\ldots,k}$ can be obtained from 
the associated empirical cumulative distribution function by 
solving the condition 
$\tilde{F}_{n}(\tilde{x}_{0.5})\stackrel{!}{=}0.5$
for $\tilde{x}_{0.5}$; cf. Eq.~(\ref{klempvert}).\footnote{From a 
mathematical point of view, this amounts to the following problem: 
consider a straight line which contains the point with coordinates 
$(x_{0},y_{0})$ and has non-zero slope $y^{\prime}(x_{0}) \neq 0$, 
i.e., $y=y_{0}+y^{\prime}(x_{0})(x-x_{0})$. Re-arranging to solve 
for the variable $x$ then yields 
$x=x_{0}+[y^{\prime}(x_{0})]^{-1}(y-y_{0})$.}
\end{itemize}
\underline{\textbf{Remark:}} Note that the value of the median of a 
univariate relative frequency distribution is reasonably 
insensitive to so-called \textbf{outliers} in a statistical sample.

\medskip
\noindent
\underline{\R:} \texttt{median(\textit{variable})} \\
\underline{EXCEL, OpenOffice:} \texttt{MEDIAN}
(dt.: \texttt{MEDIAN}) \\
\underline{SPSS:} Analyze $\rightarrow$ Descriptive Statistics
$\rightarrow$ Frequencies \ldots $\rightarrow$ Statistics
\ldots: Median

\subsection[$\alpha$--Quantile]{$\boldsymbol{\alpha}$--Quantile}
A generalisation of the median is the concept of the
$\boldsymbol{\alpha}$\textbf{--quantile} $\tilde{x}_{\alpha}$
(ord, metr) of the relative frequency distribution for an ordinally 
or metrically scaled one-dimensional variable $X$. Again, 
it is necessary to first arrange the $n$ observed values 
$\{x_{i}\}_{i=1,\ldots,n}$ in their ascending natural \textbf{rank 
order}, i.e., $x_{(1)} \leq x_{(2)} \leq \ldots \leq x_{(n)}$.

\medskip
\noindent
\underline{\textbf{Def.:}} For the ascendingly ordered $n$
observed values $\{x_{i}\}_{i=1,\ldots,n}$, and for given $\alpha$ 
with $0<\alpha<1$, at most $\alpha\times$100\% have a rank
lower of equal to resp.\ are less or equal to the 
$\alpha$--quantile $\tilde{x}_{\alpha}$, and at most 
$(1-\alpha)\times$100\% have a rank higher or equal 
to resp.\ are greater or equal to the 
$\alpha$--quantile $\tilde{x}_{\alpha}$.
\begin{itemize}

\item[(i)] Discrete data \hfill
$\fbox{$F_{n}(\tilde{x}_{\alpha})\geq \alpha$}$
\be
\tilde{x}_{\alpha}=\begin{cases}
x_{(k)} & \text{if}\ n\alpha\notin\mathbb{N}, k>n\alpha \\
\frac{1}{2}[x_{(k)}+x_{(k+1)}] & \text{if}
\ k=n\alpha\in\mathbb{N}
\end{cases} \ .
\ee

\item[(ii)] Binned data \hfill
$\fbox{$\tilde{F}_{n}(\tilde{x}_{\alpha})=\alpha$}$

The class interval $K_{i}$ contains the $\alpha$--quantile 
$\tilde{x}_{\alpha}$, if
$\displaystyle\sum_{j=1}^{i-1}h_{j}<\alpha$ and 
$\displaystyle\sum_{j=1}^{i}h_{j}\geq \alpha$. Then
\be
\tilde{x}_{\alpha}=u_{i}+\frac{b_{i}}{h_{i}}
\left(\alpha-\sum_{j=1}^{i-1}h_{j}\right) \ .
\ee
Alternatively, an $\alpha$--quantile of a statistical sample 
$\boldsymbol{S_{\Omega}}$ for a continuous variable $X$ with 
binned data $(K_{j},h_{j})_{j=1,\ldots,k}$ can be obtained from 
the associated empirical cumulative distribution function by 
solving the condition 
$\tilde{F}_{n}(\tilde{x}_{\alpha})\stackrel{!}{=}\alpha$
for $\tilde{x}_{\alpha}$; cf. Eq.~(\ref{klempvert}).

\end{itemize}
\underline{\textbf{Remark:}} The quantiles
$\tilde{x}_{0.25}$, $\tilde{x}_{0.5}$, $\tilde{x}_{0.75}$
(also denoted by $Q_{1}$, $Q_{2}$, $Q_{3}$) have special status. 
They are referred to as the \textbf{first quartile} $\rightarrow$
\textbf{second quartile (median)} $\rightarrow$
\textbf{third quartile} of a relative frequency distribution for an 
ordinally or a metrically scaled one-dimensional variable $X$ and 
form the core of the \textbf{five number summary} of this 
distribution. Occasionally, $\alpha$--quantiles are also referred 
to as \textbf{percentile values}.

\medskip
\noindent
\underline{\R:} \texttt{quantile(\textit{variable}, $\alpha$)} \\
\underline{EXCEL, OpenOffice:} \texttt{PERCENTILE.EXC}
(dt.: \texttt{QUANTIL.EXKL}, \texttt{QUANTIL}) \\
\underline{SPSS:} Analyze $\rightarrow$ Descriptive Statistics
$\rightarrow$ Frequencies \ldots $\rightarrow$ Statistics
\ldots: Percentile(s)

\subsection[Five number summary]{Five number summary}
The \textbf{five number summary} (ord, metr) of the relative 
frequency distribution for an ordinally or metrically scaled 
one-dimensional variable $X$ is a compact compilation of 
information giving the (i)~lowest rank resp.\ smallest value, 
(ii)~first quartile, (iii)~second quartile or median, (iv)~third 
quartile, and (v)~highest rank resp.\ largest value that $X$ takes 
in a univariate raw data set $\{x_{i}\}_{i=1,\ldots,n}$ from a 
statistical sample $\boldsymbol{S_{\Omega}}$, i.e.,
\be
\{x_{(1)}, \tilde{x}_{0.25}, \tilde{x}_{0.5},
\tilde{x}_{0.75}, x_{(n)}\} \ .
\ee
Alternative notation: $\{Q_{0}, Q_{1}, Q_{2}, Q_{3}, Q_{4}\}$.

\medskip
\noindent
\underline{\R:} \texttt{fivenum(\textit{variable})},
\texttt{summary(\textit{variable})} \\
\underline{EXCEL, OpenOffice:} \texttt{MIN}, \texttt{QUARTILE.INC},
\texttt{MAX} (dt.: \texttt{MIN}, \texttt{QUARTILE.INKL},
\texttt{QUARTILE}, \texttt{MAX})\\
\underline{SPSS:} Analyze $\rightarrow$ Descriptive Statistics
$\rightarrow$ Frequencies \ldots $\rightarrow$ Statistics
\ldots: Quartiles, Minimum, Maximum

\medskip
\noindent
All measures of central tendency which we will discuss hereafter 
are defined exclusively for characterising relative frequency 
distributions for \textit{metrically scaled one-dimensional 
variables}~$X$ only.

\subsection[Sample mean]{Sample mean}
The best known measure of central tendency is the dimensionful
\textbf{sample mean} $\bar{x}$ (metr) (also referred to as the 
arithmetical mean). Amongst the first to have employed the sample 
mean as a characteristic statistical measure in the systematic 
analysis of quantitative emprical data ranks the English 
physicist, mathematician, astronomer and philosopher
\href{http://www-history.mcs.st-and.ac.uk/Biographies/Newton.html}{Sir
Isaac Newton PRS MP (1643--1727)}; cf. Mlodinow 
(2008)~\ct[p~127]{mlo2008}. Given metrically scaled data, it is 
defined by:
\begin{itemize}
\item[(i)] From a raw data set:
\be
\lb{eq:arithmean1}
\fbox{$\displaystyle
\bar{x}
:= \frac{1}{n}\left(x_{1}+\ldots+x_{n}\right)
=: \frac{1}{n}\sum_{i=1}^{n}x_{i} \ .
$}
\ee
\item[(ii)] From a relative frequency distribution:
\be
\bar{x}
\lb{eq:arithmean2}
:= a_{1}h_{n}(a_{1})+\ldots+a_{k}h_{n}(a_{k})
=: \sum_{j=1}^{k}a_{j}h_{n}(a_{j}) \ .
\ee
\end{itemize}

\medskip
\noindent
\underline{\textbf{Remarks:}} (i)~The value of the sample mean is
very sensitive to \textbf{outliers}.\\ (ii)~For binned data one
selects the midpoint of each class interval $K_{i}$ to represent
the $a_{j}$ (provided the raw data set is no longer accessible).

\medskip
\noindent
\underline{\R:} \texttt{mean(\textit{variable})} \\
\underline{EXCEL, OpenOffice:} \texttt{AVERAGE} (dt.:
\texttt{MITTELWERT}) \\
\underline{SPSS:} Analyze $\rightarrow$ Descriptive Statistics
$\rightarrow$ Frequencies \ldots $\rightarrow$ Statistics
\ldots: Mean

\subsection[Weighted mean]{Weighted mean}
In practice, one also encounters the dimensionful \textbf{weighted 
mean} $\bar{x}_{w}$ (metr), defined by
\be
\fbox{$\displaystyle
\bar{x}_{w}
:= w_{1}x_{1}+\ldots+w_{n}x_{n}
=: \sum_{i=1}^{n}w_{i}x_{i} \ ;
$}
\ee
the $n$ \textbf{weight factors} $w_{1}$, \ldots, $w_{n}$ need to 
satisfy the constraints
\be
0 \leq w_{1}, \ldots, w_{n} \leq 1
\quad \text{and} \quad
w_{1}+\ldots+w_{n}=\sum_{i=1}^{n}w_{i}=1 \ .
\ee
%

\section[Measures of variability]{Measures of variability}
\lb{sec:streuung}
The idea behind the \textbf{measures of variability} is to convey a 
notion of the ``spread'' of data in a given statistical sample 
$\boldsymbol{S_{\Omega}}$, technically referred to also as the 
\textbf{dispersion} of the data. As the realisation of this
intention requires a well-defined concept of \textbf{distance}, 
the measures of variability are meaningful for data relating to 
\textit{metrically scaled one-dimensional variables}~$X$ only. One 
can distinguish two kinds of such measures: (i)~simple 
$2$-data-point measures, and (ii)~sophisticated $n$-data-point 
measures. We begin with two examples belonging to the first 
category.

\subsection[Range]{Range}
For a univariate raw data set $\{x_{i}\}_{i=1,\ldots,n}$ of $n$ 
observed values for $X$, the dimensionful \textbf{range} $R$ (metr) 
simply expresses the difference between the largest and the 
smallest value in this set, i.e.,
\be
\lb{eq:range}
R:=x_{(n)}-x_{(1)} \ .
\ee
The basis of this measure is the ascendingly ordered data set 
$x_{(1)} \leq x_{(2)} \leq \ldots \leq x_{(n)}$. Alternatively, 
the range can be denoted by $R=Q_{4}-Q_{0}$.

\medskip
\noindent
\underline{\R:} \texttt{range(\textit{variable})},
$\texttt{max(\textit{variable})}
- \texttt{min(\textit{variable})}$ \\
\underline{SPSS:} Analyze $\rightarrow$ Descriptive Statistics
$\rightarrow$ Frequencies \ldots $\rightarrow$ Statistics
\ldots: Range

\subsection[Interquartile range]{Interquartile range}
In the same spirit as the range, the dimensionful
\textbf{interquartile range} $d_{Q}$ (metr) is defined as the 
difference between the third quantile and the first quantile of 
the relative frequency distribution for some metrically scaled
$X$, i.e.,
\be
d_{Q}:=\tilde{x}_{0.75}-\tilde{x}_{0.25} \ .
\ee
Alternatively, this is $d_{Q}=Q_{3}-Q_{1}$.

\medskip
\noindent
\underline{\R:} \texttt{IQR(\textit{variable})}

\medskip
\noindent
Viewing the interquartile range~$d_{Q}$ of a univariate metrically
scaled raw data set~$\{x_{i}\}_{i=1,\ldots,n}$ as a reference
length, it is commonplace to define a specific value~$x_{i}$ to be
an
\begin{itemize}

\item \textbf{outlier}, if either $x_{i} < \tilde{x}_{0.25}
- 1.5d_{Q}$ and $x_{i} \geq \tilde{x}_{0.25}
- 3d_{Q}$, or $x_{i} > \tilde{x}_{0.75} + 1.5d_{Q}$ and
$x_{i} \leq \tilde{x}_{0.75} + 3d_{Q}$,

\item \textbf{extreme value}, if either $x_{i} < \tilde{x}_{0.25}
- 3d_{Q}$, or $x_{i} > \tilde{x}_{0.75} + 3d_{Q}$.

\end{itemize}

\medskip
\noindent
A very convenient graphical method for transparently displaying 
distributional features of metrically scaled data relating to a 
five number summary, also making explicit the interquartile
range, outliers and extreme values, is provided by a \textbf{box
plot}; see, e.g., Tukey (1977)~\ct{tuk1977}. An example of a single
box plot is depicted in Fig.~\ref{fig:1Dboxplot}, of parallel
box plots in Fig.~\ref{fig:parallelBoxplots}.

\medskip
\noindent
\underline{\R:} \texttt{boxplot(\textit{variable})},
\texttt{boxplot(\textit{variable}~\texttildelow~\textit{group
variable})}

\begin{figure}[!htb]
\begin{center}
\includegraphics[scale=0.8]{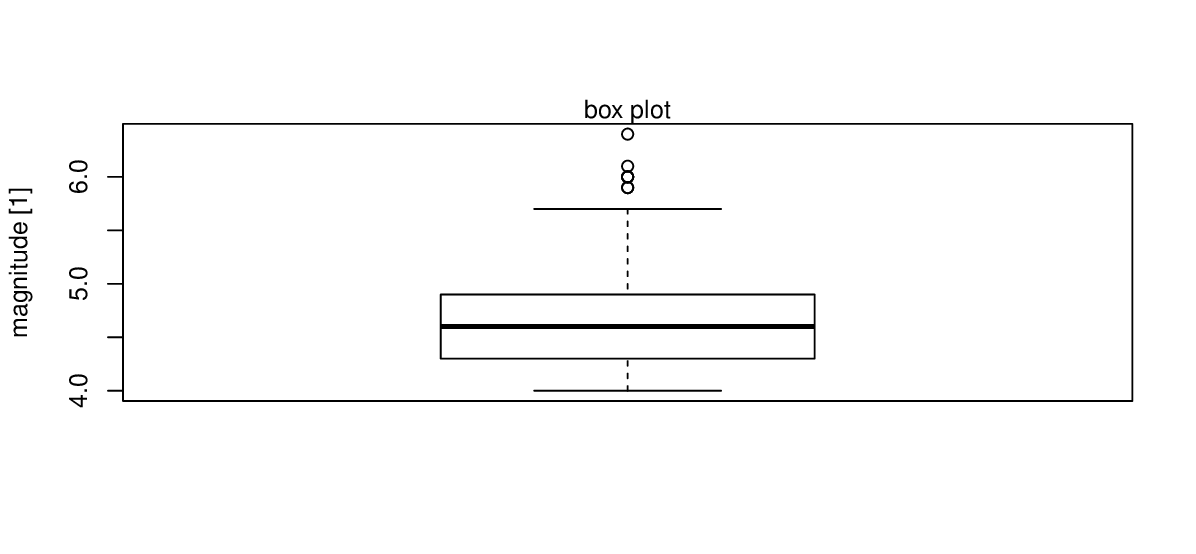}
\end{center}
\caption{Example of a box plot, representing elements of the five
number summary for the distribution of measured values for the
variable ``magnitude'' in the \R{} data set ``quakes.'' The open
circles indicate the positions of outliers. \newline
\underline{\R:} \newline
\texttt{data("quakes")} \newline
\texttt{?quakes} \newline
\texttt{boxplot( quakes\$mag )}}
\lb{fig:1Dboxplot}
\end{figure}
\begin{figure}[!htb]
\begin{center}
\includegraphics[scale=0.8]{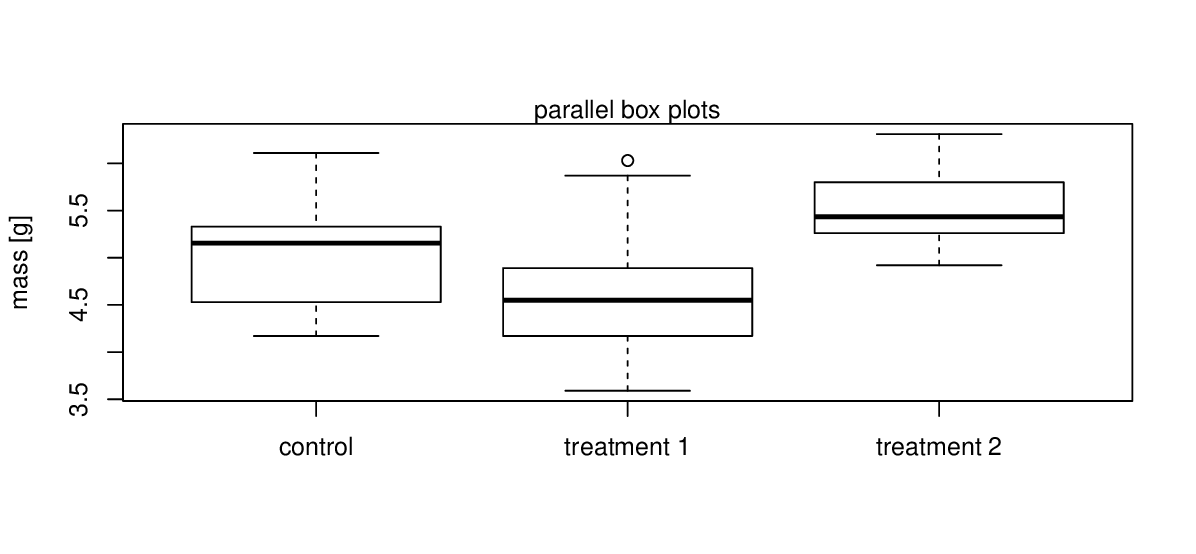}
\end{center}
\caption{Example of parallel box plots, comparing elements of
the five number summary for the distribution of measured values for
the variable ``weight'' between categories of the variable
``group'' in the \R{} data set ``PlantGrowth.'' The open
circle indicates the position of an outlier. \newline
\underline{\R:} \newline
\texttt{data("PlantGrowth")} \newline
\texttt{?PlantGrowth} \newline
\texttt{boxplot( PlantGrowth\$weight~\texttildelow~PlantGrowth\$group )}}
\lb{fig:parallelBoxplots}
\end{figure}
%

%
%
%

\subsection[Sample variance]{Sample variance}
The most frequently employed measure of variability in \textbf{
Statistics} is the dimensionful $n$-data-point \textbf{sample 
variance} $s^{2}$ (metr), and the related sample standard 
deviation to be discussed below. One of the originators of these 
concepts is the French mathematician 
\href{http://www-history.mcs.st-and.ac.uk/Biographies/De_Moivre.html}{Abraham de Moivre (1667--1754)}; cf. Bernstein (1998)~\ct[p~5]{ber1998}.
Given a univariate raw data set  
$\{x_{i}\}_{i=1,\ldots,n}$ for $X$, its spread is essentially 
quantified in terms of the sum of squared deviations of the $n$ 
data points~$x_{i}$ from their common sample mean~$\bar{x}$. Due 
to the algebraic identity
\[
\left(x_{1}-\bar{x}\right) + \ldots + \left(x_{n}-\bar{x}\right)
= \sum_{i=1}^{n}\left(x_{i}-\bar{x}\right)
= \left(\sum_{i=1}^{n}x_{i}\right) - n\bar{x}
\stackrel{\text{Eq.}~(\ref{eq:arithmean1})}{\equiv} 0 \ ,
\]
there are only $n-1$ \textbf{degrees of freedom} involved in this 
measure. The sample variance is thus defined by:
\begin{itemize}
\item[(i)] From a raw data set:
\be
\lb{eq:sampvardescr}
\fbox{$\displaystyle
s^{2} := 
\frac{1}{n-1}\left[\,(x_{1}-\bar{x})^{2}+\ldots+(x_{n}-\bar{x})^{2}
\,\right]
=: \frac{1}{n-1}\sum_{i=1}^{n}\left(x_{i}-\bar{x}\right)^{2} \ ;
$}
\ee
alternatively, by the \textbf{shift theorem}:\footnote{That is,
the algebraic identity 
$\displaystyle\sum_{i=1}^{n}\left(x_{i}-\bar{x}\right)^{2}
= \sum_{i=1}^{n}\left(x_{i}^{2}-2x_{i}\bar{x}+\bar{x}^{2}\right)
\stackrel{\text{Eq.}~(\ref{eq:arithmean1})}{\equiv} 
\sum_{i=1}^{n}x_{i}^{2}-\sum_{i=1}^{n}\bar{x}^{2}
= \sum_{i=1}^{n}x_{i}^{2}-n\bar{x}^{2}$.}
\be
\lb{eq:sampvardescr2}
s^{2}
= \frac{1}{n-1}\left[\,x_{1}^{2}+\ldots+x_{n}^{2}-n\bar{x}{}^{2}
\,\right]
= \frac{1}{n-1}\left[\,\sum_{i=1}^{n}x_{i}^{2}-n\bar{x}{}^{2}
\,\right] \ .
\ee
\item[(ii)] From a relative frequency distribution:
\begin{eqnarray}
\lb{eq:ssqurel1}
s^{2}
& := & \frac{n}{n-1}\left[\,(a_{1}-\bar{x})^{2}h_{n}(a_{1})+\ldots
+(a_{k}-\bar{x})^{2}h_{n}(a_{k})\,\right] \nonumber\\
& =: & \frac{n}{n-1}\sum_{j=1}^{k}\left(a_{j}-\bar{x}\right)^{2}
h_{n}(a_{j}) \ ;
\end{eqnarray}
alternatively:
\begin{eqnarray}
\lb{eq:ssqurel2}
s^{2}
& = & \frac{n}{n-1}\left[\,a_{1}^{2}h_{n}(a_{1})+\ldots
+a_{k}^{2}h_{n}(a_{k})-\bar{x}{}^{2}\,\right] \nonumber\\
& = & \frac{n}{n-1}\left[\,\sum_{j=1}^{k}a_{j}^{2}h_{n}(a_{j})
-\bar{x}{}^{2}\,\right] \ .
\end{eqnarray}

\end{itemize}
\underline{\textbf{Remarks:}} (i)~We point out that the alternative 
formulae for a sample variance provided here prove computationally 
more efficient.\\
(ii)~For binned data, when one selects the midpoint of each class 
interval $K_{j}$ to represent the~$a_{j}$ (given the raw data set 
is no longer accessible), a correction of Eqs.~(\ref{eq:ssqurel1}) 
and (\ref{eq:ssqurel2}) by an additional term 
$(1/12)(n/n-1)\sum_{j=1}^{k}b_{j}^{2}h_{j}$ becomes necessary,
assuming uniformly distributed data within each of the class
intervals $K_{j}$ of width $b_{j}$; cf. Eq.~(\ref{eq:varUniDistr}).

\medskip
\noindent
\underline{\R:} \texttt{var(\textit{variable})} \\
\underline{EXCEL, OpenOffice:} \texttt{VAR.S} (dt.: \texttt{VAR.S},
\texttt{VARIANZ}) \\
\underline{SPSS:} Analyze $\rightarrow$ Descriptive Statistics
$\rightarrow$ Frequencies \ldots $\rightarrow$ Statistics
\ldots: Variance

\subsection[Sample standard deviation]{Sample standard deviation}
For ease of handling dimensions associated with a metrically 
scaled one-dimensional variable $X$, one defines the dimensionful 
\textbf{sample standard deviation} $s$ (metr) simply as the positive 
square root of the sample variance~(\ref{eq:sampvardescr}), i.e.,
\be
s := +\sqrt{s^{2}} \ ,
\ee
such that a measure for the spread of data results which shares 
the dimension of $X$ and its sample mean~$\bar{x}$.

\medskip
\noindent
\underline{\R:} \texttt{sd(\textit{variable})} \\
\underline{EXCEL, OpenOffice:} \texttt{STDEV.S} (dt.:
\texttt{STABW.S}, \texttt{STABW}) \\
\underline{SPSS:} Analyze $\rightarrow$ Descriptive Statistics
$\rightarrow$ Frequencies \ldots $\rightarrow$ Statistics
\ldots: Std. deviation

\subsection[Sample coefficient of variation]{Sample coefficient of
variation}
For ratio scaled one-dimensional variables $X$, a dimensionless 
relative measure of variability is the \textbf{sample coefficient of 
variation} $v$ (metr: ratio), defined by
\be
v := \frac{s}{\bar{x}} \ ,
\quad\text{if}\quad
\bar{x} > 0 \ .
\ee
%

\subsection[Standardisation]{Standardisation}
Data for metrically scaled one-dimensional variables $X$ is 
amenable to the process of \textbf{standardisation}. By this is meant 
a linear affine transformation~$X \rightarrow Z$, which generates 
from a univariate raw data set $\{x_{i}\}_{i=1,\ldots,n}$ of $n$ 
measured values for a dimensionful variable $X$, with sample mean 
$\bar{x}$ and sample standard deviation $s_{X}>0$, data for an 
equivalent dimensionless variable $Z$ according to
\be
\lb{eq:standardisationdiscriptive}
\fbox{$\displaystyle
x_{i} \mapsto z_{i}:=\frac{x_{i}-\bar{x}}{s_{X}}
\qquad \text{for all}\ i=1,\ldots,n \ .
$}
\ee
For the resultant $Z$-data, referred to as the
\textbf{$\boldsymbol{Z}$ scores} of the original metrical $X$-data,
this has the convenient practical consequences that (i)~all 
one-dimensional metrical data is thus represented on the 
\textit{same dimensionless measurement scale}, and
(ii)~the corresponding sample mean and sample standard deviation of
the $Z$-data amount to
\[
\bar{z} = 0
\quad\text{and}\quad s_{Z}=1 \ ,
\]
respectively. Employing Z scores, specific values~$x_{i}$ of the
original metrical $X$-data will be expressed in terms of sample
standard deviation units, i.e., by how many sample standard
deviations they fall on either side of the common sample mean.
Essential information on characteristic distributional features of
one-dimensional metrical data will be preserved by the process of standardisation.

\medskip
\noindent
\underline{\R:} \texttt{scale(\textit{variable}, center = TRUE,
scale = TRUE)} \\
\underline{EXCEL, OpenOffice:} \texttt{STANDARDIZE} (dt.:
\texttt{STANDARDISIERUNG})\\
\underline{SPSS:} Analyze $\rightarrow$ Descriptive Statistics
$\rightarrow$ Descriptives \ldots $\rightarrow$ Save standardized 
values as variables

\section[Measures of relative distortion]{Measures of relative
distortion}
\lb{sec:distortion}
The third family of measures characterising relative frequency 
distributions for univariate data~$\{x_{i}\}_{i=1,\ldots,n}$ for 
metrically scaled one-dimensional variables~$X$, having specific 
sample mean~$\bar{x}$ and sample standard deviation~$s_{X}$, 
relate to the issue of the \textbf{shape} of a distribution. These 
measures take a \textbf{Gau\ss ian normal distribution} (cf. 
Sec.~\ref{sec:normverteil} below) as as a reference case, with the
values of its two free parameters equal to the given $\bar{x}$
and~$s_{X}$. With respect to this \textit{reference distribution},
one defines two kinds of dimensionless \textbf{measures of relative
distortion} as described in the following (cf., e.g., Joanes and
Gill (1998)~\ct{joagil1998}).

\subsection[Skewness]{Skewness}
\lb{subsec:skew}
The \textbf{skewness} $g_{1}$ (metr) is a dimensionless measure to 
quantify the degree of relative distortion of a given frequency
distribution in the \textit{horizontal direction}. Its 
implementation in the software package EXCEL employs the definition
\be
\lb{eq:scew}
g_{1} := 
\frac{n}{(n-1)(n-2)}\sum_{i=1}^{n}\left(\frac{x_{i}-\bar{x}}{s_{X}}
\right)^{3}
\qquad\text{for}\quad n > 2 \ ,
\ee
wherein the observed values $\{x_{i}\}_{i=1,\ldots,n}$ enter in 
their standardised form according to 
Eq.~(\ref{eq:standardisationdiscriptive}). Note that $g_{1}=0$ for 
an exact Gau\ss ian normal distribution.

\medskip
\noindent
\underline{\R:} \texttt{skewness(\textit{variable},
type = 2)} (package: \texttt{e1071}, by Meyer \textit{et al}
(2019)~\ct{meyetal2019}) \\
\underline{EXCEL, OpenOffice:} \texttt{SKEW} (dt.:
\texttt{SCHIEFE}) \\
\underline{SPSS:} Analyze $\rightarrow$ Descriptive Statistics
$\rightarrow$ Frequencies \ldots $\rightarrow$ Statistics
\ldots: Skewness

\subsection[Excess kurtosis]{Excess kurtosis}
\lb{subsec:kurt}
The \textbf{excess kurtosis} $g_{2}$ (metr) is a dimensionless 
measure to quantify the degree of relative distortion of a given 
frequency distribution in the \textit{vertical direction}. Its 
implementation in the software package EXCEL employs the definition
\be
\lb{eq:kurt}
g_{2} := 
\frac{n(n+1)}{(n-1)(n-2)(n-3)}\sum_{i=1}^{n}\left(\frac{x_{i}
-\bar{x}}{s_{X}}\right)^{4} - \frac{3(n-1)^{2}}{(n-2)(n-3)}
\qquad\text{for}\quad n > 3 \ ,
\ee
wherein the observed values $\{x_{i}\}_{i=1,\ldots,n}$ enter in
their standardised form according to 
Eq.~(\ref{eq:standardisationdiscriptive}). Note that $g_{2}=0$ for 
an exact Gau\ss ian normal distribution.

\medskip
\noindent
\underline{\R:} \texttt{kurtosis(\textit{variable},
type = 2)} (package: \texttt{e1071}, by Meyer \textit{et al}
(2019)~\ct{meyetal2019}) \\
\underline{EXCEL, OpenOffice:} \texttt{KURT} (dt.:
\texttt{KURT}) \\
\underline{SPSS:} Analyze $\rightarrow$ Descriptive Statistics
$\rightarrow$ Frequencies \ldots $\rightarrow$ Statistics
\ldots: Kurtosis

\section[Measures of concentration]{Measures of concentration}
\lb{sec:konz}
Finally, for univariate data~$\{x_{i}\}_{i=1,\ldots,n}$ relating 
to a ratio scaled one-dimensional variable $X$, which has a 
discrete spectrum of values $\{a_{j}\}_{j=1,\ldots,k}$, or which
was binned into $k$~different categories $\{K_{j}\}_{j=1,\ldots,k}$ 
with respective midpoints $a_{j}$, two kinds of \textbf{measures of 
concentration} are commonplace in \textbf{Statistics}; one 
qualitative in nature, the other quantitative.

\medskip
\noindent
Begin by defining the \textbf{total sum} for the data 
$\{x_{i}\}_{i=1,\ldots,n}$ by
\be
\lb{eq:totalsum}
S := \sum_{i=1}^{n}x_{i}
= \sum_{j=1}^{k}a_{j}o_{n}(a_{j})
\stackrel{\text{Eq.}~(\ref{eq:arithmean1})}{=} n\bar{x} \ ,
\ee
where $(a_{j},o_{n}(a_{j}))_{j=1,\ldots,k}$ is the absolute 
frequency distribution for the observed values (or categories) 
of~$X$. Then the \textbf{relative proportion} that the value
$a_{j}$ (or the category $K_{j}$) takes in~$S$ is
\be
\lb{eq:relprop}
\frac{a_{j}o_{n}(a_{j})}{S} = \frac{a_{j}h_{n}(a_{j})}{\bar{x}} \ .
\ee
%

\subsection[Lorenz curve]{Lorenz curve}
\lb{subsec:lorenz}
From the elements introduced in Eqs.~(\ref{eq:totalsum}) and 
(\ref{eq:relprop}), the US--American economist
\href{http://en.wikipedia.org/wiki/Max_O._Lorenz}{Max Otto Lorenz 
(1876--1959)} constructed cumulative relative quantities which 
constitute the coordinates of a so-called \textbf{Lorenz curve} 
representing concentration in the 
distribution for the ratio scaled one-dimensional variable $X$; cf. 
Lorenz (1905)~\ct{lor1905}. These coordinates are defined as 
follows:
\begin{itemize}
\item Horizontal axis:
\be
k_{i} := \sum_{j=1}^{i}\frac{o_{n}(a_{j})}{n}
= \sum_{j=1}^{i}h_{n}(a_{j})
\qquad (i=1,\ldots,k) \ ,
\ee
\item Vertical axis:
\be
l_{i} := \sum_{j=1}^{i}\frac{a_{j}o_{n}(a_{j})}{S}
= \sum_{j=1}^{i}\frac{a_{j}h_{n}(a_{j})}{\bar{x}}
\qquad (i=1,\ldots,k) \ .
\ee
\end{itemize}
The initial point on a Lorenz curve is generally the coordinate
system's origin, $(k_{0},l_{0})=(0,0)$, the final point 
is $(1,1)$. As a reference facility to measure concentration in 
the distribution of $X$ in qualitative terms, one defines a
\textbf{null concentration curve} as the bisecting line linking
$(0,0)$ to $(1,1)$. The Lorenz curve is interpreted as stating that
a point on the curve with coordinates $(k_{i},l_{i})$ represents
the fact that $k_{i}\times 100\%$ of the $n$ statistical units take
a share of $l_{i}\times 100\%$ in the total sum $S$ for the ratio
scaled one-dimensional variable $X$. Qualitatively, for given
univariate data~$\{x_{i}\}_{i=1,\ldots,n}$, the concentration in
the distribution of $X$ is the stronger, the larger is the dip of
the Lorenz curve relative to the null concentration curve. Note
that in addition to the null concentration curve, one can define as
a second reference facility a \textbf{maximum concentration curve}
such that only the largest value $a_{k}$ (or category $K_{k}$) in
the spectrum of values of $X$ takes the full share of $100\%$ in
the total sum $S$ for $\{x_{i}\}_{i=1,\ldots,n}$.

\subsection[Normalised Gini coefficient]{Normalised Gini 
coefficient}
\lb{subsec:gini}
The Italian statistician, demographer and sociologist
\href{http://en.wikipedia.org/wiki/Corrado_Gini}{Corrado
Gini (1884--1965)} devised a quantitative measure for 
concentration in the distribution for a ratio scaled
one-dimensional variable~$X$; cf. Gini (1921)~\ct{gin1921}. The
dimensionless \textbf{normalised Gini coefficient} $G_{+}$ (metr:
ratio) can be interpreted geometrically as the ratio of areas
\be
G_{+} := \frac{(\text{area enclosed between Lorenz and
null concentration curves})}{(\text{area enclosed between maximum 
and null concentration curves})} \ .
\ee
Its related computational definition is given by
\be
\fbox{$\displaystyle
G_{+} := \frac{n}{n-1}\left[\,\sum_{i=1}^{k}(k_{i-1}+k_{i})\,
\frac{a_{i}o_{n}(a_{i})}{S}-1\,\right] \ .
$}
\ee
Due to normalisation, the range of values is $0 \leq G_{+} \leq 1$.
Thus, null concentration amounts to $G_{+}=0$, while maximum 
concentration amounts to $G_{+}=1$.\footnote{In September 2012 it 
was reported (implicitly) in the public press that the coordinates 
underlying the Lorenz curve describing the distribution of private 
equity in Germany at the time were $(0.00,0.00)$, $(0.50,0.01)$, 
$(0.90,0.50)$, and $(1.00,1.00)$; cf. Ref.~\ct{sue2012}. Given 
that in this case $n \gg 1$, these values amount to a Gini 
coefficient of $G_{+}=0.64$. The Oxfam Report on Wealth Inequality
2019 can be found at the URL (cited on May 31, 2019):
\href{https://www.oxfam.org/en/research/public-good-or-private-wealth}{www.oxfam.org/en/research/public-good-or-private-wealth}.}


\chapter[Measures of association for bivariate
distributions]{Descriptive measures of association for bivariate
frequency distributions}
\lb{ch4}
Now we come to describe and characterise specific features of 
bivariate frequency distributions, i.e., intrinsic structures of 
bivariate raw data sets $\{(x_{i},y_{i})\}_{i=1,\ldots,n}$ 
obtained from samples~$\boldsymbol{S_{\Omega}}$ for a 
two-dimensional statistical variable~$(X,Y)$ from some target 
population of study objects~$\boldsymbol{\Omega}$. Let us suppose 
that the spectrum of values resp.\ categories of $X$ is $a_{1}, 
a_{2}, \ldots, a_{k}$, and the spectrum of values resp.\ 
categories of $Y$ is $b_{1}, b_{2}, \ldots, b_{l}$, where $k, l 
\in \mathbb{N}$. Hence, for the bivariate \textbf{joint
distribution} there exists a total of $k \times l$ possible
combinations $\{(a_{i},b_{j})\}_{i=1,\ldots,k;j=1,\ldots,l}$ of
values resp.\ categories for $(X,Y)$. In the following, we will
denote associated bivariate absolute (observed) frequencies by 
$o_{ij}:=o_{n}(a_{i},b_{j})$, and bivariate relative frequencies 
by $h_{ij}:=h_{n}(a_{i},b_{j})$.

\section[$(k \times l)$ contingency tables]{$\boldsymbol{(k \times 
l)}$ contingency tables}
\lb{sec:konttaf}
Consider a bivariate raw data set 
$\{(x_{i},y_{i})\}_{i=1,\ldots,n}$ for a two-dimensional 
statistical variable~$(X,Y)$, giving rise to $k \times l$ 
combinations of values resp.\ 
categories $\{(a_{i},b_{j})\}_{i=1,\ldots,k;j=1,\ldots,l}$. The 
bivariate joint distribution of observed \textbf{absolute 
frequencies} $o_{ij}$ may be conveniently represented in terms of 
a $\boldsymbol{(k \times l)}$ \textbf{contingency table}, or
\textbf{cross tabulation}, by
\be
\begin{array}{c|cccccc|c}
o_{ij} & b_{1} & b_{2} & \ldots & b_{j} & \ldots & b_{l} &
\Sigma_{j} \\
\hline
a_{1} & o_{11} & o_{12} & \ldots & o_{1j} & \ldots & o_{1l} &
o_{1+} \\
a_{2} & o_{21} & o_{22} & \ldots & o_{2j} & \ldots & o_{2l} &
o_{2+} \\
\vdots & \vdots & \vdots & \ddots & \vdots & \ddots & \vdots &
\vdots \\
a_{i} & o_{i1} & o_{i2} & \ldots & o_{ij} & \ldots & o_{il} &
o_{i+} \\
\vdots & \vdots & \vdots & \ddots & \vdots & \ddots & \vdots &
\vdots \\
a_{k} & o_{k1} & o_{k2} & \ldots & o_{kj} & \ldots & o_{kl} &
o_{k+} \\
\hline
\Sigma_{i} & o_{+1} & o_{+2} & \ldots & o_{+j} &
\ldots & o_{+l} & n
\end{array} \ ,
\ee
where it holds for all $i=1,\ldots,k$ and $j=1,\ldots,l$ that
\be
0 \leq o_{ij} \leq n \qquad\text{and}\qquad
\sum_{i=1}^{k}\sum_{j=1}^{l}o_{ij}=n \ .
\ee
The corresponding univariate \textbf{marginal absolute frequencies} 
of $X$ and of $Y$ are
\bea
\lb{eq:margfreq1}
o_{i+} & := & o_{i1} + o_{i2} + \ldots + o_{ij} + \ldots + o_{il}
\ =: \ \sum_{j=1}^{l}o_{ij} \\
\lb{eq:margfreq2}
o_{+j} & := & o_{1j} + o_{2j} + \ldots + o_{ij} + \ldots + o_{kj}
\ =: \ \sum_{i=1}^{k}o_{ij} \ .
\eea

\medskip
\noindent
\underline{\R:} \texttt{CrossTable(\textit{row variable},
\textit{column variable})} (package: \texttt{gmodels}, by
Warnes \textit{et al} (2018)~\ct{waretal2018}) \\
\underline{SPSS:} Analyze $\rightarrow$ Descriptive Statistics
$\rightarrow$ Crosstabs \ldots $\rightarrow$ Cells
\ldots: Observed

\medskip
\noindent
One obtains the related bivariate joint distribution of observed 
\textbf{relative frequencies} $h_{ij}$ following the systematics of 
Eq.~(\ref{eq:relfreq}) to yield
\be
\begin{array}{c|cccccc|c}
h_{ij} & b_{1} & b_{2} & \ldots & b_{j} & \ldots & b_{l} &
\Sigma_{j} \\
\hline
a_{1} & h_{11} & h_{12} & \ldots & h_{1j} & \ldots & h_{1l} &
h_{1+} \\
a_{2} & h_{21} & h_{22} & \ldots & h_{2j} & \ldots & h_{2l} &
h_{2+} \\
\vdots & \vdots & \vdots & \ddots & \vdots & \ddots & \vdots &
\vdots \\
a_{i} & h_{i1} & h_{i2} & \ldots & h_{ij} & \ldots & h_{il} &
h_{i+} \\
\vdots & \vdots & \vdots & \ddots & \vdots & \ddots & \vdots &
\vdots \\
a_{k} & h_{k1} & h_{k2} & \ldots & h_{kj} & \ldots & h_{kl} &
h_{k+} \\
\hline
\Sigma_{i} & h_{+1} & h_{+2} & \ldots & h_{+ j} &
\ldots & h_{+ l} & 1
\end{array} \ .
\ee
Again, it holds for all $i=1,\ldots,k$ and $j=1,\ldots,l$ that
\be
0 \leq h_{ij} \leq 1 \qquad\text{and}\qquad
\sum_{i=1}^{k}\sum_{j=1}^{l}h_{ij}=1 \ ,
\ee
while the univariate \textbf{marginal relative frequencies} of $X$ 
and of $Y$ are
\bea
h_{i+} & := & h_{i1} + h_{i2} + \ldots + h_{ij} + \ldots + h_{il}
\ =: \ \sum_{j=1}^{l}h_{ij} \\
h_{+j} & := & h_{1j} + h_{2j} + \ldots + h_{ij} + \ldots + h_{kj}
\ =: \ \sum_{i=1}^{k}h_{ij} \ .
\eea

\medskip
\noindent
On the basis of a $(k \times l)$ contingency table displaying the 
relative frequencies of the bivariate joint distribution for some 
two-dimensional variable~$(X,Y)$, one may define two 
kinds of related \textbf{conditional relative frequency 
distributions}, namely (i)~the conditional distribution of $X$ 
given~$Y$ by
\be
\lb{eq:condrelfreq1}
h(a_{i}|b_{j}) := \frac{h_{ij}}{h_{+j}} \ ,
\ee
and (ii)~the conditional distribution of $Y$ given~$X$ by
\be
\lb{eq:condrelfreq2}
h(b_{j}|a_{i}) := \frac{h_{ij}}{h_{i+}} \ .
\ee
Then, by means of these conditional distributions, a notion of 
\textbf{statistical independence} of variables $X$ and $Y$ is
defined to correspond to the simultaneous properties
\be
h(a_{i}|b_{j}) = h(a_{i}) = h_{i+}
\qquad\text{and}\qquad
h(b_{j}|a_{i}) = h(b_{j}) = h_{+j} \ .
\ee
Given these properties hold, it follows from 
Eqs.~(\ref{eq:condrelfreq1}) and (\ref{eq:condrelfreq2}) that
\be
\lb{eq:statindep}
h_{ij} = h_{i+}h_{+j} \ ;
\ee
the bivariate relative frequencies $h_{ij}$ in this case are 
numerically equal to the product of the corresponding univariate 
marginal relative frequencies $h_{i+}$ and $h_{+j}$.

\section[Measures of association for the metrical scale
level]{\href{https://www.youtube.com/watch?v=8o-Z2i_aOOU}{Measures
of association for the metrical scale level}}
\lb{sec:2Dmetr}
Next, specifically consider a bivariate raw data set 
$\{(x_{i},y_{i})\}_{i=1,\ldots,n}$ from a statistical sample 
$\boldsymbol{S_{\Omega}}$ for a metrically scaled two-dimensional
variable~$(X,Y)$. The bivariate joint distribution for $(X,Y)$ in 
this sample can be conveniently represented graphically in terms 
of a \textbf{scatter plot}, cf. Fig.~\ref{fig:scatterplot}, thus
uniquely locating the positions of 
$n$ sampling units in (a subset of) \textbf{Euclidian space} 
$\mathbb{R}^{2}$. Let us now introduce two kinds of 
measures for the description of specific characteristic features 
of such bivariate joint distributions.

\medskip
\noindent
\underline{\R:} \texttt{plot(\textit{variable1},
\textit{variable2})}

\begin{figure}[!htb]
\begin{center}
\includegraphics[scale=0.8]{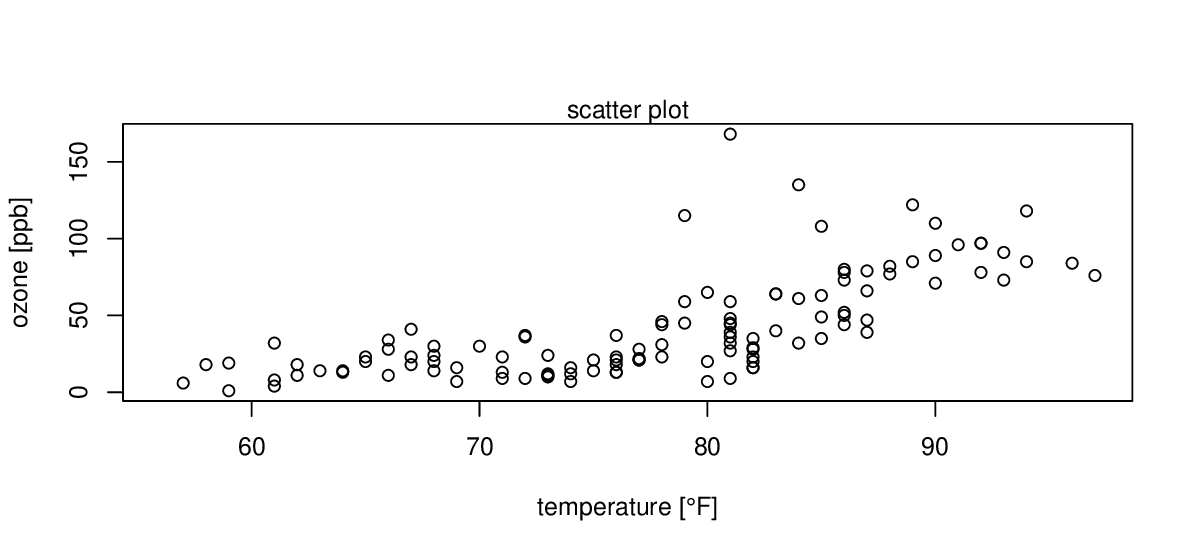}
\end{center}
\caption{Example of a scatter plot, representing the joint
distribution of measured values for the variables ``temperature''
and ``ozone'' in the \R{} data set ``airquality.'' \newline
\underline{\R:} \newline
\texttt{data("airquality")} \newline
\texttt{?airquality} \newline
\texttt{plot( airquality\$Temp , airquality\$Ozone )}}
\lb{fig:scatterplot}
\end{figure}
%

\subsection[Sample covariance]{Sample covariance}
\lb{subsec:covar}
The first standard measure describing degree of association in the 
joint distribution for a metrically scaled two-dimensional 
variable~$(X,Y)$ is the dimensionful \textbf{sample covariance} 
$s_{XY}$ (metr), defined by

\begin{itemize}
\item[(i)] From a raw data set:
\begin{eqnarray}
\lb{eq:covar}
s_{XY}
& := & \frac{1}{n-1}\left[\,(x_{1}-\bar{x})(y_{1}-\bar{y})+\ldots
+ (x_{n}-\bar{x})(y_{n}-\bar{y})\,\right]
\nonumber \\
& =: & \frac{1}{n-1}\sum_{i=1}^{n}\left(x_{i}-\bar{x}\right)
\left(y_{i}-\bar{y}\right)
\ ;
\end{eqnarray}
alternatively:
\begin{eqnarray}
\lb{eq:covar2}
s_{XY}
& = & \frac{1}{n-1}\left[\,x_{1}y_{1}+\ldots+x_{n}y_{n}
-n\bar{x}\bar{y}\,\right] \nonumber\\
& = & \frac{1}{n-1}\left[\,\sum_{i=1}^{n}x_{i}y_{i}
-n\bar{x}\bar{y}\,\right] \ .
\end{eqnarray}

\item[(ii)] From a relative frequency distribution:
\begin{eqnarray}
s_{XY}
& := & \frac{n}{n-1}\left[\,(a_{1}-\bar{x})(b_{1}-\bar{y})h_{11}
+ \ldots + (a_{k}-\bar{x})(b_{l}-\bar{y})h_{kl}
\,\right] \nonumber\\
& =: & \frac{n}{n-1}\sum_{i=1}^{k}\sum_{j=1}^{l}
\left(a_{i}-\bar{x}\right)\left(b_{j}-\bar{y}\right)h_{ij} \ ;
\end{eqnarray}
alternatively:
\begin{eqnarray}
\lb{eq:covarrelfreq2}
s_{XY}
& = & \frac{n}{n-1}\left[\,a_{1}b_{1}h_{11}+\ldots
+a_{k}b_{l}h_{kl}-\bar{x}\bar{y}\,\right] \nonumber \\
& = & \frac{n}{n-1}\left[\,\sum_{i=1}^{k}\sum_{j=1}^{l}
a_{i}b_{j}h_{ij}-\bar{x}\bar{y}\,\right] \ .
\end{eqnarray}
\end{itemize}
\underline{\textbf{Remark:}} The alternative formulae provided here 
prove computationally more efficient.

\medskip
\noindent
\underline{\R:} \texttt{cov(\textit{variable1},
\textit{variable2})} \\
\underline{EXCEL, OpenOffice:} \texttt{COVARIANCE.S} (dt.:
\texttt{KOVARIANZ.S, KOVAR})

\vspace{5mm}
\noindent
In view of its defining equation~(\ref{eq:covar}), the sample 
covariance can be given the following geometrical interpretation. 
For a total of $n$ data points $(x_{i},y_{i})$, it quantitfies the 
degree of excess of \textit{signed} rectangular areas 
$\left(x_{i}-\bar{x}\right)\left(y_{i}-\bar{y}\right)$ with 
respect to the common \textbf{centroid} ${\displaystyle 
\boldsymbol{r}_{C}:=\left(\begin{array}{c}
\bar{x} \\ \bar{y} \end{array}\right)}$ of the $n$~data points in 
favour of either positive or negative signed areas, if 
any.\footnote{The centroid is the special case of equal mass 
points, with masses ${\displaystyle m_{i}=\frac{1}{n}}$, of the 
centre of gravity of a system of $n$ discrete massive objects, 
defined by 
${\displaystyle\boldsymbol{r}_{C}:=\frac{\sum_{i=1}^{n}m_{i}
\boldsymbol{r}_{i}}{\sum_{j=1}^{n}m_{j}}}$. In two Euclidian 
dimensions the position vector is ${\displaystyle 
\boldsymbol{r}_{i}=\left(\begin{array}{c} x_{i} \\ y_{i} 
\end{array}\right)}$.}

\medskip
\noindent
It is worthwhile to point out that in the research literature it  
is standard to define for the joint distribution for a metrically 
scaled two-dimensional variable~$(X,Y)$ a dimensionful symmetric 
$\boldsymbol{(2 \times 2)}$ \textbf{sample covariance matrix} 
$\boldsymbol{S^{2}}$ according to
\be
\lb{eq:2dcovmat}
\boldsymbol{S^{2}} :=
\left(\begin{array}{cc}
s_{X}^{2} & s_{XY} \\
s_{XY} & s_{Y}^{2}
\end{array}\right) \ ,
\ee
the components of which are defined by
Eqs.~(\ref{eq:sampvardescr}) and (\ref{eq:covar}). The determinant 
of $\boldsymbol{S^{2}}$, given by
$\det(\boldsymbol{S^{2}})=s_{X}^{2}s_{Y}^{2}-s_{XY}^{2}$, is
positive as long as $s_{X}^{2}s_{Y}^{2}-s_{XY}^{2} > 0$, which
applies in most practical cases. Then $\boldsymbol{S^{2}}$ is
regular, and thus a corresponding inverse
$(\boldsymbol{S^{2}})^{-1}$ exists; cf. 
Ref.~\ct[Sec.~3.5]{hve2009}.

\medskip
\noindent
The concept of a regular sample covariance matrix
$\boldsymbol{S^{2}}$ and 
its inverse $(\boldsymbol{S^{2}})^{-1}$ generalises in a
straightforward fashion to the case of multivariate joint
distributions for metrically scaled $m$-dimensional statistical
variables~$(X,Y, \ldots, Z)$, where $\boldsymbol{S^{2}} \in
\mathbb{R}^{m \times m}$ is given by
\be
\lb{eq:mdcovmat}
\boldsymbol{S^{2}} :=
\left(\begin{array}{cccc}
s_{X}^{2} & s_{XY} & \ldots & s_{ZX} \\
s_{XY} & s_{Y}^{2} & \ldots & s_{YZ} \\
\vdots & \vdots & \ddots & \vdots \\
s_{ZX} & s_{YZ} & \ldots & s_{Z}^{2}
\end{array}\right) \ ,
\ee
and $\det(\boldsymbol{S^{2}}) \neq 0$ is required.

\subsection[Bravais and Pearson's sample correlation 
coefficient]{Bravais and Pearson's sample correlation coefficient}
The sample covariance $s_{XY}$ constitutes the basis for
the second standard measure characterising the joint distribution 
for a metrically scaled two-dimensional variable~$(X,Y)$ by 
descriptive means, which is the normalised and dimensionless
\textbf{sample correlation coefficient} $r$ (metr) devised by the
French physicist
\href{http://en.wikipedia.org/wiki/Auguste_Bravais}{Auguste
Bravais (1811--1863)} and the English mathematician and 
statistician
\href{http://www-history.mcs.st-and.ac.uk/Biographies/Pearson.html}{Karl
Pearson FRS (1857--1936)} for the purpose of analysing 
corresponding bivariate raw data 
$\{(x_{i},y_{i})\}_{i=1,\ldots,n}$ for the existence of a 
\textit{linear~(!!!)} statistical association. It is defined in 
terms of the bivariate sample covariance $s_{XY}$ and the 
univariate sample standard deviations $s_{X}$ and $s_{Y}$ by
(cf. Bravais (1846)~\ct{bra1846} and Pearson
(1901, 1920)~\ct{pea1901,pea1920})
\be
\lb{eq:correl}
\fbox{$\displaystyle
r:=\frac{s_{XY}}{s_{X}s_{Y}} \ .
$}
\ee
With Eq.~(\ref{eq:covar}) for $s_{XY}$, this becomes
\be
\lb{eq:correl2}
r = \frac{1}{n-1}\sum_{i=1}^{n}\left(\frac{x_{i}-\bar{x}}{s_{X}}
\right)\left(\frac{y_{i}-\bar{y}}{s_{Y}}\right)
= \frac{1}{n-1}\sum_{i=1}^{n}z_{i}^{X}z_{i}^{Y} \ ,
\ee
employing standardisation according to
Eq.~(\ref{eq:standardisationdiscriptive}) in the final step.
Due to its normalisation, the range of the sample correlation 
coefficient is $-1 \leq r \leq +1$. The sign of $r$ encodes the 
\textbf{direction} of a correlation. As to interpreting the
\textbf{strength} of a correlation via the magnitude $|r|$, in
practice one typically employs the following qualitative

\medskip
\noindent
\underline{\textbf{Rule of thumb:}}\\
$0.0 = |r|$: no correlation\\
$0.0 < |r| < 0.2$: very weak correlation\\
$0.2 \leq |r| < 0.4$: weak correlation\\
$0.4 \leq |r| < 0.6$: moderately strong correlation\\
$0.6 \leq |r| \leq 0.8$: strong correlation\\
$0.8 \leq |r| < 1.0$: very strong correlation\\
$1.0=|r|$: perfect correlation.

\medskip
\noindent
\underline{\R:} \texttt{cor(\textit{variable1},
\textit{variable2})} \\
\underline{EXCEL, OpenOffice:} \texttt{CORREL} (dt.:
\texttt{KORREL}) \\
\underline{SPSS:} Analyze $\rightarrow$ Correlate
$\rightarrow$ Bivariate \ldots: Pearson

\medskip
\noindent
In line with Eq.~(\ref{eq:2dcovmat}), it is convenient to define 
a dimensionless symmetric $\boldsymbol{(2 \times 2)}$
\textbf{sample correlation matrix} $\boldsymbol{R}$ by
\be
\lb{eq:2dcorrelmat}
\boldsymbol{R} :=
\left(\begin{array}{cc}
1 & r \\
r & 1
\end{array}\right) \ ,
\ee
which is regular and positive definite as long as its determinant 
$\det(\boldsymbol{R}) =1-r^{2}>0$. In this case, its 
inverse $\boldsymbol{R}^{-1}$ is given by
\be
\lb{eq:2dcorrelmatinv}
\boldsymbol{R}^{-1} =
\frac{1}{1-r^{2}}\left(\begin{array}{rr}
1 & -r \\
-r & 1
\end{array}\right) \ .
\ee
Note that for \textit{non-correlating} metrically scaled 
variables $X$ and $Y$, i.e., when $r=0$, the sample correlation
matrix degenerates to become a unit matrix, 
$\boldsymbol{R}=\boldsymbol{1}$.

\medskip
\noindent
Again, the concept of a regular and positive definite sample
correlation matrix $\boldsymbol{R}$, with inverse
$\boldsymbol{R}^{-1}$,
generalises to multivariate joint distributions for metrically 
scaled $m$-dimensional statistical variables~$(X,Y, \ldots, Z)$, 
where $\boldsymbol{R} \in \mathbb{R}^{m \times m}$ is given 
by\footnote{Given a data matrix $\boldsymbol{X} \in \mathbb{R}^{n 
\times m}$ for a metrically scaled $m$-dimensional statistical 
variable~$(X,Y, \ldots, Z)$, one can show that upon 
standardisation of the data according to 
Eq.~(\ref{eq:standardisationdiscriptive}), which 
amounts to a transformation $\boldsymbol{X} \mapsto \boldsymbol{Z} 
\in \mathbb{R}^{n \times m}$, the sample correlation matrix can be 
represented by 
$\displaystyle\boldsymbol{R}=\frac{1}{n-1}\,\boldsymbol{Z}^{T}
\boldsymbol{Z}$. The form of this relation is equivalent to
Eq.~(\ref{eq:correl2}).}
\be
\lb{eq:mdcorrelmat}
\boldsymbol{R} :=
\left(\begin{array}{cccc}
1 & r_{XY} & \ldots & r_{ZX} \\
r_{XY} & 1 & \ldots & r_{YZ} \\
\vdots & \vdots & \ddots & \vdots \\
r_{ZX} & r_{YZ} & \ldots & 1
\end{array}\right) \ ,
\ee
and $\det(\boldsymbol{R}) \neq 0$. Note that $\boldsymbol{R}$ is a 
dimensionless quantity which, hence, is \textbf{scale-invariant};
cf. Sec.~\ref{sec:paretodistr}.

\section[Measures of association for the ordinal scale
level]{\href{https://www.youtube.com/watch?v=6IJf7izTtjA}{Measures
of association for the ordinal scale level}}
\lb{sec:2Dord}
At the ordinal scale level, bivariate raw data 
$\{(x_{i},y_{i})\}_{i=1,\ldots,n}$ for a two-dimensional 
variable~$(X,Y)$ is not necessarily quantitative in nature. 
Therefore, in order to be in a position to define a sensible 
quantitative bivariate measure of statistical association for 
ordinal variables, one needs to introduce meaningful surrogate 
data which is numerical. This task is realised by means of 
defining so-called \textbf{rank numbers}, which are assigned to the 
original ordinal data according to the procedure described in the 
following.

\medskip
\noindent
Begin by establishing amongst the observed values 
$\{x_{i}\}_{i=1,\ldots,n}$ resp.\ $\{y_{i}\}_{i=1,\ldots,n}$ their 
natural ascending rank order, i.e.,
\be
\lb{eq:xyordered}
x_{(1)} \leq x_{(2)} \leq \ldots \leq x_{(n)}
\qquad\text{and}\qquad
y_{(1)} \leq y_{(2)} \leq \ldots \leq y_{(n)} \ .
\ee
Then, every individual $x_{i}$ resp.\ $y_{i}$ is assigned a 
\textbf{rank number} which corresponds to its position in the 
ordered sequences~(\ref{eq:xyordered}):
\be
x_{i} \mapsto R(x_{i}) \ , \quad
y_{i} \mapsto R(y_{i}) \ ,
\quad\quad\text{for all}\quad i=1,\ldots,n \ .
\ee
Should there be any ``tied ranks'' due to equality of some $x_{i}$ 
or $y_{i}$, one assigns the arithmetical mean of the corresponding 
rank numbers to all $x_{i}$ resp.\ $y_{i}$ involved in the 
``tie.'' Ultimately, by this procedure, the entire bivariate raw 
data undergoes a transformation
\be
\{(x_{i},y_{i})\}_{i=1,\ldots,n} \mapsto 
\{[R(x_{i}),R(y_{i})]\}_{i=1,\ldots,n} \ ,
\ee
yielding $n$ pairs of rank numbers to numerically represent the 
original bivariate ordinal data.

\medskip
\noindent
Given surrogate rank number data, the \textbf{means of rank
numbers} always amount to
\bea
\bar{R}(x) & := & \frac{1}{n}\sum_{i=1}^{n}R(x_{i})
\ = \ \frac{n+1}{2} \\
\bar{R}(y) & := & \frac{1}{n}\sum_{i=1}^{n}R(y_{i})
\ = \ \frac{n+1}{2} \ .
\eea
The \textbf{variances of rank numbers} are defined in accordance
with Eqs.~(\ref{eq:sampvardescr2}) and (\ref{eq:ssqurel2}), i.e.,
\bea
s_{R(x)}^{2} & := & 
\frac{1}{n-1}\,\left[\,\sum_{i=1}^{n}R^{2}(x_{i})
-n\bar{R}^{2}(x)\,\right]
\ = \ \frac{n}{n-1}\,\left[\,\sum_{i=1}^{k}R^{2}(a_{i})h_{i+}
-\bar{R}^{2}(x)\,\right] \\
s_{R(y)}^{2} & := & 
\frac{1}{n-1}\,\left[\,\sum_{i=1}^{n}R^{2}(y_{i})
-n\bar{R}^{2}(y)\,\right]
\ = \ \frac{n}{n-1}\,\left[\,\sum_{j=1}^{l}R^{2}(b_{j})h_{+ j}
-\bar{R}^{2}(y)\,\right] \ .
\eea
In addition, to characterise the joint distribution of rank 
numbers, a \textbf{sample covariance of rank numbers} is defined in 
line with Eqs.~(\ref{eq:covar2}) and (\ref{eq:covarrelfreq2}) by
\bea
s_{R(x)R(y)} & := & \frac{1}{n-1}\,\left[\,\sum_{i=1}^{n}
R(x_{i})R(y_{i})
-n\bar{R}(x)\bar{R}(y)\,\right] \nonumber \\
& = & \frac{n}{n-1}\,\left[\,\sum_{i=1}^{k}\sum_{j=1}^{l}
R(a_{i})R(b_{j})h_{ij} -\bar{R}(x)\bar{R}(y)\,\right] \ .
\eea

\vspace{5mm}
\noindent
On this fairly elaborate technical backdrop, the English 
psychologist and statistician
\href{http://en.wikipedia.org/wiki/Charles_Edward_Spearman}{Charles
Edward Spearman FRS (1863--1945)} defined a dimensionless
\textbf{sample rank correlation coefficient}~$r_{S}$ (ord), in
analogy to Eq.~(\ref{eq:correl}), by (cf. Spearman
(1904)~\ct{spe1904})
\be
\lb{eq:rankcorrelcoeff}
\fbox{$\displaystyle
r_{S}:=\frac{s_{R(x)R(y)}}{s_{R(x)}s_{R(y)}} \ .
$}
\ee
The range of this rank correlation coefficient is $-1 \leq r_{S} 
\leq +1$. Again, while the sign of $r_{S}$ encodes the 
\textbf{direction} of a rank correlation, in interpreting the
\textbf{strength} of a rank correlation via the magnitude $|r_{S}|$
one usually employs the qualitative

\medskip
\noindent
\underline{\textbf{Rule of thumb:}}\\
$0.0 = |r_{S}|$: no rank correlation\\
$0.0 < |r_{S}| < 0.2$: very weak rank correlation\\
$0.2 \leq |r_{S}| < 0.4$: weak rank correlation\\
$0.4 \leq |r_{S}| < 0.6$: moderately strong rank correlation\\
$0.6 \leq |r_{S}| \leq 0.8$: strong rank correlation\\
$0.8 \leq |r_{S}| < 1.0$: very strong rank correlation\\
$1.0=|r_{S}|$: perfect rank correlation.

\medskip
\noindent
\underline{\R:} \texttt{cor(\textit{variable1},
\textit{variable2}, method = "spearman")} \\
\underline{SPSS:} Analyze $\rightarrow$ Correlate
$\rightarrow$ Bivariate \ldots: Spearman

\medskip
\noindent
When \textit{no tied ranks} occur, Eq.~(\ref{eq:rankcorrelcoeff})
simplifies to (cf. Hartung \textit{et al} 
(2005)~\ct[p~554]{haretal2005})
\be
r_{S}=1-\frac{6\sum_{i=1}^{n}[R(x_{i})-R(y_{i})]^{2}}{n(n^{2}-1)}
\ .
\ee
%

\section[Measures of association for the nominal scale
level]{Measures of association for the nominal scale level}
\lb{sec:2Dnom}
Lastly, let us turn to consider the case of quantifying 
the degree of statistical association in bivariate raw data 
$\{(x_{i},y_{i})\}_{i=1,\ldots,n}$ for a nominally scaled 
two-dimensional variable~$(X,Y)$, with categories 
$\{(a_{i},b_{j})\}_{i=1,\ldots,k;j=1,\ldots,l}$. The starting 
point are the observed bivariate \textbf{absolute}
resp.\ \textbf{relative (cell) frequencies} $o_{ij}$ and $h_{ij}$
of the joint distribution for $(X,Y)$, with univariate
\textbf{marginal frequencies} $o_{i+}$ resp.\ $h_{i+}$ for $X$ and
$o_{+ j}$ resp.\ $h_{+ j}$ for $Y$. The
\textbf{$\boldsymbol{\chi}^{2}$--statistic} 
devised by the English mathematical statistician 
\href{http://www-history.mcs.st-and.ac.uk/Biographies/Pearson.html}{Karl
Pearson FRS (1857--1936)} rests on the notion of statistical 
independence of two one-dimensional variables $X$ and $Y$ in that 
it takes the corresponding formal condition provided by 
Eq.~(\ref{eq:statindep}) as a reference state. A simple algebraic 
manipulation of this condition obtains 
\be
h_{ij} = h_{i+}h_{+j}
\quad\Rightarrow\quad
\frac{o_{ij}}{n} = \frac{o_{i +}}{n}\,\frac{o_{+ j}}{n}
\quad\overbrace{\Rightarrow}^{\text{multiplication by}\ n}\quad
o_{ij} = \frac{o_{i +}o_{+ j}}{n} \ .
\ee
Pearson's descriptive $\chi^{2}$--statistic (cf. Pearson 
(1900)~\ct{pea1900}) is then defined by
\be
\fbox{$\displaystyle
\chi^{2} := \sum_{i=1}^{k}\sum_{j=1}^{l}\frac{\left(o_{ij}
-{\displaystyle\frac{o_{i+}o_{+ 
j}}{n}}\right)^{2}}{{\displaystyle\frac{o_{i+}
o_{+ j}}{n}}}
= n\sum_{i=1}^{k}\sum_{j=1}^{l}\frac{\left(h_{ij}
-h_{i+}h_{+ j}\right)^{2}}{h_{i+}h_{+ j}} \ ,
$}
\ee
whose range of values amounts to $0 \leq \chi^{2} \leq 
\max(\chi^{2})$, with $\max(\chi^{2}):=n\,[\min(k,l)-1]$.

\medskip
\noindent
\underline{\textbf{Remark:}} Provided $\displaystyle 
\frac{o_{i+}o_{+j}}{n} \geq 5$ for all $i=1, \ldots, k$ and $j=1, 
\ldots, l$, Pearson's $\chi^{2}$--statistic can be employed for 
the analysis of statistical associations amongst the components of 
a two-dimensional variable~$(X,Y)$ of almost all combinations of 
scale levels.

\medskip
\noindent
The problem with Pearson's $\chi^{2}$--statistic is that, due to 
its variable spectrum of values, it is not immediately clear how 
to use it efficiently in interpreting the \textbf{strength} of 
statistical associations. This shortcoming can, however, be 
overcome by resorting to the \textbf{measure of association}
proposed by the Swedish mathematician, actuary, and statistician 
\href{http://www-history.mcs.st-and.ac.uk/Biographies/Cramer_Harald.html}{Carl
Harald Cram\'{e}r (1893--1985)}, which basically is the 
result of a special kind of normalisation of Pearson's measure. 
Thus, \textbf{Cram\'{e}r's $\boldsymbol{V}$}, as it has come to be 
known, is defined by (cf. Cram\'{e}r (1946)~\ct{cra1946})
%
\be
\lb{eq:cramv}
\fbox{$\displaystyle
V:=\sqrt{\frac{\chi^{2}}{\max(\chi^{2})}} \ ,
$}
\ee
with range $0 \leq V \leq 1$. For the interpretation of the 
strength of statistical association in the joint distribution for a 
two-dimensional categorical variable~$(X,Y)$, one may thus employ 
the qualitative

\medskip
\noindent
\underline{\textbf{Rule of thumb:}}\\
$0.0 \leq V < 0.2$: weak association\\
$0.2 \leq V < 0.6$: moderately strong association\\
$0.6 \leq V \leq 1.0$: strong association.

\medskip
\noindent
\underline{\R:} \texttt{assocstats(\textit{contingency table})}
(package: \texttt{vcd}, by Meyer \textit{et al}
(2017)~\ct{meyetal2017}) \\
\underline{SPSS:} Analyze $\rightarrow$ Descriptive Statistics
$\rightarrow$ Crosstabs \ldots $\rightarrow$ Statistics \ldots:
Chi-square, Phi and Cramer's V


\chapter[Descriptive linear regression analysis]{\href{https://www.youtube.com/watch?v=uMpJZ2g0e3E}{Descriptive linear regression analysis}}
\lb{ch5}
For strongly correlating bivariate sample data 
$\{(x_{i},y_{i})\}_{i=1,\ldots,n}$ for a metrically scaled 
two-dimensional statistical variable~$(X,Y)$, i.e., when $0.71 
\leq |r| \leq 1.0$, it is meaningful to construct a mathematical 
model of the linear quantitative statistical association so 
diagnosed. The standard method to realise this by systematic means 
is due to the German mathematician and astronomer 
\href{http://www-groups.dcs.st-and.ac.uk/~history/Biographies/Gauss.html}{Carl Friedrich Gau\ss\ (1777--1855)} and is known by the 
name of \textbf{descriptive linear regression analysis};
cf. Gau\ss\ (1809)~\ct{gau1809}. We here restrict our attention to
the case of \textbf{simple linear regression}, which aims to
explain the variability in one \textbf{dependent variable} in terms
of the variability in a single \textbf{independent variable}.

\medskip
\noindent
To be determined is a \textbf{best-fit linear model} to given 
bivariate metrical data
$\{(x_{i},y_{i})\}_{i=1,\ldots,n}$. The linear model in question 
can be expressed in mathematical terms by
\be
\fbox{$\displaystyle
\lb{eq:descriptlinreg}
\hat{y}=a+bx \ ,
$}
\ee
with unknown regression coefficients
\textbf{$\boldsymbol{y}$-intercept} $a$ and \textbf{slope} $b$.
Gau\ss' \textbf{method of least squares} works as follows.

\section[Method of least squares]{Method of least squares}
\lb{sec:kq}
At first, one has to make a choice: assign $X$ the status of an 
\textbf{independent variable}, and $Y$ the status of a
\textbf{dependent variable} (or vice versa; usually this freedom of
choice does exist, unless one is testing a specific functional or
suspected causal relationship, $y=f(x)$). Then, considering the
measured values~$x_{i}$ for $X$ as fixed, to be minimised for the
$Y$-data is the \textbf{sum of the squared vertical deviations} of
the measured values~$y_{i}$ from the model values 
$\hat{y}_{i}=a+bx_{i}$. The latter are associated 
with an arbitrary straight line through the \textbf{cloud of data 
points} $\{(x_{i},y_{i})\}_{i=1,\ldots,n}$ in a \textbf{scatter
plot}. This sum, given by
\be
\lb{eq:sumofsquares}
S(a,b):=\sum_{i=1}^{n}(y_{i}-\hat{y}_{i})^{2}
= \sum_{i=1}^{n}(y_{i}-a-bx_{i})^{2} \ ,
\ee
constitutes a non-negative real-valued function of two 
variables, $a$ and $b$. Hence, determining its (local)
\textbf{minimum values} entails satisfying (i)~the necessary
condition of \textit{simultaneously vanishing} first partial
derivatives
\be
0 \stackrel{!}{=} \frac{\partial S(a,b)}{\partial a} \ , \qquad
0 \stackrel{!}{=} \frac{\partial S(a,b)}{\partial b} \ ,
\ee
--- this yields a well-determined $(2 \times 2)$ system of linear 
algebraic equations for the unknowns $a$ and $b$, cf. 
Ref.~\ct[Sec.~3.1]{hve2009} ---, and (ii)~the sufficient 
condition of a \textit{positive definite} \textbf{Hessian matrix} 
$H(a,b)$ of second partial derivatives,
\be
H(a,b) := \left(\begin{array}{ccc}
{\displaystyle\frac{\partial^{2} S(a,b)}{\partial a^{2}}} & &
{\displaystyle\frac{\partial^{2} S(a,b)}{\partial a\partial b}}
\\ \\
{\displaystyle\frac{\partial^{2} S(a,b)}{\partial b\partial a}} & &
{\displaystyle\frac{\partial^{2} S(a,b)}{\partial b^{2}}}
\end{array}\right) \ ,
\ee
at the candidate optimal values of $a$ and $b$. $H(a,b)$ is
referred to as positive definite when all of its
\textbf{eigenvalues} are positive; cf. Ref.~\ct[Sec.~3.6]{hve2009}.

\section[Empirical regression line]{Empirical regression line}
\lb{sec:empregg}
It is a fairly straightforward algebraic exercise (see, e.g., 
Toutenburg (2004) \ct[p~141ff]{tou2004}) to show that the 
values of the unknowns $a$ and $b$, which determine a unique 
global minimum of $S(a,b)$, amount to
\be
\lb{eq:abestimators}
\fbox{$\displaystyle
b=\frac{s_{Y}}{s_{X}}\,r \ ,\qquad
a=\bar{y}-b\bar{x} \ .
$}
\ee
These values are referred to as the \textbf{least squares
estimators} for $a$ and $b$. Note that they are exclusively
expressible in terms of familiar univariate and bivariate measures characterising the joint distribution for $(X,Y)$.

\medskip
\noindent
With the solutions $a$ and $b$ of Eq.~(\ref{eq:abestimators})
inserted in Eq.~(\ref{eq:descriptlinreg}), the resultant
\textbf{best-fit linear model} is given by
\be
\lb{eq:linregmodel}
\fbox{$\displaystyle
\hat{y}=\bar{y}+\frac{s_{Y}}{s_{X}}\,r\,(x-\bar{x}) \ .
$}
\ee
It may be employed for the purpose of generating intrapolating 
\textbf{predictions} of the kind $x \mapsto \hat{y}$, for
$x$-values confined to the empirical interval $[x_{(1)},x_{(n)}]$.
An example of a best-fit linear model obtained by the method of
least squares is shown in Fig.~\ref{fig:linReg}.

\medskip
\noindent
\underline{\R:}
\texttt{lm(\textit{variable:y}~\texttildelow~\textit{variable:x})}
\\
\underline{EXCEL, OpenOffice:} \texttt{SLOPE}, \texttt{INTERCEPT}
(dt.: \texttt{STEIGUNG}, \texttt{ACHSENABSCHNITT}) \\
\underline{SPSS:} Analyze $\rightarrow$ Regression
$\rightarrow$ Linear \ldots

\begin{figure}[!htb]
\begin{center}
\includegraphics[scale=0.8]{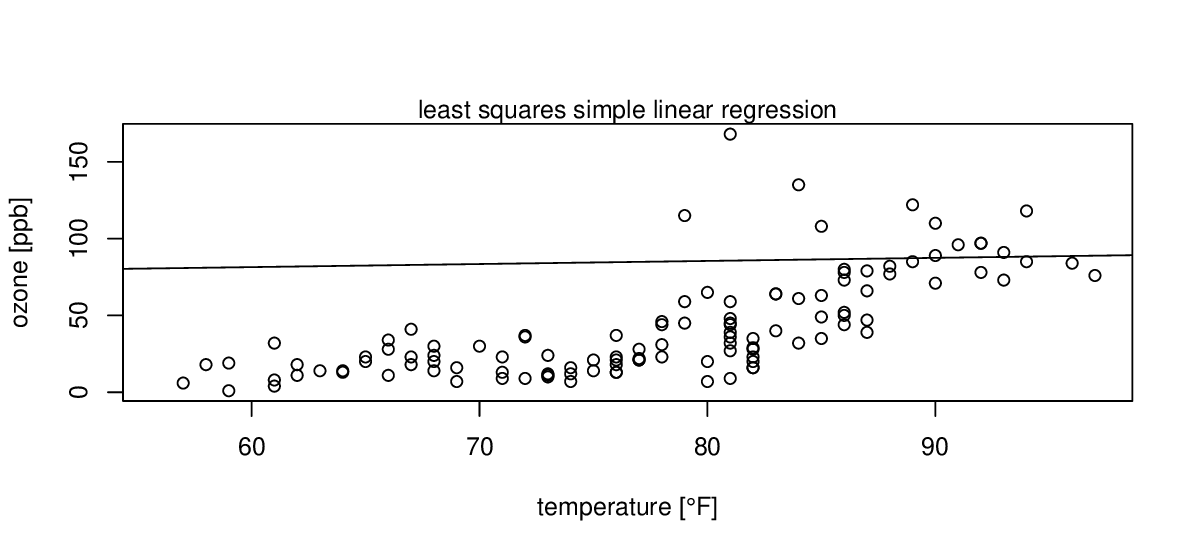}
\end{center}
\caption{Example of a best-fit linear model obtained by the method
of least squares for the case of the bivariate joint distribution
featured in Fig-~\ref{fig:scatterplot}. The least squares
estimators for the $y$-intercept and the slope take values
$a = 69.41~\text{ppb}$ and
$b = 0.20~(\text{ppb}/\text{°F})$, respectively. \newline
\underline{\R:} \newline
\texttt{data("airquality")} \newline
\texttt{?airquality} \newline
\texttt{regMod <- lm( airquality\$Temp~\texttildelow~airquality\$Ozone )} \newline
\texttt{summary(regMod)} \newline
\texttt{plot( airquality\$Temp , airquality\$Ozone )} \newline
\texttt{abline(regMod)}}
\lb{fig:linReg}
\end{figure}

\medskip
\noindent
Note that Eq.~(\ref{eq:linregmodel}) may be re-expressed in terms 
of the corresponding $Z$ scores of $X$ and $\hat{Y}$, according to
Eq.~(\ref{eq:standardisationdiscriptive}). This yields
\be
\left(\frac{\hat{y}-\bar{y}}{s_{Y}}\right)
= r\left(\frac{x-\bar{x}}{s_{X}}\right)
\qquad\Leftrightarrow\qquad
\hat{z}_{Y} = rz_{X} \ .
\ee
%

\section[Coefficient of determination]{Coefficient of
determination}
\lb{sec:bestreg}
The quality of any particular simple linear regression model, 
i.e., its \textbf{goodness-of-the-fit}, is assessed by means of the 
\textbf{coefficient of determination} $B$ (metr). This measure is 
derived by starting from the algebraic identity
\be
\sum_{i=1}^{n}(y_{i}-\bar{y})^{2}
=\sum_{i=1}^{n}(\hat{y}_{i}-\bar{y})^{2}
+\sum_{i=1}^{n}(y_{i}-\hat{y}_{i})^{2} \ ,
\ee
which, upon conveniently re-arranging, leads to defining a quantity
\be
\lb{eq:linregcoeffdetdescr}
\fbox{$\displaystyle
B:=\frac{{\displaystyle\sum_{i=1}^{n}(y_{i}-\bar{y})^{2}
-\sum_{i=1}^{n}(y_{i}-\hat{y}_{i})^{2}}}{{\displaystyle
\sum_{i=1}^{n}(y_{i}-\bar{y})^{2}}}
= \frac{{\displaystyle\sum_{i=1}^{n}(\hat{y}_{i}-\bar{y})^{2}}}{
{\displaystyle\sum_{i=1}^{n}(y_{i}-\bar{y})^{2}}} \ ,
$}
\ee
with range $0 \leq B \leq 1$. A perfect fit is signified by $B=1$, 
while no fit amounts to $B=0$. The coefficient of determination 
provides a descriptive measure for the proportion of variability 
of $Y$ in a bivariate data set $\{(x_{i},y_{i})\}_{i=1,\ldots,n}$ 
that can be accounted for as due to the association with $X$ via 
the simple linear regression model. Note that in simple linear 
regression it holds that
\be
\lb{eq:linregrsq}
B=r^{2} \ ;
\ee
see, e.g., Toutenburg (2004) \ct[p~150f]{tou2004}).

\medskip
\noindent
\underline{\R:}
\texttt{summary( lm(\textit{variable:y}~\texttildelow~\textit{variable:x}) )} \\
\underline{EXCEL, OpenOffice:} \texttt{RSQ} (dt.:
\texttt{BESTIMMTHEITSMASS}) \\
\underline{SPSS:} Analyze $\rightarrow$ Regression
$\rightarrow$ Linear \ldots $\rightarrow$ Statistics \ldots: Model 
fit

\vspace{5mm}
\noindent
This concludes Part I of these lecture notes, the introductory 
discussion on uni- and bivariate
\href{https://www.youtube.com/watch?time_continue=2&v=pTtYMdqZ1M4}{\textbf{descriptive statistical methods of data analysis}}. We wish to
encourage the interested reader to adhere to accepted scientific
standards when actively getting involved with data analysis
her/him-self. This entails, amongst other aspects, foremost the
truthful documentation of all data taken into account in a specific
analysis conducted. Features facilitating understanding such as
visualisations of empirical distributions by means of, where
appropriate, histograms, bar charts, box plots or scatter plots, or
providing the values of five number summaries, sample means, sample
standard deviations, standardised skewness and excess kurtosis
measures, or sample correlation coefficients should be commonplace
in any kind of research report. It must be a prime objective of the
researcher to empower potential readers to retrace the inferences
made by her/him.

\medskip
\noindent
To set the stage for the application of inferential statistical
methods in Part III, we now turn to review the elementary concepts
underlying \textbf{Probability Theory}, predominantly as
interpreted in the \textbf{frequentist approach} to this topic.


\chapter[Elements of probability theory]{Elements of probability
theory}
\lb{ch6}
All examples of \textbf{inferential statistical methods of data 
analysis} to be presented in Chs.~\ref{ch12} and \ref{ch13} have 
been developed in the context of the so-called \textbf{frequentist 
approach} to \textbf{Probability Theory}.\footnote{The origin of
the term ``probability'' is traced back to the Latin word 
\textit{probabilis}, which the Roman philosopher 
\href{https://en.wikipedia.org/wiki/Cicero}{Cicero (106 BC--43 
BC)} used to capture a notion of plausibility or likelihood; see 
Mlodinow (2008)~\ct[p~32]{mlo2008}.} The issue in
\textbf{Inferential Statistics} is to estimate the plausibility or
likelihood of hypotheses given the observational \textbf{evidence}
for them. The \textbf{frequentist approach} was pioneered by the
Italian mathematician, physician, astrologer, philosopher and
gambler 
\href{http://www-history.mcs.st-and.ac.uk/Biographies/Cardan.html}{Girolamo Cardano (1501--1576)}, the French lawyer and amateur mathematician  
\href{http://www-history.mcs.st-and.ac.uk/Biographies/Fermat.html}{Pierre
de Fermat (1601--1665)}, the French mathematician, 
physicist, inventor, writer and Catholic philosopher 
\href{http://www-history.mcs.st-and.ac.uk/Biographies/Pascal.html}{Blaise Pascal (1623--1662)}, the Swiss mathematician
\href{http://www-history.mcs.st-and.ac.uk/Biographies/Bernoulli_Jacob.html}{Jakob Bernoulli (1654--1705)}, and the French 
mathematician and astronomer
\href{http://www-history.mcs.st-and.ac.uk/Biographies/Laplace.html}{Marquis Pierre Simon de Laplace (1749--1827)}. It is deeply 
rooted in the two fundamental assumptions that any particular 
\textbf{random experiment} can be repeated arbitrarily often 
(i)~under the ``same conditions,'' and (ii)~completely 
``independent of one another,'' so that a theoretical basis is 
given for defining allegedly ``\textit{objective probabilities}''
for random events and hypotheses via the \textbf{relative
frequencies} of very long sequences of repetition of the same
random experiment.\footnote{A special role in the context of the
frequentist approach to Probability Theory is assumed by Jakob
Bernoulli's law of large numbers, as well as the concept of
independently and identically distributed 
(in short: ``i.i.d.'') random variables; we will discuss these 
issues in Sec.~\ref{sec:zentrgrenz} below.} This is a highly 
idealised viewpoint, however, which shares only a limited degree 
of similarity with the actual conditions pertaining to an 
observer's resp.~experimentor's reality. Renowned textbooks 
adopting the \textbf{frequentist viewpoint} of \textbf{Probability
Theory} and  \textbf{Inferential Statistics} are, e.g., Cram\'{e}r 
(1946)~\ct{cra1946} and Feller (1968)~\ct{fel1968}.

\medskip
\noindent
Not everyone in \textbf{Statistics} is entirely happy, though, with 
the philosophy underlying the \textbf{frequentist approach} to
introducing the concept of \textbf{probability}, as a number of its
central ideas rely on unobserved data (information). A
complementary viewpoint is taken by the framework which originated
from the work of the English mathematician and Presbyterian
minister \href{http://www-history.mcs.st-and.ac.uk/Biographies/Bayes.html}{Thomas Bayes (1702--1761)}, and later of Laplace, and so is
commonly referred to as the \textbf{Bayes--Laplace approach}; cf. 
Bayes (1763)~\ct{bay1763} and Laplace (1812)~\ct{lap1812}. A 
striking conceptual difference to the \textbf{frequentist approach} 
consists in its use of prior, allegedly ``\textit{subjective
probabilities}'' for random events and hypotheses, quantifying a
persons's individual reasonable \textbf{degree-of-belief} in their 
likelihood, which are subsequently updated by analysing relevant
empirical data.\footnote{Anscombe and Aumann (1963)~\ct{ansaum1963}
in their seminal paper refer to ``objective probabilities'' as
associated with ``roulette lotteries,'' and to ``subjective
probabilities'' as associated with ``horse lotteries.'' Savage
(1954)~\ct{sav1954} employs the alternative terminology of
distinguishing between ``objectivistic probabilities'' and
``personalistic probabilities.''} Renowned textbooks adopting the
\textbf{Bayes--Laplace viewpoint} of \textbf{Probability Theory}
and \textbf{Inferential Statistics} are, e.g., Jeffreys
(1939)~\ct{jef1939} and Jaynes (2003)~\ct{jay2003}, while general
information regarding the  \textbf{Bayes--Laplace approach} is
available from the website \href{http://bayes.wustl.edu/}{\texttt{bayes.wustl.edu}}. More recent textbooks, which assist in the
implementation of advanced computational routines, have been issued
by Gelman \textit{et al} (2014)~\ct{geletal2014} and by McElreath
(2016)~\ct{mce2016}. A discussion of the pros and cons of either of
these two competing approaches to \textbf{Probability Theory} can
be found, e.g., in Sivia and Skilling (2006)
\ct[p~8ff]{sivski2006}, or in Gilboa (2009) \ct[Sec.~5.3]{gil2009}.

\medskip
\noindent
A common denominator of both frameworks, \textbf{frequentist} and 
\textbf{Bayes--Laplace}, is the attempt to quantify a notion of 
\textbf{uncertainty} that can be related to in formal treatments of 
\textbf{decision-making}. In the following we turn to discuss the 
general principles on which \textbf{Probability Theory} is built.

\section[Random events]{Random events}
\lb{sec:zufall}
We begin by introducing some basic formal constructions and 
corresponding terminology used in the \textbf{frequentist 
approach} to \textbf{Probability Theory}:
\begin{itemize}

\item \textbf{Random experiments}: Random experiments are
experiments which can be repeated arbitrarily often under identical 
conditions, with \textbf{events} --- also called \textbf{outcomes}
--- that cannot be predicted with certainty. Well-known simple 
examples are found amongst games of chance such as tossing a coin, 
rolling dice, or playing roulette.

\item \textbf{Sample space} $\boldsymbol{\Omega} =\{\omega_{1},
\omega_{2},\ldots\}$: The sample space associated with a random 
experiment is constituted by the set of all possible
\textbf{elementary events} (or elementary outcomes) $\omega_{i}$
($i=1, 2, \ldots$), which are signified by their property of
\textit{mutual exclusivity}. The sample space~$\boldsymbol{\Omega}$
of a random experiment may contain either
\begin{itemize}
\item[(i)] a finite number $n$ of elementary events; then
$|\boldsymbol{\Omega}|=n$, or
\item[(ii)] countably many elementary events in the sense of a 
one-to-one correspondence with the set of natural numbers 
$\mathbb{N}$, or
\item[(iii)] uncountably may elements in the sense of a 
one-to-one correspondence with the set of real numbers 
$\mathbb{R}$, or an open or closed subset thereof.\footnote{For
reasons of definiteness, we will assume in this case
that the sample space $\boldsymbol{\Omega}$ 
associated with a random experiment is compact.}
\end{itemize}
The essential concept of the sample space associated with a random 
experiment was introduced to \textbf{Probability Theory} by the 
Italian mathematician 
\href{http://www-history.mcs.st-and.ac.uk/Biographies/Cardan.html}{Girolamo Cardano (1501--1576)}; see Cardano (1564)~\ct{car1564}, 
Mlodinow (2008)~\ct[p~42]{mlo2008}, and Bernstein 
(1998)~\ct[p~47ff]{ber1998}.

\item \textbf{Random events} $A, B, \ldots \subseteq 
\boldsymbol{\Omega}$: Random events are formally defined as all 
kinds of subsets of $\boldsymbol{\Omega}$ that can be formed from 
the elementary events $\omega_{i} \in \boldsymbol{\Omega}$.

\item \textbf{Certain event} $\boldsymbol{\Omega}$: The certain
event is synonymous with the sample space itself. When a particular 
random experiment is conducted, ``something will happen for sure.''

\item \textbf{Impossible event} $\emptyset=\{\}
= \bar{\boldsymbol{\Omega}}$: The impossible event is the natural 
complement to the certain event. When a particular random 
experiment is conducted, ``it is not possible that nothing will 
happen at all.''

\item \textbf{Event space} ${\cal P}(\boldsymbol{\Omega}) := \{A|A 
\subseteq \boldsymbol{\Omega}\}$: The event space, also referred 
to as the \textbf{power set} of~$\boldsymbol{\Omega}$, is the set
of all possible subsets (random events!) that can be formed from 
elementary events $\omega_{i} \in \boldsymbol{\Omega}$. Its size 
(or cardinality) is given by $|{\cal P}(\boldsymbol{\Omega})| = 
2^{|\boldsymbol{\Omega}|}$. The event space 
${\cal P}(\boldsymbol{\Omega})$ constitutes a so-called
\textbf{$\boldsymbol{\sigma}$--algebra} associated with the sample
space $\boldsymbol{\Omega}$; cf. Rinne (2008)~\ct[p~177]{rin2008}.
When $|\boldsymbol{\Omega}|=n$, i.e., when $\boldsymbol{\Omega}$ is 
finite, then $|{\cal P}(\boldsymbol{\Omega})|=2^{n}$.

\end{itemize}
In the formulation of probability theoretical laws and 
computational rules, the following set operations and identities 
prove useful.

\subsection*{Set operations}
%
\begin{enumerate}
\item $\bar{A} = \boldsymbol{\Omega}\backslash A$
--- \textbf{complementation} of a set (or event) $A$ (``not $A$'')

\item $A\backslash B=A\cap\bar{B}$ --- formation of the
\textbf{difference} of sets (or events) $A$ and $B$ (``$A$, but not
$B$'')

\item $A \cup B$ --- formation of the union of sets (or events) 
$A$ and $B$, otherwise referred to as the \textbf{disjunction} of
$A$ and $B$ (``$A$ or $B$'')

\item $A \cap B$ --- formation of the intersection of sets (or 
events) $A$ and $B$, otherwise referred to as the
\textbf{conjunction} of $A$ and $B$ (``$A$ and $B$'')

\item $A \subseteq B$ --- \textbf{inclusion} of a set (or event)
$A$ in a set (or event) $B$ (``$A$ is a subset of or equal
to~$B$'')

\end{enumerate}
%
\subsection*{Computational rules and identities}
%
\begin{enumerate}
\item $A \cup B=B\cup A$ and $A \cap B=B\cap A$
\hfill (\textbf{commutativity})

\item $(A \cup B)\cup C=A\cup (B\cup C)$ and \\
$(A \cap B)\cap C=A\cap (B\cap C)$
\hfill (\textbf{associativity})

\item $(A \cup B)\cap C=(A\cap C)\cup(B\cap C)$ and \\
$(A \cap B)\cup C=(A\cup C)\cap(B\cup C)$
\hfill (\textbf{distributivity})

\item $\overline{A\cup B}=\bar{A}\cap\bar{B}$ and
$\overline{A\cap B}=\bar{A}\cup\bar{B}$
\hfill (\textbf{de Morgan's laws})

\end{enumerate}

\medskip
\noindent
Before addressing the central axioms of \textbf{Probability
Theory}, we first provide the following important definition.

\medskip
\noindent
\underline{\textbf{Def.:}} Suppose given a \textit{compact} sample
space $\boldsymbol{\Omega}$ of some random experiment. Then one 
understands by a finite \textbf{complete partition} of 
$\boldsymbol{\Omega}$ a set of $n \in \mathbb{N}$ random events 
$\{A_{1}, \ldots, A_{n}\}$ such that
\begin{itemize}
\item[(i)] $A_{i} \cap A_{j} = \emptyset$ for $i \neq j$, i.e., 
they are \textbf{pairwise disjoint} (mutually exclusive), and 
\item[(ii)] $\displaystyle\bigcup_{i=1}^{n}A_{i} = 
\boldsymbol{\Omega}$, i.e., their union is identical to the full 
\textbf{sample space}.
\end{itemize}
%

\section[Kolmogorov's axioms of probability theory]{Kolmogorov's
axioms of probability theory}
\lb{sec:kolaxiom}
It took a fairly long time until, in 1933, a unanimously accepted 
basis of \textbf{Probability Theory} was established. In part the 
delay was due to problems with providing a unique definition 
of \textbf{probability}, and how it could be measured and
interpreted in practice. The situation was resolved only when the
Russian mathematician
\href{http://www-history.mcs.st-and.ac.uk/Biographies/Kolmogorov.html}{Andrey Nikolaevich Kolmogorov (1903--1987)} proposed to discard 
the intention of providing a unique definition of
\textbf{probability} altogether, and to restrict the issue instead
to merely prescribing in an axiomatic fashion a minimum set of 
properties any \textbf{probability measure} needs to possess
in order to be coherent and consistent. We now recapitulate the 
axioms that Kolmogorov put forward; cf. Kolmogoroff 
(1933)~\ct{kol1933}.

\medskip
\noindent
For a given \textbf{random experiment}, let $\boldsymbol{\Omega}$
be its \textbf{sample space} and ${\cal P}(\boldsymbol{\Omega})$
the associated \textbf{event space}. Then a mapping
\be
P: {\cal P}(\boldsymbol{\Omega}) \rightarrow \mathbb{R}_{\geq 0}
\ee
defines a \textbf{probability measure} with the following
properties:
\begin{enumerate}
\item for all \textbf{random events} $A \in {\cal 
P}(\boldsymbol{\Omega})$, 
\hfill (\textbf{non-negativity})
\be
\lb{eq:axiom1}
P(A) \geq 0 \ ,
\ee

\item for the \textbf{certain event} $\boldsymbol{\Omega} \in 
{\cal P}(\boldsymbol{\Omega})$, \hfill (\textbf{normalisability})
\be
\lb{eq:axiom2}
P(\boldsymbol{\Omega}) = 1 \ ,
\ee

\item for all \textbf{pairwise disjoint random events} $A_{1},
A_{2}, \ldots \in {\cal P}(\boldsymbol{\Omega})$, i.e., $A_{i}
\cap A_{j} = \emptyset$ for all $i \neq j$,\\
\mbox{} \hfill ($\boldsymbol{\sigma}$\textbf{--additivity})
\be
\lb{eq:axiom3}
P\left(\bigcup_{i=1}^{\infty}A_{i}\right)
= P(A_{1} \cup A_{2} \cup \ldots)
= P(A_{1}) + P(A_{2}) + \ldots
= \sum_{i=1}^{\infty}P(A_{i}) \ .
\ee

\end{enumerate}
The first two axioms imply the property
\be
\lb{eq:pleqone}
0 \leq P(A) \leq 1 \ ,
\quad\text{for all}\quad
A \in {\cal P}(\boldsymbol{\Omega}) \ ;
\ee
the expression $P(A)$ itself is referred to as the
\textbf{probability} of a random event $A \in {\cal 
P}(\boldsymbol{\Omega})$. A less strict version of the third axiom
is given by requiring only \textbf{finite additivity} of a
probability measure. This means it shall possess the property
\be
P(A_{1} \cup A_{2}) = P(A_{1}) + P(A_{2}) \ ,
\quad\text{for any two }\quad
A_{1}, A_{2} \in {\cal P}(\boldsymbol{\Omega})
\quad\text{with}\quad A_{1} \cap A_{2} = \emptyset \ .
\ee

\medskip
\noindent
The triplet
\[
\left(\boldsymbol{\Omega}, {\cal P}, P\right)
\]
constitutes a special case of a so-called \textbf{probability
space}.

\medskip
\noindent
The following consequences for random events $A, B, A_{1}, A_{2}, 
\ldots \in {\cal P}(\boldsymbol{\Omega})$ can be derived from 
Kolmogorov's three axioms of probability theory; cf., e.g., 
Toutenburg (2005) \ct[p~19ff]{tou2005}. Their implications can be 
convienently visualised by means of \textbf{Venn diagrams}, 
named in honour of the English logician and philosopher 
\href{http://www-history.mcs.st-and.ac.uk/Biographies/Venn.html}{John
Venn FRS FSA (1834--1923)}; see Venn (1880)~\ct{ven1880}, and also,
e.g., Wewel (2014)~\ct[Ch.~5]{wew2014}.

\medskip
\noindent
\textbf{Consequences}
\begin{enumerate}

\item $P(\bar{A}) = 1-P(A)$

\item $P(\emptyset) = P(\bar{\boldsymbol{\Omega}}) = 0$

\item If $A \subseteq B$, then $P(A) \leq P(B)$.

\item $P(A_{1} \cup A_{2}) = P(A_{1})+P(A_{2})-P(A_{1} \cap 
A_{2})$.

\item $\displaystyle P(B) = \sum_{i=1}^{n}P(B \cap A_{i})$,
provided the $n \in \mathbb{N}$ random events $A_{i}$ constitute a 
finite \textbf{complete partition} of the sample space 
$\boldsymbol{\Omega}$.

\item $P(A\backslash B) = P(A) - P(A \cap B)$.

\end{enumerate}

\vspace{5mm}
\noindent
Employing its \textbf{complementation}~$\bar{A}$ and the first of
the consequences stated above, one defines by the ratio
\be
\lb{eq:odds}
O(A) := \frac{P(A)}{P(\bar{A})} = \frac{P(A)}{1-P(A)}
\ee
the so-called \textbf{odds} of a random event
$A \in {\cal P}(\boldsymbol{\Omega})$.

\medskip
\noindent
The renowned Israeli--US-American experimental psychologists 
Daniel Kahneman and Amos Tversky (the latter of which deceased in 
1996, aged fifty-nine) refer to the third of the consequences
stated above as the \textbf{extension rule}; see Tversky and
Kahneman (1983)~\ct[p~294]{tvekah1983}. It provides a cornerstone
to their remarkable investigations on the ``intuitive statistics''
applied by Humans in everyday \textbf{decision-making}, which focus
in particular on the \textbf{conjunction rule},
\be
\lb{eq:conjrule}
P(A \cap B) \leq P(A)
\quad\text{and}\quad
P(A \cap B) \leq P(B) \ ,
\ee
and the associated \textbf{disjunction rule},
\be
\lb{eq:disjrule}
P(A \cup B) \geq P(A)
\quad\text{and}\quad
P(A \cup B) \geq P(B) \ .
\ee
Both may be perceived as subcases of the fourth law above, which 
is occasionally referred to as the \textbf{convexity} property of a 
probability measure; cf. Gilboa (2009) \ct[p~160]{gil2009}. By 
means of their famous ``Linda the bank teller'' example in 
particular, Tversky and Kahneman (1983)~\ct[p~297ff]{tvekah1983} 
were able to demonstrate the startling empirical fact that the 
conjunction rule is frequently violated in everyday (intuitive)
decision-making; in their view, in consequence of decision-makers 
often resorting to a so-called \textit{representativeness
heuristic} as an aid in corresponding situations; see also Kahneman
(2011)~\ct[Sec.~15]{kah2011}. In recognition of their as much
intriguing as groundbreaking work, which sparked the discipline of
\textbf{Behavioural Economics}, Daniel Kahneman was awarded the  
\href{http://www.nobelprize.org/nobel_prizes/economics/laureates/2002/}{Sveriges
Riksbank Prize in Economic Sciences in Memory of Alfred Nobel}
in 2002.

\section[Laplacian random experiments]{Laplacian random
experiments}
\lb{sec:laplace}
Games of chance with a \textit{finite} number $n$ of possible 
mutually exclusive elementary outcomes, such as tossing a single 
coin once, rolling a single dye once, or selecting a single 
playing card from a deck of 32, belong to the simplest kinds of 
random experiments. In this context, there exists a clear-cut 
frequentist notion of a unique ``\textit{objective probability}'' 
associated with any kind of possible random event (outcome) that 
may occur.
Such probabilities can be computed according to a straightforward 
prescription due to the French mathematician and astronomer
\href{http://www-history.mcs.st-and.ac.uk/Biographies/Laplace.html}{Marquis Pierre Simon de Laplace (1749--1827)}. The prescription
rests on the assumption that the device generating the random 
events is a ``fair'' (i.e., unbiased) one.

\medskip
\noindent
Consider a random experiment, the $n$ \textbf{elementary events} 
$\omega_{i}$ ($i=1,\ldots,n$) of which that constitute the 
associated sample space $\boldsymbol{\Omega}$ are supposed to 
be ``equally likely,'' meaning they are assigned \textbf{equal 
probability}:
\be
P(\omega_{i}) = \frac{1}{|\boldsymbol{\Omega}|}
= \frac{1}{n} \ ,
\quad\text{for all}\quad \omega_{i}\in\boldsymbol{\Omega}
\ (i=1,\ldots,n) \ .
\ee
All random experiments of this nature are referred to as 
\textbf{Laplacian random experiments}.

\medskip
\noindent
\underline{\textbf{Def.:}} For a Laplacian random experiment,
the probability of an arbitrary random event $A \in 
{\cal P}(\boldsymbol{\Omega})$ can be computed according to the 
rule
\be
\lb{eq:classprob}
\fbox{$\displaystyle
P(A) := \frac{|A|}{|\boldsymbol{\Omega}|}
= \frac{\text{Number of cases favourable to event}\ A}{\text{Number
of all possible cases}} \ .
$}
\ee
Any probability measure $P$ which can be constructed in this 
fashion is called a \textbf{Laplacian probability measure}.

\medskip
\noindent
The systematic counting of the numbers of possible outcomes of 
random experiments in general is the central theme of
\textbf{combinatorics}. We now briefly address its main
considerations.

\section[Combinatorics]{Combinatorics}
\lb{sec:comb}
At the heart of combinatorical considerations is the well-known 
\textbf{urn model}. This supposes given an urn containing $N \in 
\mathbb{N}$ balls that are either
\begin{itemize}
\item[(a)] all different, and thus can 
be uniquely distinguished from one another, or
\item[(b)] there are $s \in 
\mathbb{N}$ ($s \leq N$) subsets of indistinguishable
like balls, of sizes $n_{1},\ldots,n_{s}$ resp., such that 
$n_{1}+\ldots+n_{s}=N$.
\end{itemize}
The first systematic developments in \textbf{Combinatorics} date
back to the Italian astronomer, physicist, engineer, philosopher,
and mathematician 
\href{http://www-history.mcs.st-and.ac.uk/Biographies/Galileo.html}{Galileo Galilei (1564--1642)} and the French mathematician
\href{http://www-history.mcs.st-and.ac.uk/Biographies/Pascal.html}{Blaise Pascal (1623--1662)}; cf. Mlodinow 
(2008)~\ct[p~62ff]{mlo2008}.

\subsection{Permutations}
\textbf{Permutations} relate to the number of distinguishable 
possibilities of arranging $N$ balls in an ordered sequences. 
Altogether, for cases (a) resp.\ (b) one finds that there are a 
total number of

\begin{center}
    \begin{tabular}[h!]{c|c}
    (a)~all balls different & (b)~$s$ subsets of like balls \\
    \hline
     & \\
    $N!$ & $\displaystyle \frac{N!}{n_{1}!n_{2}!\cdots n_{s}!}$
    \\
     & \\
    \end{tabular}
\end{center}
different possibilities. Remember that the \textbf{factorial} of a 
natural number $N \in \mathbb{N}$ is defined by
\be
N! := N\times(N-1)\times(N-2)\times\cdots\times 3 \times 2 
\times 1 \ .
\ee
\underline{\R:} \texttt{factorial($N$)}

\subsection{Combinations and variations}
\lb{subsec:combvar}
\textbf{Combinations} and \textbf{variations} ask for the total
number of distinguishable possibilities of selecting from a
collection of $N$ balls a sample of size $n \leq N$, while
differentiating between cases when
\begin{itemize}
\item[(a)] the order in which balls were selected is either 
neglected or instead accounted for, and

\item[(b)] a ball that was selected once either cannot be selected 
again or indeed can be selected again as often as a ball is being 
drawn.

\end{itemize}
These considerations result in the following cases of different 
possibilities:
\begin{center}
    \begin{tabular}[h!]{c|c|c}
     & no repetition & with repetition \\
    \hline
     & & \\
    combinations (order neglected) & $\left(\begin{array}{c}
    N \\ n \end{array}\right)$ & $\left(\begin{array}{c}
    N+n-1 \\ n \end{array}\right)$ \\
     & & \\
    \hline
     & & \\
    variations (order accounted for) & $\left(\begin{array}{c}
    N \\ n \end{array}\right)n!$ & $N^{n}$ \\
     & & \\
    \end{tabular}
\end{center}
Note that, herein, the \textbf{binomial coefficient} for two
natural numbers $n, N \in \mathbb{N}$, $n \leq N$, introduced by
\href{http://www-history.mcs.st-and.ac.uk/Biographies/Pascal.html}{Blaise Pascal (1623--1662)}, is defined by
\be
\lb{eq:binomcoeff}
\left(\begin{array}{c}
N \\ n \end{array}\right) := \frac{N!}{n!(N-n)!} \ .
\ee
For fixed value of $N$ and running value of $n \leq N$, it 
generates the positive integer entries of Pascal's well-known 
numerical triangle; see, e.g., Mlodinow 
(2008)~\ct[p~72ff]{mlo2008}. The binomial coefficient satisfies 
the identity
\be
\left(\begin{array}{c}
N \\ n \end{array}\right) \equiv \left(\begin{array}{c}
N \\ N-n \end{array}\right) \ .
\ee
\underline{\R:} \texttt{choose($N$, $n$)}

\medskip
\noindent
To conclude this chapter, we turn to discuss the essential concept 
of \textbf{conditional probabilities} of random events.

\section[Conditional probabilities]{Conditional probabilities}
\lb{sec:bedwahr}
Consider some random experiment with sample space 
$\boldsymbol{\Omega}$, event space ${\cal 
P}(\boldsymbol{\Omega})$, and a well-defined, unique probability 
measure $P$ over ${\cal P}(\boldsymbol{\Omega})$.

\medskip
\noindent
\underline{\textbf{Def.:}} For random events $A, B \in
{\cal P}(\boldsymbol{\Omega})$, with $P(B) > 0$,
\be
\lb{condprob}
\fbox{$\displaystyle
P(A|B) := \frac{P(A \cap B)}{P(B)}
$}
\ee
defines the \textbf{conditional probability} of $A$
to occur, given that it is known that $B$ occurred before.
Analogously, one defines 
a conditional probability  $P(B|A)$ with the roles of random 
events $A$ and $B$ switched, provided $P(A) > 0$. Note that since, 
by Eq.~(\ref{eq:pleqone}), $0 \leq P(A|B), P(B|A) \leq 1$, the 
implication of definition~(\ref{condprob}) is that the conjunction 
rule~(\ref{eq:conjrule}) must \textit{always} be satisfied.

\medskip
\noindent
\underline{\textbf{Def.:}} Random events $A, B \in 
{\cal P}(\boldsymbol{\Omega})$ are called \textbf{mutually 
stochastically independent}, if, simultaneously, the conditions
\be
\lb{eq:stochindep1}
\fbox{$\displaystyle
P(A|B) \stackrel{!}{=} P(A) \ ,\quad P(B|A) \stackrel{!}{=} P(B)
\quad\stackrel{\text{Eq.~\ref{condprob}}}{\Leftrightarrow}\quad
P(A \cap B) = P(A)P(B)
$}
\ee
are satisfied, i.e., when for both random events $A$ and $B$ the 
{\bf\textit{a posteriori} probabilities} $P(A|B)$ and $P(B|A)$ 
coincide with the respective {\bf\textit{a priori} probabilities} 
$P(A)$ and $P(B)$.

\medskip
\noindent
For applications, the following two prominent laws of
\textbf{Probability Theory} prove essential.

\subsection[Law of total probability]{Law of total probability}
For a random experiment with probability space 
$\left(\boldsymbol{\Omega}, {\cal P}, P\right)$, it holds by the 
\textbf{law of total probability} that for any random event $B \in 
{\cal P}(\boldsymbol{\Omega})$
\be
\lb{eq:totalprob}
\fbox{$\displaystyle
P(B)=\sum_{i=1}^{m}P(B|A_{i})P(A_{i}) \ ,
$}
\ee
provided the random events $A_{1}, \ldots, A_{m} \in 
{\cal P}(\boldsymbol{\Omega})$ constitute a finite \textbf{complete 
partition} of $\boldsymbol{\Omega}$ into $m \in \mathbb{N}$
\textbf{pairwise disjoint events}.

\medskip
\noindent
The content of this law may be conveniently visualised by means of 
a Venn diagram.

\subsection[Bayes' theorem]{Bayes' theorem}
\lb{subsec:bayes}
This important result is due to the English mathematician and
Presbyterian minister 
\href{http://www-history.mcs.st-and.ac.uk/Biographies/Bayes.html}{Thomas Bayes (1702--1761)}; see the posthumous publication Bayes 
(1763)~\ct{bay1763}. For a random experiment with probability space
$\left(\boldsymbol{\Omega}, {\cal P}, P\right)$, it states that,
given

\begin{itemize}

\item[(i)] random events $A_{1}, \ldots, A_{m} \in 
{\cal P}(\boldsymbol{\Omega})$ which constitute a finite
\textbf{complete partition} of $\boldsymbol{\Omega}$ into $m \in
\mathbb{N}$ \textbf{pairwise disjoint events},

\item[(ii)] $P(A_{i}) > 0$
for all $i = 1, \ldots, m$, with $\displaystyle
\sum_{i=1}^{m}P(A_{i}) = 1$ by Eq.~(\ref{eq:axiom2}), and

\item[(iii)] a random event $B \in {\cal P}(\boldsymbol{\Omega})$
with $\displaystyle 
P(B)\stackrel{\text{Eq.~\ref{eq:totalprob}}}{=}\sum_{i=1}^{m}
P(B|A_{i})P(A_{i}) > 0$ that is known to have occurred,

\end{itemize}
the identity
\be
\lb{eq:bayes}
\fbox{$\displaystyle
P(A_{i}|B)=\frac{P(B|A_{i})P(A_{i})}{
{\displaystyle\sum_{j=1}^{m}P(B|A_{j})P(A_{j})}}
$}
\ee
applies. This form of the theorem was given by Laplace
(1774)~\ct{lap1774}. By Eq.~(\ref{eq:axiom2}), it necessarily
follows that $\displaystyle \sum_{i=1}^{m}P(A_{i}|B) = 1$. Again,
the content of \textbf{Bayes' theorem} may be conveniently
visualised by means of a Venn diagram.

\medskip
\noindent
Some of the different terms appearing in Eq.~(\ref{eq:bayes}) have 
been given names in their own right:
\begin{itemize}

\item $P(A_{i})$ is referred to as the \textbf{prior probability}
of random event, or hypothesis, $A_{i}$,

\item $P(B|A_{i})$ is the \textbf{likelihood} of random event,
or empirical evidence, $B$, given random event, or hypothesis,
$A_{i}$, and

\item $P(A_{i}|B)$ is called the \textbf{posterior probability} of 
random event, or hypothesis, $A_{i}$, given random event,
or empirical evidence, $B$.

\end{itemize}

\medskip
\noindent
The most common interpretation of \textbf{Bayes' theorem} is 
that it essentially provides a means for computing the
\textbf{posterior probability} of a random event, or hypothesis,
$A_{i}$, given information on the factual realisation of an
associated random event, or evidence, $B$, in terms of the product
of the \textbf{likelihood} of $B$, given $A_{i}$, and the
\textbf{prior probability} of~$A_{i}$,
\be
\lb{eq:bayes2}
P(A_{i}|B) \propto P(B|A_{i}) \times P(A_{i}) \ .
\ee
This result
is at the heart of the interpretation that \textbf{empirical 
learning} amounts to updating the prior ``\textit{subjective 
probability}'' one has assigned to a specific random 
event, or hypothesis, $A_{i}$, in order to quantify one's initial
reasonable \textbf{degree-of-belief} in its occurrence resp. in its
truth content, by means of adequate experimental or observational
data and corresponding theoretical considerations; see, e.g., Sivia
and Skilling (2006)~\ct[p~5ff]{sivski2006}, Gelman \textit{et al}
(2014)~\ct[p~6ff]{geletal2014}, or McElreath
(2016)~\ct[p~4ff]{mce2016}.

\medskip
\noindent
The \textbf{Bayes--Laplace approach} to tackling 
quantitative--statistical problems in \textbf{Econometrics} was 
pioneered by Zellner in the early 1970ies; see the 1996 reprint of 
his renowned 1971 monograph~\ct{zel1996}. A recent thorough 
introduction into its main considerations is provided by the 
graduate textbook by Greenberg (2013)~\ct{gre2013}.

\medskip
\noindent
A particularly prominent application of this framework in
\textbf{Econometrics} is given by proposals to the mathematical
modelling of economic agents' \textbf{decision-making} (in the
sense of choice behaviour) under conditions of
\textbf{uncertainty}, which, fundamentally, assume \textit{rational
behaviour} on the part of the agents; see, e.g., the graduate
textbook by Gilboa (2009)~\ct{gil2009}, and the brief reviews by
Svetlova and van Elst (2012, 2014)~\ct{svehve2012,svehve2014}, as
well as references therein. \textit{Psychological dimensions} of 
\textbf{decision-making}, on the other hand, such as the
empirically established existence of reference points, loss
aversion, and distortion of probabilities into corresponding
decision weights, have been accounted for in Kahneman and Tversky's 
(1979)~\ct{kahtve1979} \textbf{Prospect Theory}.


\chapter[Discrete and continuous random variables]{\href{https://www.youtube.com/watch?v=1rEczO3Jh_Y}{Discrete and continuous random
variables}}
\lb{ch7}
Applications of \textbf{inferential statistical methods}
rooted in the \textbf{frequentist approach} to \textbf{Probability
Theory}, some of which are to be discussed in Chs.~\ref{ch12} and
\ref{ch13} below, rest fundamentally on the concept of a
probability-dependent quantity arising in the context of
\textbf{random experiments} that is referred to as a \textbf{random
variable}. The present chapter aims to provide a basic introduction
to the general properties and characteristic features of random
variables. We begin by stating the definition of this concept.

\medskip
\noindent
\underline{\textbf{Def.:}} A real-valued one-dimensional
\textbf{random variable} is defined as a one-to-one mapping
\be
X: \boldsymbol{\Omega} \rightarrow D \subseteq \mathbb{R}
\ee
of the sample space $\boldsymbol{\Omega}$ of some random 
experiment with associated probability space 
$\left(\boldsymbol{\Omega}, {\cal P}, P\right)$ into a subset $D$ 
of the real numbers $\mathbb{R}$.

\medskip
\noindent
Depending on the nature of the \textbf{spectrum of values} of $X$,
we will distinguish in the following between random 
variables of the \textbf{discrete} and \textbf{continuous} kinds.

\section[Discrete random variables]{Discrete random variables}
\lb{sec:diskretz}
\textbf{Discrete random variables} are signified by the existence
of a finite or countably infinite

\medskip
\noindent
\textbf{Spectrum of values:}
\be
X \mapsto x \in \left\{x_{1}, \ldots, x_{n}\right\}
\subset \mathbb{R} \ ,
\quad\quad\text{with}\quad n \in \mathbb{N} \ .
\ee
All values $x_{i}$ ($i=1,\ldots,n$) in this spectrum,
referred to as possible \textbf{realisations} of $X$, are assigned 
individual probabilities $p_{i}$ by a real-valued

\medskip
\noindent
\textbf{Probability function:}
\be
\fbox{$\displaystyle
P(X=x_{i}) = p_{i} \quad\quad\text{for}\quad i=1, \dots, n \ ,
$}
\ee
with properties\\[-5mm]
\begin{center}
\begin{itemize}
\item[(i)] $0 \leq p_{i} \leq 1$, and
\hfill (\textbf{non-negativity})\\[-5mm]

\item[(ii)] $\displaystyle\sum_{i=1}^{n}p_{i}=1$.
\hfill (\textbf{normalisability})\\[-5mm]
\end{itemize}
\end{center}
Specific distributional features of a discrete random variable $X$ 
deriving from its probability function $P(X=x_{i})$ are encoded in 
the associated theoretical

\medskip
\noindent
\textbf{Cumulative distribution function (\texttt{cdf}):}
\be
\fbox{$\displaystyle
F_{X}(x) = \texttt{cdf}(x) := P(X \leq x)
= \sum_{i|x_{i}\leq x}P(X=x_{i}) \ .
$}
\ee
The \texttt{cdf} exhibits the asymptotic behaviour
\be
\lim_{x\to-\infty}F_{X}(x)=0 \ ,
\qquad
\lim_{x\to+\infty}F_{X}(x)=1 \ .
\ee
Information on the central tendency and the variability of a 
discrete random variable~$X$ is quantified in terms of its

\medskip
\noindent
\textbf{Expectation value} and \textbf{variance:}
\bea
\lb{eq:evdiscr}
\mathrm{E}(X) & := & \sum_{i=1}^{n}x_{i}P(X=x_{i}) \\
\lb{eq:vardiscr}
\mathrm{Var}(X) & := & \sum_{i=1}^{n}\left(x_{i}-\mathrm{
E}(X)\right)^{2}P(X=x_{i}) \ .
\eea
One of the first occurrences of the notion of the expectation 
value of a random variable relates to the famous ``wager'' put 
forward by the French mathematician
\href{http://www-history.mcs.st-and.ac.uk/Biographies/Pascal.html}{Blaise Pascal (1623--1662)}; cf. Gilboa 
(2009)~\ct[Sec.~5.2]{gil2009}.

\medskip
\noindent
By the so-called \textbf{shift theorem} it holds that the variance 
may alternatively be obtained from the computationally more 
efficient formula 
\be
\mathrm{Var}(X)=\mathrm{E}\left[(X-\mathrm{E}(X))^{2}\right]
=\mathrm{E}(X^{2})-\left[\mathrm{E}(X)\right]^{2} \ .
\ee
Specific values of $\mathrm{E}(X)$ and $\mathrm{ Var}(X)$ will be 
denoted throughout by the Greek letters $\mu$ and $\sigma^{2}$, 
respectively. The \textbf{standard deviation} of $X$ amounts to 
$\sqrt{\mathrm{Var}(X)}$; its specific values will be denoted by 
$\sigma$.

\medskip
\noindent
The evaluation of \textbf{event probabilities} for a discrete
random variable~$X$ with known probability function~$P(X=x_{i})$
follows from the

\medskip
\noindent
\textbf{Computational rules:}
\bea
\lb{eq:comprulesdiscr1}
P(X \leq d) & = & F_{X}(d) \\
P(X < d) & = & F_{X}(d) - P(X=d) \\
P(X \geq c) & = & 1 - F_{X}(c) + P(X=c) \\
P(X > c) & = &  1- F_{X}(c) \\
P(c \leq X \leq d) & = & F_{X}(d) - F_{X}(c) + P(X=c) \\
P(c < X \leq d) & = & F_{X}(d) - F_{X}(c) \\
P(c \leq X < d) & = & F_{X}(d) - F_{X}(c) - P(X=d) + P(X=c) \\
\lb{eq:comprulesdiscr8}
P(c < X < d) & = & F_{X}(d) - F_{X}(c) - P(X=d) \ ,
\eea
where $c$ and $d$ denote arbitrary lower and upper cut-off values 
imposed on the spectrum of $X$.

\medskip
\noindent
In applications it is frequently of interest to know the values of 
a discrete \texttt{cdf}'s

\medskip
\noindent
$\boldsymbol{\alpha}$\textbf{--quantiles:}\\
These are realisations $x_{\alpha}$ of $X$ specifically
determined by the condition that $X$ take values $x \leq 
x_{\alpha}$ at least with probability $\alpha$ (for $0<\alpha<1$),
i.e.,
\be
F_{X}(x_{\alpha}) = P(X \leq x_{\alpha}) \stackrel{!}{\geq} \alpha
\qquad\text{and}\qquad
F_{X}(x) = P(X \leq x) < \alpha
\quad\quad\text{for}\quad
x < x_{\alpha} \ .
\ee
Occasionally, $\alpha$--quantiles of a probability distribution 
are also referred to as \textbf{percentile values}.

\section[Continuous random variables]{Continuous random variables}
\lb{sec:stetigz}
\textbf{Continuous random variables} possess an uncountably
infinite

\medskip
\noindent
\textbf{Spectrum of values:}
\be
X \mapsto x \in D \subseteq \mathbb{R} \ .
\ee
It is, therefore, no longer meaningful to assign probabilities to 
individual \textbf{realisations} $x$ of $X$, but only to 
infinitesimally small intervals $\mathrm{d}x \in D$ instead, by
means of a real-valued

\medskip
\noindent
\textbf{Probability density function (\texttt{pdf}):}
\be
\fbox{$\displaystyle
f_{X}(x) = \texttt{pdf}(x) \ .
$}
\ee
Hence, approximately,
\[
P(X \in \mathrm{d}x) \approx f_{X}(\xi)\,\mathrm{d}x \ ,
\]
for some representative $\xi \in \mathrm{d}x$. The \texttt{pdf} of
an arbitrary continuous random variable~$X$ has the defining 
properties:\\[-5mm]
\begin{center}
\begin{itemize}
\item[(i)] $f_{X}(x) \geq 0$ for all $x \in D$,
\hfill (\textbf{non-negativity})\\[-5mm]

\item[(ii)] ${\displaystyle \int_{-\infty}^{+\infty}f_{X}(x)\,
\mathrm{d}x = 1}$, and \hfill (\textbf{normalisability})\\[-5mm]

\item[(iii)] $f_{X}(x) = F_{X}^{\prime}(x)$. \hfill (\textbf{link
to} \texttt{cdf})\\[-5mm]

\end{itemize}
\end{center}
The evaluation of \textbf{event probabilities} for a continuous 
random variable~$X$ rests on the associated theoretical

\medskip
\noindent
\textbf{Cumulative distribution function (\texttt{cdf}):}
\be
\fbox{$\displaystyle
F_{X}(x) = \texttt{cdf}(x) := P(X \leq x)
= \int_{-\infty}^{x}f_{X}(t)\,\mathrm{d}t \ .
$}
\ee

\pagebreak
\noindent
Event probabilities for~$X$ are then to be obtained from the

\medskip
\noindent
\textbf{Computational rules:}
\bea
\lb{eq:comprulescont1}
P(X = d) & = & 0 \\
\lb{eq:comprulescont2}
P(X \leq d) & = & F_{X}(d) \\
\lb{eq:comprulescont3}
P(X \geq c) & = & 1 - F_{X}(c) \\
\lb{eq:comprulescont4}
P(c \leq X \leq d) & = & F_{X}(d) - F_{X}(c) \ ,
\eea
where $c$ and $d$ denote arbitrary lower and upper cut-off values 
imposed on the spectrum of $X$. Note that, again, the \texttt{cdf}
exhibits the asymptotic properties
\be
\lim_{x\to-\infty}F_{X}(x)=0 \ ,
\qquad
\lim_{x\to+\infty}F_{X}(x)=1 \ .
\ee
The central tendency and the variabilty of a continuous random 
variable~$X$ are quantified by its

\medskip
\noindent
\textbf{Expectation value} and \textbf{variance:}
\bea
\lb{eq:expectcon}
\mathrm{E}(X) & := & \int_{-\infty}^{+\infty}xf_{X}(x)\,
\mathrm{d}x \\
\lb{eq:varcon}
\mathrm{Var}(X) & := & \int_{-\infty}^{+\infty}
\left(x-\mathrm{E}(X)\right)^{2}f_{X}(x)\,\mathrm{d}x \ .
\eea
Again, by the \textbf{shift theorem} the variance may alternatively 
be obtained from the computationally more efficient formula 
$\displaystyle \mathrm{Var}(X)=\mathrm{E}
\left[(X-\mathrm{E}(X))^{2}\right]
=\mathrm{E}(X^{2})-\left[\mathrm{E}(X)\right]^{2}$. Specific values
of $\mathrm{E}(X)$ and $\mathrm{Var}(X)$ will be denoted throughout
by $\mu$ and $\sigma^{2}$, respectively. The \textbf{standard
deviation} of $X$ amounts to $\sqrt{\mathrm{Var}(X)}$; its specific
values will be denoted by $\sigma$.

\medskip
\noindent
The construction of interval estimates for unknown distribution 
parameters of continuous one-dimensional random variables~$X$ in 
given target populations~$\boldsymbol{\Omega}$, and null hypothesis
significance testing (to be discussed later in 
Chs.~\ref{ch12} and \ref{ch13}), both require explicit knowledge of
the $\boldsymbol{\alpha}$\textbf{--quantiles} associated with the
\texttt{cdf}s of the~$X$s. Generally, these are defined as follows.

\medskip
\noindent
$\boldsymbol{\alpha}$\textbf{--quantiles:} \\
$X$ take values $x \leq x_{\alpha}$ with probability $\alpha$ 
(for $0<\alpha<1$), i.e.,
\be
P(X \leq x_{\alpha}) = F_{X}(x_{\alpha}) \stackrel{!}{=} 
\alpha
\qquad \overbrace{\Leftrightarrow}^{F_{X}(x)\ \text{is strictly
monotonously increasing}}\qquad
\fbox{$\displaystyle
x_{\alpha}=F_{X}^{-1}(\alpha)$} \ .
\ee
Hence, $\alpha$--quantiles of the probability distribution for a 
continuous one-dimensional random variable~$X$ are determined by 
the inverse \texttt{cdf}, $F_{X}^{-1}$. For given $\alpha$, the 
spectrum of $X$ is thus naturally partitioned into domains $x \leq 
x_{\alpha}$ and $x \geq x_{\alpha}$. Occasionally, 
$\alpha$--quantiles of a probability distribution are also 
referred to as \textbf{percentile values}.

\section[Skewness and excess kurtosis]{Skewness and excess 
kurtosis}
\lb{sec:skewkurt}
In analogy to the descriptive case of Sec.~\ref{sec:distortion},
dimensionless \textbf{measures of relative distortion}
characterising the \textbf{shape} of the probability distribution
for a discrete or a continuous one-dimensional random variable~$X$
are defined by the

\medskip
\noindent
\textbf{Skewness} and \textbf{excess kurtosis:}
\bea
\lb{eq:skew2}
\mathrm{Skew}(X) & := & \frac{\mathrm{E}\left[(X-\mathrm{
E}(X))^{3}\right]}{\left[\mathrm{Var}(X)\right]^{3/2}} \\
\lb{eq:kurt2}
\mathrm{Kurt}(X) & := & \frac{\mathrm{E}\left[(X-\mathrm{
E}(X))^{4}\right]}{\left[\mathrm{Var}(X)\right]^{2}} - 3 \ ,
\eea
given $\mathrm{Var}(X) > 0$; cf. Rinne (2008)~\ct[p~196]{rin2008}. 
Specific values of $\mathrm{Skew}(X)$ and $\mathrm{Kurt}(X)$ may be 
denoted by $\gamma_{1}$ and $\gamma_{2}$, respectively.

\section[Lorenz curve for continuous random variables]{Lorenz 
curve for continuous random variables}
For a continuous one-dimensional random variable~$X$, the
\textbf{Lorenz curve} expressing qualitatively the degree of
concentration involved in its associated probability distribution
of is defined by
\be
\lb{lorcurve}
\fbox{$\displaystyle
L(x_{\alpha})
= 
\frac{{\displaystyle\int_{-\infty}^{x_{\alpha}}tf_{X}(t)\,
\mathrm{d}t}}{{\displaystyle\int_{-\infty}^{+\infty}tf_{X}(t)\,
\mathrm{d}t}} \ ,
$}
\ee
with $x_{\alpha}$ denoting a particular $\alpha$--quantile
of the distribution in question.

\section[Linear transformations of random variables]{Linear
transformations of random variables}
\lb{sec:lintransf}
\textbf{Linear transformations} of real-valued one-dimensional
random variables $X$ are determined by the two-parameter relation
\be
\lb{lintransf}
\fbox{$\displaystyle
Y = a+bX \quad\text{with}\quad a,b \in \mathbb{R}, b \neq 0 \ ,
$}
\ee
where $Y$ denotes the resultant new random variable. 
Transformations of random variables of this kind have 
the following effects on the computation of expectation values and 
variances.

%
\subsection[Effect on expectation values]{Effect on expectation 
values}
%
\begin{enumerate}
\item $\mathrm{E}(a) = a$
\item $\mathrm{E}(bX) = b\mathrm{E}(X)$
\item $\mathrm{E}(Y) = \mathrm{E}(a+bX) = \mathrm{E}(a)
+ \mathrm{E}(bX)
= a+b\mathrm{E}(X)$.
\end{enumerate}
%

\subsection[Effect on variances]{Effect on variances}
\lb{subsec:vartrans}
%
\begin{enumerate}
\item $\mathrm{Var}(a) = 0$
\item $\mathrm{Var}(bX) = b^{2}\mathrm{Var}(X)$
\item $\mathrm{Var}(Y) = \mathrm{Var}(a+bX) = \mathrm{Var}(a)
+ \mathrm{Var}(bX) = b^{2}\mathrm{Var}(X)$.
\end{enumerate}
%

\subsection[Standardisation]{Standardisation}
\lb{subsec:standard}
\textbf{Standardisation} of an arbitrary one-dimensional random 
variable $X$, with $\sqrt{\mathrm{Var}(X)} > 0$, implies the 
determination of a special linear transformation $X \mapsto Z$ 
according to Eq.~(\ref{lintransf}) such that the expectation value 
and variance of $X$ are re-scaled to their simplest values 
possible, i.e., $\mathrm{E}(Z)=0$ and $\mathrm{Var}(Z)=1$. Hence,
the two (in part non-linear) conditions
\[
0 \stackrel{!}{=} \mathrm{E}(Z)= a+b\mathrm{E}(X)
\quad\text{and}\quad
1 \stackrel{!}{=} \mathrm{Var}(Z)= b^{2}\mathrm{Var}(X) \ ,
\]
for unknowns $a$ and $b$, need to be satisfied simultaneously. 
These are solved by, respectively,
\be
a=-\frac{\mathrm{E}(X)}{\sqrt{\mathrm{Var}(X)}}
\quad\text{and}\quad
b=\frac{1}{\sqrt{\mathrm{Var}(X)}} \ ,
\ee
and so
\be
\lb{eq:standardisation}
\fbox{$\displaystyle
X \rightarrow Z = \frac{X-\mathrm{E}(X)}{\sqrt{\mathrm{Var}(X)}}
\ , \qquad
x \mapsto z = \frac{x-\mu}{\sigma} \in \bar{\mathbb{D}}
\subseteq \mathbb{R} \ ,
$}
\ee
irrespective of whether the random variable $X$ is of the discrete 
kind (cf. Sec.~\ref{sec:diskretz}) or of the continuous kind (cf. 
Sec.~\ref{sec:stetigz}). It is essential for applications to 
realise that under the process of standardisation 
the values of event probabilities for a random variable $X$ 
remain \textbf{invariant} (unchanged), i.e.,
\be
P(X\leq x)
= P\left(\frac{X-\mathrm{E}(X)}{\sqrt{\mathrm{Var}(X)}}
\leq \frac{x-\mu}{\sigma}\right) = P(Z\leq z) \ .
\ee
%

\section[Sums of random variables and reproductivity]{Sums of
random variables and reproductivity}
\lb{sec:sumvar}
\underline{\textbf{Def.:}} For a set of $n$ additive
one-dimensional random variables~$X_{1}, \ldots, X_{n}$, one
defines a \textbf{total sum} random variable~$Y_{n}$ and an
associated \textbf{mean} random variable~$\bar{X}_{n}$ according to
\be
\lb{eq:sumnmean}
\fbox{$\displaystyle
Y_{n} := \sum_{i=1}^{n}X_{i}
\quad\quad\text{and}\quad\quad
\bar{X}_{n} := \frac{1}{n}\,Y_{n} \ .
$}
\ee

\medskip
\noindent
By \textit{linearity} of the expectation value 
operation,\footnote{That is: $\mathrm{E}(X_{1}+X_{2})
= \mathrm{E}(X_{1})+\mathrm{E}(X_{2})$.} it then holds that
\be
\lb{eq:sumexpv}
\mathrm{E}(Y_{n})
= \mathrm{E}\left(\sum_{i=1}^{n}X_{i}\right)
= \sum_{i=1}^{n}\mathrm{E}(X_{i})
\quad\quad\text{and}\quad\quad
\mathrm{E}(\bar{X}_{n}) = \frac{1}{n}\,\mathrm{E}(Y_{n}) \ .
\ee
If, in addition, the $X_{1}, \ldots, X_{n}$ are \textit{mutually 
stochastically independent} according to 
Eq.~(\ref{eq:stochindep1}) (see also 
Sec.~\ref{subsec:2dcovtheoret} below), it follows from 
Sec.~\ref{subsec:vartrans} that the variances of $Y_{n}$ and 
$\bar{X}_{n}$ are given by
\be
\lb{eq:sumvar}
\mathrm{Var}(Y_{n})
= \mathrm{Var}\left(\sum_{i=1}^{n}X_{i}\right)
= \sum_{i=1}^{n}\mathrm{Var}(X_{i})
\quad\quad\text{and}\quad\quad
\mathrm{Var}(\bar{X}_{n})
= \left(\frac{1}{n}\right)^{2}\mathrm{Var}(Y_{n}) \ ,
\ee
respectively.

\medskip
\noindent
\underline{\textbf{Def.:}} \textbf{Reproductivity} of a probability 
distribution law (\texttt{cdf}) $F(x)$ is given when the total 
sum~$Y_{n}$ of $n$ independent and identically distributed (in 
short: ``i.i.d.'') additive one-dimensional random 
variables~$X_{1}, \ldots, X_{n}$, which each individually satisfy 
distribution laws $F_{X_{i}}(x) \equiv F(x)$, inherits \textit{this 
very} distribution law $F(x)$ from its underlying $n$ random 
variables. Examples of reproductive distribution laws, to be 
discussed in the following Ch.~\ref{ch8}, are the binomial, the 
Gau\ss ian normal, and the $\chi^{2}$--distributions.

\section[Two-dimensional random variables]{Two-dimensional random 
variables}
\lb{sec:2dvar}
The \textbf{empirical tests for association} between two 
statistical variables $X$ and $Y$ of Ch.~\ref{ch13} require the 
notions of \textbf{two-dimensional random variables} and their 
bivariate \textbf{joint probability distributions}. Recommended 
introductory literature on these matters are, e.g., Toutenburg 
(2005)~\ct[p~57ff]{tou2005} and Kredler (2003)~\ct[Ch.~2]{kre2003}.

\medskip
\noindent
\underline{\textbf{Def.:}} A real-valued two-dimensional
\textbf{random variable} is defined as a one-to-one mapping
\be
\left(X,Y\right): \boldsymbol{\Omega} \rightarrow D \subseteq 
\mathbb{R}^{2}
\ee
of the sample space $\boldsymbol{\Omega}$ of some random 
experiment with associated probability space 
$\left(\boldsymbol{\Omega}, {\cal P}, P\right)$ into a subset $D$ 
of the two-dimensional Euclidian space $\mathbb{R}^{2}$.

\medskip
\noindent
We proceed by sketching some important concepts relating to 
two-dimensional random variables.

\subsection[Joint probability distributions]{Joint probability 
distributions}
\subsubsection[Discrete case]{\underline{Discrete case:}}
Two-dimensional \textbf{discrete random variables} possess a

\medskip
\noindent
\textbf{Spectrum of values:}
\be
(X,Y) \mapsto (x,y) \in \left\{x_{1}, \ldots, x_{k}\right\} \times
\left\{y_{1}, \ldots, y_{l}\right\}
\subset \mathbb{R}^{2} \ ,
\quad\quad\text{with}\quad k, l \in \mathbb{N} \ .
\ee
All pairs of values $(x_{i},y_{j})_{i=1,\ldots,k; j=1,\ldots,l}$ 
in this spectrum are assigned individual probabilities $p_{ij}$ by 
a real-valued

\medskip
\noindent
\textbf{Joint probability function:}
\be
\fbox{$\displaystyle
P(X=x_{i},Y=y_{j}) = p_{ij}
\quad\quad\text{for}\quad i=1, \dots, k;
j=1, \dots, l \ ,
$}
\ee
with properties\\[-5mm]
\begin{center}
\begin{itemize}
\item[(i)] $0 \leq p_{ij} \leq 1$, and
\hfill (\textbf{non-negativity})\\[-5mm]

\item[(ii)] ${\displaystyle\sum_{i=1}^{k}\sum_{j=1}^{l}p_{ij}=1}$.
\hfill (\textbf{normalisability})\\[-5mm]
\end{itemize}
\end{center}
By analogy to the case of one-dimensional random variables, 
specific \textbf{event probabilities} for $(X,Y)$ are obtained from 
the associated

\medskip
\noindent
\textbf{Joint cumulative distribution function (\texttt{cdf}):}
\be
\fbox{$\displaystyle
F_{XY}(x,y) = \texttt{cdf}(x,y) := P(X \leq x,Y \leq y)
= \sum_{i|x_{i}\leq x}\sum_{j|y_{j}\leq y}p_{ij} \ .
$}
\ee
%

\subsubsection[Continous case]{\underline{Continuous case:}}
For two-dimensional \textbf{continuous random variables} the range 
can be represented by the

\medskip
\noindent
\textbf{Spectrum of values:}
\be
(X,Y) \mapsto (x,y) \in D = (x_\mathrm{min},x_\mathrm{max}) \times 
(y_\mathrm{min},y_\mathrm{max}) \subseteq \mathbb{R}^{2} \ .
\ee
Probabilities are now assigned to infinitesimally small areas
$\mathrm{d}x \times \mathrm{d}y \in D$ by means of a real-valued

\medskip
\noindent
\textbf{Joint probability density function (\texttt{pdf}):}
\be
\fbox{$\displaystyle
f_{XY}(x,y) = \texttt{pdf}(x,y) \ ,
$
}
\ee
with properties:\\[-5mm]
\begin{center}
\begin{itemize}
\item[(i)] $f_{XY}(x,y) \geq 0$ for all $(x,y) \in D$, and
\hfill (\textbf{non-negativity})\\[-5mm]

\item[(ii)] ${\displaystyle\int_{-\infty}^{+\infty}
\int_{-\infty}^{+\infty}
f_{XY}(x,y)\,\mathrm{d}x\mathrm{d}y=1}$.
\hfill (\textbf{normalisability})\\[-5mm]

\end{itemize}
\end{center}
Approximately, one now has
\[
P(X \in \mathrm{d}x, Y \in \mathrm{d}y) \approx f_{XY}(\xi, \eta)\,
\mathrm{d}x\mathrm{d}y \ ,
\]
for representative $\xi \in \mathrm{d}x$ and $\eta \in
\mathrm{d}y$. Specific \textbf{event probabilities} for $(X,Y)$ are
obtained from the associated

\medskip
\noindent
\textbf{Joint cumulative distribution function (\texttt{cdf}):}
\be
\fbox{$\displaystyle
F_{XY}(x,y) = \texttt{cdf}(x,y) := P(X \leq x,Y \leq y)
= \int_{-\infty}^{x}\int_{-\infty}^{y}f_{XY}(t,u)\,
\mathrm{d}t\mathrm{d}u \ .
$}
\ee
%

\subsection[Marginal and conditional distributions]{Marginal and 
conditional probability distributions}
\subsubsection[Discrete case]{\underline{Discrete case:}}
The univariate \textbf{marginal probability functions} for $X$ and
$Y$ induced by the joint probability function $P(X=x_{i},Y=y_{j}) 
= p_{ij}$ are
\be
\lb{eq:margxdiscr}
p_{i+} := \sum_{j=1}^{l}p_{ij} = P(X=x_{i}) 
\quad\quad\text{for}\quad i=1,\ldots,k \ ,
\ee
and
\be
\lb{eq:margydiscr}
p_{+j} := \sum_{i=1}^{k}p_{ij} = P(Y=y_{j})
\quad\quad\text{for}\quad j=1,\ldots,l \ .
\ee
In addition, one defines \textbf{conditional probability functions} 
for $X$ given $Y=y_{j}$, with $p_{+j}>0$, and for $Y$ given 
$X=x_{i}$, with $p_{i+}>0$, by
\be
\lb{eq:condxdiscr}
p_{i|j} := \frac{p_{ij}}{p_{+j}} = P(X=x_{i}|Y=y_{j})
\quad\quad\text{for}\quad i=1,\ldots,k \ ,
\ee
respectively
\be
\lb{eq:condydiscr}
p_{j|i} := \frac{p_{ij}}{p_{i+}} = P(Y=y_{j}|X=x_{i})
\quad\quad\text{for}\quad j=1,\ldots,l \ .
\ee
%

\subsubsection[Continous case]{\underline{Continuous case:}}
The univariate \textbf{marginal probability density functions} for
$X$ and $Y$ induced by the joint probability density function 
$f_{XY}(x,y)$ are
\be
\lb{eq:margxcont}
f_{X}(x) = \int_{-\infty}^{+\infty}f_{XY}(x,y)\,\mathrm{d}y \ ,
\ee
and
\be
\lb{eq:margycont}
f_{Y}(y) = \int_{-\infty}^{+\infty}f_{XY}(x,y)\,\mathrm{d}x \ .
\ee
Moreover, one defines \textbf{conditional probability density 
functions} for $X$ given $Y$, and for $Y$ given $X$, by
\be
\lb{eq:condxcont}
f_{X|Y}(x|y) := \frac{f_{XY}(x,y)}{f_{Y}(y)}
\quad\quad\text{for}\quad f_{Y}(y)>0 \ ,
\ee
respectively
\be
\lb{eq:condycont}
f_{Y|X}(y|x) := \frac{f_{XY}(x,y)}{f_{X}(x)}
\quad\quad\text{for}\quad f_{X}(x)>0 \ .
\ee
%

\subsection[Bayes' theorem for two-dimensional random variables]{Bayes' theorem for two-dimensional random variables}
The concept of a bivariate joint probability distribution is at 
the heart of the formulation of Bayes' theorem, 
Eq.~(\ref{eq:bayes}), for a real-valued two-dimensional random 
variable $(X,Y)$.

\subsubsection[Discrete case]{\underline{Discrete case:}}
Let $P(X=x_{i}) = p_{i+} > 0$ be a \textbf{prior probability 
function} for a discrete random variable $X$. Then, on the grounds 
of a joint probability function $P(X=x_{i},Y=y_{j}) = p_{ij}$ and
Eqs.~(\ref{eq:condxdiscr}) and (\ref{eq:condydiscr}), the
\textbf{posterior probability function} for $X$ given $Y=y_{j}$,
with$P(Y=y_{j}) = p_{+j} > 0$, is determined by
\be
p_{i|j} = \frac{p_{j|i}}{p_{+j}}\,p_{i+}
\quad\quad\text{for}\quad i=1,\ldots,k \ .
\ee
By using Eqs.~(\ref{eq:margydiscr}) and (\ref{eq:condydiscr}) to 
re-expressed the denominator $p_{+j}$, this may be given in the 
standard form
\be
\lb{eq:bayesdiscr2drandvars}
\fbox{$\displaystyle
p_{i|j} = \frac{p_{j|i}\,p_{i+}}{\displaystyle\sum_{i=1}^{k}p_{j|i}
\,p_{i+}}
\quad\quad\text{for}\quad i=1,\ldots,k \ .
$}
\ee
%

\subsubsection[Continous case]{\underline{Continuous case:}}
Let $f_{X}(x) > 0$ be a \textbf{prior probability density function} 
for a continuous random variable $X$. Then, on the grounds 
of a joint probability density function $f_{XY}(x,y)$ and
Eqs.~(\ref{eq:condxcont}) and (\ref{eq:condycont}), the
\textbf{posterior probability density function} for $X$ given $Y$,
with $f_{Y}(y)>0$, is determined by
\be
f_{X|Y}(x|y) = \frac{f_{Y|X}(y|x)}{f_{Y}(y)}\,f_{X}(x) \ .
\ee
By using Eqs.~(\ref{eq:margycont}) and (\ref{eq:condycont}) to 
re-expressed the denominator $f_{Y}(y)$, this may be stated in the 
standard form
\be
\lb{eq:bayescont2drandvars}
\fbox{$\displaystyle
f_{X|Y}(x|y) = 
\frac{f_{Y|X}(y|x)\,f_{X}(x)}{\displaystyle\int_{-\infty}^{+\infty}
f_{Y|X}(y|x)\,f_{X}(x)\,\mathrm{d}x} \ .
$}
\ee

\medskip
\noindent
In practical applications, evaluation of the, at times intricate, 
single and double integrals contained in this representation of 
Bayes' theorem is managed by employing sophisticated numerical 
approxi\-mation techniques; cf. Saha (2002)~\ct{sah2002},
Sivia and Skilling (2006)~\ct{sivski2006}, Greenberg
(2013)~\ct{gre2013}, Gelman \textit{et al} (2014)~\ct{geletal2014},
or McElreath (2016)~\ct{mce2016}.

\subsection[Covariance and correlation]{Covariance and correlation}
\lb{subsec:2dcovtheoret}
We conclude this section by reviewing the standard measures for 
characterising the degree of \textbf{stochastic association}
between two random variables $X$ and $Y$.

\medskip
\noindent
The \textbf{covariance} of $X$ and $Y$ is defined by
\be
\lb{eq:2dvarcov}
\mathrm{Cov}(X,Y) := \mathrm{E}\left[\left(X-\mathrm{E}(X)\right)
\left(Y-\mathrm{E}(Y)\right)\right] \ .
\ee
It constitutes the off-diagonal component of the symmetric 
$\boldsymbol{(2 \times 2)}$ \textbf{covariance matrix}
\be
\lb{eq:2dcovmattheoret}
\boldsymbol{\Sigma}(X,Y) :=
\left(\begin{array}{cc}
\mathrm{Var}(X) & \mathrm{Cov}(X,Y) \\
\mathrm{Cov}(X,Y) & \mathrm{Var}(Y)
\end{array}\right) \ ,
\ee
which is regular and thus invertible as long as 
$\det[\boldsymbol{\Sigma}(X,Y)] \neq 0$.

\medskip
\noindent
By a suitable normalisation procedure, one defines from 
Eq.~(\ref{eq:2dvarcov}) the \textbf{correlation coefficient} of $X$ 
and $Y$ as
\be
\lb{eq:2dvarcorrel}
\rho(X,Y) := \frac{\mathrm{Cov}(X,Y)}{\sqrt{\mathrm{Var}(X)}
\sqrt{\mathrm{Var}(Y)}} \ .
\ee
This features as the off-diagonal component in the symmetric 
$\boldsymbol{(2 \times 2)}$ \textbf{correlation matrix}
\be
\lb{eq:2dcorrelmattheoret}
\boldsymbol{R}(X,Y) :=
\left(\begin{array}{cc}
1 & \rho(X,Y) \\
\rho(X,Y) & 1
\end{array}\right) \ ,
\ee
which is positive definite and thus invertible for $0 < 
\det[\boldsymbol{R}(X,Y)] = 1-\rho^{2} \leq 1$.

\medskip
\noindent
\underline{\textbf{Def.:}} Two random variables $X$ and $Y$ are 
referred to as \textbf{mutually stochastically independent}
provided that
\be
\lb{eq:stochindep2}
\mathrm{Cov}(X,Y) = 0
\qquad\Leftrightarrow\qquad
\rho(X,Y) = 0 \ .
\ee
It then follows that
\be
\lb{eq:stochindep3}
P(X \leq x,Y \leq y) = P(X \leq x) \times P(Y\leq y)
\quad\Leftrightarrow\quad
F_{XY}(x,y) = F_{X}(x) \times F_{Y}(y)
\ee
for $(x,y) \in D \subseteq \mathbb{R}^{2}$. Moreover, in this case 
(i)~$\mathrm{E}(X \times Y) = \mathrm{E}(X) \times \mathrm{E}(Y)$,
and (ii)~$\mathrm{Var}(aX+bY) = a^{2}\mathrm{Var}(X)+b^{2}
\mathrm{Var}(Y)$.

\vspace{5mm}
\noindent
In the next chapter we will highlight a number of standard 
univariate probability distributions for discrete and continuous 
one-dimensional random variables.


\chapter[Standard univariate probability distributions]{Standard 
univariate probability distributions for discrete and continuous 
random variables}
\lb{ch8}
In this chapter, we review (i)~the univariate probability 
distributions for one-dimensional random variables which one
typically encounters as \textbf{theoretical probability
distributions} in the context of frequentist \textbf{null
hypothesis significance testing} (cf. Chs. \ref{ch12}
and \ref{ch13}), but we also include
(ii)~cases of well-established pedagogical merit, and (iii)~a few
examples of rather specialised univariate probability
distributions, which, nevertheless, prove to be of interest in the description and modelling of various theoretical market situations
in \textbf{Economics}. We split our considerations into two main
parts according to whether a one-dimensional random variable~$X$
underlying a particular distribution law varies discretely or
continuously. For each of the cases to be presented, we list the
\textbf{spectrum of values} of $X$, its \textbf{probability
function} (for discrete $X$) or \textbf{probability density
function} (\texttt{pdf}) (for continuous $X$), its
\textbf{cumulative distribution function} (\texttt{cdf}), its
\textbf{expectation value} and its \textbf{variance}, and, in some
continuous cases, also its \textbf{skewness}, \textbf{excess
kurtosis} and $\boldsymbol{\alpha}$\textbf{--quantiles}. Additional information, e.g., commands in \R, on a GDC, in EXCEL, or in
OpenOffice, by which a specific distribution function may be
activated for computational purposes or be plotted, is included
where available.

\section[Discrete uniform distribution]{Discrete uniform
distribution}
\lb{sec:dgleichverteil}
One of the simplest probability distributions for a discrete 
one-dimensional random variable $X$ is given by the one-parameter 
\textbf{discrete uniform distribution},
\be
X \sim L(n) \ ,
\ee
which is characterised by the number $n$ of different values in 
$X$'s

\medskip
\noindent
Spectrum of values:
\be
X \mapsto x \in \left\{x_{1}, \ldots, x_{n}\right\}
\subset \mathbb{R} \ ,
\quad\text{with}\quad n \in \mathbb{N} \ .
\ee
Probability function:
\be
\lb{eq:lprob}
\fbox{$\displaystyle
P(X=x_{i}) = \frac{1}{n} \quad\quad\text{for}\quad
i=1, \dots, n \ ;
$}
\ee
its graph is shown in Fig.~\ref{fig:lprob} below for 
$n=6$.
\begin{figure}[!htb]
\begin{center}
\includegraphics[scale=0.8]{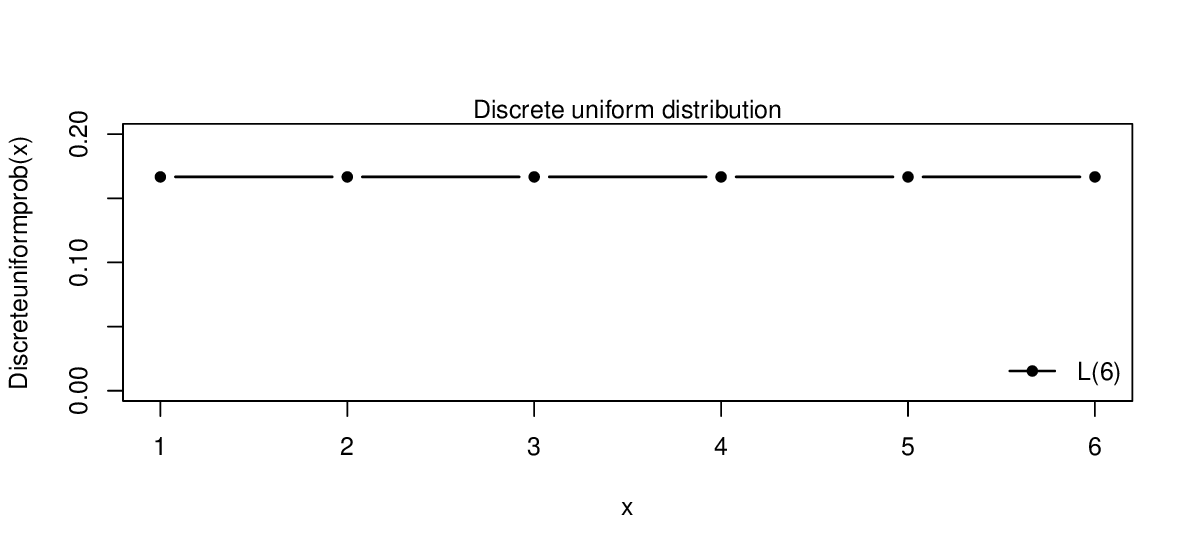}
\end{center}
\caption{Probability function of the discrete uniform distribution 
according to Eq.~(\ref{eq:lprob}) for the case $L(6)$. An
enveloping line is also shown.}
\lb{fig:lprob}
\end{figure}

\medskip
\noindent
Cumulative distribution function (\texttt{cdf}):
\be
\fbox{$\displaystyle
F_{X}(x) = P(X \leq x) = \sum_{i|x_{i}\leq x}\frac{1}{n} \ .
$}
\ee
Expectation value and variance:
\bea
\mathrm{E}(X) & = & \sum_{i=1}^{n}x_{i} \times \frac{1}{n}
= \mu \\
\mathrm{Var}(X) & = & \left(\sum_{i=1}^{n}x_{i}^{2} \times
\frac{1}{n}\right) - \mu^{2} \ .
\eea
For skewness and excess kurtosis, see, e.g., Rinne
(2008)~\ct[p~372f]{rin2008}.

\medskip
\noindent
The discrete uniform distribution is identical to a Laplacian 
probability measure; cf. Sec.~\ref{sec:laplace}. This is 
well-known from games of chance such as tossing a fair coin once, 
selecting a single card from a deck of cards, rolling a fair dye 
once, or the fair roulette lottery.

\medskip
\noindent
\underline{\R:} $\texttt{ddunif}(x,x_{1},x_{n})$,
$\texttt{pdunif}(x,x_{1},x_{n})$, 
$\texttt{qdunif}(\alpha,x_{1},x_{n})$,
$\texttt{rdunif}(n_{\mathrm{simulations}},x_{1},x_{n})$ (package:
{\tt extraDistr}, by Wolodzko (2018)~\ct{wol2018})

\section[Binomial distribution]{Binomial distribution}
\lb{sec:binomverteil}
\subsection[Bernoulli distribution]{Bernoulli distribution}
Another simple probability distribution, for a discrete 
one-dimensional random variable $X$ with only two possible values, 
$0$ and $1$,\footnote{Any one-dimensional random variable of this 
kind is referred to as dichotomous.} is due to the Swiss 
mathematician 
\href{http://www-history.mcs.st-and.ac.uk/Biographies/Bernoulli_Jacob.html}{Jakob Bernoulli (1654--1705)}. The \textbf{Bernoulli 
distribution},
\be
X \sim B(1;p) \ ,
\ee
depends on a single free parameter, the probability $p \in [0;1]$
for the event $X = x = 1$.

\medskip
\noindent
Spectrum of values:
\be
X \mapsto x \in \left\{0, 1\right\} \ .
\ee
Probability function:
\be
\lb{eq:bprob}
\fbox{$\displaystyle
P(X=x) = \left(\begin{array}{c}
               1 \\
               x
               \end{array}\right)p^{x}(1-p)^{1-x} \ ,
\quad\text{with}\quad 0 \leq p \leq 1 \ ;
$}
\ee
its graph is shown in Fig.~\ref{fig:bprob} below for 
$\displaystyle p=\frac{1}{3}$.
\begin{figure}[!htb]
\begin{center}
\includegraphics[scale=0.8]{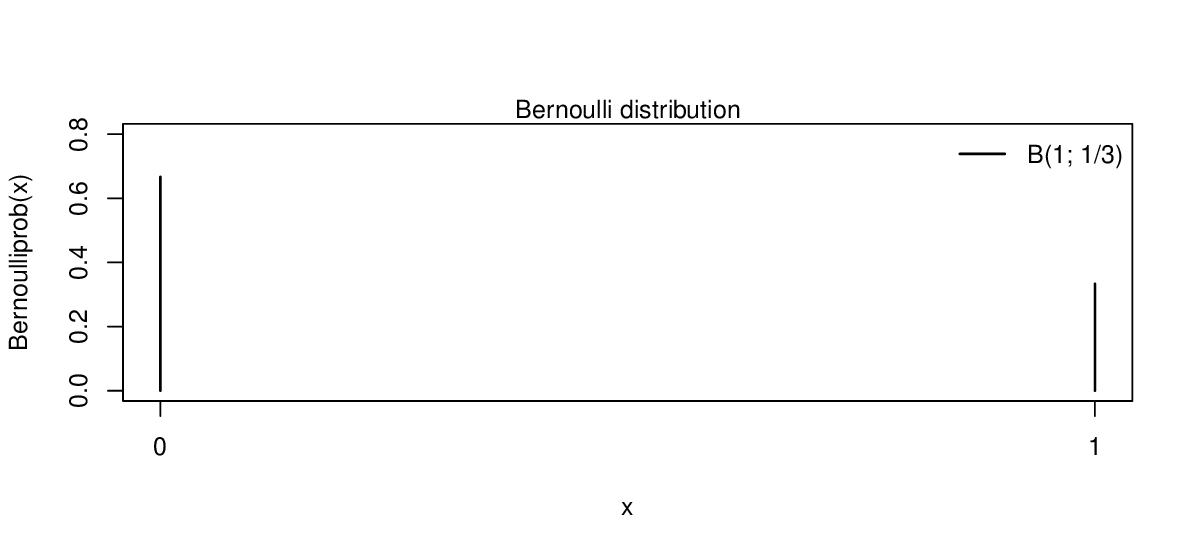}
\end{center}
\caption{Probability function of the Bernoulli distribution 
according to Eq.~(\ref{eq:bprob}) for the case $\displaystyle 
B\left(1;\frac{1}{3}\right)$.}
\lb{fig:bprob}
\end{figure}

\medskip
\noindent
Cumulative distribution function (\texttt{cdf}):
\be
\fbox{$\displaystyle
F_{X}(x) = P(X \leq x)
= \sum_{k=0}^{\left\lfloor x\right\rfloor}
\left(\begin{array}{c}
1 \\
k
\end{array}\right)p^{k}(1-p)^{1-k} \ .
$}
\ee
Expectation value and variance:
\bea
\mathrm{E}(X) & = & 0\times(1-p)+1\times p \ = \ p \\
\mathrm{Var}(X) & = & 0^{2}\times(1-p)+1^{2}\times p - p^{2}
\ = \ p(1-p) \ .
\eea
%

\subsection[General binomial distribution]{General binomial
distribution}
A direct generalisation of the Bernoulli distribution is the 
case of a discrete one-dimensional random variable $X$ which is 
the \textit{sum} of $n$ mutually stochastically independent, 
identically Bernoulli-distributed (``i.i.d.'') one-dimensional 
random variables $X_{i}\sim B(1;p)$ ($i=1,\ldots,n$), i.e.,
\be
X:=\sum_{i=1}^{n}X_{i}=X_{1}+\ldots +X_{n} \ ,
\ee
which yields the reproductive two-parameter \textbf{binomial 
distribution}
\be
X \sim B(n;p) \ ,
\ee
again with $p \in [0;1]$ the probability for a single event 
$X_{i} = x_{i} = 1$.

\medskip
\noindent
Spectrum of values:
\be
X \mapsto x \in \left\{0, \ldots, n\right\} \ , 
\quad\quad\text{with}\quad n \in \mathbb{N} \ .
\ee
Probability function:\footnote{In the context of an urn model with 
$M$ black balls and $N-M$ white balls, and the random selection of 
$n$ balls from a total of $N$, with repetition, this probability 
function can be derived from Laplace's principle of forming the 
ratio between the ``number of favourable cases'' and the ``number 
of all possible cases,'' cf. Eq.~(\ref{eq:classprob}). Thus,
$\displaystyle P(X=x) = \frac{\left(\begin{array}{c}
n \\
x
\end{array}\right)M^{x}(N-M)^{n-x}}{N^{n}}$,
where $x$ denotes the number of black balls drawn, and one 
substitutes accordingly from the definition $p:=M/N$.}
\be
\lb{eq:binomialprob}
\fbox{$\displaystyle
P(X=x) = \left(\begin{array}{c}
               n \\
               x
               \end{array}\right)p^{x}(1-p)^{n-x} \ ,
\quad\quad\text{with}\quad 0 \leq p \leq 1 \ ;
$}
\ee
its graph is shown in Fig.~\ref{fig:biprob} below for 
$n=10$ and $\displaystyle p=\frac{3}{5}$. Recall that 
$\left(\begin{array}{c} n \\ x \end{array}\right)$ denotes the 
binomial coefficient defined in Eq.~(\ref{eq:binomcoeff}), which
generates the positive integer entries of Pascal's triangle.
\begin{figure}[!htb]
\begin{center}
\includegraphics[scale=0.8]{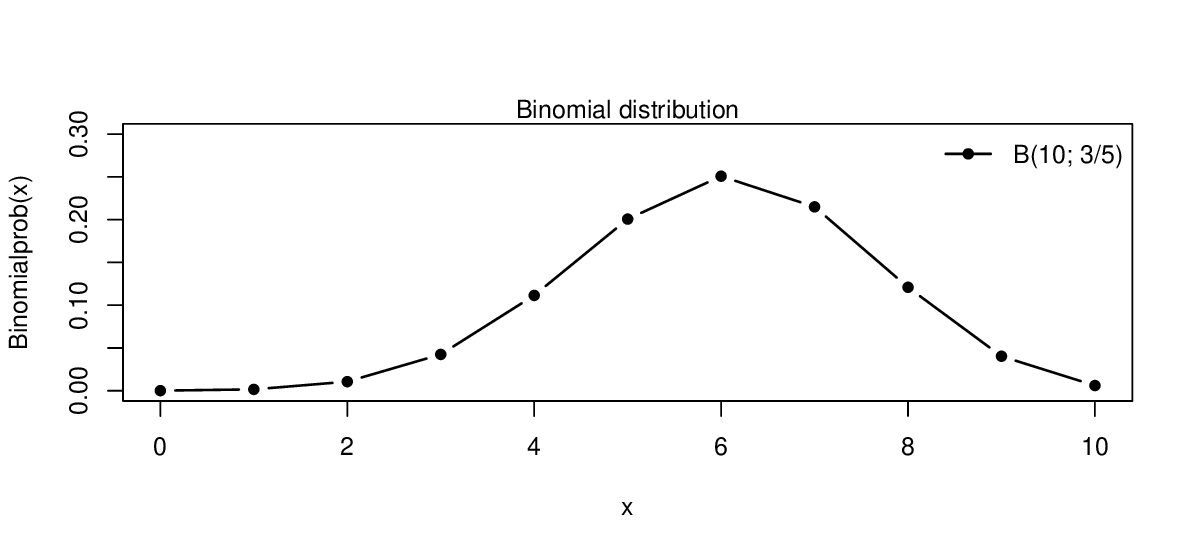}
\end{center}
\caption{Probability function of the binomial distribution 
according to Eq.~(\ref{eq:binomialprob}) for the case 
$\displaystyle B\left(10;\frac{3}{5}\right)$. An enveloping line is
also shown.}
\lb{fig:biprob}
\end{figure}

\medskip
\noindent
Cumulative distribution function (\texttt{cdf}):
\be
\fbox{$\displaystyle
F_{X}(x) = P(X \leq x)
= \sum_{k=0}^{\left\lfloor x\right\rfloor}
\left(\begin{array}{c}
n \\
k
\end{array}\right)p^{k}(1-p)^{n-k} \ .
$}
\ee
Expectation value, variance, skewness and excess kurtosis (cf. 
Rinne (2008)~\ct[p~260]{rin2008}):
\bea
\mathrm{E}(X) & = & \sum_{i=1}^{n}p \ = \ np \\
\mathrm{Var}(X) & = & \sum_{i=1}^{n}p(1-p) \ = \ np(1-p) \\
\mathrm{Skew}(X) & = & \frac{1-2p}{\sqrt{np(1-p)}} \\
\mathrm{Kurt}(X) & = & \frac{1-6p(1-p)}{np(1-p)} \ .
\eea
The results for $\mathrm{E}(X)$ and $\mathrm{Var}(X)$ are based on
the rules (\ref{eq:sumexpv}) and (\ref{eq:sumvar}), the latter of 
which applies to a set of mutually stochastically independent
random variables.

\medskip
\noindent
\underline{\R:} $\texttt{dbinom}(x,n,p)$, $\texttt{pbinom}(x,n,p)$, 
$\texttt{qbinom}(\alpha,n,p)$,
$\texttt{rbinom}(n_{\mathrm{simulations}},n,p)$ \\
\underline{GDC:} \texttt{binompdf}$(n,p,x)$,
\texttt{binomcdf}$(n,p,x)$ \\
\underline{EXCEL, OpenOffice:} \texttt{BINOM.DIST} (dt.:
\texttt{BINOM.VERT}, \texttt{BINOMVERT}), \texttt{BINOM.INV} (for 
$\alpha$--quantiles)

\section[Hypergeometric distribution]{Hypergeometric distribution}
\lb{sec:hypgeomverteil}
The \textbf{hypergeometric distribution} for a discrete 
one-dimensional random variable $X$ derives from an urn model with 
$M$ black balls and $N-M$ white balls, and the random selection of 
$n$ balls from a total of $N$ ($n \leq N$), without repetition. If 
$X$ represents the number of black balls amongst the $n$ selected 
balls, it is subject to the three-parameter probability 
distribution
\be
X \sim H(n,M,N) \ .
\ee
In particular, this model forms the mathematical basis of the 
internationally popular National Lottery ``6 out of 49,'' in 
which case there are $M=6$ winning numbers amongst a total of 
$N=49$ numbers, and $X \in \left\{0, 1, \ldots, 6\right\}$ counts 
the total of correctly guessed winning numbers on an individual 
gambler's lottery ticket.

\medskip
\noindent
Spectrum of values:
\be
X \mapsto x \in \left\{\max(0,n-(N-M)), \ldots,
\min(n,M)\right\} \ .
\ee
Probability function:
\be
\lb{eq:hypergeomprob}
\fbox{$\displaystyle
P(X=x) = \frac{\left(\begin{array}{c}
               M \\
               x
               \end{array}\right)
         \left(\begin{array}{c}
               N-M \\
               n-x
               \end{array}\right)}{
               \left(\begin{array}{c}
               N \\
               n
               \end{array}\right)} \ ;
$}
\ee
its graph is shown in Fig.~\ref{fig:hypergeomprob} below for 
the National Lottery example, so $n=6$, $M=6$ and $N=49$.
\begin{figure}[!htb]
\begin{center}
\includegraphics[scale=0.8]{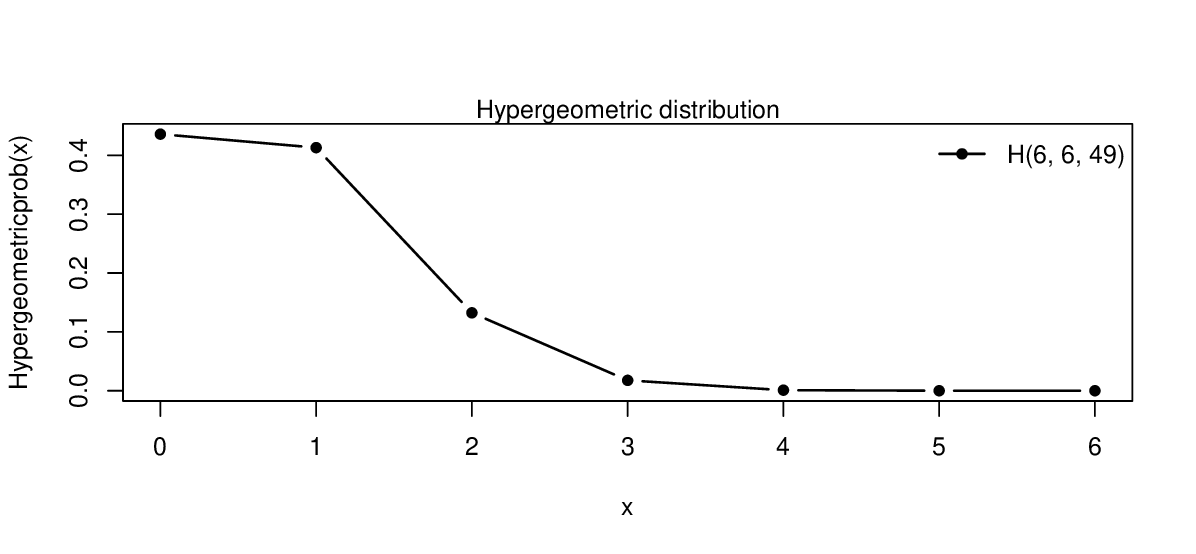}
\end{center}
\caption{Probability function of the hypergeometric distribution 
according to Eq.~(\ref{eq:hypergeomprob}) for the case 
$H\left(6,6,49\right)$. An enveloping line is also shown.}
\lb{fig:hypergeomprob}
\end{figure}

\medskip
\noindent
Cumulative distribution function (\texttt{cdf}):
\be
\fbox{$\displaystyle
F_{X}(x) = P(X \leq x)
= \sum_{k=\max(0,n-(N-M))}^{\left\lfloor x\right\rfloor}
\frac{\left(\begin{array}{c}
               M \\
               k
               \end{array}\right)
         \left(\begin{array}{c}
               N-M \\
               n-k
               \end{array}\right)}{
               \left(\begin{array}{c}
               N \\
               n
               \end{array}\right)} \ .
$}
\ee
Expectation value and variance:
\bea
\mathrm{E}(X) & = & n\,\frac{M}{N} \\
\mathrm{Var}(X) & = & n\,\frac{M}{N}\left(1-\frac{M}{N}\right)
\left(\frac{N-n}{N-1}\right) \ .
\eea
For skewness and excess kurtosis, see, e.g., Rinne
(2008)~\ct[p~270]{rin2008}.

\medskip
\noindent
\underline{\R:} $\texttt{dhyper}(x,M,N-M,n)$,
$\texttt{phyper}(x,M,N-M,n)$,
$\texttt{qhyper}(\alpha,M,N-M,n)$, \\
$\texttt{rhyper}(n_{\mathrm{simulations}},M,N-M,n)$ \\
\underline{EXCEL, OpenOffice:} \texttt{HYPGEOM.DIST}
(dt.: \texttt{HYPGEOM.VERT}, \texttt{HYPGEOMVERT})

\section[Poisson distribution]{Poisson distribution}
\lb{sec:poissonverteil}
The one-parameter \textbf{Poisson distribution} for a discrete 
one-dimensional random variable $X$,
\be
X \sim Pois(\lambda) \ .
\ee
plays a major role in analysing \textbf{count data} when the
maximum number of possible counts associated with a corresponding
data-generating process is unknown. This distribution is named
after the French mathematician, engineer, and physicist 
\href{http://www-history.mcs.st-and.ac.uk/Biographies/Poisson.html}{Baron
Sim\'{e}on Denis Poisson FRSFor HFRSE MIF (1781--1840)} and can be
considered a special case of the binomial distribution, discussed
in Sec.~\ref{sec:binomverteil}, when $n$ is very large ($n \gg 1$)
and $p$ is very small ($0 < p \ll 1$); cf. Sivia and Skilling
(2006)~\ct[Sec.~5.4]{sivski2006}.

\medskip
\noindent
Spectrum of values:
\be
X \mapsto x \in \left\{0, \ldots, n\right\} \ , 
\quad\quad\text{with}\quad n \in \mathbb{N} \ . \ .
\ee
Probability function:
\be
\lb{eq:poissonprob}
\fbox{$\displaystyle
P(X=x) = \frac{\lambda^{x}}{x!}\exp\left(-\lambda\right) \ ,
\quad\text{with}\quad \lambda \in \mathbb{R}_{>0}\ ;
$}
\ee
$\lambda$ is a dimensionless rate parameter. It is also  referred
to as the intensity parameter. The graph of the probability
function is shown in Fig.~\ref{fig:poissonprob} below for the case
$\displaystyle\lambda=\frac{3}{2}$.
\begin{figure}[!htb]
\begin{center}
\includegraphics[scale=0.8]{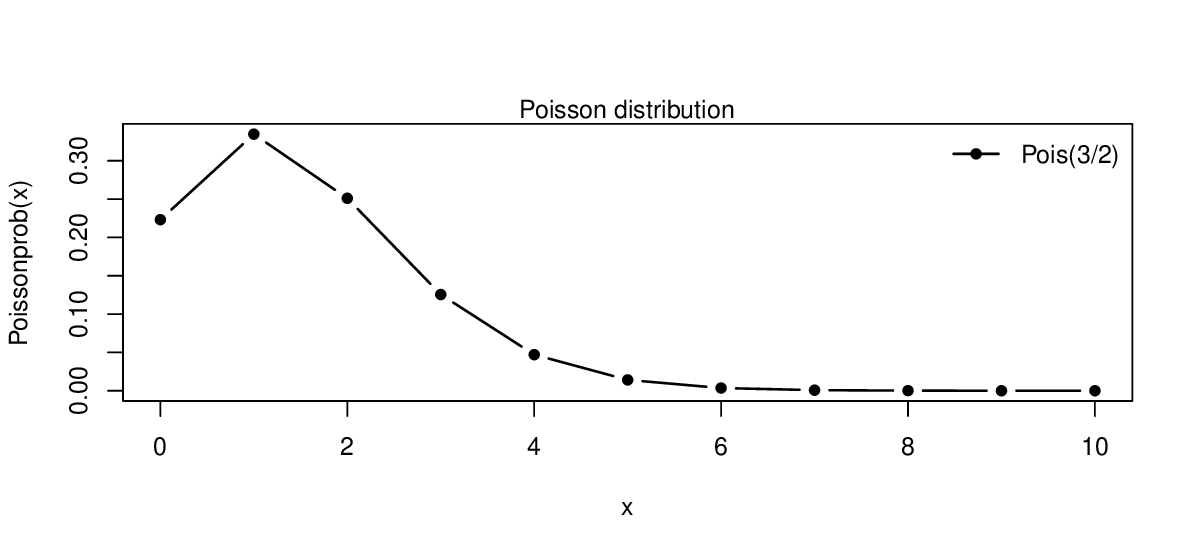}
\end{center}
\caption{Probability function of the Poisson distribution 
according to Eq.~(\ref{eq:poissonprob}) for the case 
$\displaystyle Pois\left(\frac{3}{2}\right)$. An enveloping line is
also shown.}
\lb{fig:poissonprob}
\end{figure}

\medskip
\noindent
Cumulative distribution function (\texttt{cdf}):
\be
\fbox{$\displaystyle
F_{X}(x) = P(X \leq x)
= \left(\sum_{k=0}^{\left\lfloor x\right\rfloor}
\frac{\lambda^{k}}{k!}\right)\exp\left(-\lambda\right) \ .
$}
\ee
Expectation value, variance, skewness and excess kurtosis (cf. 
Rinne (2008)~\ct[p~285f]{rin2008}):\footnote{Note that for a
binomial distribution, cf. Sec.~\ref{sec:binomverteil}, in the
limit that $n \gg 1$ while simultaneously $0 < p \ll 1$ it holds
that $np \approx np(1-p)$, and so the corresponding expectation
value and variance become more and more equal.}
\bea
\mathrm{E}(X) & = & \lambda \\
\mathrm{Var}(X) & = & \lambda \\
\mathrm{Skew}(X) & = & \frac{1}{\sqrt{\lambda}} \\
\mathrm{Kurt}(X) & = & \frac{1}{\lambda} \ .
\eea

\medskip
\noindent
\underline{\R:} $\texttt{dpois}(x,\lambda)$,
$\texttt{ppois}(x,\lambda)$, $\texttt{qpois}(\alpha,\lambda)$,
$\texttt{rpois}(n_{\mathrm{simulations}},\lambda)$ \\
\underline{EXCEL, OpenOffice:} \texttt{POISSON.DIST}
(dt.: \texttt{POISSON.VERT}), \texttt{POISSON}

\section[Continuous uniform distribution]{Continuous uniform
distribution}
\lb{sec:sgleichverteil}
The simplest example of a probability distribution for a 
continuous one-dimensional random variable $X$ is the
\textbf{continuous uniform distribution},
\be
X \sim U(a;b) \ ,
\ee
also referred to as the \textbf{rectangular distribution}. Its two 
free parameters, $a$ and $b$, denote the limits of $X$'s

\medskip
\noindent
Spectrum of values:
\be
X \mapsto x \in \left[a,b\right]
\subset \mathbb{R} \ .
\ee
Probability density function (\texttt{pdf}):\footnote{It is a nice 
and instructive little exercise, strongly recommended to the 
reader, to go through the details of explicitly computing from 
this simple \texttt{pdf} the corresponding \texttt{cdf}, expectation 
value, variance, skewness and excess kurtosis of $X \sim U(a;b)$.}
\be
\lb{eq:recpdf}
\fbox{$\displaystyle
f_{X}(x) =
\begin{cases}
{\displaystyle \frac{1}{b-a}} &
\text{for}\quad x \in \left[a,b\right] \\
\\
0 & \text{otherwise}
\end{cases} \ ;
$}
\ee
its graph is shown in Fig.~\ref{fig:unifpdf} below for four
different combinations of the parameters $a$ and $b$.
\begin{figure}[!htb]
\begin{center}
\includegraphics[scale=0.8]{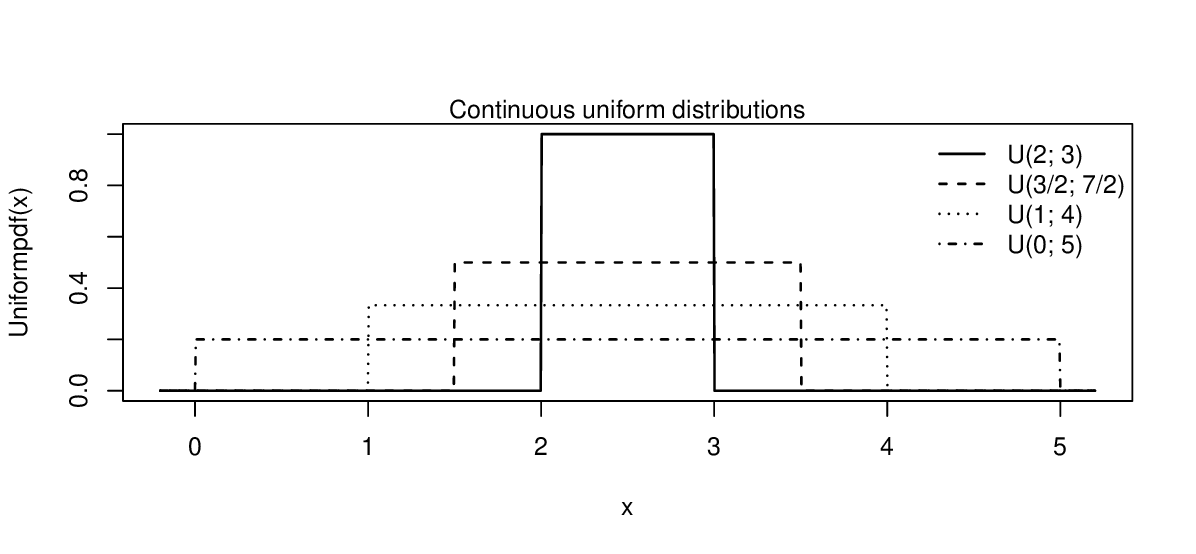}
\end{center}
\caption{\texttt{pdf} of the continuous uniform distribution 
according to Eq.~(\ref{eq:recpdf}) for the cases $U(0;5)$, 
$U(1;4)$, $U(3/2;7/2)$ and $U(2;3)$.}
\lb{fig:unifpdf}
\end{figure}

\medskip
\noindent
Cumulative distribution function (\texttt{cdf}):
\be
\lb{eq:reccdf}
\fbox{$\displaystyle
F_{X}(x) = P(X \leq x) =
\begin{cases}
0 & \text{for}\quad x < a \\ \\
{\displaystyle \frac{x-a}{b-a}} &
\text{for}\quad x \in \left[a,b\right] \\ \\
1 & \text{for}\quad x > b
\end{cases} \ .
$}
\ee
Expectation value, variance, skewness and excess kurtosis:
\bea
\mathrm{E}(X) & = & \frac{a+b}{2} \\
\lb{eq:varUniDistr}
\mathrm{Var}(X) & = & \frac{(b-a)^{2}}{12} \\
\mathrm{Skew}(X) & = & 0 \\
\mathrm{Kurt}(X) & = & -\,\frac{6}{5} \ .
\eea
Using some of these results, as well as Eq.~(\ref{eq:reccdf}), one 
finds that for all continuous uniform distributions the event 
probability
\bea
P(|X-\mathrm{E}(X)| \leq \sqrt{\mathrm{Var}(X)})
& = & P\left(\frac{\sqrt{3}(a+b)-(b-a)}{2\sqrt{3}}
\leq X \leq
\frac{\sqrt{3}(a+b)+(b-a)}{2\sqrt{3}}\right) \nonumber \\
& = & \frac{1}{\sqrt{3}} \ \approx\ 0.5773 \ ,
\eea
i.e., the event probability that $X$ falls within one standard 
deviation (``$1\sigma$'') of $\mathrm{E}(X)$ is $1/\sqrt{3}$. 
$\alpha$--quantiles of continuous uniform distributions are 
obtained by straightforward inversion, i.e., for $0 < \alpha < 1$,
\be
\alpha \stackrel{!}{=} F_{X}(x_{\alpha})
= \frac{x_{\alpha}-a}{b-a}
\qquad\Leftrightarrow\qquad
x_{\alpha} = F_{X}^{-1}(\alpha) = a + \alpha(b-a) \ .
\ee

\medskip
\noindent
\underline{\R:} $\texttt{dunif}(x,a,b)$,
$\texttt{punif}(x,a,b)$, $\texttt{qunif}(\alpha,a,b)$,
$\texttt{runif}(n_{\mathrm{simulations}},a,b)$

\medskip
\noindent
Standardisation of $X \sim U(a;b)$ according to
Eq.~(\ref{eq:standardisation}) yields a one-dimensional random 
variable $Z \sim U(-\sqrt{3};\sqrt{3})$ by
\be
X \rightarrow Z = \sqrt{3}\,\frac{2X-(a+b)}{b-a}
\mapsto z \in \left[-\sqrt{3}, \sqrt{3}\right] \ ,
\ee
with $\texttt{pdf}$
\be
f_{Z}(z) =
\begin{cases}
{\displaystyle \frac{1}{2\sqrt{3}}} & \text{for}\quad z \in
\left[-\sqrt{3}, \sqrt{3}\right] \\ \\
0 & \text{otherwise}
\end{cases} \ ,
\ee
and $\texttt{cdf}$
\be
F_{Z}(z) = P(Z \leq z) =
\begin{cases}
0 & \text{for}\quad z < -\sqrt{3} \\ \\
{\displaystyle \frac{z+\sqrt{3}}{2\sqrt{3}}} &
\text{for}\quad z \in \left[-\sqrt{3},\sqrt{3}\right] \\ \\
1 & \text{for}\quad z > \sqrt{3}
\end{cases} \ .
\ee
%

\section[Gau\ss ian normal distribution]{\href{https://www.youtube.com/watch?v=8fjDkBT641o}{Gau\ss ian normal distribution}}
\lb{sec:normverteil}
The best-known probability distribution for a continuous
one-dimensional random variable $X$, 
which proves ubiquitous in \textbf{Inferential Statistics} (see
Chs. \ref{ch12}  and \ref{ch13} below), is due to 
\href{http://www-groups.dcs.st-and.ac.uk/~history/Biographies/Gauss.html}{Carl Friedrich Gau\ss\ (1777--1855)}; cf. Gau\ss\ 
(1809)~\ct{gau1809}. This is the reproductive two-parameter
\textbf{normal distribution}
\be
X \sim N(\mu;\sigma^{2}) \ ;
\ee
the meaning of the parameters $\mu$ and $\sigma^{2}$ will be 
explained shortly. The extraordinary status of the \textbf{normal 
distribution} in \textbf{Probability Theory} and
\textbf{Statistics} was cemented through the discovery of the
\textbf{central limit theorem} by the French mathematician and
astronomer
\href{http://www-history.mcs.st-and.ac.uk/Biographies/Laplace.html}{Marquis Pierre Simon de Laplace (1749--1827)}, cf. Laplace 
(1809)~\ct{lap1809}; see Sec.~\ref{sec:zentrgrenz} below.

\medskip
\noindent
Spectrum of values:
\be
X \mapsto x \in D \subseteq \mathbb{R} \ .
\ee
Probability density function (\texttt{pdf}):
\be
\lb{glockenpdf}
\fbox{$\displaystyle
f_{X}(x) = \frac{1}{\sqrt{2\pi}\sigma}\exp\left[-\frac{1}{2}
\left(\frac{x-\mu}{\sigma}\right)^{2}\right] \ ,
\quad\text{with}\quad \sigma \in \mathbb{R}_{>0} \ .
$}
\ee
This normal--\texttt{pdf} defines a reflection-symmetric 
characteristic bell-shaped curve, the analytical properties of 
which were first discussed by the French mathematician 
\href{http://www-history.mcs.st-and.ac.uk/Biographies/De_Moivre.html}{Abraham de Moivre (1667--1754)}. The $x$--position of this 
curve's (global) maximum is specified by $\mu$, while the 
$x$--positions of its two points of inflection are given by 
$\mu-\sigma$ resp.~$\mu+\sigma$. The effects of different values 
of the parameters $\mu$ and $\sigma$ on the bell-shaped curve are 
illustrated in Figs.~\ref{fig:norm1pdf} and~\ref{fig:norm2pdf}
below.
\begin{figure}[!htb]
\begin{center}
\includegraphics[scale=0.8]{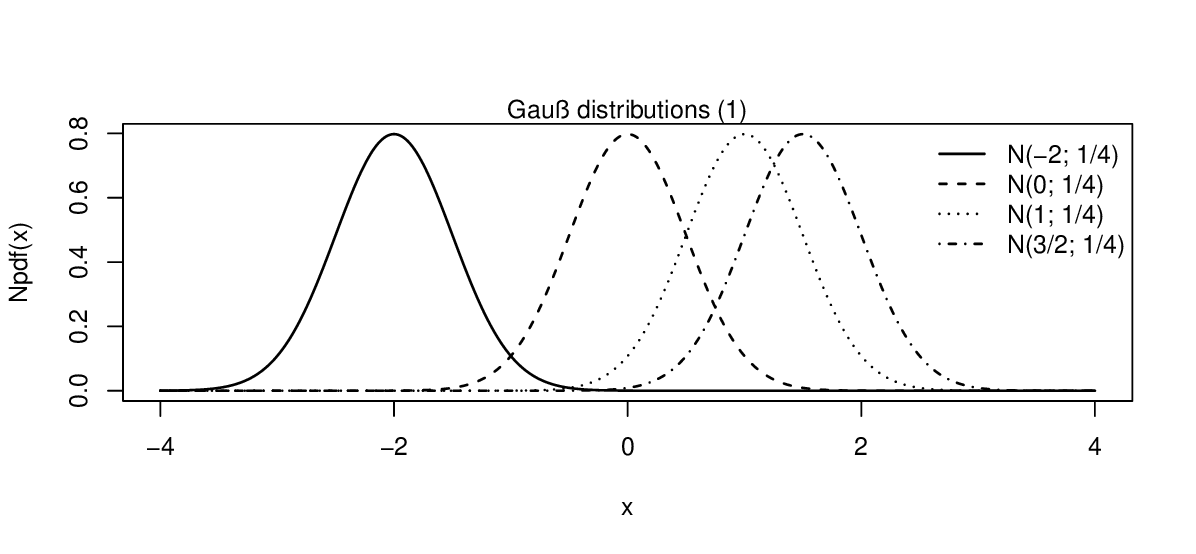}
\end{center}
\caption{\texttt{pdf} of the Gau\ss ian normal distribution
according to Eq.~(\ref{glockenpdf}). Cases $N(-2;1/4)$,
$N(0;1/4)$, $N(1;1/4)$ and $N(3/2;1/4)$, which have
constant~$\sigma$.}
\lb{fig:norm1pdf}
\end{figure}
\begin{figure}[!htb]
\begin{center}
\includegraphics[scale=0.8]{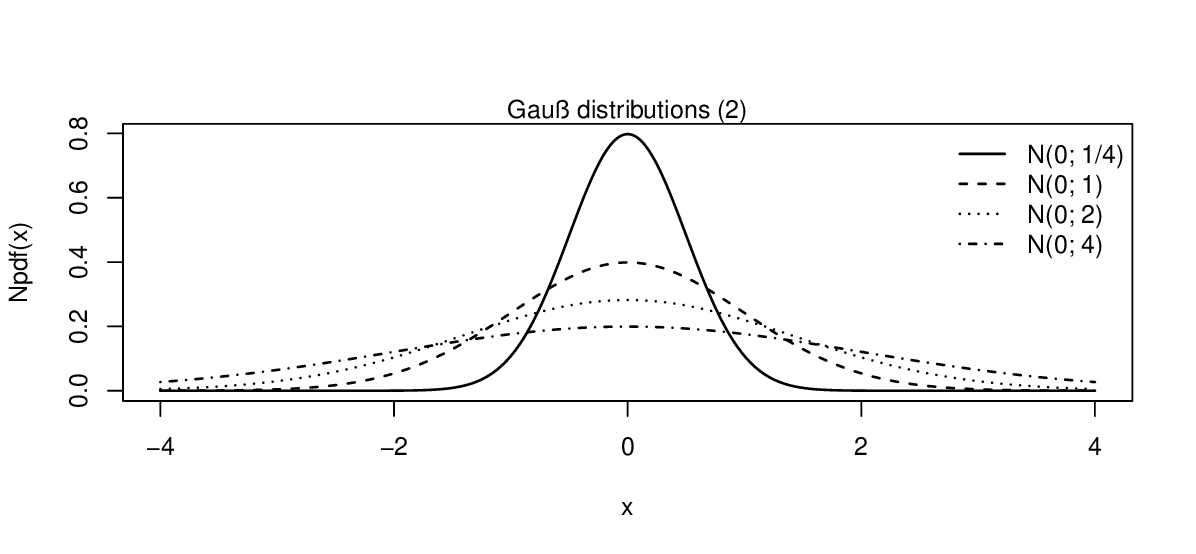}
\end{center}
\caption{\texttt{pdf} of the Gau\ss ian normal distribution
according to Eq.~(\ref{glockenpdf}). Cases $N(0;1/4)$, $N(0;1)$,
$N(0;2)$ and $N(0;4)$, which have constant~$\mu$.}
\lb{fig:norm2pdf}
\end{figure}

\medskip
\noindent
Cumulative distribution function (\texttt{cdf}):
\be
\lb{eq:gaussiancdf}
\fbox{$\displaystyle
F_{X}(x) = P (X \leq x) = \int_{-\infty}^{x}
\frac{1}{\sqrt{2\pi}\sigma}\exp\left[-\frac{1}{2}
\left(\frac{t-\mu}{\sigma}\right)^{2}\right]\mathrm{d}t \ .
$}
\ee
We emphasise the fact that the normal--\texttt{cdf} \textit{cannot}
be expressed in terms of elementary mathematical functions.

\medskip
\noindent
Expectation value, variance, skewness and excess kurtosis (cf. 
Rinne (2008)~\ct[p~301]{rin2008}):
\bea
\mathrm{E}(X) & = & \mu \\
\mathrm{Var}(X) & = & \sigma^{2} \\
\mathrm{Skew}(X) & = & 0 \\
\mathrm{Kurt}(X) & = & 0 \ .
\eea

\medskip
\noindent
\underline{\R:} $\texttt{dnorm}(x,\mu,\sigma)$,
$\texttt{pnorm}(x,\mu,\sigma)$,
$\texttt{qnorm}(\alpha,\mu,\sigma)$,
$\texttt{rnorm}(n_{\mathrm{simulations}},\mu,\sigma)$ \\
\underline{GDC:} \texttt{normalpdf}$(x,\mu,\sigma)$,
\texttt{normalcdf}$(-\infty,x,\mu,\sigma)$ \\
\underline{EXCEL, OpenOffice:} \texttt{NORM.DIST} (dt.:
\texttt{NORM.VERT}, \texttt{NORMVERT})

\vspace{5mm}
\noindent
Upon standardisation of a normally distributed one-dimensional 
random variable $X$ according to Eq.~(\ref{eq:standardisation}), 
the cor\-responding normal distribution $N(\mu;\sigma^{2})$ is 
transformed into the unique \textbf{standard normal distribution}, 
$N(0;1)$, with

\medskip
\noindent
Probability density function (\texttt{pdf}):
\be
\lb{sglockenpdf}
\fbox{$\displaystyle
\varphi(z) := \frac{1}{\sqrt{2\pi}}\exp\left[-\frac{1}{2}\,z^{2}
\right] \quad\text{for}\quad
z \in \mathbb{R} \ ;
$}
\ee
its graph is shown in Fig.~\ref{fig:snormpdf} below.
\begin{figure}[!htb]
\begin{center}
\includegraphics[scale=0.8]{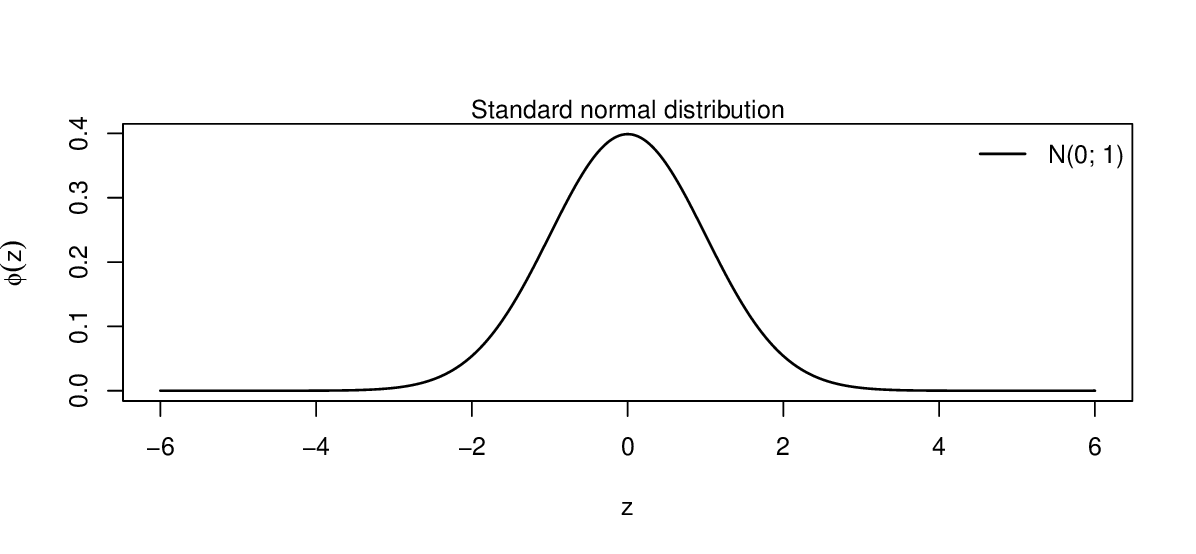}
\end{center}
\caption{\texttt{pdf} of the standard normal distribution according 
to Eq.~(\ref{sglockenpdf}).}
\lb{fig:snormpdf}
\end{figure}

\medskip
\noindent
Cumulative distribution function (\texttt{cdf}):
\be
\fbox{$\displaystyle
\Phi(z) := P(Z \leq z) = \int_{-\infty}^{z}
\frac{1}{\sqrt{2\pi}}\exp\left[-\frac{1}{2}\,t^{2}\right]
\mathrm{d}t \ .
$}
\ee

\medskip
\noindent
\underline{\R:} $\texttt{dnorm}(z)$,
$\texttt{pnorm}(z)$, $\texttt{qnorm}(\alpha)$,
$\texttt{rnorm}(n_{\mathrm{simulations}})$ \\
\underline{EXCEL:} \texttt{NORM.S.DIST} (dt.: \texttt{NORM.S.VERT})

\vspace{5mm}
\noindent
The resultant random variable $Z \sim N(0;1)$ satisfies the

\medskip
\noindent
Computational rules:
\bea
P(Z \leq b) & = & \Phi(b) \\
P(Z \geq a) & = & 1-\Phi(a) \\
P(a \leq Z \leq b) & = & \Phi(b) - \Phi(a) \\
\Phi(-z) & = & 1 - \Phi(z) \\
\lb{eq:symzint}
P(-z \leq Z \leq z) & = & 2\Phi(z) - 1 \ .
\eea
The event probability that a (standard) normally distributed 
one-dimensional random variable 
takes values inside an interval of length $k$ times two standard 
deviations, 
centred on its expectation value, is given by the important
$\boldsymbol{k}\boldsymbol{\sigma}$\textbf{--rule}. This states
that 
\be
P(|X-\mu| \leq k\sigma) 
\overbrace{=}^{{\text{Eq.~(\ref{eq:standardisation})}}}
P(-k \leq Z \leq +k)
\overbrace{=}^{{\text{Eq.~(\ref{eq:symzint})}}}
2\Phi(k) - 1 \quad \text{for}\quad k>0 \ .
\ee
According to this rule, the event probability of a normally 
distributed one-dimensional random variable to deviate from its 
mean by \textit{more than six standard deviations} amounts to
\be
P(|X-\mu| > 6\sigma) = 2\left[1-\Phi(6)\right]
\approx 1.97 \times 10^{-9} \ ,
\ee
i.e., about two parts in one billion. Thus, in this scenario 
the occurrence of extreme \textbf{outliers} for~$X$ is practically 
impossible. In turn, the persistent occurrence of so-called 
$\boldsymbol{6\sigma}$\textbf{--events}, or larger deviations from
the mean, in quantitative statistical surveys can be interpreted as
evidence \textit{against} the assumption of an underlying
Gau\ss ian random process; cf. Taleb (2007)~\ct[Ch.~15]{tal2007}.

\medskip
\noindent
The rapid, accelerated decline in the event probabilities for 
deviations from the mean of a  Gau\ss ian normal distribution can 
be related to the fact that the elasticity of the 
standard normal--\texttt{pdf} is given by (cf. 
Ref.~\ct[Sec.~7.6]{hve2009})
\be
\lb{eq:phielasts}
\varepsilon_{\varphi}(z) = -\,z^{2} \ .
\ee
Manifestly this is negative for all $z \neq 0$ and increases
non-linearly in absolute value as one moves away from $z=0$.

\medskip
\noindent
$\alpha$--quantiles associated with $Z \sim N(0;1)$ are obtained
from the inverse standard normal--\texttt{cdf} according to
\be
\alpha \stackrel{!}{=} P(Z \leq z_{\alpha}) = \Phi(z_{\alpha})
\qquad\Leftrightarrow\qquad
z_{\alpha} = \Phi^{-1}(\alpha)
\quad\text{for\ all}\quad 0 < \alpha < 1 \ .
\ee
Due to the reflection symmetry of $\varphi(z)$ with respect to the
vertical axis at $z=0$, it holds that
\be
z_{\alpha} = -z_{1-\alpha} \ .
\ee
For this reason, one typically finds $z_{\alpha}$-values listed in 
textbooks on \textbf{Statistics} only for $\alpha \in [1/2,1)$. 
Alternatively, a particular $z_{\alpha}$ may be obtained from \R,
a GDC, EXCEL, or from OpenOffice. The backward transformation from
a particular $z_{\alpha}$ of the standard normal distribution to
the corresponding~$x_{\alpha}$ of a given normal distribution
follows from Eq.~(\ref{eq:standardisation}) and amounts to
$x_{\alpha} = \mu+z_{\alpha}\sigma$.

\medskip
\noindent
\underline{\R:} $\texttt{qnorm}(\alpha)$ \\
\underline{GDC:} \texttt{invNorm}$(\alpha)$ \\
\underline{EXCEL, OpenOffice:} \texttt{NORM.S.INV} (dt.:
\texttt{NORM.S.INV}, \texttt{NORMINV})

\vspace{5mm}
\noindent
At this stage, a few historical remarks are in order. The
Gau\ss ian normal distribution gained a prominent, though in 
parts questionable status in the \textbf{Social Sciences} through
the highly influential work of the Belgian astronomer, 
mathematician, statistician and sociologist 
\href{http://www-history.mcs.st-and.ac.uk/Biographies/Quetelet.html}{Lambert Adolphe Jacques Quetelet (1796--1874)} during the 
$19^\mathrm{th}$ Century. In particular, his research programme on 
the generic properties of \textit{l'homme moyen} (engl.: the
average man), see Quetelet (1835)~\ct{que1835}, an ambitious and to
some extent obsessive attempt to quantify and classify
physiological and sociological human characteristics according to
the principles of a normal distribution, left a lasting impact on
the field, with repercussions to this day. Quetelet, by the way,
co-founded the \href{http://www.rss.org.uk}{Royal Statistical
Society (\texttt{rss.org.uk})} in 1834. Further visibility was
given to Quetelet's ideas at the time by a contemporary, the
English empiricist 
\href{http://www-history.mcs.st-and.ac.uk/Biographies/Galton.html}{Sir Francis Galton FRS (1822--1911)}, whose intense studies on 
heredity in Humans, see Galton (1869)~\ct{gal1869}, which he later 
subsumed under the term ``eugenics,'' complemented Quetelet's 
investigations, and profoundly shaped subsequent developments in 
social research; cf. Bernstein (1998)~\ct[Ch.~9]{ber1998}. 
Incidently, amongst many other contributions to the field, 
Galton's activities helped to pave the way for making
\textbf{questionnaires} and \textbf{surveys} a commonplace for
collecting statistical data from Humans.

\vspace{5mm}
\noindent
The (standard) normal distribution, as well as the next three 
examples of probability distributions for a continuous
one-dimensional random variable $X$, are commonly referred to as
the \textbf{test distributions}, due to the central roles they play
in null hypothesis significance testing (cf. Chs. \ref{ch12} and
\ref{ch13}).

\section[$\chi^{2}$--distribution]{
$\boldsymbol{\chi}^{2}$--distribution with $n$ degrees of freedom}
\lb{sec:chi2verteil}
The reproductive one-parameter
$\boldsymbol{\chi}^{2}$\textbf{--distribution with}
$\boldsymbol{n}$ \textbf{degrees of freedom} was
devised by the English mathematical statistician 
\href{http://www-history.mcs.st-and.ac.uk/Biographies/Pearson.html}{Karl Pearson FRS (1857--1936)}; cf. Pearson (1900)~\ct{pea1900}. 
The underlying continuous one-dimensional random variable
\be
\fbox{$\displaystyle
X \sim \chi^{2}(n) \ ,
$}
\ee
is perceived of as the sum of squares of $n$ stochastically 
independent, identically standard normally distributed 
(``i.i.d.'') random variables $Z_{i} \sim N(0;1)$ 
($i=1,\ldots,n$), i.e.,
\be
X:=\sum_{i=1}^{n}Z_{i}^{2}
= Z_{1}^{2} + \ldots + Z_{n}^{2} \ ,
\quad\text{with}\quad n \in \mathbb{N} \ .
\ee

\medskip
\noindent
Spectrum of values:
\be
X \mapsto x \in D \subseteq \mathbb{R}_{\geq 0} \ .
\ee
The probability density function (\texttt{pdf}) of a 
$\chi^{2}$--distribution with $df=n$ degrees of freedom is a 
fairly complicated mathematical expression; see Rinne 
(2008)~\ct[p~319]{rin2008} or Ref.~\ct[Eq.~(3.26)]{hve2018} for the
explicit representation of the $\chi^{2}$\texttt{pdf}. Plots are
shown for four different values of the parameter $n$ in
Fig.~\ref{fig:chi2pdf}. The $\chi^{2}$\texttt{cdf} \textit{cannot}
be expressed in terms of elementary mathematical functions.
\begin{figure}[!htb]
\begin{center}
\includegraphics[scale=0.8]{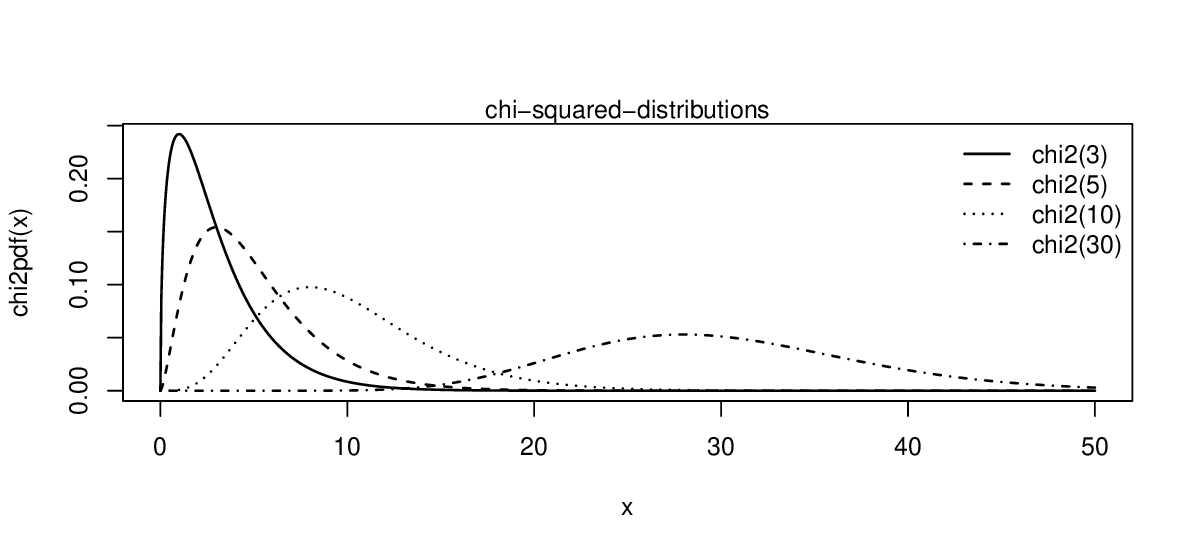}
\end{center}
\caption{\texttt{pdf} of the $\chi^{2}$--distribution for $df=n \in 
\{3, 5, 10, 30\}$ degrees of freedom.}
\lb{fig:chi2pdf}
\end{figure}

\medskip
\noindent
Expectation value, variance, skewness and excess kurtosis (cf. 
Rinne (2008)~\ct[p~320f]{rin2008}):
\bea
\mathrm{E}(X) & = & n \\
\mathrm{Var}(X) & = & 2n \\
\mathrm{Skew}(X) & = & \sqrt{\frac{8}{n}} \\
\mathrm{Kurt}(X) & = & \frac{12}{n} \ .
\eea
$\alpha$--quantiles, $\chi^{2}_{n;\alpha}$, of 
$\chi^{2}$--distributions are generally tabulated in textbooks on 
\textbf{Statistics}. Alternatively, they may be obtained from \R,
EXCEL, or from OpenOffice.

\medskip
\noindent
Note that for $n \geq 50$ a $\chi^{2}$--distribution may be 
approximated reasonably well by a normal distribution, $N(n,2n)$. 
This is a reflection of the \textbf{central limit theorem}, to be 
discussed in Sec.~\ref{sec:zentrgrenz} below.

\medskip
\noindent
\underline{\R:} $\texttt{dchisq}(x,n)$, $\texttt{pchisq}(x,n)$,
$\texttt{qchisq}(\alpha,n)$,
$\texttt{rchisq}(n_{\mathrm{simulations}},n)$ \\
\underline{GDC:} $\chi^{2}\texttt{pdf}(x,n)$,
$\chi^{2}\texttt{cdf}(0,x,n)$ \\
\underline{EXCEL, OpenOffice:} \texttt{CHISQ.DIST},
\texttt{CHISQ.INV} (dt.: \texttt{CHIQU.VERT}, \texttt{CHIQVERT}, \\
\texttt{CHIQU.INV}, \texttt{CHIQINV})

\section[$t$--distribution]{$\boldsymbol{t}$--distribution with
$n$ degrees of freedom}
\lb{sec:tverteil}
The non-reproductive one-parameter
$\boldsymbol{t}$\textbf{--distribution with} $\boldsymbol{n}$
\textbf{degrees of freedom} was discovered by the English
statistician
\href{http://www-history.mcs.st-and.ac.uk/Biographies/Gosset.html}{William Sealy Gosset (1876--1937)}. Intending to some extent to irritate
the scientific community, he published his findings under the 
pseudonym of ``Student;'' cf. Student (1908)~\ct{stu1908}. 
Consider two stochastically independent one-dimensional random 
variables, $Z \sim N(0;1)$ and $X \sim \chi^{2}(n)$, satisfying 
the indicated distribution laws. Then the quotient random variable 
defined by
\be
\fbox{$\displaystyle
T:=\frac{Z}{\sqrt{X/n}} \sim t(n) \ ,
\quad\text{with}\quad n \in \mathbb{N} \ ,
$}
\ee
is $t$--distributed with $df=n$ degrees of freedom.

\medskip
\noindent
Spectrum of values:
\be
T \mapsto t \in D \subseteq \mathbb{R} \ .
\ee
The probability density function (\texttt{pdf}) of a 
$t$--distribution, which exhibits a reflection symmetry with 
respect to the vertical axis at $t=0$, is a fairly complicated 
mathematical expression; see Rinne (2008)~\ct[p~326]{rin2008}
or Ref.~\ct[Eq.~(2.26)]{hve2018} for the explicit representation of
the $t$\texttt{pdf}. Plots are shown for four different values of
the parameter $n$ in Fig.~\ref{fig:tpdf}. The $t$\texttt{cdf}
\textit{cannot} be expressed in terms of elementary mathematical
functions.
\begin{figure}[!htb]
\begin{center}
\includegraphics[scale=0.8]{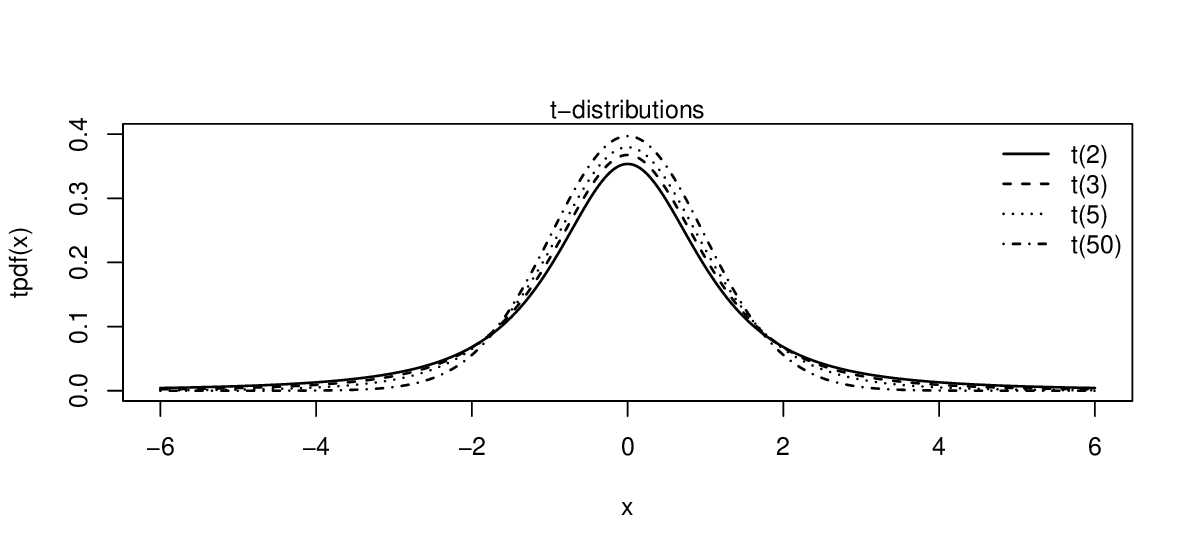}
\end{center}
\caption{\texttt{pdf} of the $t$--distribution for $df=n \in \{2,
3, 5, 50\}$ degrees of freedom. For the case $t(50)$, the
$t$\texttt{pdf} is essentially equivalent to the standard
normal~\texttt{pdf}. Notice the fatter tails of the 
$t$\texttt{pdf} for small values of $n$.}
\lb{fig:tpdf}
\end{figure}

\medskip
\noindent
Expectation value, variance, skewness and excess kurtosis (cf. 
Rinne (2008)~\ct[p~327]{rin2008}):
\bea
\mathrm{E}(X) & = & 0 \\
\mathrm{Var}(X) & = & \frac{n}{n-2}
\quad\text{for}\quad n > 2 \\
\mathrm{Skew}(X) & = & 0
\quad\text{for}\quad n > 3 \\
\mathrm{Kurt}(X) & = & \frac{6}{n-4}
\quad\text{for}\quad n > 4 \ .
\eea
$\alpha$--quantiles, $t_{n;\alpha}$, of $t$--distributions, for 
which, due to the reflection symmetry of the $t$\texttt{pdf}, the 
identity $t_{n;\alpha}=-t_{n;1-\alpha}$ holds, are generally 
tabulated in textbooks on \textbf{Statistics}. Alternatively, they 
may be obtained from \R, some GDCs, EXCEL, or from OpenOffice.

\medskip
\noindent
Note that for $n \geq 50$ a $t$--distribution may be 
approximated reasonably well by the standard normal distribution, 
$N(0;1)$. Again, this is a manifestation of the \textbf{central
limit theorem}, to be discussed in Sec.~\ref{sec:zentrgrenz} below.
For $n=1$, a $t$--distribution amounts to the special case
$a=1$, $b=0$ of the Cauchy distribution; cf.
Sec.~\ref{sec:cauchyverteil}.

\medskip
\noindent
\underline{\R:} $\texttt{dt}(x,n)$, $\texttt{pt}(x,n)$,
$\texttt{qt}(\alpha,n)$,
$\texttt{rt}(n_{\mathrm{simulations}},n)$ \\
\underline{GDC:} $\texttt{tpdf}(t,n)$, $\texttt{tcdf}(-10,t,n)$,
$\texttt{invT}(\alpha,n)$ \\
\underline{EXCEL, OpenOffice:} \texttt{T.DIST}, \texttt{T.INV}
(dt.: \texttt{T.VERT}, \texttt{TVERT}, \texttt{T.INV},
\texttt{TINV})

\section[$F$--distribution]{$\boldsymbol{F}$--distribution with
$n_{1}$ and $n_{2}$ degrees of freedom}
\lb{sec:fverteil}
The reproductive two-parameter
$\boldsymbol{F}$\textbf{--distribution with}
$\boldsymbol{n_{1}}$ \textbf{and} $\boldsymbol{n_{2}}$
\textbf{degrees of freedom} was made prominent in
\textbf{Statistics} by the English statistician, evolutionary
biologist, eugenicist and geneticist
\href{http://www-history.mcs.st-and.ac.uk/Biographies/Fisher.html}{Sir
Ronald Aylmer Fisher FRS (1890--1962)}, and the
US-American mathematician and statistician
\href{http://en.wikipedia.org/wiki/George_W._Snedecor}{George
Waddel Snedecor (1881--1974)}; cf. Fisher (1924)~\ct{fis1924} and 
Snedecor (1934)~\ct{sne1934}. Consider two sets of stochastically 
independent, identically standard normally distributed 
(``i.i.d.'') one-dimensional random variables, $X_{i} \sim N(0;1)$
($i=1,\ldots,n_{1}$), and $Y_{j} \sim N(0;1)$
($j=1,\ldots,n_{2}$). Define the sums
\be
X:=\sum_{i=1}^{n_{1}}X_{i}^{2}
\qquad\text{and}\qquad
Y:=\sum_{j=1}^{n_{2}}Y_{j}^{2} \ ,
\ee
each of which satisfies a $\chi^{2}$--distribution with $n_{1}$ 
resp.~$n_{2}$ degrees of freedom. Then the quotient random variable
\be
\fbox{$\displaystyle
F_{n_{1},n_{2}} := \frac{X/n_{1}}{Y/n_{2}} \sim F(n_{1},n_{2}) \ ,
\quad\text{with}\quad n_{1},n_{2} \in \mathbb{N} \ ,
$}
\ee
is $F$--distributed with $df_{1}=n_{1}$ and $df_{2}=n_{2}$ degrees 
of freedom.

\medskip
\noindent
Spectrum of values:
\be
F_{n_{1},n_{2}} \mapsto f_{n_{1},n_{2}}
\in D \subseteq \mathbb{R}_{\geq 0} \ .
\ee
The probability density function (\texttt{pdf}) of an 
$F$--distribution is quite a complicated 
mathematical expression; see Rinne (2008)~\ct[p~330]{rin2008} for
the explicit representation of the $F$\texttt{pdf}. Plots are 
shown for four different combinations of the parameters $n_{1}$ 
and $n_{2}$ in Fig.~\ref{fig:Fpdf}. The $F$\texttt{cdf}
\textit{cannot} be expressed in terms of elementary mathematical
functions.
\begin{figure}[!htb]
\begin{center}
\includegraphics[scale=0.8]{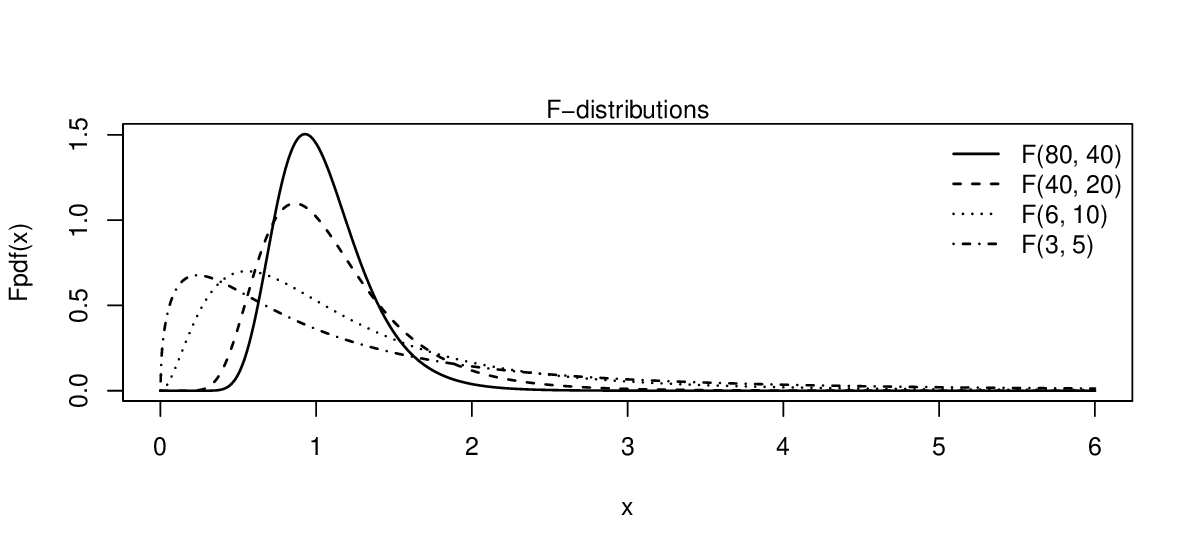}
\end{center}
\caption{\texttt{pdf} of the $F$--distribution for four
combinations of degrees of freedom $(df_{1}=n_{1}, df_{2}=n_{2})$.
The curves correspond to the cases $F(80,40)$, $F(40,20)$,
$F(6,10)$ and $F(3,5)$, respectively.}
\lb{fig:Fpdf}
\end{figure}

\medskip
\noindent
Expectation value, variance, skewness and excess kurtosis (cf. 
Rinne (2008)~\ct[p~332]{rin2008}):
\bea
\mathrm{E}(X) & = & \frac{n_{2}}{n_{2}-2} \quad\text{for}\quad
n_{2} > 2\\
\mathrm{Var}(X) & = & \frac{2n_{2}^{2}(n_{1}+n_{2}-2)}{
n_{1}(n_{2}-2)^{2}(n_{2}-4)}
\quad\text{for}\quad n_{2} > 4 \\
\mathrm{Skew}(X) & = & \frac{(2n_{1}+n_{2}-2)\sqrt{8(n_{2}-4)}}{
(n_{2}-6)\sqrt{n_{1}(n_{1}+n_{2}-2)}}
\quad\text{for}\quad n_{2} > 6 \\
\mathrm{Kurt}(X) & = & 
12\,\frac{n_{1}(5n_{2}-22)(n_{1}+n_{2}-2)+(n_{2}-2)^{2}(n_{2}-4)}{
n_{1}(n_{2}-6)(n_{2}-8)(n_{1}+n_{2}-2)}
\quad\text{for}\quad n_{2} > 8 \ .
\eea
$\alpha$--quantiles, $f_{n_{1},n_{2};\alpha}$, of 
$F$--distributions are tabulated in advanced textbooks on 
\textbf{Statistics}. Alternatively, they may be obtained from \R,
EXCEL, or from OpenOffice.

\medskip
\noindent
\underline{\R:} $\texttt{df}(x,n_{1}, n_{2})$,
$\texttt{pf}(x,n_{1}, n_{2})$, $\texttt{qf}(\alpha,n_{1}, n_{2})$,
$\texttt{rf}(n_{\mathrm{simulations}},n_{1}, n_{2})$ \\
\underline{GDC:} $F\texttt{pdf}(x,n_{1},n_{2})$,
$F\texttt{cdf}(0,x,n_{1},n_{2})$ \\
\underline{EXCEL, OpenOffice:} \texttt{F.DIST}, \texttt{F.INV}
(dt.: \texttt{F.VERT}, \texttt{FVERT}, \texttt{F.INV},
\texttt{FINV})

\section[Pareto distribution]{Pareto distribution}
\lb{sec:paretodistr}
When studying the distribution of wealth and income of people in 
Italy towards the end of the $19^\mathrm{th}$ Century, the Italian 
engineer, sociologist, economist, political scientist and 
philosopher
\href{http://en.wikipedia.org/wiki/Vilfredo_Pareto}{Vilfredo
Federico Damaso Pareto (1848--1923)} discovered a certain type of 
quantitative regularity which he could model mathematically in 
terms of a simple power-law function involving only two free 
parameters; cf. Pareto (1896)~\ct{par1896}. The one-dimensional 
random variable~$X$ underlying such a \textbf{Pareto distribution},
\be
X \sim Par(\gamma,x_\mathrm{min}) \ ,
\ee
has a

\medskip
\noindent
Spectrum of values:
\be
X \mapsto x \in \{x|x \geq x_\mathrm{min}\} \subset \mathbb{R}_{>0}
\ ,
\ee
and a

\medskip
\noindent
Probability density function (\texttt{pdf}):
\be
\lb{paretopdf}
\fbox{$\displaystyle
f_{X}(x)=
\begin{cases}
0 & \text{for}\quad x < x_\mathrm{min} \\
\\
{\displaystyle \frac{\gamma}{x_\mathrm{min}}
\left(\frac{x_\mathrm{min}}{x}\right)^{\gamma+1}} \ ,
\quad \gamma \in \mathbb{R}_{>0}
& \text{for}\quad x \geq x_\mathrm{min}
\end{cases} \ ;
$}
\ee
its graph is shown in Fig.~\ref{fig:parpdf} below for four 
different values of the dimensionless exponent~$\gamma$.
\begin{figure}[!htb]
\begin{center}
\includegraphics[scale=0.8]{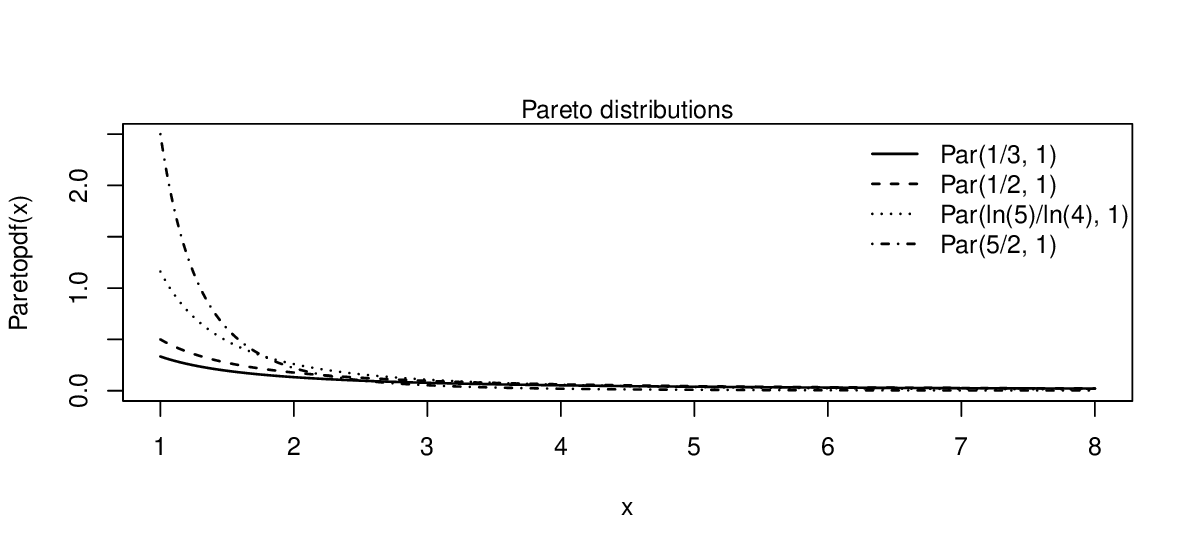}
\end{center}
\caption{\texttt{pdf} of the Pareto distribution according 
to Eq.~(\ref{paretopdf}) for $x_\mathrm{min}=1$ and $\displaystyle 
\gamma \in \left\{\frac{1}{3}, \frac{1}{2}, \frac{\ln(5)}{\ln(4)}, 
\frac{5}{2}\right\}$.}
\lb{fig:parpdf}
\end{figure}

\medskip
\noindent
Cumulative distribution function (\texttt{cdf}):
\be
\fbox{$\displaystyle
F_{X}(x) = P(X \leq x)
=
\begin{cases}
0 & \text{for}\quad x < x_\mathrm{min} \\
\\
{\displaystyle 1-\left(\frac{x_\mathrm{min}}{x}\right)^{\gamma}}
& \text{for}\quad x \geq x_\mathrm{min}
\end{cases} \ .
$}
\lb{paretocdf}
\ee
Expectation value, variance, skewness and excess kurtosis (cf. 
Rinne (2008)~\ct[p~362]{rin2008}):
\bea
\mathrm{E}(X) & = & \frac{\gamma}{\gamma-1}\,x_\mathrm{min}
\qquad\text{for}\quad \gamma > 1 \\
\lb{eq:paretovar}
\mathrm{Var}(X) & = & \frac{\gamma}{(\gamma-1)^{2}(\gamma-2)}\,
x_\mathrm{min}^{2}
\qquad\text{for}\quad \gamma > 2 \\
\mathrm{Skew}(X) & = & 
\frac{2(1+\gamma)}{\gamma-3}\,\sqrt{\frac{\gamma-2}{\gamma}}
\qquad\text{for}\quad \gamma > 3 \\
\mathrm{Kurt}(X) & = & 
\frac{6(\gamma^{3}+\gamma^{2}-6\gamma-2)}{\gamma(\gamma-3)
(\gamma-4)}
\qquad\text{for}\quad \gamma > 4 \ .
\eea
It is important to realise that $\mathrm{E}(X)$, $\mathrm{Var}(X)$, 
$\mathrm{Skew}(X)$ and $\mathrm{Kurt}(X)$ are \textit{well-defined
only} for the values of $\gamma$ indicated; otherwise these
measures do not exist.

\medskip
\noindent
$\alpha$--quantiles:
\be
\lb{alphapareto}
\alpha \stackrel{!}{=} F_{X}(x_{\alpha})
= 1-\left(\frac{x_\mathrm{min}}{x_{\alpha}}\right)^{\gamma}
\ \Leftrightarrow\ 
x_{\alpha} = F_{X}^{-1}(\alpha)
= \sqrt[\gamma]{\frac{1}{1-\alpha}}\,x_\mathrm{min}
\quad\text{for\ all}\quad 0 < \alpha < 1 \ .
\ee

\medskip
\noindent
\underline{\R:}
$\texttt{dpareto}(x, \gamma, x_{\mathrm{min}})$,
$\texttt{ppareto}(x, \gamma, x_{\mathrm{min}})$,
$\texttt{qpareto}(\alpha, \gamma, x_{\mathrm{min}})$, \\
$\texttt{rpareto}(n_{\mathrm{simulations}}, \gamma,
x_{\mathrm{min}})$ (package: {\tt extraDistr}, by Wolodzko
(2018)~\ct{wol2018})

\medskip
\noindent
Note that it follows from Eq.~(\ref{paretocdf}) that the 
probability of a Pareto-distributed continuous one-dimensional 
random variable $X$ to exceed a certain threshold value $x$ is 
given by the simple power-law rule
\be
P(X > x)
= 1 - P(X \leq x)
= \left(\frac{x_\mathrm{min}}{x}\right)^{\gamma} \ .
\ee
Hence, the ratio of probabilities
\be
\frac{P(X>kx)}{P(X>x)}
=\frac{{\displaystyle \left(\frac{x_\mathrm{min}}{kx}
\right)^{\gamma}}}{{\displaystyle
\left(\frac{x_\mathrm{min}}{x}\right)^{\gamma}}}
= \left(\frac{1}{k}\right)^{\gamma} \ ,
\ee
with $k \in \mathbb{R}_{>0}$, is \textbf{scale-invariant}, meaning
independent of a particular scale $x$ at which one observes $X$ 
(cf. Taleb (2007)~\ct[p~256ff and p~326ff]{tal2007}). This 
behaviour is a direct consequence of a special mathematical 
property of Pareto distributions which is technically referred to 
as \textbf{self-similarity}. It is determined by the fact that 
a Pareto--\texttt{pdf} (\ref{paretopdf}) has \textit{constant}
elasticity, i.e. (cf. Ref.~\ct[Sec.~7.6]{hve2009})
\be
\varepsilon_{f_{X}}(x) = -(\gamma+1)
\quad\text{for}\quad x \geq x_\mathrm{min} \ ,
\ee
which contrasts with the case of the standard normal 
distribution; cf. Eq.~(\ref{eq:phielasts}). This feature implies
that in the present scenario the occurrence of extreme
\textbf{outliers} for~$X$ is not entirely unusual.

\medskip
\noindent
Further interesting examples, in various fields of applied science,
of distributions of quantities which also feature the 
scale-invariance of scaling laws are described in 
Wiesenfeld (2001)~\ct{wie2001}. Nowadays, Pareto distributions 
play an important role in the quantitative modelling of financial 
risk; see, e.g., Bouchaud and Potters (2003)~\ct{boupot2003}.

\medskip
\noindent
Working out the equation of the Lorenz curve associated with a 
Pareto distribution according to Eq.~(\ref{lorcurve}), using
Eq.~(\ref{alphapareto}), yields a particularly simple result
given by
\be
L(\alpha;\gamma) = 1 - (1-\alpha)^{1-(1/\gamma)} \ .
\ee
This result forms the basis of Pareto's famous \textbf{80/20 rule} 
concerning concentration in the distribution of various assets of 
general importance in a given population. According to Pareto's 
empirical findings, typically 80\% of such an asset are owned by 
just 20\% of the population considered (and vice versa); cf. 
Pareto (1896)~\ct{par1896}.\footnote{See also footnote 2 in 
Sec.~\ref{subsec:gini}.} The \textbf{80/20 rule} applies exactly
for a value of the power-law index of ${\displaystyle\gamma = 
\frac{\ln(5)}{\ln(4)}} \approx 1.16$. It is a prominent example of 
the phenomenon of \textbf{universality}, frequently observed in the 
mathematical modelling of quantitative--empirical relationships 
between variables in a wide variety of scientific disciplines; cf. 
Gleick (1987)~\ct[p~157ff]{gle1987}.

\medskip
\noindent
For purposes of numerical simulation it is useful to work with a 
\textbf{truncated Pareto distribution}, for which the
one-dimensional random variable~$X$ takes values in an interval
$\left[x_\mathrm{min},x_\mathrm{cut}\right] \subset
\mathbb{R}_{>0}$. Samples of 
random values for such an $X$ can be easily generated from a 
one-dimensional random variable $Y$ that is uniformly distributed 
on the interval $\left[0,1\right]$. The sample values of the latter
are subsequently transformed according to the formula; cf. 
Ref.~\ct{wol2015}:
\be
x(y) = \frac{x_\mathrm{min}x_\mathrm{cut}}{\left[x_\mathrm{
cut}^{\gamma}-\left(x_\mathrm{cut}^{\gamma}-x_\mathrm{
min}^{\gamma}\right)y\right]^{1/\gamma}} \ .
\ee
The required uniformly distributed random numbers $y \in 
\left[0,1\right]$ can be obtained, e.g., from \R\ by means of
$\texttt{runif}(n_{\mathrm{simulations}},0,1)$, or from the random
number generator \texttt{RAND()} (dt.: \texttt{ZUFALLSZAHL()})
in EXCEL or in OpenOffice.

\section[Exponential distribution]{Exponential distribution}
\lb{sec:expdistr}
The \textbf{exponential distribution} for a continuous 
one-dimensional random variable $X$,
\be
X \sim Ex(\lambda) \ ,
\ee
depends on a single free parameter, $\lambda \in \mathbb{R}_{>0}$, 
which represents an inverse scale.

\medskip
\noindent
Spectrum of values:
\be
X \mapsto x \in \mathbb{R}_{\geq 0} \ .
\ee
Probability density function (\texttt{pdf}):
\be
\lb{eq:exppdf}
\fbox{$\displaystyle
f_{X}(x) = 
\begin{cases}
0 &
\text{for}\quad x < 0 \\ \\
\lambda\exp\left[-\lambda x\right]  \ ,
\quad \lambda \in \mathbb{R}_{>0} &
\text{for}\quad x \geq 0
\end{cases} \ ;
$}
\ee
its graph is shown in Fig.~\ref{fig:exppdf} below.
\begin{figure}[!htb]
\begin{center}
\includegraphics[scale=0.8]{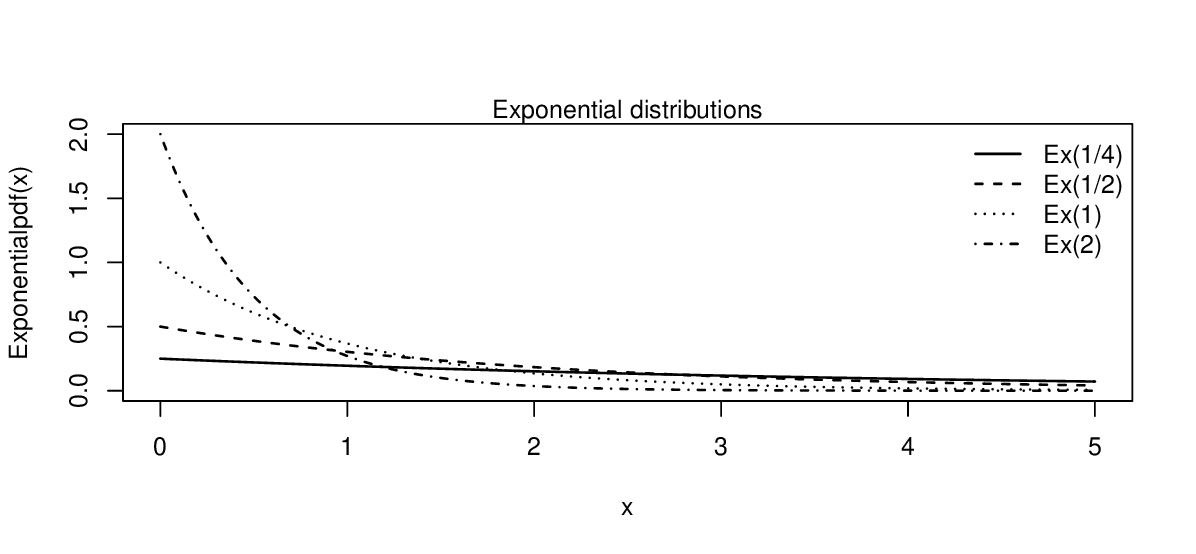}
\end{center}
\caption{\texttt{pdf} of the exponential distribution 
according to Eq.~(\ref{eq:exppdf}). Displayed are the cases 
$Ex(1/4)$, $Ex(1/2)$, $Ex(1)$ and $Ex(2)$.}
\lb{fig:exppdf}
\end{figure}

\medskip
\noindent
Cumulative distribution function (\texttt{cdf}):
\be
\fbox{$\displaystyle
F_{X}(x) = P(X \leq x)
=
\begin{cases}
0 & \text{for}\quad x < 0 \\ \\
1 - \exp\left[-\lambda x\right] &
\text{for}\quad x \geq 0
\end{cases} \ .
$}
\ee
Expectation value, variance, skewness and excess 
kurtosis:\footnote{The derivation of these results entails 
integration by parts for a number of times; see, e.g., 
Ref.~\ct[Sec.~8.1]{hve2009}.}
\bea
\mathrm{E}(X) & = & \frac{1}{\lambda} \\
\mathrm{Var}(X) & = & \frac{1}{\lambda^{2}} \\
\mathrm{Skew}(X) & = & 2 \\
\mathrm{Kurt}(X) & = & 6 \ .
\eea
$\alpha$--quantiles:
\be
\alpha \stackrel{!}{=} F_{X}(x_{\alpha})
= 1 - \exp\left[-\lambda x_{\alpha}\right]
\ \Leftrightarrow\ 
x_{\alpha} = F_{X}^{-1}(\alpha)
= -\,\frac{\ln(1-\alpha)}{\lambda}
\quad\text{for\ all}\quad 0 < \alpha < 1 \ .
\ee

\medskip
\noindent
\underline{\R:} $\texttt{dexp}(x,\lambda)$,
$\texttt{pexp}(x,\lambda)$, $\texttt{qexp}(\alpha,\lambda)$,
$\texttt{rexp}(n_{\mathrm{simulations}},\lambda)$

\section[Logistic distribution]{Logistic distribution}
\lb{sec:logisticdistr}
The \textbf{logistic distribution} for a continuous one-dimensional 
random variable $X$,
\be
X \sim Lo(\mu;s) \ ,
\ee
depends on two free parameters: a location parameter $\mu \in \mathbb{R}$ and a scale parameter $s \in \mathbb{R}_{>0}$.

\medskip
\noindent
Spectrum of values:
\be
X \mapsto x \in \mathbb{R} \ .
\ee
Probability density function (\texttt{pdf}):
\be
\lb{eq:logisticpdf}
\fbox{$\displaystyle
f_{X}(x) = \frac{\exp\left[{\displaystyle 
-\frac{x-\mu}{s}}\right]}{s\left(1+
\exp\left[{\displaystyle -\frac{x-\mu}{s}}\right]\right)^{2}} \ ,
\quad \mu \in\mathbb{R}\ , s \in \mathbb{R}_{>0} \ ;
$}
\ee
its graph is shown in Fig.~\ref{fig:logisticpdf} below.
\begin{figure}[!htb]
\begin{center}
\includegraphics[scale=0.8]{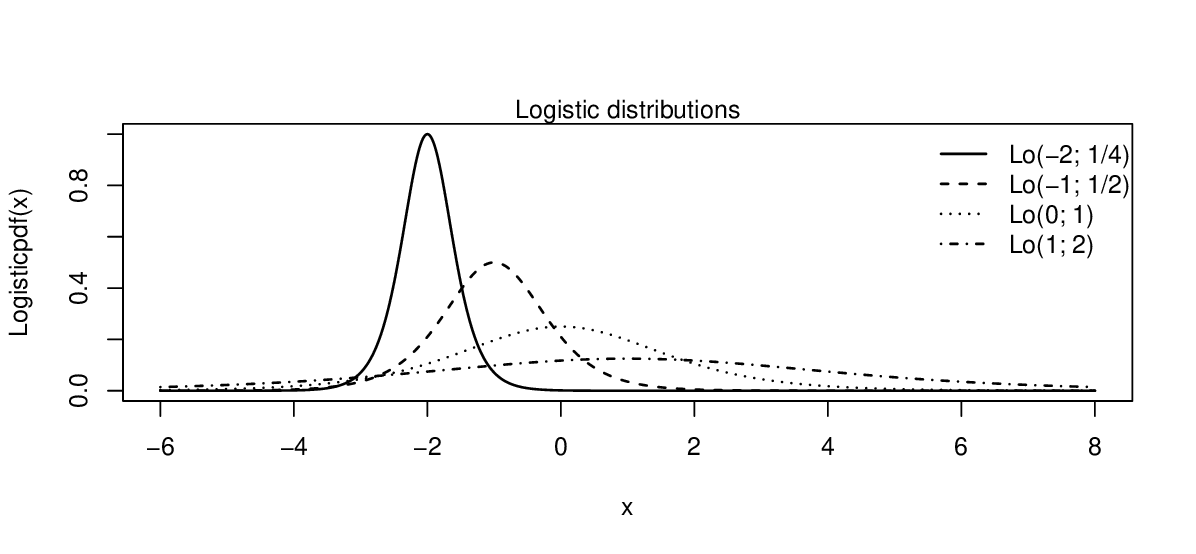}
\end{center}
\caption{\texttt{pdf} of the logistic distribution 
according to Eq.~(\ref{eq:logisticpdf}). Displayed are the cases 
$Lo(-2;1/4)$, $Lo(-1;1/2)$, $Lo(0;1)$ and $Lo(1;2)$.}
\lb{fig:logisticpdf}
\end{figure}

\medskip
\noindent
Cumulative distribution function (\texttt{cdf}):
\be
\fbox{$\displaystyle
F_{X}(x) = P(X \leq x)
= \frac{1}{1+\exp\left[{\displaystyle -\frac{x-\mu}{s}}\right]} \ .
$}
\ee
Expectation value, variance, skewness and excess kurtosis (cf. 
Rinne (2008)~\ct[p~359]{rin2008}):
\bea
\mathrm{E}(X) & = & \mu \\
\mathrm{Var}(X) & = & \frac{s^{2}\pi^{2}}{3} \\
\mathrm{Skew}(X) & = & 0 \\
\mathrm{Kurt}(X) & = & \frac{6}{5} \ .
\eea
$\alpha$--quantiles:
\be
\alpha \stackrel{!}{=} F_{X}(x_{\alpha})
= \frac{1}{1+\exp\left[{\displaystyle 
-\frac{x_{\alpha}-\mu}{s}}\right]}
\ \Leftrightarrow\ 
x_{\alpha} = F_{X}^{-1}(\alpha)
= \mu + s\ln\left(\frac{\alpha}{1-\alpha}\right)
\quad\text{for\ all}\quad 0 < \alpha < 1 \ .
\ee

\medskip
\noindent
\underline{\R:} $\texttt{dlogis}(x,\mu,s)$,
$\texttt{plogis}(x,\mu,s)$, $\texttt{qlogis}(\alpha,\mu,s)$,
$\texttt{rlogis}(n_{\mathrm{simulations}},\mu,s)$

\section[Special hyperbolic distribution]{Special hyperbolic
distribution}
\lb{sec:hyperbelverteil}
The complex dynamics associated with the formation of generic 
singularities in relativistic cosmology can be perceived as a 
random process. In this context, the following \textbf{special 
hyperbolic distribution} for a continuous one-dimensional random 
variable $X$,
\be
X \sim sHyp \ ,
\ee
which does not depend on any free parameters, was introduced by 
Khalatnikov {\em et al\/} (1985)~\ct{khaetal1985} to aid a 
simplified dynamical description of singularity formation; see 
also Heinzle {\em et al\/} (2009)~\ct[Eq.~(50)]{heietal2009}.

\medskip
\noindent
Spectrum of values:
\be
X \mapsto x \in \left[0,1\right]
\subset \mathbb{R}_{\geq 0} \ .
\ee
Probability density function (\texttt{pdf}):
\be
\lb{eq:sHyppdf}
\fbox{$\displaystyle
f_{X}(x) = 
\begin{cases}
{\displaystyle \frac{1}{\ln(2)}\,\frac{1}{1+x}} &
\text{for}\quad x \in \left[0,1\right] \\ \\
0 & \text{otherwise}
\end{cases} \ ;
$}
\ee
its graph is shown in Fig.~\ref{fig:sHyppdf} below.
\begin{figure}[!htb]
\begin{center}
\includegraphics[scale=0.8]{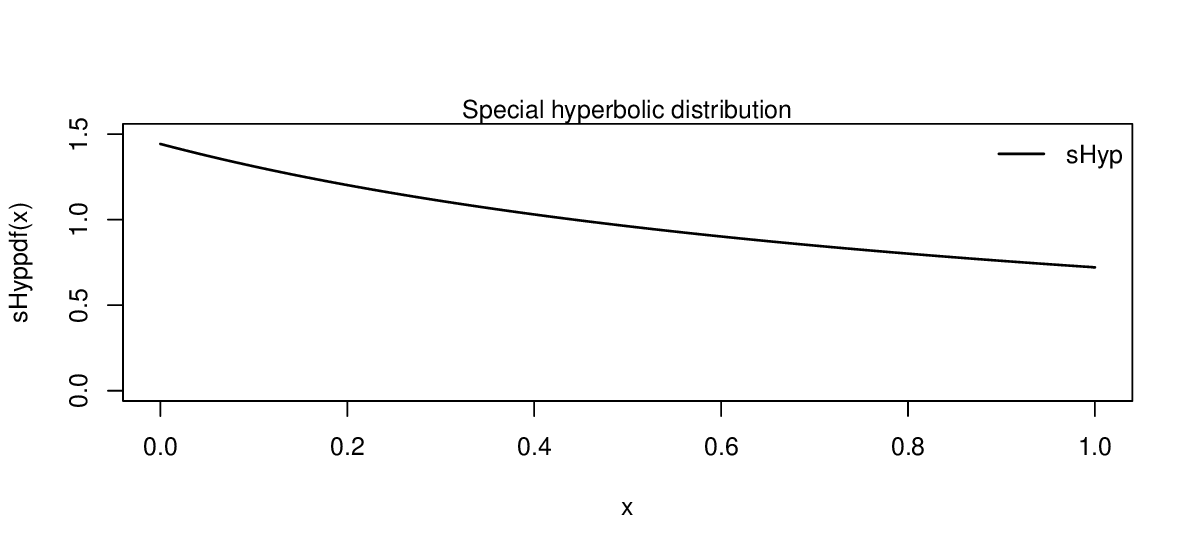}
\end{center}
\caption{\texttt{pdf} of the special hyperbolic distribution 
according to Eq.~(\ref{eq:sHyppdf}).}
\lb{fig:sHyppdf}
\end{figure}

\medskip
\noindent
Cumulative distribution function (\texttt{cdf}):
\be
\fbox{$\displaystyle
F_{X}(x) = P(X \leq x)
=
\begin{cases}
0 & \text{for}\quad x < 0 \\ \\
{\displaystyle \frac{1}{\ln(2)}\,\ln(1+x)} &
\text{for}\quad x \in \left[0,1\right] \\ \\
1 & \text{for}\quad x > 1
\end{cases} \ .
$}
\ee
Expectation value, variance, skewness and excess 
kurtosis:\footnote{Use polynomial division to simplify the 
integrands in the ensuing moment integrals when verifying these 
results.}
\bea
\mathrm{E}(X) & = & \frac{1-\ln(2)}{\ln(2)} \\
\mathrm{Var}(X) & = & \frac{3\ln(2)-2}{2\left[
\ln(2)\right]^{2}} \\
\mathrm{Skew}(X) & = & \frac{7\left[
\ln(2)\right]^{2}-\frac{27}{2}\ln(2)+6}{3\left(\frac{1}{2}
\right)^{3/2}\left[3\ln(2)-2\right]^{3/2}} \\
\mathrm{Kurt}(X) & = & \frac{15\left[
\ln(2)\right]^{3}-\frac{193}{3}\left[
\ln(2)\right]^{2}+72\ln(2)-24}{\left[
3\ln(2)-2\right]^{2}} \ .
\eea
$\alpha$--quantiles:
\be
\alpha \stackrel{!}{=} F_{X}(x_{\alpha})
= \frac{1}{\ln(2)}\,\ln(1+x_{\alpha})
\ \Leftrightarrow\ 
x_{\alpha} = F_{X}^{-1}(\alpha)
= e^{\alpha\ln(2)}-1
\quad\text{for\ all}\quad 0 < \alpha < 1 \ .
\ee
%

\section[Cauchy distribution]{Cauchy distribution}
\lb{sec:cauchyverteil}
The French mathematician
\href{http://www-history.mcs.st-and.ac.uk/Biographies/Cauchy.html} {Augustin Louis Cauchy (1789--1857)} is credited with the inception into
\textbf{Statistics} of the continuous two-parameter 
distribution law
\be
X \sim Ca(b;a) \ ,
\ee
with properties

\medskip
\noindent
Spectrum of values:
\be
X \mapsto x \in \mathbb{R} \ .
\ee
Probability density function (\texttt{pdf}):
\be
\lb{cauchypdf}
\fbox{$\displaystyle
f_{X}(x) = \frac{1}{\pi}\,\frac{a}{a^{2}+(x-b)^{2}} \ ,
\qquad\text{with}\quad a \in \mathbb{R}_{>0},\ b \in \mathbb{R} \ ;
$}
\ee
its graph is shown in Fig.~\ref{fig:capdf} below for four 
particular cases.
\begin{figure}[!htb]
\begin{center}
\includegraphics[scale=0.8]{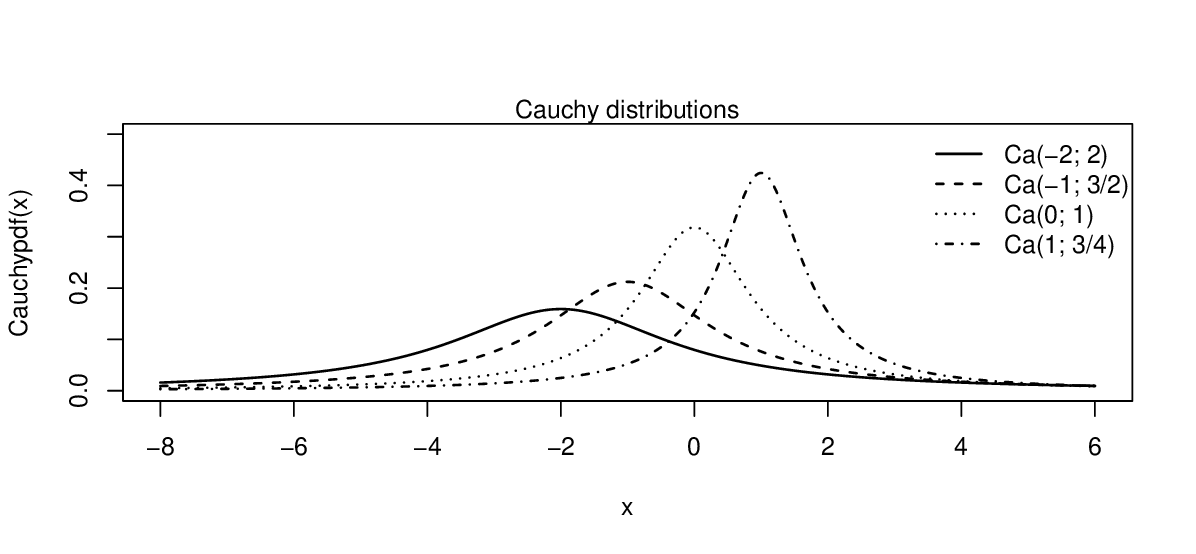}
\end{center}
\caption{\texttt{pdf} of the Cauchy distribution according to 
Eq.~(\ref{cauchypdf}). Displayed are the cases $Ca(-2;2)$,
$Ca(-1;3/2)$, $Ca(0;1)$ and $Ca(1;3/4)$. The case $Ca(0;1)$
corresponds to a $t$--distribution with $df = 1$ degree of freedom;
cf. Sec.~\ref{sec:tverteil}.}
\lb{fig:capdf}
\end{figure}

\medskip
\noindent
Cumulative distribution function (\texttt{cdf}):
\be
\lb{cauchycdf}
\fbox{$\displaystyle
F_{X}(x) = P(X \leq x)
= \frac{1}{2}
+ \frac{1}{\pi}\,\arctan\left(\frac{x-b}{a}\right) \ .
$}
\ee
Expectation value, variance, skewness and excess 
kurtosis:\footnote{In the case of a Cauchy distribution the 
fall-off in the tails of the \texttt{pdf} is not sufficiently fast 
for the expectation value and variance integrals, 
Eqs.~(\ref{eq:expectcon}) and (\ref{eq:varcon}), to 
converge to finite values. Consequently, this also concerns the 
skewness and excess kurtosis given in Eqs.~(\ref{eq:skew2}) and 
(\ref{eq:kurt2}).}
\bea
\mathrm{E}(X): & \quad & \text{does NOT exist due to a
diverging integral} \\
\mathrm{Var}(X): & \quad & \text{does NOT exist due to a
diverging integral} \\
\mathrm{Skew}(X): & \quad & \text{does NOT exist due to a
diverging integral} \\
\mathrm{Kurt}(X): & \quad & \text{does NOT exist due to a
diverging integral} \ .
\eea
See, e.g., Sivia and Skilling (2006) \ct[p~34]{sivski2006}.

\medskip
\noindent
$\alpha$--quantiles:
\be
\alpha \stackrel{!}{=} F_{X}(x_{\alpha})
\quad\Leftrightarrow\quad
x_{\alpha} = F_{X}^{-1}(\alpha)
= b + a\tan\left[\pi\left(\alpha-\frac{1}{2}\right)\right]
\quad\text{for\ all}\quad 0 < \alpha < 1 \ .
\ee

\medskip
\noindent
\underline{\R:} $\texttt{dcauchy}(x,b,a)$,
$\texttt{pcauchy}(x,b,a)$, $\texttt{qcauchy}(\alpha,b,a)$,
$\texttt{rcauchy}(n_{\mathrm{simulations}},b,a)$

\section[Central limit theorem]{Central limit theorem}
\lb{sec:zentrgrenz}
The first systematic derivation and presentation of the paramount
\textbf{central limit theorem} of \textbf{Probability Theory} is
due to the French mathematician and astronomer 
\href{http://www-history.mcs.st-and.ac.uk/Biographies/Laplace.html}{Marquis Pierre Simon de Laplace (1749--1827)}, cf. Laplace 
(1809)~\ct{lap1809}.

\medskip
\noindent
Consider a set of $n$ \textbf{mutually stochastically independent} 
[cf. Eqs.~(\ref{eq:stochindep2}) and~(\ref{eq:stochindep3})],
\textbf{additive} one-dimensional random variables
$X_{1},\ldots, X_{n}$, with
\begin{itemize}

\item[(i)] \textit{finite} expectation values 
$\mu_{1}, \ldots, \mu_{n}$,

\item[(ii)] \textit{finite} variances 
$\sigma_{1}^{2}, \ldots, \sigma_{n}^{2}$,
which are not too different from one another, and 

\item[(iii)] corresponding \texttt{cdf}s 
$F_{1}(x), \ldots, F_{n}(x)$.
\end{itemize}
Introduce for this set a \textbf{total sum}~$Y_{n}$ according to 
Eq.~(\ref{eq:sumnmean}), and, by standardisation 
via~Eq.~(\ref{eq:standardisation}), a related \textbf{standardised 
summation random variable}
\be
\displaystyle
Z_{n} := 
\frac{Y_{n}-{\displaystyle\sum_{i=1}^{n}\mu_{i}}}{\sqrt{{\displaystyle\sum_{j=1}^{n}\sigma_{j}^{2}}}} \ .
\ee
Let ${\cal F}_{n}(z_{n})$ denote the \texttt{cdf} associated with
$Z_{n}$.

\medskip
\noindent
Then, subject to the convergence condition
\be
\lim_{n \to \infty}\max_{1 \leq i \leq 
n}\frac{\sigma_{i}}{\sqrt{{\displaystyle\sum_{j=1}^{n}\sigma_{j}^{2}}}} = 0 \ ,
\ee
i.e., that asymptotically the standard deviation of the total sum 
dominates the standard deviations of any of the individual $X_{i}$,
and certain additional regularity requirements (see, e.g., Rinne
(2008)~\ct[p~427 f]{rin2008}), the \textbf{central limit theorem}
in its general form according to the Finnish mathematician 
\href{http://en.wikipedia.org/wiki/Jarl_Waldemar_Lindeberg}{Jarl 
Waldemar Lindeberg (1876--1932)} and the Croatian--American 
mathematician 
\href{http://www-history.mcs.st-and.ac.uk/Biographies/Feller.html}{William Feller (1906--1970)} states that in the asymptotic limit 
of infinitely many $X_{i}$ contributing to $Y_{n}$ (and so to
$Z_{n}$), it holds that
\be
\lim_{n \to \infty}{\cal F}_{n}(z_{n}) = \Phi(z) \ ,
\ee
i.e., the limit of the sequence of probability distributions ${\cal 
F}_{n}(z_{n})$ for the standardised summation random variables 
$Z_{n}$ is constituted by the \textbf{standard normal distribution} 
$N(0;1)$, discussed in Sec.~\ref{sec:normverteil}; cf. Lindeberg 
(1922)~\ct{lin1922} and Feller (1951)~\ct{fel1951}. Earlier 
results on the asymptotic distributional properties of a 
sum of independent additive one-dimensional random variables were 
obtained by the Russian mathematician, mechanician and physicist 
\href{http://www-history.mcs.st-and.ac.uk/Biographies/Lyapunov.html}{Aleksandr Mikhailovich Lyapunov (1857--1918)}; cf. Lyapunov 
(1901)~\ct{lya1901}.

\medskip
\noindent
Thus, under fairly general conditions, the normal distribution 
acts as a stable \textbf{attractor distribution} for the sum of $n$ 
mutually stochastically independent, additive random variables 
$X_{i}$.\footnote{Put differently, for 
increasingly large $n$ the \texttt{cdf} of the total sum 
$Y_{n}$ approximates a normal distribution with expectation value 
$\displaystyle\sum_{i=1}^{n}\mu_{i}$ and variance 
$\displaystyle\sum_{i=1}^{n}\sigma_{i}^{2}$ to an increasingly 
accurate degree. In particular, all reproductive distributions 
may be approximated by a normal distribution as $n$ becomes 
large.} In oversimplified terms: this result bears a certain 
economical convenience for most practical purposes in that, given 
favourable conditions, when the size of a random sample is 
sufficiently large (in practice, a typical rule of thumb is $n 
\geq 50$), one essentially needs to know the characteristic 
features of only a single continuous univariate probability 
distribution to perform, e.g., null hypothesis significance testing
within the frequentist framework; cf. Ch.~\ref{ch11}. 
As will become apparent in subsequent chapters, 
the central limit theorem has profound ramifications for 
applications in all empirical scientific disciplines.

\medskip
\noindent
Note that for \textit{finite} $n$ the central limit theorem makes 
\textit{no} statement as to the nature of the \textit{tails} of the 
probability distribution for $Z_{n}$ (or for $Y_{n}$), where, in 
principle, it can be very different from a normal distribution; 
cf. Bouchaud and Potters (2003) \ct[p~25f]{boupot2003}.

\medskip
\noindent
A direct consequence of the central limit theorem and its 
preconditions is the fact that for the \textbf{sample mean} 
$\bar{X}_{n}$, defined in Eq.~(\ref{eq:sumnmean}) above, both
\[
\lim_{n \to \infty}\mathrm{E}(\bar{X}_{n})
= \lim_{n \to \infty}\frac{{\displaystyle\sum_{i=1}^{n}\mu_{i}}}{n}
\quad\quad\text{and}\quad\quad
\lim_{n \to \infty}\mathrm{Var}(\bar{X}_{n})
= \lim_{n \to 
\infty}\frac{{\displaystyle\sum_{i=1}^{n}\sigma_{i}^{2}}}{n^{2}}
\]
converge to finite values. This property is most easily recognised
in the special case of $n$ \textbf{mutually stochastically 
independent and identically distributed} (in short: ``i.i.d.'') 
additive one-dimensional random variables $X_{1}, \ldots, X_{n}$, 
which have common finite expectation value $\mu$, common finite 
variance $\sigma^{2}$, and common \texttt{cdf}
$F(x)$.\footnote{These conditions lead to the central limit theorem
in the special form according to Jarl Waldemar Lindeberg
(1876--1932) and the French mathematician 
\href{http://en.wikipedia.org/wiki/Paul_Pierre_Levy}{Paul Pierre 
L\'{e}vy (1886--1971)}.} Then,
\bea
\lim_{n \to \infty}\mathrm{E}(\bar{X}_{n})
& = & \lim_{n \to \infty}\frac{n\mu}{n}
\ = \ \mu \\
\lim_{n \to \infty}\mathrm{Var}(\bar{X}_{n})
& = & \lim_{n \to \infty}\frac{n\sigma^{2}}{n^{2}}
\ = \ \lim_{n \to \infty}\frac{\sigma^{2}}{n}
\ = \ 0 \ .
\eea
This result is known as the \textbf{law of large numbers} according 
to the Swiss mathematician 
\href{http://www-history.mcs.st-and.ac.uk/Biographies/Bernoulli_Jacob.html}{Jakob Bernoulli (1654--1705)}; the sample mean 
$\bar{X}_{n}$ \textbf{converges stochastically} to its expectation 
value~$\mu$.

\medskip
\noindent
We point out that a counter-example to the central limit theorem 
is given by a set of $n$ i.i.d. Pareto-distributed with exponent 
$\gamma \leq 2$ one-dimensional random variables~$X_{i}$, since in 
this case the variance of the $X_{i}$ is undefined; cf. 
Eq.~(\ref{eq:paretovar}).

\vspace{5mm}
\noindent
This ends Part II of these lecture notes, and we now turn to Part 
III in which we focus on a number of useful applications of 
\textbf{inferential statistical methods of data analysis} 
within the \textbf{frequentist framework}. Data analysis techniques
within the conceptually compelling Bayes--Laplace framework have
been reviewed, e.g., in the online lecture notes by Saha
(2002)~\ct{sah2002}, in the textbooks by Sivia and Skilling (2006)
\ct{sivski2006}, Gelman \textit{et al} (2014)~\ct{geletal2014} and
McElreath (2016)~\ct{mce2016}, and in the lecture notes of
Ref.~\ct{hve2018}.


\chapter[Likert's scaling method of summated item 
ratings]{\href{https://www.youtube.com/watch?v=_IJJ_D6UOaQ}{Operationalisation 
of latent variables: Likert's scaling method of summated item
ratings}}
\lb{ch9}
A sound \textbf{operationalisation} of ones's portfolio of
\textbf{statistical variables} in quantitative--empirical research
is key to a successful and effective application of
\textbf{statistical methods of data analysis}, particularly in the
\textbf{Social Sciences} and \textbf{Humanities}. The most
frequently practiced method to date for operationalising
\textbf{latent variables} (such as unobservable ``social
constructs'') is due to the US-American psychologist 
\href{http://en.wikipedia.org/wiki/Rensis_Likert}{Rensis Likert's 
(1903--1981)}. In his 1932 paper \ct{lik1932}, which completed his 
thesis work for a Ph.D., he expressed the idea that \textbf{latent 
statistical variables}~$X_{L}$, when they may be perceived as 
\textit{one-dimensional} in nature, can be rendered measurable in a 
\textit{quasi-metrical} fashion by means of the \textbf{summated 
ratings} over an extended set of suitable and observable
\textbf{indicator items} $X_{i}$ ($i=1,2,\ldots$), which, in order
to ensure effectiveness, ought to be (i)~\textit{highly
interdependent} and possess (ii)~\textit{high discriminatory
power}. Such indicator items are often formulated as specific
statements relating to the theoretical concept a particular
one-dimensional latent variable~$X_{L}$ is supposed to capture,
with respect to which test persons need to express their subjective
level of agreement or, in different settings, indicate a specific
subjective degree of intensity. A typical \textbf{item rating
scale} for the indicator items~$X_{i}$, 
providing the necessary item ratings, is given for instance by the 
5--level ordinally ranked attributes of agreement\\[-5mm]
\begin{itemize}
\item[1:] strongly disagree/strongly unfavourable\\[-5mm]
\item[2:] disagree/unfavourable\\[-5mm]
\item[3:] undecided\\[-5mm]
\item[4:] agree/favourable\\[-5mm]
\item[5:] strongly agree/strongly favourable.\\[-5mm]
\end{itemize}
In the research literature, one also encounters 7--level or 
10--level item rating scales, which offer more flexibility. Note 
that it is \textit{assumed (!)} fom the outset that the
items~$X_{i}$, and thus their ratings, can be treated as
\textbf{additive}, so that the conceptual principles of
Sec.~\ref{sec:sumvar} relating to 
sums of random variables can be relied upon. When forming the sum 
over the ratings of all the indicator items one selected, it is 
essential to carefully pay attention to the \textbf{polarity} of
the items involved. For the resultant \textbf{total sum} 
${\displaystyle\sum_{i}X_{i}}$ to be consistent, the polarity of 
all items used needs to be uniform.\footnote{For a questionnaire, 
however, it is strongly recommended to include also indicator 
items of reversed polarity. This will improve the overall 
construct validity of the measurement tool.}

\medskip
\noindent
The construction of a consistent and coherent \textbf{Likert scale} 
for a one-dimensional latent statistical variable~$X_{L}$ involves 
four basic steps (see, e.g., Trochim (2006)~\ct{tro2006}):\\[-5mm]
\begin{itemize}
\item[(i)] the compilation of an initial list of 80 to 100 
potential \textbf{indicator items} $X_{i}$ for the one-dimensional 
latent variable of interest,\\[-5mm]

\item[(ii)] the draw of a \textbf{gauge random sample} from the 
target population~$\boldsymbol{\Omega}$,\\[-5mm]

\item[(iii)] the computation of the \textbf{total sum} 
${\displaystyle\sum_{i}X_{i}}$ of item ratings, and, most 
importantly, \\[-5mm]

\item[(iv)] the performance of an \textbf{item analysis} based on
the sample data and the associated total sum 
${\displaystyle\sum_{i}X_{i}}$ of item ratings.\\[-5mm]
\end{itemize}
The item analysis, in particular, consists of the consequential 
application of two exclusion criteria, which aim at establishing 
the scientific quality of the final \textbf{Likert scale}. Items
are being discarded from the list when either\\[-6mm]
\begin{itemize}
\item[(a)] they show a weak \textbf{item-to-total 
correlation} with the total sum ${\displaystyle\sum_{i}X_{i}}$ (a 
rule of thumb is to exclude items with correlations less than 
$0.5$), or\\[-6mm]
  
\item[(b)] it is possible to increase the value of
\textbf{Cronbach's}\footnote{Named 
after the US-American educational psychologist
\href{http://en.wikipedia.org/wiki/Lee_Cronbach}{Lee Joseph 
Cronbach (1916--2001)}. The range of the normalised real-valued
$\alpha$--coefficient is the interval $[0,1]$.} 
$\boldsymbol{\alpha}$\textbf{--coefficient} (see Cronbach (1951) 
\ct{cro1951}), a measure of the scale's \textbf{internal
consistency reliability}, by excluding a particular item from the
list (the objective being to attain $\alpha$-values greater than
$0.8$).\\[-6mm]
\end{itemize}
For a set of $m \in \mathbb{N}$ indicator items $X_{i}$, 
Cronbach's $\alpha$--coefficient is defined by
\be
\lb{eq:cronbachalpha}
\fbox{$\displaystyle
\alpha := \left(\frac{m}{m-1}\right)\left(1 - 
\frac{{\displaystyle\sum_{i=1}^{m}S_{i}^{2}}}{S_\mathrm{total}^{2}}
\right) \ ,
$}
\ee
where $S_{i}^{2}$ denotes the sample variance associated with the 
$i$th indicator item (perceived as being metrically scaled),
and $S_\mathrm{total}^{2}$ is the sample variance of the total sum 
${\displaystyle\sum_{i}X_{i}}$.

\medskip
\noindent
\underline{\R:} \texttt{alpha({\it items\/})} (package:
\texttt{psych}, by Revelle (2019)~\ct{rev2019}) \\
\underline{SPSS:} Analyze $\rightarrow$ Scale $\rightarrow$ 
Reliability Analysis \ldots (Model: Alpha) $\rightarrow$ 
Statistics \ldots: Scale if item deleted

\medskip
\noindent
The outcome of the item analysis is a drastic reduction of the 
initial list to a set of just $k \in \mathbb{N}$ indicator items 
$X_{i}$ ($i=1,\ldots,k$) of high discriminatory power, where $k$ 
is typically in the range of $10$ to $15$.\footnote{However, in 
many research papers one finds Likert scales with a minimum of 
just four indicator items.} The associated \textbf{total sum}
\be 
X_{L} := \sum_{i=1}^{k}X_{i}
\ee
thus operationalises the one-dimensional latent statistical 
variable~$X_{L}$ in a quasi-metrical fashion, since it is to be 
measured on an \textbf{interval scale} with a \textit{discrete} 
spectrum of values given (for a 5--level item rating scale) by
\be
X_{L} \mapsto \sum_{i=1}^{k}x_{i} \in \left[1k,5k\right] \ .
\ee
The structure of a finalised discrete $k$-indicator-item Likert 
scale for some one-dimensional latent statistical variable~$X_{L}$ 
with an equidistant graphical 5--level item rating scale is 
displayed in Tab.~\ref{tab:likert}.
\begin{table}

\underline{\textbf{One-dimensional latent statistical
variable}~$\boldsymbol{X_{L}}$:}
\begin{center}
\begin{tabular}[!htb]{cccccccc}
$\bullet$\ \textbf{Item} $X_{1}$: &
strongly disagree & $\bigcirc$\quad\quad\mbox{} 
& $\bigcirc$\quad\quad\mbox{} & $\bigcirc$\quad\quad\mbox{} & 
$\bigcirc$\quad\quad\mbox{} & $\bigcirc$ & strongly agree \\ \\
$\bullet$\ \textbf{Item} $X_{2}$: &
strongly disagree & $\bigcirc$\quad\quad\mbox{} 
& $\bigcirc$\quad\quad\mbox{} & $\bigcirc$\quad\quad\mbox{} & 
$\bigcirc$\quad\quad\mbox{} & $\bigcirc$ & strongly agree \\ \\
\vdots & \vdots & \vdots\quad\quad\mbox{} & 
\vdots\quad\quad\mbox{} & \vdots\quad\quad\mbox{} & 
\vdots\quad\quad\mbox{} & \vdots & \vdots\\ \\
$\bullet$\ \textbf{Item} $X_{k}$: & 
strongly disagree & $\bigcirc$\quad\quad\mbox{} 
& $\bigcirc$\quad\quad\mbox{} & $\bigcirc$\quad\quad\mbox{} & 
$\bigcirc$\quad\quad\mbox{} & $\bigcirc$ & strongly agree
\end{tabular}
\end{center}
\caption{Structure of a discrete $k$-indicator-item Likert scale 
for some one-dimensional latent statistical variable~$X_{L}$, 
based on a visualised equidistant 5--level item rating scale.}
\lb{tab:likert}
\end{table}

\medskip
\noindent
Likert's scaling method of aggregating information from a set of  
$k$ highly interdependent ordinally scaled items to form an 
effectively quasi-metrical, one-dimensional total sum 
${\displaystyle X_{L}=\sum_{i}X_{i}}$ draws its legitimisation to 
a large extent from a generalised version of the \textbf{central 
limit theorem} (cf. Sec.~\ref{sec:zentrgrenz}), wherein the 
precondition of mutually stochastically independent variables 
contributing to the sum is relaxed. In practice it is found that 
for many cases of interest in the samples one has available for 
research the total sum  ${\displaystyle X_{L}=\sum_{i}X_{i}}$ is 
normally distributed in to a very good approximation. 
Nevertheless, the normality property of Likert scale data needs to 
be established on a case-by-case basis. The main shortcoming of 
Likert's approach is its dependency of the gauging process of the 
scale on the target population.

\medskip
\noindent
In the \textbf{Social Sciences} there is available a broad variety
of operationalisation procedures alternative to the discrete
\textbf{Likert scale}. We restrict ourselves here to mention but
one example, namely the \textit{continuous} psychometric
\textbf{visual analogue scale (VAS)} developed by Hayes and
Paterson (1921)~\ct{haypat1921} and by Freyd (1923)~\ct{fre1923}.
Further measurement scales for latent statistical variables can be 
obtained from the websites 
\href{http://zis.gesis.org/ZisApplication/}{\texttt{zis.gesis.org}},
German Social Sciences measurement scales (ZIS), 
and \href{http://www.ssrn.com}{\texttt{ssrn.com}}, Social Science 
Research Network (SSRN). On a historical note: one of the first 
systematically designed \textbf{questionnaires} as a measurement
tool for collecting socio-economic data (from workers on strike at
the time in Britain) was published by the Statistical Society of 
London in 1838; see Ref.~\ct{ssl1838}.


\chapter[Random sampling of target populations]{\href{https://www.youtube.com/watch?v=ApHx0GE3zQI}{Random sampling of target populations}}
\lb{ch10}
\textbf{Quantitative--empirical research methods} may be employed
for \textbf{exploratory} as well as for \textbf{confirmatory data
analysis}. Here we will focus on the latter, in the context of a
\textbf{frequentist viewpoint} of \textbf{Probability Theory} and
\textbf{statistical inference}. To investigate \textbf{research
questions} systematically by statistical means, with the objective
to make inferences about the distributional properties of a set of
\textbf{statistical variables} in a specific \textbf{target 
population}~$\boldsymbol{\Omega}$ of study objects, on the basis 
of analysis of data from just a few units in a \textbf{sample} 
$\boldsymbol{S_{\Omega}}$, the following three issues have to be 
addressed in a clearcut fashion:
\begin{itemize}
\item[(i)] the \textbf{target population} $\boldsymbol{\Omega}$ of 
the research activity needs to be defined in an unambiguous way,

\item[(ii)] an adequate \textbf{random 
sample}~$\boldsymbol{S_{\Omega}}$ needs to be drawn from an 
underlying \textbf{sampling frame}~$\boldsymbol{L_{\Omega}}$ 
associated with $\boldsymbol{\Omega}$, and

\item[(iii)] a reliable mathematical procedure for
\textbf{estimating quantitative population parameters} from random
sample data needs to be employed.
\end{itemize}
We will briefly discuss these issues in turn, beginning with a 
review in Tab.~\ref{tab:notation} of conventional \textbf{notation} 
for distinguishing specific statistical measures relating to target
populations $\boldsymbol{\Omega}$ on the one-hand side from the 
corresponding ones relating to random samples 
$\boldsymbol{S_{\Omega}}$ on the other.
\begin{table}
\begin{center}
\begin{tabular}[!h]{c|c}
    	\hline \\
      \textbf{Target population} $\boldsymbol{\Omega}$ &
      \textbf{Random sample} $\boldsymbol{S_{\Omega}}$ \\ \\
      \hline \\
      population size $N$ & sample size $n$ \\ \\
      arithmetical mean $\mu$ & sample mean $\bar{X}_{n}$
      \\ \\
      standard deviation $\sigma$ & sample standard deviation
      $S_{n}$ \\ \\
      median $\tilde{x}_{0.5}$ & sample median
      $\tilde{X}_{0.5,n}$ \\ \\
      correlation coefficient $\rho$ & sample correlation
      coefficient $r$ \\ \\
      rank correlation coefficient $\rho_{S}$ & sample rank
      correl. coefficient $r_{S}$ \\ \\
      regression coefficient (intercept) $\alpha$ & sample 
      regression intercept $a$ 
      \\ \\
      regression coefficient (slope) $\beta$ & sample regression 
      slope $b$ 
      \\ \\
      \hline
\end{tabular}
\end{center}
\caption{Notation for distinguishing between statistical 
measures relating to a target population~$\boldsymbol{\Omega}$ on 
the one-hand side, and to the corresponding quantities and 
unbiased maximum likelihood point estimator functions obtained 
from a random sample~$\boldsymbol{S_{\Omega}}$ on the other.}
\lb{tab:notation}
\end{table}

\medskip
\noindent
One-dimensional \textbf{random variables} in a target population 
$\boldsymbol{\Omega}$ (of size $N$), as what \textbf{statistical 
variables} will be understood to constitute subsequently, will be
denoted by capital Latin letters such as $X$, $Y$, \ldots, $Z$,
while their \textbf{realisations} in random samples
$\boldsymbol{S_{\Omega}}$ (of size $n$) will be denoted by lower
case Latin letters such as $x_{i}$, $y_{i}$, \ldots, $z_{i}$
($i=1,\ldots,n$). In addition, one denotes \textbf{population
parameters} by lower case Greek letters, while for their
corresponding \textbf{point estimator functions} relating to random
samples, which are also perceived as random variables, again
capital Latin letters are used for representation. The ratio $n/N$
will be referred to as the \textbf{sampling fraction}. As is
standard in the statistical literature, we will denote a
particular \textbf{random sample} of size $n$ for a one-dimensional
random variable $X$ by a set $\boldsymbol{S_{\Omega}}$:~$(X_{1},
\ldots, X_{n})$, with $X_{i}$ representing any arbitrary random
variable associated with $X$ in this sample.

\medskip
\noindent
In actual practice, it is often not possible to acquire access for 
the purpose of enquiry to every single statistical unit belonging 
to an identified target population $\boldsymbol{\Omega}$, not 
even in principle. For example, this could be due to the fact that 
$\boldsymbol{\Omega}$'s size $N$ is far too large to be determined 
accurately. In this case, to ensure a reliable investigation, one 
needs to resort to using a \textbf{sampling frame} 
$\boldsymbol{L_{\Omega}}$ for $\boldsymbol{\Omega}$. By this one 
understands a representative list of  elements in 
$\boldsymbol{\Omega}$ to which access can actually be obtained one
way or another. Such a list will have to be compiled by some
authority of scientific integrity. In an attempt to avoid a
notational overflow in the following, we will continue to use~$N$
to denote \textit{both}: the size of the \textbf{target 
population}~$\boldsymbol{\Omega}$ and the  size of its associated 
\textbf{sampling frame}~$\boldsymbol{L_{\Omega}}$ (even though this 
is not entirely accurate). As regards the specific sampling
process, one may distinguish \textbf{cross-sectional} one-off
sampling at a fixed instant from \textbf{longitudinal} multiple
sampling over a finite time interval.\footnote{In a sense,
cross-sectional sampling will yield a ``snapshot'' of a target
population of interest in a particular state, while longitudinal
sampling is the basis for producing a ``film'' featuring a
particular evolutionary aspect of a target population of interest.}

\medskip
\noindent
We now proceed to introduce the three most commonly practiced 
methods of drawing \textbf{random samples} from given fixed target
populations~$\boldsymbol{\Omega}$ of statistical units.

\section[Random sampling methods]{Random sampling methods}
\lb{sec:sampling}
\subsection[Simple random sampling]{Simple random sampling}
\lb{subsec:einfzustich}
The \textbf{simple random sampling} technique can be best
understood in terms of the \textbf{urn model} of
\textbf{combinatorics} introduced in Sec.~\ref{sec:comb}. Given a
target population~$\boldsymbol{\Omega}$ (or sampling frame 
$\boldsymbol{L_{\Omega}}$) of $N$ distinguishable statistical 
units, there is a total of $\left(\begin{array}{c} N \\ n 
\end{array}\right)$ distinct possibilities of drawing samples of 
size $n$ from $\boldsymbol{\Omega}$ (or 
$\boldsymbol{L_{\Omega}}$), given the order of selection is 
\textit{not} being accounted for and \textit{excluding
repetitions}, see Sec.~\ref{subsec:combvar}. A \textbf{simple
random sample} is then defined by the property that its
probability of selection is equal to
\be
\frac{1}{\left(\begin{array}{c} N \\ n \end{array}\right)} \ ,
\ee
according to the Laplacian principle of Eq.~(\ref{eq:classprob}). 
This has the immediate consequence that the \textit{a priori} 
probability of selection of any single statistical unit is given 
by\footnote{In the statistical literature this particular property 
of a random sample is referred to as ``epsem'': equal probability 
of selection method.}
\be
1 - \frac{\left(\begin{array}{c} N-1 \\ n \end{array}\right)}{
\left(\begin{array}{c} N \\ n \end{array}\right)}
= 1 - \frac{N-n}{N} = \frac{n}{N} \ .
\ee
On the other hand, the probability that two statistical units $i$ 
and $j$ will be members of the \textit{same} sample of size~$n$
amounts to
\be
\frac{n}{N} \times \frac{n-1}{N-1} \ .
\ee
As such, by Eq.~(\ref{eq:stochindep1}), this type of a selection 
procedure of two statistical units proves \textit{not} to yield two 
stochastically independent units (in which case the joint 
probability of selection would be $n/N \times n/N$). However, for 
sampling fractions $n/N \leq 0.05$, stochastic independence of the 
selection of statistical units generally holds to a reasonably 
good approximation. When, in addition, the sample size is $n
\geq 50$, the conditions for the \textbf{central limit theorem} in
the variant of Lindeberg and L\'{e}vy (cf.
Sec.~\ref{sec:zentrgrenz}) to apply often hold to a fairly good
degree.

\subsection[Stratified random sampling]{Stratified random sampling}
\lb{subsec:stratsampling}
\textbf{Stratified random sampling} adapts the sampling process to
a known intrinsic structure of the target 
population~$\boldsymbol{\Omega}$ (and its associated sampling 
frame $\boldsymbol{L_{\Omega}}$), as provided by the $k$ mutually 
exclusive and exhaustive categories of some qualitative (nominal 
or ordinal) variable; these thus define a set of $k$
\textbf{strata} (layers) of $\boldsymbol{\Omega}$ (or
$\boldsymbol{L_{\Omega}}$). By construction, there are $N_{i}$
statistical units belonging to the $i$th stratum ($i=1, \ldots,
k$). Simple random samples of sizes $n_{i}$ are drawn from each
stratum according to the principles outlined in
Sec.~\ref{subsec:einfzustich}, yielding a total sample of size
$n=n_{1}+\ldots+n_{k}$. Frequently applied variants of this
sampling technique are (i)~\textbf{proportionate allocation} of
statistical units, defined by the condition\footnote{Note that,
thus, this also has the ``epsem'' property.}
\be
\frac{n_{i}}{n} \stackrel{!}{=} \frac{N_{i}}{N}
\qquad\Rightarrow\qquad
\frac{n_{i}}{N_{i}} = \frac{n}{N} \ ;
\ee
in particular, this allows for a fair representation of minorities 
in $\boldsymbol{\Omega}$, and (ii)~\textbf{optimal allocation} of 
statistical units which aims at a minimisation of the resultant 
sampling errors of the variables investigated. Further details on 
the stratified random sampling technique can be found, e.g., in 
Bortz and D\"{o}ring (2006)~\ct[p~425ff]{bordoe2006}.

\subsection[Cluster random sampling]{Cluster random sampling}
\lb{subsec:clustersampling}
When the target population~$\boldsymbol{\Omega}$ (and its 
associated sampling frame $\boldsymbol{L_{\Omega}}$) naturally 
subdivides into an exhaustive set of $K$ mutually exclusive
\textbf{clusters} of statistical units, a convenient sampling
strategy is given by selecting $k < K$ clusters from this set at
random and perform complete surveys within each of the chosen
clusters. The probability of selection of any particular
statistical unit from $\boldsymbol{\Omega}$ (or
$\boldsymbol{L_{\Omega}}$) thus amounts to $k/K$. This
\textbf{cluster random sampling} method has the practical advantage
of being less contrived. However, in general it entails sampling
errors that are greater than for the previous two sampling methods.
Further details on the cluster random sampling technique can be
found, e.g., in Bortz and D\"{o}ring
(2006)~\ct[p~435ff]{bordoe2006}.

\vspace{5mm}
\noindent
We emphasise at this point that empirical data gained from
\textbf{convenience samples} (in contrast to random samples) is
\textit{not} amenable to \textbf{statistical inference}, in that
its information content \textit{cannot} be generalised to the
target population~$\boldsymbol{\Omega}$ from which it was drawn;
see, e.g., Bryson (1976)~\ct[p~185]{bry1976}, or Schnell \textit{et
al} (2013)~\ct[p~289]{schetal2013}.

\section[Point estimator functions]{Point estimator functions}
Many \textbf{inferential statistical methods of data analysis} in
the \textbf{frequentist framework} revolve around the
\textbf{estimation} of unknown \textbf{distribution parameters}
$\theta$ with respect to some target
population~$\boldsymbol{\Omega}$ by means of corresponding
\textbf{maximum likelihood point estimator functions}
$\hat{\theta}_{n}(X_{1},\ldots,X_{n})$ (or: 
\textbf{statistics}), the values of which are computed from the
data of \textbf{random samples}~$\boldsymbol{S_{\Omega}}$:~$(X_{1}, 
\ldots, X_{n})$. Owing to the stochastic nature of the random 
sampling process, any point estimator function 
$\hat{\theta}_{n}(X_{1},\ldots,X_{n})$ is subject to a
\textbf{random sampling error}. One can show that this estimation
procedure becomes reliable provided that a point estimator function 
satisfies the following two important criteria of quality:
\begin{itemize}
\item[(i)] \textbf{Unbiasedness:}
$\mathrm{E}(\hat{\theta}_{n})=\theta$, and

\item[(ii)] \textbf{Consistency:}
$\displaystyle\lim_{n\to\infty}\mathrm{Var}(\hat{\theta}_{n})=0$.
\end{itemize}
For metrically scaled one-dimensional random variables $X$, 
defining for a given random sample 
$\boldsymbol{S_{\Omega}}$:~$(X_{1}, \ldots, X_{n})$ of size $n$ a 
\textbf{sample total sum} by
\be
\lb{eq:samptot}
Y_{n} := \sum_{i=1}^{n}X_{i} \ ,
\ee
the two most prominent \textbf{maximum likelihood point estimator 
functions} satisfying the \textbf{unbiasedness} and
\textbf{consistency} conditions are the \textbf{sample mean} and
\textbf{sample variance}, defined by
\bea
\lb{eq:sampmean}
\bar{X}_{n} & := & \frac{1}{n}\,Y_{n} \\
\lb{eq:sampvar}
S_{n}^{2} & := & 
\frac{1}{n-1}\,\sum_{i=1}^{n}(X_{i}-\bar{X}_{n})^{2} \ .
\eea
These will be frequently employed in subsequent considerations in 
Ch.~\ref{ch12} for point-estimating the values of the
\textbf{location} and \textbf{scale parameters}~$\mu$ and 
$\sigma^{2}$ of the distribution for a one-dimensional random 
variable~$X$ in a target population~$\boldsymbol{\Omega}$.
\textbf{Sampling theory} in the \textbf{frequentist framework}
holds it that the \textbf{standard errors (SE)} associated with the 
maximum likelihood point estimator functions $\bar{X}_{n}$ and 
$S_{n}^{2}$, defined in Eqs.~(\ref{eq:sampmean}) and 
(\ref{eq:sampvar}), amount to the standard deviations of the 
underlying theoretical \textbf{sampling distributions} for these 
functions; see, e.g., Cram\'{e}r (1946)~\ct[Chs.~27 to 
29]{cra1946}. For a given target population~$\boldsymbol{\Omega}$ 
(or sampling frame $\boldsymbol{L_{\Omega}}$) of size $N$, imagine 
drawing all possible $\left(\begin{array}{c} N \\ n 
\end{array}\right)$ mutually independent random samples of a fixed 
size $n$ (no order accounted for and repetitions excluded), from 
each of which individual realisations of $\bar{X}_{n}$ and 
$S_{n}^{2}$ are obtained. The theoretical distributions 
for all such realisations of $\bar{X}_{n}$ resp.~$S_{n}^{2}$ for 
given $N$ and $n$ are referred to as their corresponding
\textbf{sampling distributions}. A useful simulation illustrating
the concept of a sampling distribution is available at the website 
\href{http://onlinestatbook.com/stat_sim/sampling_dist/index.html}
{\texttt{onlinestatbook.com}}. In the limit that $N \to \infty$
while keeping $n$ fixed, the theoretical \textbf{sampling
distributions} of $\bar{X}_{n}$ and $S_{n}^{2}$ become normal (cf. 
Sec.~\ref{sec:normverteil}) resp.~$\chi^{2}$ with $n-1$ degrees of 
freedom (cf. Sec.~\ref{sec:chi2verteil}), with standard deviations
\bea
\lb{eq:sesammean}
\text{SE}\bar{X}_{n} & := & \frac{S_{n}}{\sqrt{n}} \\
\lb{eq:sesamvar}
\text{SE}S_{n}^{2} & := & \sqrt{\frac{2}{n-1}}\,S_{n}^{2} \ ;
\eea
cf., e.g., Lehman and Casella (1998)~\ct[p~91ff]{lehcas1998},
and Levin \textit{et al} (2010)~\ct[Ch.~6]{levetal2009}. 
Thus, for a \textit{finite} sample standard deviation $S_{n}$, 
these two \textbf{standard errors} decrease with the sample 
size~$n$ in proportion to the inverse of $\sqrt{n}$ resp.~the 
inverse of $\sqrt{n-1}$. It is a main criticism of proponents of 
the \textbf{Bayes--Laplace approach} to \textbf{Probability Theory}
and \textbf{statistical inference} that the concept of a
\textbf{sampling distribution} for a maximum likelihood point
estimator function is based on \textit{unobserved data}; cf.
Greenberg (2013)~\ct[p~31f]{gre2013}.

\medskip
\noindent
There are likewise unbiased maximum likelihood point estimators 
for the \textbf{shape} parameters $\gamma_{1}$ and $\gamma_{2}$ of 
the probability distribution for a one-dimensional random 
variable~$X$ in a target population~$\boldsymbol{\Omega}$, as 
given in Eqs.~(\ref{eq:skew2}) and (\ref{eq:kurt2}). For $n>2$ 
resp.~$n>3$, the \textbf{sample skewness} and \textbf{sample excess 
kurtosis} in, e.g., their implementation in the software packages 
\R\ (package: \texttt{e1071}, by Meyer \textit{et al}
(2019)~\ct{meyetal2019}) or SPSS are defined by (see, e.g.,
Joanes and Gill (1998)~\ct[p~184]{joagil1998})
\bea
\lb{eq:sampskew}
G_{1} & := & 
\frac{\sqrt{(n-1)n}}{n-2}\,\frac{\frac{1}{n}\,\sum_{i=1}^{n}(X_{i}
-\bar{X}_{n})^{3}}{\left(\frac{1}{n}\,\sum_{j=1}^{n}(X_{j}
-\bar{X}_{n})^{2}\right)^{3/2}}\\
\lb{eq:sampkurt}
G_{2} & := & 
\frac{n-1}{(n-2)(n-3)}\,\left[(n+1)\left(\frac{\frac{1}{n}
\,\sum_{i=1}^{n}(X_{i}-\bar{X}_{n})^{4}}{\left(\frac{1}{n}
\,\sum_{j=1}^{n}(X_{j}-\bar{X}_{n})^{2}\right)^{2}}-3\right)
+6\right] \ ,
\eea
with associated standard errors (cf. Joanes and Gill 
(1998)~\ct[p~185f]{joagil1998})
\bea
\lb{eq:sesamskew}
\text{SE}G_{1} & := & \sqrt{\frac{6(n-1)n}{(n-2)(n+1)(n+3)}} \\
\lb{eq:sesamkurt}
\text{SE}G_{2} & := & 
2\,\sqrt{\frac{6(n-1)^{2}n}{(n-3)(n-2)(n+3)(n+5)}} \ .
\eea
%



\chapter[Null hypothesis significance testing]{\href{https://www.youtube.com/watch?v=jnLHSyYQ1LE}{Null hypothesis significance testing}}
\lb{ch11}
\textbf{Null hypothesis significance testing} by means of
observable quantities is the centrepiece of the current body of
inferential statistical methods in the \textbf{frequentist
framework}. Its logic of an ongoing routine of systematic
\textbf{falsification} of null hypotheses by empirical means is
firmly rooted in the ideas of \textbf{critical rationalism} and
\textbf{logical positivism}. The latter were expressed most
emphatically by the Austro--British 
philosopher \href{http://en.wikipedia.org/wiki/Karl_Popper}{Sir 
Karl Raimund Popper CH FRS FBA (1902--1994)}; see, e.g., Popper 
(2002)~\ct{pop2002}. The systematic procedure for \textbf{null
hypothesis significance testing} on the grounds of observational
\textbf{evidence}, as practiced today within the
\textbf{frequentist framework} as a standardised method of
probability-based \textbf{decision-making}, was developed during
the first half of the $20^\mathrm{th}$ Century, predominantly by
the English statistician, evolutionary biologist, 
eugenicist and geneticist
\href{http://www-history.mcs.st-and.ac.uk/Biographies/Fisher.html}{Sir
Ronald Aylmer Fisher FRS (1890--1962)}, the 
Polish--US-American mathematician and statistician
\href{http://www-history.mcs.st-and.ac.uk/Biographies/Neyman.html}{Jerzy Neyman (1894--1981)}, the English mathematician and statistician
\href{http://www-history.mcs.st-and.ac.uk/Biographies/Pearson.html}{Karl Pearson FRS (1857--1936)}, and his son, the English statistician
\href{http://www-history.mcs.st-and.ac.uk/Biographies/Pearson_Egon.html}{Egon Sharpe Pearson CBE FRS (1895--1980)}; cf. Fisher
(1935)~\ct{fis1935}, Neyman and Pearson (1933)~\ct{neypea1933},
and Pearson (1900)~\ct{pea1900}.
We will describe the main steps of the systematic test procedure 
in the following.

\section[General procedure]{General procedure}
\lb{sec:testgen}
The central aim of \textbf{null hypothesis significance testing} is
to separate, as reliably as possible, \textbf{true effects} in a
\textbf{target population} $\boldsymbol{\Omega}$ of statistical
units concerning distributional properties of, or relations
between, selected \textbf{statistical variables} $X, Y, \ldots, Z$
from \textbf{chance effects} potentially injected by the sampling
approach to probing the nature of $\boldsymbol{\Omega}$. The
sampling approach results in a, generally unavoidable, state of
\textbf{incomplete information} on the part of the researcher.

\medskip
\noindent
In an inferential statistical context, (null and/or research)
\textbf{hypotheses} are formulated as assumptions on\\[-5mm]
\begin{itemize}
\item[(i)] the \textbf{probability distribution function} $F$ of
one or more \textbf{random variables} $X, Y, \ldots, Z$ in 
$\boldsymbol{\Omega}$, or on\\[-5mm]

\item[(ii)] one or more \textbf{parameters} $\theta$ of this
probability distribution function.\\[-5mm]
\end{itemize}
Generically, statistical hypotheses need to be viewed as 
probabilistic statements. As such the researcher will always have 
to deal with a fair amount of \textbf{uncertainty} in deciding 
whether an observed, potentially only apparent effect is
\textbf{statistically significant} and/or \textbf{practically
significant} in $\boldsymbol{\Omega}$ or not. Bernstein
(1998)~\ct[p~207]{ber1998} summarises the circumstances relating to
the test of a specific hypothesis as follows:
\begin{quotation}
``Under conditions of uncertainty, the choice is not between 
rejecting a hypothesis and accepting it, but between reject and 
not--reject.''
\end{quotation}

\medskip
\noindent
The question arises as to \textit{which kinds of quantitative 
problems can be efficiently settled by statistical means}?
With respect to a given target population $\boldsymbol{\Omega}$, 
in the simplest kinds of applications of \textbf{null hypothesis
significance testing}, one may (a)~\textbf{test for differences} in
the distributional properties of a single one-dimensional
statistical variable $X$ between a number of subgroups of
$\boldsymbol{\Omega}$, necessitating \textbf{univariate methods} of
data analysis, or one may (b)~\textbf{test for association} for a
two-dimensional statistical variable $(X,Y)$, thus requiring
\textbf{bivariate methods} of data analysis. The standardised
procedure for \textbf{null hypothesis significance testing},
practiced within the \textbf{frequentist framework} for the purpose
of assessing statistical significance of an observed, potentially
apparent effect, takes the following six steps on the way to making
a \textbf{decision}:

\medskip
\noindent
\textbf{Six-step procedure for null hypothesis significance
testing}
\begin{enumerate}

\item Formulation, with respect to the target population 
$\boldsymbol{\Omega}$, of a pair of mutually exclusive
\textbf{hypotheses}:
\begin{enumerate}
\item the \textbf{null hypothesis} $H_{0}$ conjectures that ``there 
exists \textit{no} effect in $\boldsymbol{\Omega}$ of the kind 
envisaged by the researcher,'' while

\item the \textbf{research hypothesis} $H_{1}$ conjectures  that 
``there \textit{does} exist a true effect in $\boldsymbol{\Omega}$ 
of the kind envisaged by the researcher.''
\end{enumerate}
The starting point of the test procedure is the \textit{assumption 
(!)} that it is the content of the $H_{0}$ conjecture which is 
realised in $\boldsymbol{\Omega}$. The objective is to try to 
refute $H_{0}$ empirically on the basis of random sample data 
drawn from $\boldsymbol{\Omega}$, to a level of significance which 
needs to be specified in advance. In this sense it is $H_{0}$ 
which is being subjected to a statistical test.\footnote{Bernstein 
(1998)~\ct[p~209]{ber1998} refers to the statistical test of a 
(null) hypothesis as a ``mathematical stress test.''} The striking 
\textit{asymmetry} regarding the roles of $H_{0}$ and $H_{1}$ in 
the test procedure embodies the notion of a \textbf{falsification}
of hypotheses, as advocated by critical rationalism.

\item Specification of a \textbf{significance level} $\alpha$ prior
to the performance of the test, where, by convention, $\alpha \in 
[0.01,0.05]$. The parameter $\alpha$ is synonymous with the 
probability of committing a Type~I error (to be defined below) in 
making a test decision.

\item Construction of a suitable continuous real-valued measure 
for quantifying deviations of the data in a random sample 
$\boldsymbol{S_{\Omega}}$:~$(X_{1}, \ldots, X_{n})$ of size~$n$ 
from the initial ``no effect in $\boldsymbol{\Omega}$'' conjecture
of $H_{0}$, a \textbf{test statistic} $T_{n}(X_{1},\ldots, X_{n})$ 
that is perceived as a one-dimensional random variable with (under 
the $H_{0}$ assumption) \textit{known (!)} associated
\textbf{theoretical probability distribution} for computing related
event probabilities. The latter is referred to as the \textbf{test
distribution}.\footnote{Within the frequentist framework of
null hypothesis significance testing the test statistic and its
partner test distribution form an intimate pair of decision-making
devices.}

\item Determination of the \textbf{rejection region}
$B_{\alpha}$ for $H_{0}$ within the spectrum of values of 
the test statistic $T_{n}(X_{1},\ldots, X_{n})$ from re-arranging 
the conditional probability condition
\be
P\left(T_{n}(X_{1}, \ldots, X_{n}) \in B_{\alpha}|H_{0}\right)
\stackrel{!}{\leq} \alpha \ ,
\ee
where $P(\ldots)$ and the threshold
$\alpha$--quantile(s)~$P^{-1}(\alpha)$ demarking the boundary(ies)
of $B_{\alpha}$ are to be calculated from the assumed (continuous)
test distribution.

\item Computation of a specific \textbf{realisation}
$t_{n}(x_{1}, \ldots, x_{n})$ of the test statistic
$T_{n}(X_{1}, \ldots, X_{n})$ from the data $x_{1}, 
\ldots, x_{n}$ in a \textbf{random sample} 
$\boldsymbol{S_{\Omega}}$:~$(X_{1}, \ldots, X_{n})$, the latter of 
which constitutes the required observational
\textbf{evidence}.

\item Derivation of a \textbf{test decision} on the basis of the 
following alternative criteria: when for the realisation 
$t_{n}(x_{1}, \ldots, x_{n})$ of the test statistic $T_{n}(X_{1}, 
\ldots, X_{n})$, resp.~the $p$--value (to be defined in 
Sec.~\ref{sec:pvalue} below) associated with this 
realisation,\footnote{The statistical software packages \R{}\ and
SPSS provide $p$--values as a means for making decisions in null
hypothesis significance testing.} it holds that
\begin{itemize}
\item[(i)]\ $t_{n} \in B_{\alpha}$, resp.
$\boldsymbol{p}\text{{\bf--value}} < \alpha, \text{then}
\quad\Rightarrow\quad
\text{reject}~H_{0}$,\\[-5mm]
\item[(ii)]\ $t_{n} \notin B_{\alpha}$, resp.
$\boldsymbol{p}\text{{\bf--value}} \geq \alpha, \text{then}
\quad\Rightarrow\quad\text{not\ reject}~H_{0}$.
\end{itemize}
\end{enumerate}

\medskip
\noindent
A fitting metaphor for the six-step procedure for \textbf{null
hypothesis significance testing} just described is that of a
statistical long jump competition. The issue here is to find out
whether actual empirical data deviates sufficiently strongly from
the ``no effect'' reference state conjectured in the
given \textbf{null hypothesis}~$H_{0}$, so as to land in the 
corresponding \textbf{rejection region}~$B_{\alpha}$ within the 
spectrum of values of the \textbf{test 
statistic}~$T_{n}(X_{1}, \ldots, X_{n})$. Steps~1 to~4 prepare the 
long jump facility (the test stage), while the evaluation of the 
outcome of the jump attempt takes place in steps~5 and~6. Step~4 
necessitates the direct application of \textbf{Probability Theory}
within the \textbf{frequentist framework} in that the determination
of the \textbf{rejection region}~$B_{\alpha}$ for~$H_{0}$ entails
the calculation of a conditional event probability from an
\textit{assumed} \textbf{test distribution}.

\medskip
\noindent
When an effect observed on the basis of random sample data proves 
to possess \textbf{statistical signifi\-cance} (to a predetermined 
significance level), this means that most likely it has come about 
\textit{not by chance} due to the sampling methodology. A different 
matter altogether is whether such an effect also possesses
\textbf{practical significance}, so that, for instance, management 
decisions ought to be adapted to it. \textbf{Practical
significance} of an observed effect can be evaluated, e.g., with
the standardised and scale-invariant \textbf{effect size} measures
proposed by Cohen (1992, 2009)~\ct{coh1992,coh2009}. Addressing
the \textbf{practical significance} of an observed effect should be 
commonplace in any report on inferential statistical data analysis;
see also Sullivan and R Feinn (2012)~\ct{sulfei2012}.

\medskip
\noindent
When performing \textbf{null hypothesis significance testing}, the
researcher is always at \textbf{risk} of making a wrong 
decision. Hereby, one distinguishes between the following two 
kinds of potential error:
\begin{itemize}
\item \textbf{Type I error:} reject an $H_{0}$ which, however,
is true, with conditional probability $P(H_{1}|H_{0}\ 
\text{true})=\alpha$; this case is also referred to as a ``false 
positive,'' and 

\item \textbf{Type II error:} not reject an $H_{0}$ which, however,
is false, with conditional probability $P(H_{0}|H_{1}\ 
\text{true})=\beta$; this case is also referred to as a ``false 
negative.''

\end{itemize}
By fixing the significance level $\alpha$ prior to running 
a statistical test, one controls the risk of committing a Type~I 
error in the decision process. We condense the different possible 
outcomes when making a test decision in
Tab.~\ref{tab:decerrors}.
%
\begin{table}
%
\begin{center}
    \begin{tabular}[!h]{c|ccc}
    	\hline
    	 & & & \\
    	 & $H_{0}$: \textit{no effect} & \fbox{Decision for:} & 
    	 $H_{1}$: \textit{effect} \\
    	 & & & \\
    	\hline
    	 & & & \\
    	$H_{0}$: \textit{no effect} & correct decision: & &
			\textbf{Type I error}: \\
    	true & $P(H_{0}|H_{0}\ \text{true})=1-\alpha$ & & 
    	$P(H_{1}|H_{0}\ \text{true})=\alpha$ \\
    	 & & & \\
    	\fbox{Reality / $\boldsymbol{\Omega}$:} & & & \\
    	 & & & \\
    	$H_{1}$: \textit{effect} & \textbf{Type II error}: & &
			correct decision: \\
    	true & $P(H_{0}|H_{1}\ \text{true})=\beta$ & & 
    	$P(H_{1}|H_{1}\ \text{true})=1-\beta$ \\
    	 & & & \\
      \hline
    \end{tabular}
\end{center}
\caption{Consequences of test decisions in null hypothesis
significance testing.}
\lb{tab:decerrors}
\end{table}

\medskip
\noindent
While the probability $\alpha$ is required to be specified
\textit{a priori} to a statistical test, the probability $\beta$ is 
typically computed \textit{a posteriori}. One refers to the 
probability $1-\beta$ associated with the latter as the
\textbf{power} of a statistical test. Its magnitude is determined
in particular by the parameters \textbf{sample size}~$n$,
\textbf{significance level}~$\alpha$, and the \textbf{effect size}
of the phenomenon to be investigated; see, e.g., Cohen 
(2009)~\ct{coh2009} and Hair \textit{et al} 
(2010)~\ct[p~9f]{haietal2010}.


\medskip
\noindent
As emphasised at the beginning of this chapter, \textbf{null
hypothesis significance testing} is at the heart of
quantitative--empirical research rooted in the
\textbf{frequentist framework}. To foster scientific progress in
this context, it is essential that the scientific community, in an
act of self-control, aims at repeated \textbf{replication} of
specific test results in independent investigations. An interesting
article in this respect was published by the weekly magazine
\texttt{The Economist} on Oct 19, 2013, see Ref.~\ct{eco2013},
which points out that, when subjected to such scrutiny, in general
negative empirical results ($H_{0}$ not rejected) prove much more
reliable than positive ones ($H_{0}$ rejected), though scientific
journals tend to have a bias towards publication of the latter.
A similar viewpoint is expressed in the paper by Nuzzo
(2014)~\ct{nuz2014}. Rather critical accounts of the conceptual
foundations of null hypothesis significance testing are given in
the works by Gill (1999)~\ct{gil1999} and by Kruschke and Liddell
(2017)~\ct{krulid2017}.

\medskip
\noindent
The complementary \textbf{Bayes--Laplace approach} to 
\textbf{statistical data analysis} (cf. Sec. \ref{subsec:bayes}) 
does neither require the prior specification of a significance 
level~$\alpha$, nor the introduction of a test statistic 
$T_{n}(X_{1}, \ldots, X_{n})$ with a partner test distribution for
the empirical testing of a (null) hypothesis. As described in
detail by Jeffreys (1939)~\ct{jef1939}, Jaynes (2003)~\ct{jay2003},
Sivia and Skilling (2006)~\ct{sivski2006}, Gelman \textit{et al}
(2014)~\ct{geletal2014} or McElreath (2016)~\ct{mce2016}, here
\textbf{statistical inference} is practiced entirely on the basis
of a \textbf{posterior probability distribution}
$P(\text{hypothesis}|\text{data}, I)$ for
the (research) hypothesis to be tested, conditional on the
empirical data that was analysed for this purpose, and on the
``relevant background information~$I$'' available to the researcher 
beforehand. By employing \textbf{Bayes' theorem} [cf. 
Eq.~(\ref{eq:bayes})], this \textbf{posterior probability 
distribution} is computed in particular from the product between 
the \textbf{likelihood function} $P(\text{data}|\text{hypothesis}, 
I)$ of the data, given the hypothesis and $I$, and the
\textbf{prior probability distribution} $P(\text{hypothesis}, I)$ 
encoding the researcher's initial reasonable
\textbf{degree-of-belief} in the truth content of the hypothesis on
the backdrop of $I$. That is (see Sivia and Skilling
(2006)~\ct[p~6]{sivski2006}),
\be
P(\text{hypothesis}|\text{data}, I) \propto 
P(\text{data}|\text{hypothesis}, I) \times P(\text{hypothesis}, I) 
\ .
\ee
%

\medskip
\noindent
The \textbf{Bayes--Laplace approach} can be viewed as a proposal to 
the formalisation of the process of \textbf{learning}. Note that
the posterior probability distribution of one round of data
generation and analysis can serve as the prior probability
distribution for a subsequent round of generation and analysis of
new data. Further details on the principles within the
\textbf{Bayes--Laplace framework} underlying the estimation of
distribution parameters, the optimal curve-fitting to a given set
of empirical data points, and the related selection of an adequate mathematical model are given in, e.g., Greenberg
(2013)~\ct[Chs.~3~and~4]{gre2013}, Saha (2002)~\ct[p~8ff]{sah2002},
Lupton (1993)~\ct[p~50ff]{lup1993}, and in Ref.~\ct{hve2018}.

\section[Definition of a $p$--value]{Definition of a 
$\boldsymbol{p}$--value}
\lb{sec:pvalue}
\underline{\textbf{Def.:}} Let $T_{n}(X_{1}, \ldots, X_{n})$ be
the \textbf{test statistic} of a particular \textbf{null hypothesis
significance test} in the \textbf{frequentist framework}. The
\textbf{test distribution} associated with $T_{n}(X_{1}, \ldots,
X_{n})$ \textit{be known} under the assumption that the null
hypothesis $H_{0}$ holds true in the target
population~$\boldsymbol{\Omega}$. 
The $\boldsymbol{p}${\bf--value} associated with a \textbf{
realisation} $t_{n}(x_{1}, \ldots, x_{n})$ of the test statistic 
$T_{n}(X_{1}, \ldots, X_{n})$ is defined as the conditional 
probability of finding a value for $T_{n}(X_{1}, \ldots, X_{n})$  
which is \textit{equal to or more extreme} than the actual 
realisation $t_{n}(x_{1}, \ldots, x_{n})$, given that the null 
hypothesis $H_{0}$ applies in the target 
population~$\boldsymbol{\Omega}$. This conditional probability is
to be computed from the test distribution.

\vspace{5mm}
\noindent
Specifically, using the computational rules 
(\ref{eq:comprulescont2})--(\ref{eq:comprulescont4}), one obtains 
for a
\begin{itemize}
\item two-sided statistical test,
\begin{eqnarray}
\lb{eq:pvaluetwo}
p & := & P(T_{n}<-\left.|t_{n}|\right|H_{0})
+ P(T_{n}>\left.|t_{n}|\right|H_{0}) \nonumber\\
& = & P(T_{n}<-\left.|t_{n}|\right|H_{0})
+ 1 - P(T_{n}\leq\left.|t_{n}|\right|H_{0}) \nonumber\\
& = & F_{T_{n}}(-|t_{n}|) + 1 - F_{T_{n}}(|t_{n}|) \ .
\end{eqnarray}
This result specialises to $p=2\left[1 - F_{T_{n}}(|t_{n}|)\right]$
if the respective \texttt{pdf} of the test distribution exhibits
\textbf{reflection symmetry} with respect to a vertical axis at
$t_{n}=0$, i.e., when 
$F_{T_{n}}(-|t_{n}|)=1 - F_{T_{n}}(|t_{n}|)$ holds.

\item left-sided statistical test,
\be
\lb{eq:pvalueleft}
p := P(T_{n}<t_{n}|H_{0})
= F_{T_{n}}(t_{n}) \ ,
\ee

\item right-sided statistical test,
\be
\lb{eq:pvalueright}
p := P(T_{n}>t_{n}|H_{0})
= 1 - P(T_{n}\leq t_{n}|H_{0})
= 1 - F_{T_{n}}(t_{n}) \ .
\ee
\end{itemize}
With respect to the \textbf{test decision criterion} of rejecting 
an $H_{0}$ whenever $p < \alpha$, one refers to (i)~cases with 
$p<0.05$ as \textbf{significant} test results, and to (ii)~cases
with $p<0.01$ as \textbf{highly significant} test
results.\footnote{Lakens (2017)~\ct{lak2017} posted a stimulating
blog entry on the potential traps associated with the
interpretation of a $p$--value in statistical data analysis. His
remarks come along with illustrative demonstrations in \R{},
including the underlying codes.}

\medskip
\noindent
\underline{Remark:} User-friendly routines for the computation of 
$p$--values are available in \R{}, SPSS, EXCEL and OpenOffice, and 
also on some GDCs.

\vspace{5mm}
\noindent
In the following two chapters, we will turn to discuss a number of 
standard problems in \textbf{Inferential Statistics} within the 
\textbf{frequentist framework}, in association with the 
quantitative--empirical tools that have been developed in this
context to tackle them. In Ch.~\ref{ch12} we will be concerned with
problems of a \textbf{univariate} nature, in particular,
\textbf{testing for statistical differences} in the distributional
properties of a single one-dimensional statistical variable~$X$
between two of more subgroups of some target
population~$\boldsymbol{\Omega}$, while in Ch.~\ref{ch13} the
problems at hand will be of a \textbf{bivariate} nature,
\textbf{testing for statistical association} in
$\boldsymbol{\Omega}$ for a two-dimensional statistical
variable~$(X,Y)$. An entertaining exhaustive account of the history
of statistical methods of data analysis prior to the year 1900 is
given by Stigler (1986)~\ct{sti1986}.


\chapter[Univariate methods of statistical data
analysis]{Univariate methods of statistical data analysis: 
confidence intervals and testing for differences}
\lb{ch12}
In this chapter we present a selection of standard inferential 
statistical techniques within the \textbf{frequentist framework}
that, based upon the random sampling of some target 
population~$\boldsymbol{\Omega}$, were developed for the purpose 
of (a)~range-estimating unknown distribution parameters by means 
of \textbf{confidence intervals}, (b)~\textbf{testing for
differences} between a given empirical distribution of a
one-dimensional statistical variable and its \textit{a priori}
assumed theoretical distribution, and (c)~\textbf{comparing}
distributional properties and parameters of a one-dimensional
statistical variable between two or more subgroups of
$\boldsymbol{\Omega}$. Since the methods to be introduced relate to considerations on distributions of a single one-dimensional
statistical variable only, they are thus referred to as
\textbf{univariate}.

\section[Confidence intervals]{Confidence intervals}
\lb{sec:konfintv}
Assume given a continuous one-dimensional statistical variable~$X$ 
which satisfies in some target population~$\boldsymbol{\Omega}$ a 
\textbf{Gau\ss ian normal distribution} with \textit{unknown}
\textbf{distribution parameters} $\theta \in \{\mu, \sigma^{2}\}$
(cf. Sec.~\ref{sec:normverteil}). The issue is to determine, using 
empirical data from a random sample 
$\boldsymbol{S_{\Omega}}$:~$(X_{1}, \ldots, X_{n})$, a two-sided 
 \textbf{confidence interval} estimate for any one of these unknown distribution parameters~$\theta$
at (as one says) a \textbf{confidence level} $1-\alpha$, where, by
convention, $\alpha \in [0.01,0.05]$.

\medskip
\noindent
Centred on a suitable unbiased and consistent maximum likelihood 
point estimator function~$\hat{\theta}_{n}(X_{1},\ldots,X_{n})$
for $\theta$, the aim of the estimation process is to explicitly
account for the \textbf{sampling error} $\delta_{K}$ arising due to
the random selection process. This approach yields a two-sided
confidence interval
\be
K_{1-\alpha}(\theta) = \left[\hat{\theta}_{n}-\delta_{K},
\hat{\theta}_{n}+\delta_{K}\right] \ ,
\ee
such that $P(\theta \in K_{1-\alpha}(\theta))=1-\alpha$ applies.
The interpretation of the confidence interval $K_{1-\alpha}$ is
that upon arbitrarily many independent repetitions of the
random sampling process, in $(1-\alpha)\times$100\% of all cases
the unknown distribution parameter~$\theta$ will fall inside the
boundaries of $K_{1-\alpha}$ and in $\alpha\times$100\% of all
cases it will not.\footnote{In actual reality, for a given
fixed confidence interval $K_{1-\alpha}$, the unknown distribution
parameter~$\theta$ either takes its value inside $K_{1-\alpha}$, or
not, but the researcher cannot say which case applies.} In the
following we will consider the two cases which result when choosing
$\theta \in \{\mu, \sigma^{2}\}$.

\subsection[Confidence intervals for a mean]{Confidence
intervals for a population mean}
When $\theta=\mu$, and $\hat{\theta}_{n}=\bar{X}_{n}$ by 
Eq.~(\ref{eq:sampmean}), the \textbf{two-sided confidence interval 
for a population mean} $\mu$ at significance level $1-\alpha$ 
becomes
\be
K_{1-\alpha}(\mu) = \left[\bar{X}_{n}-\delta_{K},
\bar{X}_{n}+\delta_{K}\right] \ ,
\ee
with a \textbf{sampling error} amounting to
\be
\lb{eq:samplerrcimu}
\delta_{K} = t_{n-1;1-\alpha/2}\,\frac{S_{n}}{\sqrt{n}} \ ,
\ee
where $S_{n}$ is the positive square root of the \textbf{sample 
variance} $S_{n}^{2}$ according to Eq.~(\ref{eq:sampvar}), and
$t_{n-1;1-\alpha/2}$ denotes the value of the 
$(1-\alpha/2)$--quantile of a $t$--distribution with $df=n-1$ 
degrees of freedom; cf. Sec.~\ref{sec:tverteil}. The ratio 
$\displaystyle\frac{S_{n}}{\sqrt{n}}$ represents the
\textbf{standard error}~$\text{SE}\bar{X}_{n}$ associated with
$\bar{X}_{n}$; cf. Eq.~(\ref{eq:sesammean}).

\medskip
\noindent
\underline{GDC:} mode \texttt{STAT} $\rightarrow$ \texttt{TESTS} 
$\rightarrow$ \texttt{TInterval}

\medskip
\noindent
Equation~(\ref{eq:samplerrcimu}) may be inverted to obtain the 
\textbf{minimum sample size} necessary to construct a two-sided 
confidence interval for $\mu$ to a prescribed accuracy 
$\delta_\mathrm{max}$, maximal sample variance
$\sigma_\mathrm{max}^{2}$, and fixed confidence level $1-\alpha$.
Thus,
\be
n \geq \left(\frac{t_{n-1;1-\alpha/2}}{\delta_\mathrm{max}}
\right)^{2}\sigma_\mathrm{max}^{2} \ .
\ee
%

\subsection[Confidence intervals for a variance]{Confidence
intervals for a population variance}
When $\theta=\sigma^{2}$, and $\hat{\theta}_{n}=S_{n}^{2}$ by 
Eq.~(\ref{eq:sampvar}), the associated point estimator function
\be
\frac{(n-1)S_{n}^{2}}{\sigma^{2}} \sim
\chi^{2}(n-1) \ ,
\quad\text{with}\quad n \in \mathbb{N} \ ,
\ee
satisfies a $\chi^{2}$--distribution with $df=n-1$ degrees of 
freedom; cf. Sec.~\ref{sec:chi2verteil}. By inverting the 
condition
\be
P\left(\chi^{2}_{n-1;\alpha/2} \leq
\frac{(n-1)S_{n}^{2}}{\sigma^{2}} \leq
\chi^{2}_{n-1;1-\alpha/2}\right) \stackrel{!}{=} 1-\alpha \ ,
\ee
one derives a \textbf{two-sided confidence interval for a
population variance} $\sigma^{2}$ at significance level $1-\alpha$
given by
\be
\left[\frac{(n-1)S_{n}^{2}}{\chi^{2}_{n-1;1-\alpha/2}},
\frac{(n-1)S_{n}^{2}}{\chi^{2}_{n-1;\alpha/2}}\right] \ .
\ee
$\chi^{2}_{n-1;\alpha/2}$ and $\chi^{2}_{n-1;1-\alpha/2}$ again 
denote the values of particular quantiles of a 
$\chi^{2}$--distribution.

\section[One-sample $\chi^{2}$--goodness--of--fit--test]{\href{https://www.youtube.com/watch?v=EjMZdii62Fk}{One-sample
$\boldsymbol{\chi}^{2}$--goodness--of--fit--test}}
A standard research question in quantitative--empirical 
investigations deals with the issue whether or not, with respect 
to some target population~$\boldsymbol{\Omega}$ of sample units, 
the \textbf{distribution law} for a specific one-dimensional 
statistical variable~$X$ may be assumed to comply with a 
particular theoretical reference distribution. This question can 
be formulated in terms of the corresponding \texttt{cdf}s,
$F_{X}(x)$ and $F_{0}(x)$, presupposing that for practical reasons
the spectrum of values of $X$ is subdivided into a set of $k$
mutually exclusive \textbf{categories} (or \textbf{bins}), with $k$
a judiciously chosen positive integer which depends in the first
place on the size $n$ of the random sample
$\boldsymbol{S_{\Omega}}$:~$(X_{1}, \ldots, X_{n})$ to be
investigated.

\medskip
\noindent
The non-parametric \textbf{one-sample 
$\boldsymbol{\chi}^{2}$--goodness--of--fit--test} takes 
as its starting point the pair of

\medskip
\noindent
\textbf{Hypotheses:}
\be
\begin{cases}
H_{0}: F_{X}(x) = F_{0}(x)
\quad\Leftrightarrow\quad
O_{i}-E_{i} = 0 \\
H_{1}: F_{X}(x) \neq F_{0}(x)
\quad\Leftrightarrow\quad
O_{i}-E_{i} \neq 0
\end{cases} \ ,
\ee
where $O_{i}$ ($i=1,\ldots,k$) denotes the actually
\textbf{observed frequency} of category $i$ in a random sample of
size $n$, $E_{i}:=np_{i}$ denotes the, under $H_{0}$ (and so
$F_{0}(x)$), theoretically \textbf{expected frequency} of
category~$i$ in the same random sample, and $p_{i}$ is the
\textbf{probability} of finding a value of $X$ in category~$i$
under $F_{0}(x)$.

\medskip
\noindent
The present procedure, devised by Pearson (1900)~\ct{pea1900}, 
employs the \textbf{residuals} $O_{i}-E_{i}$ ($i=1\ldots,k$) to 
construct a suitable 

\medskip
\noindent
\textbf{Test statistic:}
\be
\lb{eq:chisqgofteststat}
\fbox{$\displaystyle
T_{n}(X_{1}, \ldots, X_{n})
= \sum_{i=1}^{k}\frac{(O_{i}-E_{i})^{2}}{E_{i}}
\ \stackrel{H_{0}}{\approx}\ \chi^{2}(k-1-r)
$}
\ee
in terms of a sum of rescaled squared residuals 
$\displaystyle\frac{(O_{i}-E_{i})^{2}}{E_{i}}$,\footnote{As the
$E_{i}$ ($i=1\ldots,k$) amount to count data with unknown maximum
counts, the probability distribution relevant to model variation is
the Poisson distribution discussed in
Sec.~\ref{sec:poissonverteil}. Hence, the standard deviations are
equal to~$\sqrt{E_{i}}$, and so the variances equal to~$E_{i}$; cf.
Jeffreys (1939)~\ct[p~106]{jef1939}.} which, under $H_{0}$,
approximately follows a $\boldsymbol{\chi^{2}}$\textbf{--test
distribution} with $df=k-1-r$ degrees of freedom (cf.
Sec.~\ref{sec:chi2verteil}); $r$~denotes the number of free
parameters of the reference distribution $F_{0}(x)$ which need to
be estimated from the random sample data. For this test procedure
to be reliable, it is \textit{important (!)} that the size $n$ of
the random sample be chosen such that the condition
\be
E_{i} \stackrel{!}{\geq} 5
\ee
holds for all categories $i=1,\ldots,k$, due to the fact that the 
$E_{i}$ appear in the denominator of the test statistic in 
Eq.~(\ref{eq:chisqgofteststat}) (and so would artifically inflate 
the magnitudes of the summed ratios when the denominators become 
too small).

\medskip
\noindent
\textbf{Test decision:} The rejection region for $H_{0}$ at 
significance level $\alpha$ is given by (right-sided test)
\be
t_{n}>\chi^{2}_{k-1-r;1-\alpha} \ .
\ee
By Eq.~(\ref{eq:pvalueright}), the $p$--value associated with a 
realisation $t_{n}$ of the \textbf{test
statistic}~(\ref{eq:chisqgofteststat}), which is to be calculated
from the $\boldsymbol{\chi^{2}}$\textbf{--test distribution},
amounts to
\be
p = P(T_{n}>t_{n}|H_{0}) = 1-P(T_{n}\leq t_{n}|H_{0})
= 1-\chi^{2}\texttt{cdf}(0,t_{n},k-1-r) \ .
\ee

\medskip
\noindent
\underline{\R:} \texttt{chisq.test(table(\textit{variable}))} \\
\underline{SPSS:} Analyze $\rightarrow$ Nonparametric Tests
$\rightarrow$ Legacy Dialogs $\rightarrow$ Chi-square \ldots

\medskip
\noindent
\textbf{Effect size:} In the present context, the practical
significance of the phenomenon investigated can be estimated
from the realisation $t_{n}$ and the sample size~$n$ by
\be
\fbox{$\displaystyle
\lb{eq:eschisq}
w := \sqrt{\frac{t_{n}}{n}} \ .
$}
\ee
For the interpretation of its strength Cohen 
(1992)~\ct[Tab.~1]{coh1992} recommends the

\medskip
\noindent
\underline{\textbf{Rule of thumb:}}\\
$0.10 \leq w < 0.30$: small effect\\
$0.30 \leq w < 0.50$: medium effect\\
$0.50 \leq w$: large effect.

\medskip
\noindent
Note that in the spirit of \textbf{critical rationalism} the 
one-sample $\chi^{2}$--goodness--of--fit--test provides a tool for 
empirically \textit{excluding} possibilities of distribution laws 
for $X$.

\section[One-sample $t$-- and $Z$--tests for a population 
mean]{One-sample $\boldsymbol{t}$-- and $\boldsymbol{Z}$--tests 
for a population mean}
\lb{sec:onesampttest}
The idea here is to test whether the unknown population mean $\mu$ 
of some continuous one-dimensional statistical variable~$X$ is 
equal to, less than, or greater than some reference value 
$\mu_{0}$, to a given significance level~$\alpha$. To this end, it 
is required that $X$ satisfy in the target 
population~$\boldsymbol{\Omega}$ a \textbf{Gau\ss ian normal 
distribution}, i.e., $X \sim N(\mu;\sigma^{2})$; cf. 
Sec.~\ref{sec:normverteil}. The quantitative--analytical tool to 
be employed in this case is the parametric \textbf{one-sample 
$\boldsymbol{t}$--test for a population mean} developed by Student 
[Gosset] (1908)~\ct{stu1908}, or, when the sample size $n \geq 
50$, in consequence of the \textbf{central limit theorem} discussed 
in Sec.~\ref{sec:zentrgrenz}, the corresponding
\textbf{one-sample $\boldsymbol{Z}$--test}.

\medskip
\noindent
For a random sample~$\boldsymbol{S_{\Omega}}$:~$(X_{1}, \ldots, 
X_{n})$ of size $n \geq 50$, the validity of the \textit{assumption 
(!)} of \textbf{normality} for the $X$-distribution can be tested
by a procedure due to the Russian mathematicians
\href{http://www-history.mcs.st-and.ac.uk/Biographies/Kolmogorov.html}{Andrey Nikolaevich Kolmogorov (1903--1987)} 
and 
\href{http://en.wikipedia.org/wiki/Nikolai_Smirnov_(mathematician)}
{Nikolai Vasilyevich Smirnov (1900--1966)}. This tests the null 
hypothesis $H_{0}$: ``There is no difference between the 
distribution of the sample data and the associated reference 
normal distribution'' against the alternative $H_{1}$: ``There is 
a difference between the distribution of the sample data and the 
associated reference normal distribution;''
cf. Kolmogorov (1933)~\ct{kol1933b} and Smirnov 
(1939)~\ct{smi1939}. This procedure is referred to as the
\textbf{Kolmogorov--Smirnov--test} (or, for short, the KS--test).
The associated test statistic evaluates the strength of the
deviation of the empirical cumulative distribution function [cf. 
Eq.~(\ref{klempvert})] of given random sample data, with sample 
mean $\bar{x}_{n}$ and sample variance $s_{n}^{2}$, from the
\texttt{cdf} of a reference Gau\ss ian normal distribution with
parameters $\mu$ and $\sigma^{2}$ equal to these sample values [cf. 
Eq.~(\ref{eq:gaussiancdf})].

\medskip
\noindent
\underline{\R:} \texttt{ks.test(\textit{variable}, "pnorm")} \\
\underline{SPSS:} Analyze $\rightarrow$ Nonparametric Tests
$\rightarrow$ Legacy Dialogs $\rightarrow$ 1-Sample K-S \ldots: 
Normal

\medskip
\noindent
For sample sizes $n < 50$, however, the validity of the normality 
assumption for the $X$-distribution may be estimated in terms of 
the magnitudes of the \textbf{standardised skewness and excess
kurtosis measures},
\be
\lb{eq:g1g2ratios}
\left|\frac{G_{1}}{\text{SE}G_{1}}\right|
\quad\quad\text{and}\quad\quad
\left|\frac{G_{2}}{\text{SE}G_{2}}\right| \ ,
\ee
which are constructed from the quantities defined in
Eqs.~(\ref{eq:sampskew})--(\ref{eq:sesamkurt}).
At a significance level $\alpha = 0.05$, the normality assumption 
may be maintained as long as \textit{both} measures are smaller
than the \textbf{critical value} of~$1.96$; cf. Hair \textit{et al} 
(2010)~\ct[p~72f]{haietal2010}.

\medskip
\noindent
Formulated in a non-directed or a directed fashion, the starting 
point of the $t$--test resp.~$Z$--test procedures are the

\medskip
\noindent
\textbf{Hypotheses:}
\be
\begin{cases}
H_{0}: \mu=\mu_{0}
\quad\text{or}\quad
\mu \geq \mu_{0}
\quad\text{or}\quad
\mu \leq \mu_{0} \\
H_{1}: \mu \neq \mu_{0}
\quad\text{or}\quad
\mu < \mu_{0}
\quad\text{or}\quad
\mu > \mu_{0}
\end{cases} \ .
\ee
To measure the deviation of the sample data from the state 
conjectured to hold in the null hypothesis $H_{0}$, the
difference between the sample mean $\bar{X}_{n}$ and the 
hypothesised population mean $\mu_{0}$, normalised in analogy to 
Eq.~(\ref{eq:standardisation}) by the \textbf{standard error}
\be
\text{SE}\bar{X}_{n} := \frac{S_{n}}{\sqrt{n}}
\ee
of $\bar{X}_{n}$ given in Eq.~(\ref{eq:sesammean}), serves as the 
$\mu_{0}$--dependent

\medskip
\noindent
\textbf{Test statistic:}
\be
\lb{eq:ztteststat}
\fbox{$\displaystyle
T_{n}(X_{1}, \ldots, X_{n})
= \frac{\bar{X}_{n}-\mu_{0}}{\text{SE}\bar{X}_{n}}
\ \stackrel{H_{0}}{\sim}\ 
\begin{cases}
t(n-1) & \text{for}\quad n < 50 \\
 & \\
N(0;1) & \text{for}\quad n \geq 50
\end{cases} \ ,
$}
\ee
which, under $H_{0}$, follows a $\boldsymbol{t}$\textbf{--test
distribution} with $df=n-1$ degrees of freedom (cf.
Sec.~\ref{sec:tverteil}) resp.~a \textbf{standard normal
test distribution} (cf. Sec.~\ref{sec:normverteil}). 

\medskip
\noindent
\textbf{Test decision:} Depending on the kind of test to be 
performed, the rejection region for $H_{0}$ at significance level 
$\alpha$ is given by
\begin{center}
\begin{tabular}[h]{c|c|c|c}
 & & & \\
\textbf{Kind of test} & $\boldsymbol{H_{0}}$ &
$\boldsymbol{H_{1}}$ &
\textbf{Rejection region for} $\boldsymbol{H_{0}}$ \\
 & & & \\
\hline
 & & & \\
(a)~two-sided & $\mu=\mu_{0}$ & $\mu\neq\mu_{0}$ &
$|t_{n}|>
\begin{cases}
t_{n-1;1-\alpha/2} & (t\text{--test}) \\
z_{1-\alpha/2} & (Z\text{--test})
\end{cases}$ \\
 & & & \\
\hline
 & & & \\
(b)~left-sided & $\mu\geq\mu_{0}$ & $\mu<\mu_{0}$ &
$t_{n}<
\begin{cases}
t_{n-1;\alpha}=-t_{n-1;1-\alpha} & (t\text{--test}) \\
z_{\alpha}=-z_{1-\alpha} & (Z\text{--test})
\end{cases}$ \\
 & & & \\
\hline
 & & & \\
(c)~right-sided & $\mu\leq\mu_{0}$ & $\mu>\mu_{0}$ &
$t_{n}>
\begin{cases}
t_{n-1;1-\alpha} & (t\text{--test}) \\
z_{1-\alpha} & (Z\text{--test})
\end{cases}$ \\
 & & &
\end{tabular}
\end{center}
$p$--values associated with realisations $t_{n}$ of the
\textbf{test statistic}~(\ref{eq:ztteststat}) can be obtained from
Eqs.~(\ref{eq:pvaluetwo})--(\ref{eq:pvalueright}), using
the relevant $\boldsymbol{t}$\textbf{--test distribution} resp. the
\textbf{standard normal test distribution}.

\medskip
\noindent
\underline{\R:} \texttt{t.test(\textit{variable}, mu = $\mu_{0}$)},
\\
\texttt{t.test(\textit{variable}, mu = $\mu_{0}$,
alternative = "less")}, \\
\texttt{t.test(\textit{variable}, mu = $\mu_{0}$,
alternative = "greater")} \\
\underline{GDC:}
mode \texttt{STAT} $\rightarrow$ \texttt{TESTS} $\rightarrow$
\texttt{T-Test\ldots} when $n < 50$,
resp.~mode \texttt{STAT} $\rightarrow$ \texttt{TESTS}
$\rightarrow$ \texttt{Z-Test\ldots} when $n \geq 50$. \\
\underline{SPSS:} Analyze $\rightarrow$ Compare Means
$\rightarrow$ One-Sample T Test \ldots

\medskip
\noindent
\underline{Note:} Regrettably, SPSS provides no option for 
selecting between a ``one-tailed'' (left-/right-sided) and a 
``two-tailed'' (two-sided) $t$--test. The default setting is for a 
two-sided test. For the purpose of one-sided tests the $p$--value 
output of SPSS needs to be divided by $2$.

\medskip
\noindent
\textbf{Effect size:} The practical significance of the phenomenon investigated can be estimated from the sample mean~$\bar{x}_{n}$,
the sample standard deviation~$s_{n}$, and the reference
value~$\mu_{0}$ by the scale-invariant ratio
\be
\fbox{$\displaystyle
d := \frac{\left|\bar{x}_{n}-\mu_{0}\right|}{s_{n}} \ .
$}
\ee
For the interpretation of its strength Cohen 
(1992)~\ct[Tab.~1]{coh1992} recommends the

\medskip
\noindent
\underline{\textbf{Rule of thumb:}}\\
$0.20 \leq d < 0.50$: small effect\\
$0.50 \leq d < 0.80$: medium effect\\
$0.80 \leq d$: large effect.

\medskip
\noindent
We remark that the statistical software package~\R\ holds 
available a routine \texttt{power.t.test(power, sig.level, delta,
sd, $n$, alternative, type = "one.sample")} for the purpose of
calculating any one of the parameters \texttt{power},
\texttt{delta} or $n$ (provided all remaining parameters have been
specified) in the context of empirical investigations employing the
one-sample $t$--test for a population mean. One-sided tests are
specified via the parameter setting
\texttt{alternative = "one.sided"}.

\section[One-sample $\chi^{2}$--test for a population 
variance]{One-sample $\boldsymbol{\chi^{2}}$--test for a population
variance}
\lb{sec:onesampchi2vartest}
In analogy to the statistical significance test described in the 
previous section \ref{sec:onesampttest}, one may likewise test 
hypotheses on the value of an unknown population variance 
$\sigma^{2}$ with respect to a reference value $\sigma_{0}^{2}$ 
for a continuous one-dimensional statistical variable~$X$ which 
satisfies in $\boldsymbol{\Omega}$ a \textbf{Gau\ss ian normal 
distribution}, i.e., $X \sim N(\mu;\sigma^{2})$; cf. 
Sec.~\ref{sec:normverteil}. The hypotheses may also be formulated 
in a non-directed or directed fashion according to

\medskip
\noindent
\textbf{Hypotheses:}
\be
\begin{cases}
H_{0}: \sigma^{2}=\sigma_{0}^{2}
\quad\text{or}\quad
\sigma^{2} \geq \sigma_{0}^{2}
\quad\text{or}\quad
\sigma^{2} \leq \sigma_{0}^{2} \\
H_{1}: \sigma^{2} \neq \sigma_{0}^{2}
\quad\text{or}\quad
\sigma^{2} < \sigma_{0}^{2}
\quad\text{or}\quad
\sigma^{2} > \sigma_{0}^{2}
\end{cases} \ .
\ee
In the \textbf{one-sample $\boldsymbol{\chi^{2}}$--test for a 
population variance}, the underlying $\sigma_{0}^{2}$--dependent

\medskip
\noindent
\textbf{Test statistic:}
\be
\lb{eq:chisqteststat}
\fbox{$\displaystyle
T_{n}(X_{1}, \ldots, X_{n})
= \frac{(n-1)S_{n}^{2}}{\sigma_{0}^{2}}
\ \stackrel{H_{0}}{\sim}\ \chi^{2}(n-1)
$}
\ee
is chosen to be proportional to the sample variance defined by 
Eq.~(\ref{eq:sampvar}), and so, under $H_{0}$, follows a
$\boldsymbol{\chi^{2}}$\textbf{--test distribution} with
$df=n-1$ degrees of freedom; cf. Sec.~\ref{sec:chi2verteil}.

\medskip
\noindent
\textbf{Test decision:} Depending on the kind of test to be 
performed, the rejection region for $H_{0}$ at significance level 
$\alpha$ is given by
\begin{center}
\begin{tabular}[h]{c|c|c|c}
 & & & \\
\textbf{Kind of test} & $\boldsymbol{H_{0}}$ &
$\boldsymbol{H_{1}}$ &
\textbf{Rejection region for} $\boldsymbol{H_{0}}$ \\
 & & & \\
\hline
 & & & \\
(a)~two-sided & $\sigma^{2}=\sigma_{0}^{2}$ &
$\sigma^{2}\neq\sigma_{0}^{2}$ &
$t_{n}\begin{cases}<\chi^{2}_{n-1;\alpha/2} \\
>\chi^{2}_{n-1;1-\alpha/2}\end{cases}$ \\
 & & & \\
\hline
 & & & \\
(b)~left-sided & $\sigma^{2}\geq\sigma_{0}^{2}$ &
$\sigma^{2}<\sigma_{0}^{2}$ &
$t_{n}<\chi^{2}_{n-1;\alpha}$ \\
 & & & \\
\hline
 & & & \\
(c)~right-sided & $\sigma^{2}\leq\sigma_{0}^{2}$ &
$\sigma^{2}>\sigma_{0}^{2}$ &
$t_{n}>\chi^{2}_{n-1;1-\alpha}$ \\
 & & &
\end{tabular}
\end{center}
$p$--values associated with realisations $t_{n}$ of the
\textbf{test statistic}~(\ref{eq:chisqteststat}), which are to be
calculated from the
$\boldsymbol{\chi^{2}}$\textbf{--test distribution}, can be
obtained from Eqs.~(\ref{eq:pvaluetwo})--(\ref{eq:pvalueright}).

\medskip
\noindent
\underline{\R:} \texttt{varTest(\textit{variable},
sigma.squared = $\sigma_{0}^{2}$)} (package: \texttt{EnvStats},
by Millard (2013)~\ct{mil2013}), \\
\texttt{varTest(\textit{variable},
sigma.squared = $\sigma_{0}^{2}$, alternative = "less")}, \\
\texttt{varTest(\textit{variable},
sigma.squared = $\sigma_{0}^{2}$, alternative = "greater")}

\medskip
\noindent
Regrettably, the one-sample $\chi^{2}$--test for a population 
variance does not appear to have been implemented in the SPSS 
software package.


\section[Independent samples $t$--test for a mean]{\href{https://www.youtube.com/watch?v=3alSVL8oVMM}{Two independent samples
$\boldsymbol{t}$--test for a population mean}}
\lb{sec:ttestindep}
Quantitative--empirical studies are frequently interested 
in the question as to what extent there exist significant 
differences between two subgroups of some target 
population~$\boldsymbol{\Omega}$ in 
the distribution of a metrically scaled one-dimensional 
statistical variable $X$. Given that $X$ is \textit{normally 
distributed} in $\boldsymbol{\Omega}$ 
(cf. Sec.~\ref{sec:normverteil}), the parametric \textbf{two 
independent samples $\boldsymbol{t}$--test 
for a population mean} originating from work by Student [Gosset] 
(1908)~\ct{stu1908} provides an efficient and powerful 
investigative tool.

\medskip
\noindent
For independent random samples of sizes $n_{1}, n_{2} \geq 50$,
the issue of whether there exists empirical evidence in the 
samples \textit{against} the assumption of a normally 
distributed~$X$ in  $\boldsymbol{\Omega}$ can again be tested for 
by means of the \textbf{Kolmogorov--Smirnov--test}; cf. 
Sec.~\ref{sec:onesampttest}.

\medskip
\noindent
\underline{\R:} \texttt{ks.test(\textit{variable}, "pnorm")} \\
\underline{SPSS:} Analyze $\rightarrow$ Nonparametric Tests
$\rightarrow$ Legacy Dialogs $\rightarrow$ 1-Sample K-S \ldots: 
Normal

\medskip
\noindent
For $n_{1}, n_{2} < 50$, one may resort to a consideration of the 
magnitudes of the \textbf{standardised skewness and excess kurtosis
measures}, Eqs.~(\ref{eq:g1g2ratios}), to check for the validity of
the normality assumption for the $X$-distributions. 

\medskip
\noindent
In addition, prior to the $t$--test procedure, one needs to 
establish whether or not the variances of~$X$ have to be viewed as 
significantly different in the two random samples selected.
\textbf{Levene's test} provides an empirical method to test
$H_{0}:~\sigma_{1}^{2}=\sigma_{2}^{2}$
against $H_{1}:~\sigma_{1}^{2}\neq\sigma_{2}^{2}$; cf. Levene 
(1960)~\ct{lev1960}.

\medskip
\noindent
\underline{\R:}
\texttt{leveneTest(\textit{variable}, \textit{group
variable})} (package: \texttt{car}, by Fox and Weisberg
(2011)~\ct{foxwei2011})

\medskip
\noindent
The hypotheses of a $t$--test may be formulated in a non-directed 
fashion or in a directed one. Hence, the different kinds of 
possible conjectures are

\medskip
\noindent
\textbf{Hypotheses:} \hfill (test for differences)
\be
\begin{cases}
H_{0}: \mu_{1}-\mu_{2}=0
\quad\text{or}\quad
\mu_{1}-\mu_{2} \geq 0
\quad\text{or}\quad
\mu_{1}-\mu_{2} \leq 0 \\
H_{1}: \mu_{1}-\mu_{2} \neq 0
\quad\text{or}\quad
\mu_{1}-\mu_{2} < 0
\quad\text{or}\quad
\mu_{1}-\mu_{2} > 0
\end{cases} \ .
\ee
A test statistic is constructed from the difference of sample 
means, $\bar{X}_{n_{1}}-\bar{X}_{n_{2}}$,
standardised by the \textbf{standard error}
\be
\text{SE}(\bar{X}_{n_{1}}-\bar{X}_{n_{2}})
:= \sqrt{\frac{S_{n_{1}}^{2}}{n_{1}}+\frac{S_{n_{2}}^{2}}{n_{2}}} 
\ ,
\ee
which derives from the associated theoretical \textbf{sampling 
distribution} for $\bar{X}_{n_{1}}-\bar{X}_{n_{2}}$. Thus, one
obtains the

\medskip
\noindent
\textbf{Test statistic:} 
\be
\lb{eq:indeptteststat}
\fbox{$\displaystyle
T_{n_{1},n_{2}} := \frac{\bar{X}_{n_{1}}-\bar{X}_{n_{2}}
}{\text{SE}(\bar{X}_{n_{1}}-\bar{X}_{n_{2}})}
\ \stackrel{H_{0}}{\sim}\ t(df) \ ,
$}
\ee
which, under $H_{0}$, satisfies a
$\boldsymbol{t}$\textbf{--test distribution} (cf. 
Sec.~\ref{sec:tverteil}) with a number of degrees of freedom 
determined by the relations
\be
\displaystyle
df := \begin{cases}
n_{1}+n_{2}-2\ , &
\text{when}\quad \sigma_{1}^{2}=\sigma_{2}^{2} \\
 & \\
{\displaystyle\frac{\left(\frac{S_{n_{1}}^{2}}{n_{1}}
+\frac{S_{n_{2}}^{2}}{n_{2}}\right)^{2}}{\frac{(S_{n_{1}}^{2}/n_{1})^{2}
}{n_{1}-1}+\frac{(S_{n_{2}}^{2}/n_{2})^{2}}{n_{2}-1}}} \ ,
& \text{when}\quad \sigma_{1}^{2} \neq \sigma_{2}^{2}
\end{cases} \ .
\ee

\medskip
\noindent
\textbf{Test decision:} Depending on the kind of test to be 
performed, the rejection region for $H_{0}$ at significance level 
$\alpha$ is given by
\begin{center}
\begin{tabular}[h!]{c|c|c|c}
 & & & \\
\textbf{Kind of test} & $\boldsymbol{H_{0}}$ &
$\boldsymbol{H_{1}}$ &
\textbf{Rejection region for} $\boldsymbol{H_{0}}$ \\
 & & & \\
\hline
 & & & \\
(a)~two-sided & $\mu_{1}-\mu_{2}=0$ & $\mu_{1}-\mu_{2}\neq 0$ &
$|t_{n_{1},n_{2}}|>t_{df;1-\alpha/2}$ \\
 & & & \\
\hline
 & & & \\
(b)~left-sided & $\mu_{1}-\mu_{2}\geq 0$ & $\mu_{1}-\mu_{2}<0$ &
$t_{n_{1},n_{2}}<t_{df;\alpha}=-t_{df;1-\alpha}$ \\
 & & & \\
\hline
 & & & \\
(c)~right-sided & $\mu_{1}-\mu_{2}\leq 0$ & $\mu_{1}-\mu_{2}>0$ &
$t_{n_{1},n_{2}}>t_{df;1-\alpha}$ \\
 & & &
\end{tabular}
\end{center}
$p$--values associated with realisations $t_{n_{1},n_{2}}$ of the
\textbf{test statistic}~(\ref{eq:indeptteststat}), which are to be
calculated from the
$\boldsymbol{t}$\textbf{--test distribution}, can be obtained from 
Eqs.~(\ref{eq:pvaluetwo})--(\ref{eq:pvalueright}).

\medskip
\noindent
\underline{\R:}
\texttt{t.test(\textit{variable}\texttildelow\textit{group
variable})}, \\
\texttt{t.test(\textit{variable}\texttildelow\textit{group
variable}, alternative = "less")}, \\
\texttt{t.test(\textit{variable}\texttildelow\textit{group
variable}, alternative = "greater")} \\
\underline{GDC:} mode \texttt{STAT} $\rightarrow$ \texttt{TESTS}
$\rightarrow$ \texttt{2-SampTTest\ldots} \\
\underline{SPSS:} Analyze $\rightarrow$ Compare Means
$\rightarrow$ Independent-Samples T Test \ldots

\medskip
\noindent
\underline{Note:} Regrettably, SPSS provides no option for 
selecting between a one-sided and a two-sided $t$--test. The 
default setting is for a two-sided test. For the purpose of 
one-sided tests the $p$--value output of SPSS needs to be divided 
by $2$.

\medskip
\noindent
\textbf{Effect size:} The practical significance of the phenomenon investigated can be estimated from the sample
means~$\bar{x}_{n_{1}}$ and $\bar{x}_{n_{2}}$ and the pooled sample
standard deviation
\be
s_{\mathrm{pooled}} := \sqrt{\frac{(n_{1}-1)s_{n_{1}}^{2}
+(n_{2}-1)s_{n_{2}}^{2}}{n_{1}+n_{2}-2}}
\ee
by the scale-invariant ratio
\be
\fbox{$\displaystyle
d := \frac{\left|\bar{x}_{n_{1}}-\bar{x}_{n_{2}}
\right|}{s_{\mathrm{pooled}}} \ .
$}
\ee
For the interpretation of its strength Cohen 
(1992)~\ct[Tab.~1]{coh1992} recommends the

\medskip
\noindent
\underline{\textbf{Rule of thumb:}}\\
$0.20 \leq d < 0.50$: small effect\\
$0.50 \leq d < 0.80$: medium effect\\
$0.80 \leq d$: large effect.

\medskip
\noindent
\underline{\R:}
\texttt{cohen.d(\textit{variable}, \textit{group
variable}, pooled = TRUE)} (package: \texttt{effsize}, by
Torchiano (2018)~\ct{tor2018})

\medskip
\noindent
We remark that the statistical software package~\R\ holds 
available a routine \texttt{power.t.test(power, sig.level, delta,
sd, $n$, alternative)} for the purpose of calculation of any one 
of the parameters \texttt{power}, \texttt{delta} or $n$ (provided 
all remaining parameters have been specified) in the context of 
empirical investigations employing the independent samples 
$t$--test for a population mean. Equal values of $n$ are 
required here. One-sided tests are addressed via the parameter 
setting \texttt{alternative = "one.sided"}.

\medskip
\noindent
When the necessary conditions for the application of the
independent samples $t$--test are \textit{not} satisfied, the 
following alternative test procedures (typically of a weaker test 
power, though) for comparing two subgroups of 
$\boldsymbol{\Omega}$ with respect to the distribution of a 
metrically scaled variable~$X$ exist:
\begin{itemize}
\item[(i)] at the \textbf{nominal} scale level, provided $E_{ij}
\geq 5$ for all $i,j$, the \textbf{$\boldsymbol{\chi}^{2}$--test
for homogeneity}; cf.  Sec.~\ref{sec:chisqhomo} below, and
\item[(ii)] at the \textbf{ordinal} scale level, provided $n_{1}, 
n_{2} \geq 8$, the two independent samples 
\textbf{Mann--Whitney--$\boldsymbol{U}$--test} for a  median; cf. 
the following Sec.~\ref{sec:utest}.
\end{itemize}
%

\section[Independent samples Mann--Whitney--$U$--test]{\href{https://www.youtube.com/watch?v=bG6xXoyEgB8}{Two independent samples
Mann--Whitney--$\boldsymbol{U}$--test for a population median}}
\lb{sec:utest}
The non-parametric \textbf{two independent samples 
Mann--Whitney--$\boldsymbol{U}$--test for a population median},
devised by the Austrian--US-American mathematician and statistician
\href{http://en.wikipedia.org/wiki/Henry_Mann}{Henry Berthold
Mann~(1905--2000)} and the US-American statistician
Donald Ransom Whitney~(1915--2001) in 1947~\ct{manwhi1947}, can be 
applied to random sample data for ordinally scaled one-dimensional 
statistical variables~$X$, or for metrically scaled 
one-dimensional statistical variables~$X$ which may \textit{not} be 
reasonably assumed to be normally distributed in the target 
population~$\boldsymbol{\Omega}$. In both situations, the method 
employs \textbf{rank number data} (cf. Sec.~\ref{sec:2Dord}), which 
faithfully represents the original random sample data, to 
effectively compare the medians of~$X$ (or, rather, the mean 
rank numbers) between two independent groups. It aims to test 
empirically the null hypothesis $H_{0}$ of one of the following 
pairs of non-directed or directed

\medskip
\noindent
\textbf{Hypotheses:} \hfill (test for differences)
\be
\begin{cases}
H_{0}: \tilde{x}_{0.5}(1) = \tilde{x}_{0.5}(2)
\quad\text{or}\quad
\tilde{x}_{0.5}(1) \geq \tilde{x}_{0.5}(2)
\quad\text{or}\quad
\tilde{x}_{0.5}(1) \leq \tilde{x}_{0.5}(2) \\
H_{1}: \tilde{x}_{0.5}(1) \neq \tilde{x}_{0.5}(2)
\quad\text{or}\quad
\tilde{x}_{0.5}(1) < \tilde{x}_{0.5}(2)
\quad\text{or}\quad
\tilde{x}_{0.5}(1) > \tilde{x}_{0.5}(2)
\end{cases} \ .
\ee

\medskip
\noindent
Given two independent sets of random sample data for $X$,
\textbf{ranks} are being introduced on the basis of an ordered
\textbf{joint random sample} of size $n=n_{1}+n_{2}$ according to
$x_{i}(1) \mapsto R[x_{i}(1)]$ and $x_{i}(2) \mapsto R[x_{i}(2)]$. 
From the ranks thus assigned to the elements of each of the two 
sets of data, one computes the

\medskip
\noindent
$\boldsymbol{U}$\textbf{--values:}
\begin{eqnarray}
U_{1} & := & n_{1}n_{2} + \frac{n_{1}(n_{1}+1)}{2}
- \sum_{i=1}^{n_{1}}R[x_{i}(1)] \\
U_{2} & := & n_{1}n_{2} + \frac{n_{2}(n_{2}+1)}{2}
- \sum_{i=1}^{n_{2}}R[x_{i}(2)] \ ,
\end{eqnarray}
for which the identity $U_{1}+U_{2}=n_{1}n_{2}$ applies. Choose 
$U:=\min(U_{1},U_{2})$.\footnote{Since the $U$--values are tied to 
each other by the identity $U_{1}+U_{2}=n_{1}n_{2}$, it makes no 
difference to this method when one chooses $U:=\max(U_{1},U_{2})$ 
instead.} For independent random samples of sizes $n_{1}, n_{2} 
\geq 8$ (see, e.g., Bortz (2005) \ct[p~151]{bor2005}), the 
standardised $U$--value serves as the

\medskip
\noindent
\textbf{Test statistic:} 
\be
\lb{eq:uteststat}
\fbox{$\displaystyle
T_{n_{1},n_{2}} := \frac{U-\mu_{U}}{\text{SE}U}
\ \stackrel{H_{0}}{\approx}\ N(0;1) \ ,
$}
\ee
which, under $H_{0}$, approximately satisfies a \textbf{standard
normal test distribution}; cf. Sec.~\ref{sec:normverteil}. Here,
$\mu_{U}$ denotes the mean of the $U$--value expected under
$H_{0}$; it is defined in terms of the sample sizes by
\be
\mu_{U}:=\frac{n_{1}n_{2}}{2} \ ;
\ee
$\text{SE}U$ denotes the \textbf{standard error} of the $U$--value 
and can be obtained, e.g., from Bortz (2005) 
\ct[Eq.~(5.49)]{bor2005}.

\medskip
\noindent
\textbf{Test decision:} Depending on the kind of test to be 
performed, the rejection region for $H_{0}$ at significance level 
$\alpha$ is given by
\begin{center}
\begin{tabular}[h]{c|c|c|c}
 & & & \\
\textbf{Kind of test} & $\boldsymbol{H_{0}}$ &
$\boldsymbol{H_{1}}$ &
\textbf{Rejection region for} $\boldsymbol{H_{0}}$ \\
 & & & \\
\hline
 & & & \\
(a)~two-sided & $\tilde{x}_{0.5}(1) = \tilde{x}_{0.5}(2)$ & 
$\tilde{x}_{0.5}(1) \neq \tilde{x}_{0.5}(2)$ &
$|t_{n_{1},n_{2}}|>z_{1-\alpha/2}$ \\
 & & & \\
\hline
 & & & \\
(b)~left-sided & $\tilde{x}_{0.5}(1) \geq \tilde{x}_{0.5}(2)$ & 
$\tilde{x}_{0.5}(1) < \tilde{x}_{0.5}(2)$ &
$t_{n_{1},n_{2}}<z_{\alpha}=-z_{1-\alpha}$ \\
 & & & \\
\hline
 & & & \\
(c)~right-sided & $\tilde{x}_{0.5}(1) \leq \tilde{x}_{0.5}(2)$ & 
$\tilde{x}_{0.5}(1) > \tilde{x}_{0.5}(2)$ &
$t_{n_{1},n_{2}}>z_{1-\alpha}$ \\
 & & &
\end{tabular}
\end{center}
$p$--values associated with realisations $t_{n_{1},n_{2}}$ of the
\textbf{test statistic}~(\ref{eq:uteststat}), which are to be
calculated from the
\textbf{standard normal test distribution}, can be obtained from 
Eqs.~(\ref{eq:pvaluetwo})--(\ref{eq:pvalueright}).

\medskip
\noindent
\underline{\R:}
\texttt{wilcox.test(\textit{variable}~\texttildelow~\textit{group
variable})}, \\
\texttt{wilcox.test(\textit{variable}~\texttildelow~\textit{group
variable}, alternative = "less")}, \\
\texttt{wilcox.test(\textit{variable}~\texttildelow~\textit{group
variable}, alternative = "greater")} \\
\underline{SPSS:} Analyze $\rightarrow$ Nonparametric Tests
$\rightarrow$ Legacy Dialogs $\rightarrow$ 2 Independent
Samples \ldots: Mann-Whitney U

\medskip
\noindent
\underline{Note:} Regrettably, SPSS provides no option for 
selecting between a one-sided and a two-sided $U$--test. The 
default setting is for a two-sided test. For the purpose of 
one-sided tests the $p$--value output of SPSS needs to be divided 
by $2$.

\section[Independent samples $F$--test for a variance]{Two
independent samples $\boldsymbol{F}$--test for a population
variance}
\lb{sec:Ftestindep}
In analogy to the independent samples $t$--test for a 
population mean of Sec.~\ref{sec:ttestindep}, one may likewise 
investigate for a metrically scaled one-dimensional statistical 
variable~$X$, which can be assumed to satisfy a Gau\ss ian normal 
distribution in $\boldsymbol{\Omega}$ (cf. 
Sec.~\ref{sec:normverteil}), whether there exists a significant 
difference in the values of the population variance between two 
independent random samples.\footnote{Run the 
Kolmogorov--Smirnov--test to check 
whether the assumption of normality of the distribution of $X$ in 
the two random samples drawn needs to be rejected.} The parametric 
\textbf{two independent samples $\boldsymbol{F}$--test for a 
population variance} empirically evaluates the plausibility of the 
null hypothesis $H_{0}$ in the non-directed resp.~directed pairs of

\medskip
\noindent
\textbf{Hypotheses:} \hfill (test for differences)
\be
\begin{cases}
H_{0}: \sigma_{1}^{2}=\sigma_{2}^{2}
\quad\text{or}\quad
\sigma_{1}^{2} \geq \sigma_{2}^{2}
\quad\text{or}\quad
\sigma_{1}^{2} \leq \sigma_{2}^{2} \\
H_{1}: \sigma_{1}^{2} \neq \sigma_{2}^{2}
\quad\text{or}\quad
\sigma_{1}^{2} < \sigma_{2}^{2}
\quad\text{or}\quad
\sigma_{1}^{2} > \sigma_{2}^{2}
\end{cases} \ .
\ee
Dealing with independent random samples of sizes $n_{1}$ and 
$n_{2}$, the ratio of the corresponding sample variances serves as 
a

\medskip
\noindent
\textbf{Test statistic:}
\be
\lb{eq:fteststat}
\fbox{$\displaystyle
T_{n_{1},n_{2}} := \frac{S_{n_{1}}^{2}}{S_{n_{2}}^{2}}
\ \stackrel{H_{0}}{\sim}\ F(n_{1}-1,n_{2}-1) \ ,
$}
\ee
which, under $H_{0}$, satisfies an $\boldsymbol{F}$\textbf{--test
distribution} with $df_{1}=n_{1}-1$ and $df_{2}=n_{2}-1$ degrees of
freedom; cf. Sec.~\ref{sec:fverteil}.

\medskip
\noindent
\textbf{Test decision:} Depending on the kind of test to be 
performed, the rejection region for $H_{0}$ at significance level 
$\alpha$ is given by
\begin{center}
\begin{tabular}[h]{c|c|c|c}
 & & & \\
\textbf{Kind of test} & $\boldsymbol{H_{0}}$ &
$\boldsymbol{H_{1}}$ &
\textbf{Rejection region for} $\boldsymbol{H_{0}}$ \\
 & & & \\
\hline
 & & & \\
(a)~two-sided & $\sigma_{1}^{2}=\sigma_{2}^{2}$ & $\sigma_{1}^{2} 
\neq \sigma_{2}^{2}$ &
$t_{n_{1},n_{2}}\begin{cases} < 1/f_{n_{2}-1,n_{1}-1;1-\alpha/2} \\
> f_{n_{1}-1,n_{2}-1;1-\alpha/2} \end{cases}$ \\
 & & & \\
\hline
 & & & \\
(b)~left-sided & $\sigma_{1}^{2} \geq \sigma_{2}^{2}$ & 
$\sigma_{1}^{2} < \sigma_{2}^{2}$ &
$t_{n_{1},n_{2}} < 1/f_{n_{2}-1,n_{1}-1;1-\alpha}$ \\
 & & & \\
\hline
 & & & \\
(c)~right-sided & $\sigma_{1}^{2} \leq \sigma_{2}^{2}$ & 
$\sigma_{1}^{2} > \sigma_{2}^{2}$ &
$t_{n_{1},n_{2}} > f_{n_{1}-1,n_{2}-1;1-\alpha}$ \\
 & & &
\end{tabular}
\end{center}
$p$--values associated with realisations $t_{n_{1},n_{2}}$ of the
\textbf{test statistic}~(\ref{eq:fteststat}), which are to be
calculated from the
$\boldsymbol{F}$\textbf{--test distribution}, can be obtained from 
Eqs.~(\ref{eq:pvaluetwo})--(\ref{eq:pvalueright}).

\medskip
\noindent
\underline{\R:}
\texttt{var.test(\textit{variable}~\texttildelow~\textit{group
variable})}, \\
\texttt{var.test(\textit{variable}~\texttildelow~\textit{group
variable}, alternative = "less")}, \\
\texttt{var.test(\textit{variable}~\texttildelow~\textit{group
variable}, alternative = "greater")} \\
\underline{GDC:} mode \texttt{STAT} $\rightarrow$ \texttt{TESTS}
$\rightarrow$ \texttt{2-SampFTest\ldots}

\medskip
\noindent
Regrettably, the two-sample $F$--test for a population 
variance does not appear to have been implemented in the SPSS 
software package. Instead, to address quantitative issues of the 
kind raised here, one may resort to \textbf{Levene's test}; cf. 
Sec.~\ref{sec:ttestindep}.

\section[Dependent samples $t$--test for a mean]{\href{https://www.youtube.com/watch?v=XqaTOsBSvg0}{Two dependent samples $\boldsymbol{t}$--test
for a population mean}}
\lb{sec:ttestdep}
Besides investigating for significant differences in the 
distribution of a single one-dimensional statistical variable~$X$ 
in two or more independent subgroups of some target 
population~$\boldsymbol{\Omega}$, many research projects are 
interested in finding out (i)~how the distributional properties of 
a one-dimensional statistical variable $X$ have changed within one 
and the same random sample of $\boldsymbol{\Omega}$ in an 
experimental before--after situation, or (ii)~how the distribution 
of a one-dimensional statistical variable~$X$ differs between two 
subgroups of $\boldsymbol{\Omega}$, the sample units of which 
co-exist in a natural pairwise one-to-one correspondence to one 
another.

\medskip
\noindent
When the one-dimensional statistical variable~$X$ in question is 
metrically scaled and can be assumed to satisfy a Gau\ss ian 
normal distribution in $\boldsymbol{\Omega}$, significant 
differences can be tested for by means of the parametric
\textbf{two dependent samples $\boldsymbol{t}$--test for a
population mean}. Denoting by $A$ and $B$ either temporal before
and after instants, or partners in a set of natural pairs $(A,B)$,
define for $X$ the metrically scaled \textbf{difference variable}
\be
D:=X(A)-X(B) \ .
\ee
An \textit{important test prerequisite} demands that $D$ itself may 
be assumed \textit{normally distributed} in $\boldsymbol{\Omega}$; 
cf. Sec.~\ref{sec:normverteil}. Whether this property holds true, 
can be checked for $n \geq 50$ via the
\textbf{Kolmogorov--Smirnov--test}; cf.
Sec.~\ref{sec:onesampttest}. When $n < 50$, one may resort to a
consideration of the magnitudes of the \textbf{standardised
skewness and excess kurtosis measures}, Eqs.~(\ref{eq:g1g2ratios}). 

\medskip
\noindent
With $\mu_{D}$ denoting the population mean of the difference 
variable $D$, the

\medskip
\noindent
\textbf{Hypotheses:} \hfill (test for differences)
\be
\begin{cases}
H_{0}: \mu_{D}=0
\quad\text{or}\quad
\mu_{D} \geq 0
\quad\text{or}\quad
\mu_{D} \leq 0 \\
H_{1}: \mu_{D} \neq 0
\quad\text{or}\quad
\mu_{D} < 0
\quad\text{or}\quad
\mu_{D} > 0
\end{cases}
\ee
can be given in a non-directed or a directed formulation. From the 
sample mean $\bar{D}$ and its associated \textbf{standard error},
\be
\text{SE}\bar{D} := \frac{S_{D}}{\sqrt{n}} \ ,
\ee
which derives from the theoretical \textbf{sampling distribution}
for $\bar{D}$, one obtains by means of standardisation according
to Eq.~(\ref{eq:standardisation}) the

\medskip
\noindent
\textbf{Test statistic:}
\be
\lb{eq:deptteststat}
\fbox{$\displaystyle
T_{n} := \frac{\bar{D}}{\text{SE}\bar{D}}
\ \stackrel{H_{0}}{\sim}\ t(n-1) \ ,
$}
\ee
which, under $H_{0}$, satisfies a $\boldsymbol{t}$\textbf{--test
distribution} with $df=n-1$ degrees of freedom; cf.
Sec.~\ref{sec:tverteil}.

\medskip
\noindent
\textbf{Test decision:} Depending on the kind of test to be 
performed, the rejection region for $H_{0}$ at significance level 
$\alpha$ is given by
\begin{center}
\begin{tabular}[h]{c|c|c|c}
 & & & \\
\textbf{Kind of test} & $\boldsymbol{H_{0}}$ &
$\boldsymbol{H_{1}}$ &
\textbf{Rejection region for} $\boldsymbol{H_{0}}$ \\
 & & & \\
\hline
 & & & \\
(a)~two-sided & $\mu_{D}=0$ & $\mu_{D}\neq 0$ &
$|t_{n}|>t_{n-1;1-\alpha/2}$ \\
 & & & \\
\hline
 & & & \\
(b)~left-sided & $\mu_{D} \geq 0$ & $\mu_{D}<0$ &
$t_{n}<t_{n-1;\alpha}=-t_{n-1;1-\alpha}$ \\
 & & & \\
\hline
 & & & \\
(c)~right-sided & $\mu_{D} \leq 0$ & $\mu_{D}>0$ &
$t_{n}>t_{n-1;1-\alpha}$ \\
 & & &
\end{tabular}
\end{center}
$p$--values associated with realisations $t_{n}$ of the
\textbf{test statistic}~(\ref{eq:deptteststat}), which are to be
calculated from the
$\boldsymbol{t}$\textbf{--test distribution}, can be obtained from 
Eqs.~(\ref{eq:pvaluetwo})--(\ref{eq:pvalueright}).

\medskip
\noindent
\underline{\R:}
\texttt{t.test(\textit{variableA}, \textit{variableB}, 
paired = "T")}, \\
\texttt{t.test(\textit{variableA}, \textit{variableB},
paired = "T", alternative = "less")}, \\
\texttt{t.test(\textit{variableA}, \textit{variableB},
paired = "T", alternative = "greater")} \\
\underline{SPSS:} Analyze $\rightarrow$ Compare Means
$\rightarrow$ Paired-Samples T Test \ldots

\medskip
\noindent
\underline{Note:} Regrettably, SPSS provides no option for 
selecting between a one-sided and a two-sided $t$--test. The 
default setting is for a two-sided test. For the purpose of 
one-sided tests the $p$--value output of SPSS needs to be divided 
by $2$.

\medskip
\noindent
\textbf{Effect size:} The practical significance of the phenomenon investigated can be estimated from the sample mean~$\bar{D}$ and
the sample standard deviation~$s_{D}$ by the scale-invariant ratio
\be
\fbox{$\displaystyle
d := \frac{\left|\bar{D}\right|}{s_{D}} \ .
$}
\ee
For the interpretation of its strength Cohen 
(1992)~\ct[Tab.~1]{coh1992} recommends the

\medskip
\noindent
\underline{\textbf{Rule of thumb:}}\\
$0.20 \leq d < 0.50$: small effect\\
$0.50 \leq d < 0.80$: medium effect\\
$0.80 \leq d$: large effect.

\medskip
\noindent
\underline{\R:}
\texttt{cohen.d(\textit{variable}, \textit{group
variable}, paired = TRUE)} (package: \texttt{effsize}, by
Torchiano (2018)~\ct{tor2018})

\medskip
\noindent
We remark that the statistical software package~\R\ holds 
available a routine \texttt{power.t.test(power, sig.level, delta,
sd, $n$, alternative, type = "paired")} for the purpose of
calculation of any one of the parameters \texttt{power},
\texttt{delta} or $n$ (provided all remaining parameters have been
specified) in the context of empirical investigations employing the
dependent samples $t$--test for a population mean. One-sided tests
are addressed via the parameter setting
\texttt{alternative = "one.sided"}.

\section[Dependent samples Wilcoxon--test]{Two dependent
samples Wilcoxon--test for a population median}
When the test prerequisites of the dependent samples $t$--test 
\textit{cannot} be met, i.e., a given metrically scaled 
one-dimensional statistical variable~$X$ cannot be assumed to 
satisfy a Gau\ss ian normal distribution in $\boldsymbol{\Omega}$, 
or $X$ is an ordinally scaled one-dimensional statistical variable 
in the first place, the non-parametric \textbf{signed ranks test} 
published by the US-American chemist and statistician 
\href{http://en.wikipedia.org/wiki/Frank_Wilcoxon}{Frank Wilcoxon
(1892--1965)} in 1945 \ct{wil1945} constitutes a 
quantitative--empirical tool for comparing the distributional 
properties of $X$ between two dependent random samples drawn from 
$\boldsymbol{\Omega}$. Like Mann and Whitney's $U$--test discussed 
in Sec.~\ref{sec:utest}, it is built around the idea of \textbf{rank 
number data} faithfully representing the original random sample 
data; cf. Sec.~\ref{sec:2Dord}. Defining again a variable
\be
D:=X(A)-X(B) \ ,
\ee
with associated median $\tilde{x}_{0.5}(D)$, the null hypothesis 
$H_{0}$ in the non-directed or directed pairs of

\medskip
\noindent
\textbf{Hypotheses:} \hfill (test for differences)
\be
\begin{cases}
H_{0}: \tilde{x}_{0.5}(D)=0
\quad\text{or}\quad
\tilde{x}_{0.5}(D) \geq 0
\quad\text{or}\quad
\tilde{x}_{0.5}(D) \leq 0 \\
H_{1}: \tilde{x}_{0.5}(D) \neq 0
\quad\text{or}\quad
\tilde{x}_{0.5}(D) < 0
\quad\text{or}\quad
\tilde{x}_{0.5}(D) > 0
\end{cases}
\ee
needs to be subjected to a suitable significance test.

\medskip
\noindent
For realisations $d_{i}$ ($i=1,\ldots,n$) of $D$, introduce
\textbf{rank numbers} according to $d_{i} \mapsto R[|d_{i}|]$  for
the ordered \textbf{absolute values} $|d_{i}|$, while keeping a 
record of the \textbf{sign} of each $d_{i}$. Exclude from the data
set all null differences $d_{i}=0$, leading to a sample of reduced
size $n \mapsto n_\mathrm{red}$ . Then form the \textbf{sums of
rank numbers} $W^{+}$ for the $d_{i}>0$ and $W^{-}$ for the
$d_{i}<0$, respectively, which are linked to one another by the
identity $W^{+}+W^{-}=n_\mathrm{red}(n_\mathrm{red}+1)/2$. 
Choose~$W^{+}$.\footnote{Due to the identity
$W^{+}+W^{-}=n_\mathrm{red}(n_\mathrm{red}+1)/2$, choosing 
instead $W^{-}$ would make no qualitative difference to the 
subsequent test procedure.} For reduced sample sizes $n_\mathrm{
red}>20$ (see, e.g., Rinne (2008)~\ct[p~552]{rin2008}), one 
employs the

\medskip
\noindent
\textbf{Test statistic:}
\be
\lb{eq:wilcteststat}
\fbox{$\displaystyle
T_{n_\mathrm{red}} := \frac{W^{+}-\mu_{W^{+}}}{\text{SE}W^{+}}
\ \stackrel{H_{0}}{\approx}\ N(0;1) \ ,
$}
\ee
which, under $H_{0}$, approximately satisfies a \textbf{standard
normal test distribution}; cf. Sec.~\ref{sec:normverteil}. Here,
the mean $\mu_{W^{+}}$ expected under $H_{0}$ is defined in terms
of $n_\mathrm{red}$ by
\be
\mu_{W^{+}}:=\frac{n_\mathrm{red}(n_\mathrm{red}+1)}{4} \ ,
\ee
while the \textbf{standard error} $\text{SE}W^{+}$ can be computed 
from, e.g., Bortz (2005) \ct[Eq.~(5.52)]{bor2005}.

\medskip
\noindent
\textbf{Test decision:} Depending on the kind of test to be 
performed, the rejection region for $H_{0}$ at significance level 
$\alpha$ is given by
\begin{center}
\begin{tabular}[h]{c|c|c|c}
 & & & \\
\textbf{Kind of test} & $\boldsymbol{H_{0}}$ &
$\boldsymbol{H_{1}}$ &
\textbf{Rejection region for} $\boldsymbol{H_{0}}$ \\
 & & & \\
\hline
 & & & \\
(a)~two-sided & $\tilde{x}_{0.5}(D) = 0$ & $\tilde{x}_{0.5}(D)
\neq 0$ & $|t_{n_\mathrm{red}}|>z_{1-\alpha/2}$ \\
 & & & \\
\hline
 & & & \\
(b)~left-sided & $\tilde{x}_{0.5}(D) \geq 0$ & $\tilde{x}_{0.5}(D)
< 0$ & $t_{n_\mathrm{red}}<z_{\alpha}=-z_{1-\alpha}$ \\
 & & & \\
\hline
 & & & \\
(c)~right-sided & $\tilde{x}_{0.5}(D) \leq 0$ & $\tilde{x}_{0.5}(D)
> $ & $t_{n_\mathrm{red}}>z_{1-\alpha}$ \\
 & & &
\end{tabular}
\end{center}
$p$--values associated with realisations $t_{n_\mathrm{red}}$ of 
the \textbf{test statistic}~(\ref{eq:wilcteststat}), which are to
be calculated from the
\textbf{standard normal test distribution}, can be obtained from 
Eqs.~(\ref{eq:pvaluetwo})--(\ref{eq:pvalueright}).

\medskip
\noindent
\underline{\R:}
\texttt{wilcox.test(\textit{variableA}, \textit{variableB},
paired = "T")}, \\
\texttt{wilcox.test(\textit{variableA}, \textit{variableB},
paired = "T", alternative = "less")}, \\
\texttt{wilcox.test(\textit{variableA}, \textit{variableB},
paired = "T", alternative = "greater")} \\
\underline{SPSS:} Analyze $\rightarrow$ Nonparametric Tests
$\rightarrow$ Legacy Dialogs $\rightarrow$ 2 Related
Samples \ldots: Wilcoxon

\medskip
\noindent
\underline{Note:} Regrettably, SPSS provides no option for 
selecting between a one-sided and a two-sided Wilcoxon--test. The 
default setting is for a two-sided test. For the purpose of 
one-sided tests the $p$--value output of SPSS needs to be divided 
by $2$.

\section[$\chi^{2}$--test for homogeneity]{\href{https://www.youtube.com/watch?v=Lp2M_0_OhWM}{$\boldsymbol{\chi}^{2}$--test
for homogeneity}}
\lb{sec:chisqhomo}
Due to its independence of scale levels of measurement, the 
non-parametric $\boldsymbol{\chi^{2}}$\textbf{--test for
homogeneity} constitutes the most generally applicable statistical
test for significant differences in the distributional properties
of a particular one-dimensional statistical variable~$X$ between
$k \in \mathbb{N}$ different independent subgroups of some
population $\boldsymbol{\Omega}$. By assumption, the
one-dimensional variable~$X$ may take values in a total of $l \in
\mathbb{N}$ different \textbf{categories} $a_{j}$ ($j=1,\ldots,l$).
Begin by formulating the

\medskip
\noindent
\textbf{Hypotheses:} \hfill (test for differences)
\be
\begin{cases}
H_{0}: X\ \text{satisfies the same distribution
in all}\ k\ \text{subgroups of}\ \boldsymbol{\Omega} \\
H_{1}: X\ \text{satisfies a different distribution
in at least one subgroup of}\ \boldsymbol{\Omega}
\end{cases} \ .
\ee
With $O_{ij}$ denoting the \textbf{observed frequency} of category 
$a_{j}$ in subgroup~$i$ ($i=1,\ldots,k$), and $E_{ij}$ the, under 
$H_{0}$, \textbf{expected frequency} of category $a_{j}$ in 
subgroup~$i$, the sum of rescaled squared \textbf{residuals} 
$\displaystyle\frac{(O_{ij}-E_{ij})^{2}}{E_{ij}}$ provides a useful

\medskip
\noindent
\textbf{Test statistic:}
\be
\lb{eq:chisqhomteststat}
\fbox{$\displaystyle
T_{n} := \sum_{i=1}^{k}\sum_{j=1}^{l}\frac{(O_{ij}-E_{ij}
)^{2}}{E_{ij}}
\ \stackrel{H_{0}}{\approx}\ \chi^{2}[(k-1)\times(l-1)] \ .
$}
\ee
Under $H_{0}$, this test statistic satisfies approximately a 
$\boldsymbol{\chi^{2}}$\textbf{--test distribution} with $df=(k-1)
\times (l-1)$ degrees of 
freedom; cf. Sec.~\ref{sec:chi2verteil}. The $E_{ij}$ are defined 
as \textbf{projections} of the \textbf{observed 
proportions}~$\displaystyle\frac{O_{+j}}{n}$ in the total sample  
of size~$n:=O_{1+}+\ldots+O_{k+}$ of each of the $l$~categories 
$a_{j}$ of $X$ into each of the $k$~subgroups of size $O_{i+}$ 
by [cf. Eqs.~(\ref{eq:margfreq1}) and (\ref{eq:margfreq2})]
\be
E_{ij} := O_{i+}\,\frac{O_{+j}}{n} \ .
\ee
Note the \textit{important (!) test prerequisite} that the total 
sample size~$n$ be such that
\be
E_{ij} \stackrel{!}{\geq} 5
\ee
applies for all categories $a_{j}$ and subgroups~$i$.

\medskip
\noindent
\textbf{Test decision:} The rejection region for $H_{0}$ at 
significance level $\alpha$ is given by (right-sided test)
\be
t_{n}>\chi^{2}_{(k-1)\times(l-1);1-\alpha} \ .
\ee
By Eq.~(\ref{eq:pvalueright}), the $p$--value associated with a 
realisation $t_{n}$ of the \textbf{test
statistic}~(\ref{eq:chisqhomteststat}), which is to
be calculated from the $\boldsymbol{\chi^{2}}$\textbf{--test
distribution}, amounts to
\be
p = P(T_{n}>t_{n}|H_{0}) = 1-P(T_{n}\leq t_{n}|H_{0})
= 1-\chi^{2}\texttt{cdf}\left(0,t_{n},(k-1)\times(l-1)\right) \ .
\ee

\medskip
\noindent
\underline{\R:} \texttt{chisq.test(\textit{group variable},
\textit{variable})} \\
\underline{GDC:} mode \texttt{STAT} $\rightarrow$ \texttt{TESTS}
$\rightarrow$ \texttt{$\chi^{2}$-Test\ldots}\\
\underline{SPSS:} Analyze $\rightarrow$ Descriptive Statistics
$\rightarrow$ Crosstabs \ldots $\rightarrow$ Statistics \ldots:
Chi-square

\medskip
\noindent
Typically the power of a $\chi^{2}$--test for homogeneity is 
weaker than for the related two procedures of comparing three or
more independent subgroups of $\boldsymbol{\Omega}$, which will be 
discussed in the subsequent Secs.~\ref{sec:anova} and 
\ref{sec:kruskalwallis}.

\medskip
\noindent
\textbf{Effect size:} The practical significance of the phenomenon investigated can be estimated and interpreted by means of the
effect size measure~$w$ defined in Eq.~(\ref{eq:eschisq});
cf. Cohen (1992)~\ct[Tab.~1]{coh1992}.

\section[One-way analysis of variance (ANOVA)]{\href{https://www.youtube.com/watch?v=viiFi8zfLX0}{One-way analysis of variance (ANOVA)}}
\lb{sec:anova}
This powerful quantitative--analytical tool has been developed in 
the context of investigations on biometrical genetics by 
the English statistician
\href{http://www-history.mcs.st-and.ac.uk/Biographies/Fisher.html}{Sir
Ronald Aylmer Fisher FRS (1890--1962)} (see Fisher 
(1918)~\ct{fis1918}), and later extended by the
US-American statistician
\href{http://en.wikipedia.org/wiki/Henry_Scheffe}{Henry
Scheff\'{e} (1907--1977)} (see Scheff\'{e} (1959)~\ct{sch1959}). 
It is of a parametric nature and can be interpreted alternatively 
as a method for\footnote{Only experimental designs with fixed 
effects are considered here.}
\begin{itemize}
\item[(i)] investigating the influence of a qualitative 
one-dimensional statistical variable~$Y$ with $k \geq 3$ 
categories $a_{i}$ ($i=1,\ldots,k$), generally referred to as a 
``factor,'' on a quantitative one-dimensional statistical 
variable~$X$, or\\[-0.5cm]
\item[(ii)] testing for differences of the mean of a quantitative 
one-dimensional statistical variable~$X$ between $k \geq 3$ 
different subgroups of some target 
population~$\boldsymbol{\Omega}$.\\[-0.5cm]
\end{itemize}
A necessary condition for the application of the \textbf{one-way 
analysis of variance (ANOVA)} test procedure is that the 
quantitative one-dimensional statistical variable~$X$ to be 
investigated may be reasonably assumed to be (a)~\textit{normally 
distributed} (cf. Sec.~\ref{sec:normverteil}) in the $k \geq 3$ 
subgroups of the target population~$\boldsymbol{\Omega}$ 
considered, with, in addition, (b)~\textit{equal variances}. Both
of these conditions also have to hold for each of a set of $k$ 
mutually stochastically independent random variables $X_{1}, 
\ldots, X_{k}$ representing $k$ random samples drawn independently 
from the identified $k$ subgroups of $\boldsymbol{\Omega}$, of 
sizes $n_{1}, \ldots, n_{k} \in \mathbb{N}$, respectively. In the 
following, the element $X_{ij}$ of the underlying $(n \times 2)$ 
data matrix $\boldsymbol{X}$ represents the $j$th value of $X$ in 
the random sample drawn from the $i$th subgroup of 
$\boldsymbol{\Omega}$, with $\bar{X}_{i}$ the corresponding
\textbf{subgroup sample mean}. The $k$ independent random samples
can be understood to form a \textbf{total random sample} of size 
$\displaystyle n:=n_{1}+\ldots+n_{k} =\sum_{i=1}^{k}n_{i}$, with 
\textbf{total sample mean} $\bar{X}_{n}$; cf.
Eq.~(\ref{eq:sampmean}).

\medskip
\noindent
The intention of the ANOVA procedure in the variant~(ii) stated 
above is to empirically test the null hypothesis $H_{0}$ in the 
set of

\medskip
\noindent
\textbf{Hypotheses:} \hfill (test for differences)
\be
\lb{anovahyp1}
\begin{cases}
H_{0}: \mu_{1}=\ldots=\mu_{k}=\mu_{0} \\
H_{1}: \mu_{i} \neq \mu_{0}\ \text{at least for one}
\ i=1,\ldots,k
\end{cases} \ .
\ee
The necessary test prerequisites can be checked by (a)~the 
\textbf{Kolmogorov--Smirnov--test} for normality of the 
$X$-distribution in each of the $k$ subgroups of 
$\boldsymbol{\Omega}$ (cf. Sec.~\ref{sec:onesampttest}) when 
$n_{i} \geq 50$, or, when $n_{i} < 50$, by a consideration of the 
magnitudes of the \textbf{standardised skewness and excess kurtosis
measures}, Eqs.~(\ref{eq:g1g2ratios}), and likewise by
(b)~\textbf{Levene's test} for $H_{0}: \sigma_{1}^{2} = 
\ldots=\sigma_{k}^{2}=\sigma_{0}^{2}$ against $H_{1}$: 
``$\sigma_{i}^{2} \neq \sigma_{0}^{2}$ at least for one 
$i=1,\ldots,k$'' to test for equality of the variances in these
$k$ subgroups (cf. Sec.~\ref{sec:ttestindep}).

\medskip
\noindent
\underline{\R:}
\texttt{leveneTest(\textit{variable}, \textit{group
variable})} (package: \texttt{car}, by Fox and Weisberg
(2011)~\ct{foxwei2011})

\medskip
\noindent
The starting point of the ANOVA procedure is a simple
algebraic decomposition of the \textbf{random sample values}
$X_{ij}$ into three additive components according to
\be
\lb{eq:xijdecompo}
X_{ij} = \bar{X}_{n} + (\bar{X}_{i}-\bar{X}_{n})
+ (X_{ij}-\bar{X}_{i}) \ .
\ee
This expresses the $X_{ij}$ in terms of the sum of the total 
sample mean, $\bar{X}_{n}$, the deviation of the subgroup sample 
means from the total sample mean, $(\bar{X}_{i}-\bar{X}_{n})$, and 
the residual deviation of the sample values from their respective 
subgroup sample means, $(X_{ij}-\bar{X}_{i})$. The decomposition 
of the $X_{ij}$ motivates a \textbf{linear stochastic model} for
the target population~$\boldsymbol{\Omega}$ of the 
form\footnote{Formulated in the context of this linear stochastic 
model, the null and research hypotheses are $H_{0}: 
\alpha_{1}=\ldots=\alpha_{k}=0$ and $H_{1}$: at least one 
$\alpha_{i}\neq 0$, respectively.}
\be
\text{in}\ \boldsymbol{\Omega}: \quad
X_{ij}=\mu_{0}+\alpha_{i}+\varepsilon_{ij}
\ee
in order to quantify, via the $\alpha_{i}$ ($i=1,\ldots,k$), the 
potential influence of the qualitative one-dimensional 
variable~$Y$ on the quantitative one-dimensional variable~$X$. 
Here $\mu_{0}$ is the \textbf{population mean} of $X$, it holds
that $\sum_{i=1}^{k}n_{i}\alpha_{i}=0$, and it is assumed for the
\textbf{random errors} $\varepsilon_{ij}$ that $\varepsilon_{ij} 
\stackrel{\text{i.i.d.}}{\sim} N(0;\sigma_{0}^{2})$, i.e., that 
they are identically normally distributed and mutually 
stochastically independent.

\medskip
\noindent
Having established the decomposition (\ref{eq:xijdecompo}), one 
next turns to consider the associated set of \textbf{sums of
squared deviations}, defined by
\bea
\lb{eq:bss}
\text{BSS} & := &
\sum_{i=1}^{k}\sum_{j=1}^{n_{i}}\left(\bar{X}_{i}-\bar{X}_{n}
\right)^{2}
\ = \ \sum_{i=1}^{k}n_{i}\left(\bar{X}_{i}-\bar{X}_{n}\right)^{2} 
\\
\lb{eq:rss}
\text{RSS} & := &
\sum_{i=1}^{k}\sum_{j=1}^{n_{i}}\left(X_{ij}-\bar{X}_{i}\right)^{2}
\\
\lb{eq:tss}
\text{TSS} & := &
\sum_{i=1}^{k}\sum_{j=1}^{n_{i}}\left(X_{ij}-\bar{X}_{n}\right)^{2}
\ ,
\eea
where the summations are (i)~over all $n_{i}$ sample units within 
a subgroup, and (ii)~over all of the $k$ subgroups themselves.
The sums are referred to as, resp., (a)~the sum of squared 
deviations between the subgroup samples (BSS), (b)~the residual 
sum of squared deviations within the subgroup samples (RSS), and 
(c)~the total sum of squared deviations (TSS) of the individual 
$X_{ij}$ from the total sample mean $\bar{X}_{n}$. It is a fairly 
elaborate though straightforward algebraic exercise to show that 
these three squared deviation terms relate to one another 
according to the strikingly simple and elegant identity (cf. Bosch 
(1999)~\ct[p~220f]{bos1999})
\be
\text{TSS} = \text{BSS} + \text{RSS} \ .
\ee

\medskip
\noindent
Now, from the sums of squared deviations 
(\ref{eq:bss})--(\ref{eq:tss}), one defines, resp., the 
\textbf{total sample variance},
\be
S_\mathrm{total}^{2} := 
\frac{1}{n-1}\,\sum_{i=1}^{k}\sum_{j=1}^{n_{i}}
\left(X_{ij}-\bar{X}_{n}\right)^{2}
= \frac{\text{TSS}}{n-1} \ ,
\ee
involving $df=n-1$ degrees of freedom, the \textbf{sample variance 
between subgroups},
\be
S_\mathrm{between}^{2} := 
\frac{1}{k-1}\,\sum_{i=1}^{k}n_{i}\left(\bar{X}_{i}-\bar{X}_{n}
\right)^{2}
= \frac{\text{BSS}}{k-1} \ ,
\ee
with $df=k-1$, and the \textbf{mean sample variance within
subgroups},
\be
S_\mathrm{within}^{2} := 
\frac{1}{n-k}\,\sum_{i=1}^{k}\sum_{j=1}^{n_{i}}
\left(X_{ij}-\bar{X}_{i}\right)^{2}
= \frac{\text{RSS}}{n-k} \ ,
\ee
for which $df=n-k$.

\medskip
\noindent
Employing the latter two subgroup-specific dispersion measures, 
the set of hypotheses~(\ref{anovahyp1}) may be recast into the 
alternative form

\medskip
\noindent
\textbf{Hypotheses:} \hfill (test for differences)
\be
\lb{anovahyp2}
\begin{cases}
H_{0}: {\displaystyle\frac{S_\mathrm{between}^{2}}{S_\mathrm{
within}^{2}}} \leq 1 \\ \\
H_{1}: {\displaystyle\frac{S_\mathrm{between}^{2}}{S_\mathrm{
within}^{2}}} > 1 
\end{cases} \ .
\ee

\medskip
\noindent
Finally, as a test statistic for the ANOVA procedure one chooses 
this very ratio of variances\footnote{This ratio is sometimes 
given as $\displaystyle T_{n,k} := \frac{(\text{explained 
variance})}{(\text{unexplained variance})}$, in analogy to 
expression~(\ref{eq:linregcoeffdet}) below. Occasionally, one also 
considers the coefficient $\displaystyle\eta^{2} := 
\frac{\text{BSS}}{\text{TSS}}$, which, however, does not account 
for the degrees of freedom involved. In this respect, the modified 
coefficient $\displaystyle\tilde{\eta}{}^{2}:=\frac{S_\mathrm{
between}^{2}}{S_\mathrm{total}^{2}}$ would constitute a more 
sophisticated measure.} we just employed,
\[
T_{n,k} := \frac{(\text{sample variance between 
subgroups})}{(\text{mean sample variance within subgroups})}
= \frac{\text{BSS}/(k-1)}{\text{RSS}/(n-k)}\ ,
\]
expressing the size of the ``sample variance between 
subgroups'' in terms of multiples of the ``mean sample variance 
within subgroups''; it thus constitutes a relative measure.
A real effect of difference between subgroups is thus given when 
the non-negative numerator turns out to be significantly larger 
than the non-negative denominator. Mathematically, this 
statistical measure of deviations between the data and the null 
hypothesis is captured by the

\medskip
\noindent
\textbf{Test statistic:}\footnote{Note the one-to-one
correspondence to the test statistic (\ref{eq:fteststat}) employed
in the independent samples $F$--test for a population variance.}
\be
\lb{eq:anovateststat}
\fbox{$\displaystyle
T_{n,k} := 
\frac{S_\mathrm{between}^{2}}{S_\mathrm{within}^{2}}
\ \stackrel{H_{0}}{\sim}\ F(k-1,n-k) \ .
$}
\ee
Under $H_{0}$, it satisfies an $\boldsymbol{F}$\textbf{--test
distribution} with $df_{1}=k-1$ 
and $df_{2}=n-k$ degrees of freedom; cf. Sec.~\ref{sec:fverteil}.

\medskip
\noindent
It is a well-established standard in practical applications of the 
one-way ANOVA procedure to display the results of the data 
analysis in the form of a \textbf{summary table}, here given in 
Tab.~\ref{tab:anovatab}.
\begin{table}

\begin{center}
\begin{tabular}[h]{c|c|c|c|c}
\underline{\textbf{ANOVA}} & sum of & df & mean & test \\
variability & squares &   & square & statistic \\
\hline
between groups & $\text{BSS}$ & $k-1$ &
$S_\mathrm{between}^{2}$ & $t_{n,k}$ \\
within groups & $\text{RSS}$ & $n-k$ & $S_\mathrm{within}^{2}$ & \\
\hline
total & \text{TSS} & $n-1$ & &
\end{tabular}
\end{center}
\caption{ANOVA summary table.}
\lb{tab:anovatab}
\end{table}

\medskip
\noindent
\textbf{Test decision:} The rejection region for $H_{0}$ at 
significance level $\alpha$ is given by (right-sided test)
\be
t_{n,k} > f_{k-1,n-k;1-\alpha} \ .
\ee
With Eq.~(\ref{eq:pvalueright}), the $p$--value associated with a 
specific realisation $t_{n,k}$ of the \textbf{test
statistic}~(\ref{eq:anovateststat}), which
is to be calculated from the $\boldsymbol{F}$\textbf{--test
distribution}, amounts to
\be
p = P(T_{n,k}>t_{n,k}|H_{0}) = 1-P(T_{n,k}\leq t_{n,k}|H_{0})
= 1-F\texttt{cdf}(0,t_{n,k},k-1,n-k) \ .
\ee

\medskip
\noindent
\underline{\R:}
\texttt{anova( lm(\textit{variable}~\texttildelow~\textit{group
variable}) )} (variances equal), \\
\texttt{oneway.test(\textit{variable}~\texttildelow~\textit{group
variable})} (variances not equal) \\
\underline{GDC:} mode \texttt{STAT} $\rightarrow$ \texttt{TESTS}
$\rightarrow$ \texttt{ANOVA(} \\
\underline{SPSS:} Analyze $\rightarrow$ Compare Means
$\rightarrow$ One-Way ANOVA \ldots

\medskip
\noindent
\textbf{Effect size:} The practical significance of the phenomenon investigated can be estimated from the sample sums of squared 
deviations $\text{BSS}$ and $\text{RSS}$ according to
\be
\fbox{$\displaystyle
f := \sqrt{\frac{\text{BSS}}{\text{RSS}}} \ .
$}
\ee
For the interpretation of its strength Cohen 
(1992)~\ct[Tab.~1]{coh1992} recommends the

\medskip
\noindent
\underline{\textbf{Rule of thumb:}}\\
$0.10 \leq f < 0.25$: small effect\\
$0.25 \leq f < 0.40$: medium effect\\
$0.40 \leq f$: large effect.

\medskip
\noindent
We remark that the statistical software package~\R\ holds 
available a routine \texttt{power.anova.test(groups, $n$, 
between.var, within.var, sig.level, power)} for the purpose of 
calculation of any one of the parameters \texttt{power} or $n$
(provided all remaining parameters have been specified) in the 
context of empirical investigations employing the one-way ANOVA. 
Equal values of $n$ are required here.

\medskip
\noindent
When a one-way ANOVA yields a statistically significant result, 
so-called \textbf{post-hoc tests} need to be run subsequently in 
order to identify those subgroups $i$ whose means $\mu_{i}$ differ 
most drastically from the reference value $\mu_{0}$. The
\textbf{Student--Newman--Keuls--test} (Newman (1939)~\ct{new1939}
and Keuls (1952)~\ct{keu1952}), e.g., successively subjects the
pairs of subgroups with the largest differences in sample means to 
independent samples $t$--tests; cf. Sec.~\ref{sec:ttestindep}. 
Other useful post-hoc tests are those developed by \textbf{
Holm--Bonferroni} (Holm (1979)~\ct{hol1979}), \textbf{Tukey} (Tukey 
(1977)~\ct{tuk1977}), or by \textbf{Schef\-f\'{e}} (Schef\-f\'{e} 
(1959)~\ct{sch1959}).

\medskip
\noindent
\underline{\R:} \texttt{pairwise.t.test(\textit{variable},
\textit{group variable}, p.adj = "bonferroni")} \\
\underline{SPSS:} Analyze $\rightarrow$ Compare Means
$\rightarrow$ One-Way ANOVA \ldots $\rightarrow$
Post Hoc \ldots

\section[Kruskal--Wallis--test]{Kruskal--Wallis--test for a
population median}
\lb{sec:kruskalwallis}
Finally, a feasible alternative to the one-way ANOVA, when the 
conditions for the latter's legitimate application cannot be met,
or one is interested in the distributional properties of a specific 
ordinally scaled one-dimensional statistical variable~$X$, is 
given by the non-parametric significance test devised by the 
US-American mathematician and statistician
\href{http://www-history.mcs.st-and.ac.uk/Biographies/Kruskal_William.html}{William Henry Kruskal (1919--2005)} and the
US-American economist and statistician
\href{http://en.wikipedia.org/wiki/W._Allen_Wallis}{Wilson
Allen Wallis (1912--1998)} in 1952 \ct{kruwal1952}. The \textbf{
Kruskal--Wallis--test} effectively serves to detect significant 
differences for a population median of an ordinally or metrically 
scaled one-dimensional statistical variable~$X$ between $k \geq 3$ 
independent subgroups of some target 
population~$\boldsymbol{\Omega}$. To be investigated empirically 
is the null hypothesis $H_{0}$ in the pair of mutually exclusive

\medskip
\noindent
\textbf{Hypotheses:} \hfill (test for differences)
\be
\begin{cases}
H_{0}: \tilde{x}_{0.5}(1) =
\ldots = \tilde{x}_{0.5}(k) \\
H_{1}: \text{at least one}\ \tilde{x}_{0.5}(i)\ 
(i=1,\ldots,k)\ \text{is different from
the other group medians}
\end{cases} \ .
\ee
Introduce \textbf{rank numbers} according to $x_{j}(1) \mapsto 
R[x_{j}(1)]$, \ldots, and $x_{j}(k) \mapsto R[x_{j}(k)]$ within 
the random samples drawn independently from each of the $k \geq 3$ 
subgroups of $\boldsymbol{\Omega}$ on the basis of an ordered
\textbf{joint random sample} of size $\displaystyle 
n:=n_{1}+\ldots+n_{k}=\sum_{i=1}^{k}n_{i}$; cf. 
Sec.~\ref{sec:2Dord}. Then form the \textbf{sum of rank numbers}
for each random sample separately, i.e.,
\be
R_{+i} := \sum_{j=1}^{n_{i}}R[x_{j}(i)]
\qquad (i=1,\ldots,k) \ .
\ee
Provided the sample sizes satisfy the condition $n_{i} \geq 5$ for 
all $k \geq 3$ independent random samples (hence, $n \geq 
15$), the test procedure can be based on the

\medskip
\noindent
\textbf{Test statistic:}
\be
\lb{eq:kruwalteststat}
\fbox{$\displaystyle
T_{n,k} := \left[\frac{12}{n(n+1)}
\sum_{i=1}^{k}\frac{R_{+i}^{2}}{n_{i}}\right]
- 3(n+1)
\ \stackrel{H_{0}}{\approx}\ \chi^{2}(k-1) \ ,
$}
\ee
which, under $H_{0}$, approximately satisfies a 
$\boldsymbol{\chi^{2}}$\textbf{--test distribution} with $df=k-1$
degrees of freedom (cf. Sec.~\ref{sec:chi2verteil}); see, e.g.,
Rinne (2008)~\ct[p~553]{rin2008}.

\medskip
\noindent
\textbf{Test decision:}  The rejection region for $H_{0}$ at 
significance level $\alpha$ is given by (right-sided test)
\be
t_{n,k}>\chi^{2}_{k-1;1-\alpha} \ .
\ee
By Eq.~(\ref{eq:pvalueright}), the $p$--value associated with a
realisation $t_{n,k}$ of the \textbf{test
statistic}~(\ref{eq:kruwalteststat}), which is to be
calculated from the $\boldsymbol{\chi^{2}}$\textbf{--test
distribution}, amounts to
\be
p = P(T_{n,k}>t_{n,k}|H_{0}) = 1-P(T_{n,k}\leq t_{n,k}|H_{0})
= 1-\chi^{2}\texttt{cdf}(0,t_{n,k},k-1) \ .
\ee

\medskip
\noindent
\underline{\R:}
\texttt{kruskal.test(\textit{variable}~\texttildelow~\textit{group
variable})} \\
\underline{SPSS:} Analyze $\rightarrow$ Nonparametric Tests
$\rightarrow$ Legacy Dialogs $\rightarrow$ K Independent Samples
\ldots: Kruskal-Wallis H


\chapter[Bivariate methods of statistical data analysis]{Bivariate
methods of statistical data analysis: testing for association}
\lb{ch13}
Recognising patterns of regularity in the variability of data sets 
for given (observable) statistical variables, and explaining them 
in terms of \textbf{causal relationships} in the context of a 
suitable \textbf{theoretical model}, is one of the main objectives
of any empirical scientific discipline, and thus motivation for 
corresponding \textbf{research}; see, e.g., Penrose 
(2004)~\ct{pen2004}. Causal relationships are intimately related 
to \textbf{interactions} between objects or agents of the physical 
or/and of the social kind. A \textit{necessary} (though not 
sufficient) \textit{condition} on the way to theoretically
fathoming causal relationships is to establish empirically the
existence of significant \textbf{statistical associations} between
the variables in question. \textbf{Replication} of positive
observational or experimental results of this kind, when
accomplished, yields strong support in favour of this idea.
Regrettably, however, the existence of causal relationships between
two statistical variables \textit{cannot} be established with
absolute certainty by empirical means; compelling theoretical
arguments need to stand in. Causal relationships between
statistical variables imply an unambiguous distinction between
\textbf{independent variables} and \textbf{dependent variables}. In
the following, we will discuss the principles of the simplest three inferential statistical methods within the \textbf{frequentist
framework}, each associated with specific \textbf{null hypothesis
significance tests}, that provide empirical checks of the
aforementioned necessary condition in the \textbf{bivariate case}.

\section[Correlation analysis and linear regression]{\href{https://www.youtube.com/watch?v=aLcTqycIXgU}{Correlation analysis and
linear regression}}
\lb{sec:correl}
\subsection[$t$--test for a correlation]{$\boldsymbol{t}$--test 
for a correlation}
\lb{subsec:correlttest}
The parametric \textbf{correlation analysis} presupposes a
metrically scaled two-dimensional statistical variable $(X,Y)$ 
that can be assumed to satisfy a \textbf{bivariate normal 
distribution} in some target population~$\boldsymbol{\Omega}$. Its 
aim is to investigate whether or not the components $X$ and $Y$ 
feature a quantitative--statistical association of a
\textit{linear} nature, given a data matrix $\boldsymbol{X} \in 
\mathbb{R}^{n \times 2}$ obtained from a random sample of 
size~$n$. Formulated in terms of the \textbf{population correlation 
coefficient} $\rho$ according to 
\href{http://en.wikipedia.org/wiki/Auguste_Bravais}{Auguste
Bravais (1811--1863)} and
\href{http://www-history.mcs.st-and.ac.uk/Biographies/Pearson.html}
{Karl Pearson FRS (1857--1936)}, the method tests $H_{0}$ against 
$H_{1}$ in one of the alternative pairs of

\medskip
\noindent
\textbf{Hypotheses:} \hfill (test for association)
\be
\lb{eq:correlanahypo}
\begin{cases}
H_{0}: \rho = 0
\quad\text{or}\quad
\rho \geq 0
\quad\text{or}\quad
\rho \leq 0 \\
H_{1}: \rho \neq 0
\quad\text{or}\quad
\rho < 0
\quad\text{or}\quad
\rho > 0
\end{cases} \ ,
\ee
with $-1 \leq \rho \leq +1$.

\medskip
\noindent
For sample sizes~$n \geq 50$, the assumption of normality of the 
marginal $X$- and $Y$-distributions in a given random sample 
$\boldsymbol{S_{\Omega}}$:~$(X_{1}, \ldots, X_{n};\-Y_{1}, \ldots, 
Y_{n})$ drawn from $\boldsymbol{\Omega}$ can be tested by 
means of the \textbf{Kolmogorov--Smirnov--test}; cf. 
Sec.~\ref{sec:onesampttest}. For sample sizes $n < 50$, on the 
other hand, the magnitudes of the \textbf{standardised skewness and
excess kurtosis measures}, Eqs.~(\ref{eq:g1g2ratios}), can be
considered instead. A \textbf{scatter plot} of the bivariate raw
sample data $\{(x_{i},y_{i})\}_{i=1,\ldots,n}$ displays
characteristic features of the \textbf{joint
$\boldsymbol{(X,Y)}$-distribution}.

\medskip
\noindent
\underline{\R:} \texttt{ks.test(\textit{variable}, "pnorm")} \\
\underline{SPSS:} Analyze $\rightarrow$ Nonparametric Tests
$\rightarrow$ Legacy Dialogs $\rightarrow$ 1-Sample K-S \ldots: 
Normal

\medskip
\noindent
Normalising the \textbf{sample correlation coefficient} $r$ of 
Eq.~(\ref{eq:correl}) by its \textbf{standard error},
\be
\text{SE}r :=\sqrt{\frac{1-r^{2}}{n-2}} \ ,
\ee
the latter of which can be derived from the corresponding 
theoretical \textbf{sampling distribution} for~$r$, presently
yields the (see, e.g., Toutenburg (2005)~\ct[Eq.~(7.18)]{tou2005})

\medskip
\noindent
\textbf{Test statistic:}
\be
\lb{eq:correlteststat}
\fbox{$\displaystyle
T_{n} := \frac{r}{\text{SE}r} \ \stackrel{H_{0}}{\sim}\ t(n-2) \ ,
$}
\ee
which, under $H_{0}$, satisfies a $\boldsymbol{t}$\textbf{--test
distribution} with $df=n-2$ degrees of freedom; cf.
Sec.~\ref{sec:tverteil}.

\medskip
\noindent
\textbf{Test decision:} Depending on the kind of test to be 
performed, the rejection region for $H_{0}$ at significance level 
$\alpha$ is given by
\begin{center}
\begin{tabular}[h]{c|c|c|c}
 & & & \\
\textbf{Kind of test} & $\boldsymbol{H_{0}}$ &
$\boldsymbol{H_{1}}$ &
\textbf{Rejection region for} $\boldsymbol{H_{0}}$ \\
 & & & \\
\hline
 & & & \\
(a)~two-sided & $\rho=0$ & $\rho \neq 0$ &
$|t_{n}|>t_{n-2;1-\alpha/2}$ \\
 & & & \\
\hline
 & & & \\
(b)~left-sided & $\rho \geq 0$ & $\rho<0$ &
$t_{n}<t_{n-2;\alpha}=-t_{n-2;1-\alpha}$ \\
 & & & \\
\hline
 & & & \\
(c)~right-sided & $\rho \leq 0$ & $\rho>0$ &
$t_{n}>t_{n-2;1-\alpha}$ \\
 & & &
\end{tabular}
\end{center}
$p$--values associated with realisations $t_{n}$ of the
\textbf{test statistic}~(\ref{eq:correlteststat}), which are to be
calculated from the
$\boldsymbol{t}$\textbf{--test distribution}, can  be obtained from 
Eqs.~(\ref{eq:pvaluetwo})--(\ref{eq:pvalueright}).

\medskip
\noindent
\underline{\R:} \texttt{cor.test(\textit{variable1},
\textit{variable2})}, \\
\texttt{cor.test(\textit{variable1}, \textit{variable2}, 
alternative = "less")}, \\
\texttt{cor.test(\textit{variable1}, \textit{variable2}, 
alternative = "greater")} \\
\underline{SPSS:} Analyze $\rightarrow$ Correlate
$\rightarrow$ Bivariate \ldots: Pearson

\medskip
\noindent
\textbf{Effect size:} The practical significance of the phenomenon investigated can be estimated directly from the absolute value of
the scale-invariant sample correlation coefficient~$r$ according to
Cohen's (1992)~\ct[Tab.~1]{coh1992}

\medskip
\noindent
\underline{\textbf{Rule of thumb:}}\\
$0.10 \leq |r| < 0.30$: small effect\\
$0.30 \leq |r| < 0.50$: medium effect\\
$0.50 \leq |r|$: large effect.

\medskip
\noindent
It is generally recommended to handle significant test results of 
\textbf{correlation analyses} for metrically scaled two-dimensional
statistical variables $(X,Y)$ with some care, due to the 
possibility of \textbf{spurious correlations} induced by additional 
\textbf{control variables} $Z, \ldots$, acting hidden in the 
background. To exclude this possibility, a correlation analysis 
should, e.g., be repeated for homogeneous subgroups of the 
sample~$\boldsymbol{S_{\Omega}}$. Some rather curious and 
startling cases of spurious correlations have been collected at 
the website
\href{http://www.tylervigen.com/}{\texttt{www.tylervigen.com}}.

\subsection[$F$--test of a regression 
model]{$\boldsymbol{F}$--test of a regression model}
\lb{subsec:correlFtest}
When a correlation in the joint distribution of a metrically 
scaled two-dimensional statistical variable~$(X,Y)$, significant 
in~$\boldsymbol{\Omega}$ at level~$\alpha$, proves to be 
\textit{strong}, i.e., when the magnitude of $\rho$ takes a value
in the interval
\[
0.71 \leq |\rho| \leq 1.0 \ ,
\]
it is meaningful to ask which \textit{linear} quantitative model 
best represents the detected linear statistical association; 
cf.~Pearson (1903)~\ct{pea1903}. To this end, \textbf{simple linear 
regression} seeks to devise a \textbf{linear stochastic regression 
model} for the target population~$\boldsymbol{\Omega}$ of the form
\be
\lb{eq:linregmod}
\text{in}\ \boldsymbol{\Omega}: \quad
Y_{i} = \alpha + \beta x_{i} + \varepsilon_{i}
\qquad (i=1, \ldots,n) \ ,
\ee
which, for instance, assigns $X$ the role of an \textbf{independent 
variable} (and so its values $x_{i}$ can be considered prescribed 
by the modeller) and $Y$ the role of a \textbf{dependent variable}; 
such a model is essentially \textbf{univariate} in nature. The
\textbf{regression coefficients} $\alpha$ and $\beta$ denote the 
unknown $\boldsymbol{y}$\textbf{--intercept} and \textbf{slope} of
the model in $\boldsymbol{\Omega}$. For the \textbf{random errors} 
$\varepsilon_{i}$ it is assumed that
\be
\lb{eq:linregassump}
\varepsilon_{i} \stackrel{\text{i.i.d.}}{\sim} N(0;\sigma^{2}) \ ,
\ee
meaning they are identically normally distributed (with zero mean 
and constant variance~$\sigma^{2}$) and mutually stochastically 
independent. With respect to the bivariate random sample 
$\boldsymbol{S_{\Omega}}$:~$(X_{1}, \ldots, X_{n};\-Y_{1}, \ldots, 
Y_{n})$, the supposed linear relationship between $X$ and $Y$ is 
expressed by
\be
\text{in}\ \boldsymbol{S_{\Omega}}: \quad
y_{i} = a + b x_{i} + e_{i} \qquad (i=1, \ldots,n) \ .
\ee
So-called \textbf{residuals} are then defined according to
\be
e_{i} := y_{i}-\hat{y}_{i} = y_{i} - a - b x_{i} \qquad (i=1, 
\ldots,n) \ ,
\ee
which, for given values of $x_{i}$, encode the differences between 
the observed realisations $y_{i}$ of $Y$ and the corresponding (by
the linear regression model) predicted values $\hat{y}_{i}$
of $Y$. Given the assumption expressed in
Eq.~(\ref{eq:linregassump}), the residuals must satisfy the
condition $\displaystyle\sum_{i=1}^{n}e_{i}=0$.

\medskip
\noindent
Next, introduce \textbf{sums of squared deviations} for the
$Y$-data, in line with the ANOVA procedure of Sec.~\ref{sec:anova},
i.e.,
\bea
\lb{eq:tssreg}
\text{TSS} & := & \sum_{i=1}^{n}(y_{i}-\bar{y})^{2} \\
\lb{eq:rssreg}
\text{RSS} & := & \sum_{i=1}^{n}(y_{i}-\hat{y}_{i})^{2}
\ = \ \sum_{i=1}^{n}e_{i}^{2} \ .
\eea
In terms of these quantities, the \textbf{coefficient of 
determination} of Eq.~(\ref{eq:linregcoeffdetdescr}) for assessing 
the \textbf{goodness-of-the-fit} of a regression model can be
expressed by
\be
\lb{eq:linregcoeffdet}
B = \frac{\text{TSS}-\text{RSS}}{\text{TSS}}
= \frac{(\text{total variance of}\ Y)-(\text{unexplained variance 
of}\ Y)}{(\text{total variance of}\ Y)} \ .
\ee
This normalised measure expresses the proportion of variability in 
a data set of $Y$ which can be explained by the corresponding 
variability of $X$ through the \textbf{best-fit regression model}. 
The range of $B$ is $0 \leq B \leq 1$.

\medskip
\noindent
In the methodology of a \textbf{regression analysis} within the
\textbf{frequentist framework}, the first issue to be addressed is
to test the significance of the overall \textbf{simple linear
regression model}~(\ref{eq:linregmod}), i.e., to test $H_{0}$
against $H_{1}$ in the set of

\medskip
\noindent
\textbf{Hypotheses:} \hfill (test for differences)
\be
\lb{linreganovahyp1}
\begin{cases}
H_{0}: \beta = 0 \\
H_{1}: \beta \neq 0
\end{cases} \ .
\ee
Exploiting the goodness-of-the-fit aspect of the regression model 
as quantified by $B$ in Eq.~(\ref{eq:linregcoeffdet}), one 
arrives via division by the \textbf{standard error} of $B$,
\be
\lb{eq:sersq}
\text{SE}B := \frac{1-B}{n-2} \ ,
\ee
which derives from the theoretical \textbf{sampling distribution}
for $B$, at the (see, e.g., Hatzinger and Nagel 
(2013)~\ct[Eq.~(7.8)]{hatnag2013})

\medskip
\noindent
\textbf{Test statistic:}\footnote{Note that with the identity 
$B=r^{2}$ of Eq.~(\ref{eq:linregrsq}), which applies in simple 
linear regression, this is just the square of the test 
statistic~(\ref{eq:correlteststat}).}
\be
\lb{eq:linregteststat}
\fbox{$\displaystyle
T_{n} := \frac{B}{\text{SE}B}\ \stackrel{H_{0}}{\sim}\ F(1,n-2) \ .
$}
\ee
Under $H_{0}$, this satisfies an $\boldsymbol{F}$\textbf{--test
distribution} with $df_{1}=1$ and $df_{2}=n-2$ degrees of freedom;
cf. Sec.~\ref{sec:fverteil}.

\medskip
\noindent
\textbf{Test decision:} The rejection region for $H_{0}$ at 
significance level $\alpha$ is given by (right-sided test)
\be
t_{n} > f_{1,n-2;1-\alpha} \ .
\ee
With Eq.~(\ref{eq:pvalueright}), the $p$--value associated with a 
specific realisation $t_{n}$ of the
\textbf{test statistic}~(\ref{eq:linregteststat}), which is
to be calculated from the $\boldsymbol{F}$\textbf{--test
distribution}, amounts 
to
\be
p = P(T_{n}>t_{n}|H_{0}) = 1-P(T_{n}\leq t_{n}|H_{0})
= 1-F\texttt{cdf}(0,t_{n},1,n-2) \ .
\ee
%

\subsection[$t$--test for the regression 
coefficients]{$\boldsymbol{t}$--test for the regression 
coefficients}
The second issue to be addressed in a systematic \textbf{regression 
analysis} within the \textbf{frequentist framework} is to test
statistically which of the regression coefficients in the
model~(\ref{eq:linregmod}) are significantly different from zero.
In the case of simple linear regression, though, the matter for the coefficient $\beta$ is settled already 
by the \textbf{$\boldsymbol{F}$--test} of the regression model just 
outlined, resp.~the \textbf{$\boldsymbol{t}$--test} for $\rho$ 
described in Sec.~\ref{subsec:correlttest}; see, e.g., Levin 
\textit{et al} (2010)~\ct[p~389f]{levetal2009}. In this sense, a 
further test of statistical significance is redundant in the case 
of simple linear regression. However, when extending the concept 
of \textbf{regression analysis} to the more involved case of
\textbf{multivariate data}, a quantitative approach frequently
employed in the research literature of the \textbf{Social Sciences}
and \textbf{Economics}, this question attains relevance in its own
right. In this context, the \textbf{linear stochastic regression
model} for the dependent variable $Y$ to be assessed is of the
general form (cf. Yule (1897)~\ct{yul1897})
\be
\lb{eq:multlinregmod}
\text{in}\ \boldsymbol{\Omega}: \quad
Y_{i} = \alpha + \beta_{1}x_{i1} + \ldots
+ \beta_{k}x_{ik} + \varepsilon_{i}
\qquad (i=1, \ldots,n) \ ,
\ee
containing a total of $k$ uncorrelated independent variables and 
$k+1$ regression coefficients, as well as a random error term. A 
\textbf{multiple linear regression model} to be estimated from data 
of a corresponding random sample from $\boldsymbol{\Omega}$ of 
size~$n$ thus entails $n-k-1$ degrees of freedom; cf. 
Hair \textit{et al} (2010)~\ct[p~176]{haietal2010}. In view of this 
prospect, we continue with our methodological considerations.

\medskip
\noindent
First of all, \textbf{unbiased maximum likelihood point estimators} 
for the regression coefficients $\alpha$ and $\beta$ in 
Eq.~(\ref{eq:linregmod}) are obtained from application to the data 
of Gau\ss' method of \textbf{minimising the sum of squared
residuals} (RSS) (cf. Gau\ss~(1809)~\ct{gau1809} and
Ch.~\ref{ch5}),
\[
\text{minimise}\left(\text{RSS}=\sum_{i=1}^{n}e_{i}^{2}\right) \ ,
\]
yielding solutions
\be
\lb{eq:linregab}
b = \frac{S_{Y}}{s_{X}}\,r
\qquad\text{and}\qquad
a=\bar{Y}-b\bar{x} \ .
\ee
The equation of the \textbf{best-fit simple linear regression
model} is thus given by
\be
\fbox{$\displaystyle
\lb{eq:linregmodelstand}
\hat{y}=\bar{Y}+\frac{S_{Y}}{s_{X}}\,r\,(x-\bar{x}) \ ,
$}
\ee
and can be employed for purposes of predicting values of~$Y$ from 
given values of $X$ in the empirical interval $[x_{(1)},x_{(n)}]$.

\medskip
\noindent
Next, the \textbf{standard errors} associated with the values of
the maximum likelihood point estimators $a$ and $b$ in 
Eq.~(\ref{eq:linregab}) are derived from the corresponding 
theoretical \textbf{sampling distributions} and amount to (cf.,
e.g., Hartung \textit{et al} (2005)~\ct[p~576ff]{haretal2005})
\bea
\lb{eq:seregcoeffa}
\text{SE}a & := & 
\sqrt{\frac{1}{n}+\frac{\bar{x}}{(n-1)s_{X}^{2}}}\,\text{SE}e \\
\lb{eq:seregcoeffb}
\text{SE}b & := & 
\frac{\text{SE}e}{\sqrt{n-1}\,s_{X}} \ ,
\eea
where the \textbf{standard error of the residuals} $e_{i}$ is
defined by
\be
\text{SE}e := 
\sqrt{\frac{{\displaystyle\sum_{i=1}^{n}
(Y_{i}-\hat{Y}_{i})^{2}}}{n-2}} \ .
\ee

\medskip
\noindent
We now describe the test procedure for the \textbf{regression 
coefficient}~$\beta$. To be tested is $H_{0}$ against $H_{1}$ in 
one of the alternative pairs of

\medskip
\noindent
\textbf{Hypotheses:} \hfill (test for differences)
\be
\begin{cases}
H_{0}: \beta=0
\quad\text{or}\quad
\beta \geq 0
\quad\text{or}\quad
\beta \leq 0 \\
H_{1}: \beta \neq 0
\quad\text{or}\quad
\beta < 0
\quad\text{or}\quad
\beta > 0
\end{cases} \ .
\ee
Dividing the \textbf{sample regression slope} $b$ by its
\textbf{standard error}~(\ref{eq:seregcoeffb}) yields the

\medskip
\noindent
\textbf{Test statistic:}
\be
\lb{eq:linregbteststat}
\fbox{$\displaystyle
T_{n} := \frac{b}{\text{SE}b} \ \stackrel{H_{0}}{\sim}\ t(n-2) \ ,
$}
\ee
which, under $H_{0}$, satisfies a $\boldsymbol{t}$\textbf{--test
distribution} with $df=n-2$ degrees of freedom; cf.
Sec.~\ref{sec:tverteil}.

\medskip
\noindent
\textbf{Test decision:} Depending on the kind of test to be 
performed, the rejection region for $H_{0}$ at significance level 
$\alpha$ is given by
\begin{center}
\begin{tabular}[h]{c|c|c|c}
 & & & \\
\textbf{Kind of test} & $\boldsymbol{H_{0}}$ &
$\boldsymbol{H_{1}}$ &
\textbf{Rejection region for} $\boldsymbol{H_{0}}$ \\
 & & & \\
\hline
 & & & \\
(a)~two-sided & $\beta = 0$ & $\beta \neq 0$ &
$|t_{n}|>t_{n-2;1-\alpha/2}$ \\
 & & & \\
\hline
 & & & \\
(b)~left-sided & $\beta \geq 0$ & $\beta < 0$ &
$t_{n}<t_{n-2;\alpha}=-t_{n-2;1-\alpha}$ \\
 & & & \\
\hline
 & & & \\
(c)~right-sided & $\beta \leq 0$ & $\beta > 0$ &
$t_{n}>t_{n-2;1-\alpha}$ \\
 & & &
\end{tabular}
\end{center}
$p$--values associated with realisations $t_{n}$ of the
\textbf{test statistic}~(\ref{eq:linregbteststat}), which are to be 
calculated from the
$\boldsymbol{t}$\textbf{--test distribution}, can be obtained from 
Eqs.~(\ref{eq:pvaluetwo})--(\ref{eq:pvalueright}). We emphasise 
once more that for simple linear regression the test procedure 
just described is equivalent to the \textbf{correlation analysis}
of Sec.~\ref{subsec:correlttest}.

\medskip
\noindent
An analogous \textbf{$\boldsymbol{t}$--test} needs to be run to
check whether the \textbf{regression coefficient} $\alpha$ is
non-zero, too, using the ratio $\displaystyle\frac{a}{\text{SE}a}$
as a test statistic. However, in particular when the origin of $X$
is \textit{not} contained in the empirical interval 
$[x_{(1)},x_{(n)}]$, the null hypothesis $H_{0}: \alpha=0$ is a 
meaningless statement.

\medskip
\noindent
\underline{\R:}
\texttt{regMod <- lm(\textit{variable:y}~\texttildelow~\textit{variable:x})} \\
\texttt{summary(regMod)} \\
\underline{GDC:} mode \texttt{STAT} $\rightarrow$ \texttt{TESTS}
$\rightarrow$ \texttt{LinRegTTest\ldots} \\
\underline{SPSS:} Analyze $\rightarrow$ Regression
$\rightarrow$ Linear \ldots\ldots

\medskip
\noindent
\underline{Note:} Regrettably, SPSS provides no option for 
selecting between a one-sided and a two-sided $t$--test. The 
default setting is for a two-sided test. For the purpose of 
one-sided tests the $p$--value output of SPSS needs to be divided 
by $2$.

\medskip
\noindent
The extent to which the prerequisites of a regression analysis as 
stated in Eq.~(\ref{eq:linregassump}) are satisfied can be 
assessed by means of an \textbf{analysis of the residuals}:
\begin{itemize}

\item[(i)] for $n \geq 50$, \textbf{normality} of the distribution
of \textbf{residuals} $e_{i}$ ($i=1, \ldots, n$) can be checked by
means of a \textbf{Kolmogorov--Smirnov--test}; cf. 
Sec.~\ref{sec:onesampttest}; otherwise, when $n < 50$, resort to a 
consideration of the magnitudes of the \textbf{standardised
skewness and excess kurtosis measures}, Eqs.~(\ref{eq:g1g2ratios});

\item[(ii)] \textbf{homoscedasticity} of the $e_{i}$ ($i=1, \ldots, 
n$), i.e., whether or not they can be assumed to have constant 
variance, can be investigated qualitatively in terms of a
\textbf{scatter plot} that marks the standardised~$e_{i}$ (along
the vertical axis) against the corresponding predicted 
$Y$-values~$\hat{y}_{i}$ ($i=1, \ldots, n$) (along the horizontal 
axis). An elliptically shaped envelope of the cloud of data points 
thus obtained indicates that homoscedasticity applies.
\end{itemize}

\medskip
\noindent
Simple linear regression analysis can be easily modified to 
provide a tool to test bivariate empirical data 
$\{(x_{i},y_{i})\}_{i=1,\ldots,n}$ for positive metrically scaled 
statistical variables~$(X,Y)$ for an association in the form of a 
\textbf{Pareto distribution}; cf. Sec.~\ref{sec:paretodistr}. To
begin with, the original data is subjected to logarithmic 
transformations in order to obtain data for the \textbf{logarithmic 
quantities} $\ln(y_{i})$ resp.~$\ln(x_{i})$. Subsequently, a 
correlation analysis can be performed on the transformed data. 
Given there exists a functional relationship between the original 
$Y$ and $X$ of the form $y=Kx^{-(\gamma+1)}$, the logarithmic 
quantities are related by
\be
\ln(y) = \ln(K) - (\gamma+1)\times\ln(x) \ ,
\ee
i.e., one finds a \textit{straight line relationship} between 
$\ln(y)$ and $\ln(x)$ with negative slope equal to $-(\gamma+1)$.

\medskip
\noindent
We like to draw the reader's attention 
to a remarkable statistical phenomenon that was discovered, and 
emphatically publicised, by the English empiricist 
\href{http://www-history.mcs.st-and.ac.uk/Biographies/Galton.html}{Sir
Francis Galton FRS (1822--1911)}, following years of intense 
research during the late $19^\mathrm{th}$ Century; see Galton 
(1886)~\ct{gal1886}, and also Kahneman (2011)~\ct[Ch.~17]{kah2011}.
\textbf{Regression toward the mean} is best demonstrated on the
basis of the standardised version of the best-fit simple linear 
regression model of Eq.~(\ref{eq:linregmodelstand}), namely
\be
\lb{eq:regmean}
\hat{z}_{Y} = rz_{X} \ .
\ee
For bivariate metrically scaled random sample data that exhibits a 
non-perfect positive correlation (i.e., $0 < r < 1$), one observes
that, on average, large (small) $z_{X}$-values (i.e., values that
are far from their mean; that are, perhaps, even outliers) pair
with smaller (larger) $z_{Y}$-values (i.e., values that are closer
to their mean; that are more mediocre). Since this phenomenon 
persists after the roles of $X$ and $Y$ in the regression model 
have been switched, this is clear evidence that \textbf{regression 
toward the mean} is a manifestation of \textbf{randomness}, and 
\textit{not} of \textbf{causality} (which requires an unambiguous 
temporal order between a cause and an effect). Incidently,
\textbf{regression toward the mean} ensures that many physical and
social processes cannot become unstable.

\medskip
\noindent
Ending this section we point out that in reality a lot of the
processes studied in the \textbf{Natural Sciences} and in the
\textbf{Social Sciences} prove to be of an inherently
\textbf{non-linear nature};
see e.g. Gleick (1987)~\ct{gle1987}, Penrose (2004)~\ct{pen2004},
and Smith (2007)~\ct{smi2007}. On the one hand, this increases the
level of complexity involved in the analysis of data, on the other, 
non-linear processes offer the reward of a plethora of interesting 
and intriguing (dynamical) phenomena.

\section[Rank correlation analysis]{Rank correlation analysis}
When the two-dimensional statistical variable~$(X,Y)$ is 
metrically scaled but may \textit{not} be assumed bivariate
normally distributed in the target
population~$\boldsymbol{\Omega}$, or when
$(X,Y)$ is ordinally scaled in the first place, the standard tool 
for testing for a statistical association between the 
components~$X$ and $Y$ is the parametric \textbf{rank correlation 
analysis} developed by the English psychologist and statistician
\href{http://en.wikipedia.org/wiki/Charles_Edward_Spearman}{Charles
Edward Spearman FRS (1863--1945)} in 1904~\ct{spe1904}. This 
approach, like the univariate test procedures of Mann and Whitney, 
Wilcoxon, and Kruskal and Wallis discussed in Ch.~\ref{ch12}, is 
again fundamentally rooted in the concept of \textbf{rank numbers} 
representing statistical data which possess a natural order, 
introduced in Sec.~\ref{sec:2Dord}.

\medskip
\noindent
Following the translation of the original data pairs into 
corresponding \textbf{rank number pairs},
\be
(x_{i},y_{i}) \mapsto [R(x_{i}),R(y_{i})] \qquad (i=1,\ldots,n) \ ,
\ee
the objective is to subject $H_{0}$ in the alternative sets of

\medskip
\noindent
\textbf{Hypotheses:} \hfill (test for association)
\be
\begin{cases}
H_{0}: \rho_{S} = 0
\quad\text{or}\quad
\rho_{S} \geq 0
\quad\text{or}\quad
\rho_{S} \leq 0 \\
H_{1}: \rho_{S} \neq 0
\quad\text{or}\quad
\rho_{S} < 0
\quad\text{or}\quad
\rho_{S} > 0
\end{cases} \ ,
\ee
with $\rho_{S}$ ($-1 \leq \rho_{S} \leq +1$) the \textbf{population 
rank correlation coefficient}, to a test of statistical 
significance at level $\alpha$. Provided the size of the random 
sample is such that $n \geq 30$ (see, e.g., Bortz 
(2005)~\ct[p~233]{bor2005}), by dividing the \textbf{sample rank 
correlation coefficient}~$r_{S}$ of Eq.~(\ref{eq:rankcorrelcoeff}) 
by its \textbf{standard error}
\be
\lb{eq:sers}
\text{SE}r_{S} := \sqrt{\frac{1-r_{S}^{2}}{n-2}}
\ee
derived from the theoretical \textbf{sampling distribution}
for~$r_{S}$, one obtains a suitable

\medskip
\noindent
\textbf{Test statistic:}
\be
\lb{eq:rankcorrelteststat}
\fbox{$\displaystyle
T_{n} := \frac{r_{S}}{\text{SE}r_{S}} \ \stackrel{H_{0}}{\approx}\ 
t(n-2) \ .
$}
\ee
Under $H_{0}$, this approximately satisfies a
$\boldsymbol{t}$\textbf{--test distribution} 
with $df=n-2$ degrees of freedom; cf. Sec.~\ref{sec:tverteil}.

\medskip
\noindent
\textbf{Test decision:} Depending on the kind of test to be 
performed, the rejection region for $H_{0}$ at significance level 
$\alpha$ is given by
\begin{center}
\begin{tabular}[h]{c|c|c|c}
 & & & \\
\textbf{Kind of test} & $\boldsymbol{H_{0}}$ &
$\boldsymbol{H_{1}}$ &
\textbf{Rejection region for} $\boldsymbol{H_{0}}$ \\
 & & & \\
\hline
 & & & \\
(a)~two-sided & $\rho_{S}=0$ & $\rho_{S} \neq 0$ &
$|t_{n}|>t_{n-2;1-\alpha/2}$ \\
 & & & \\
\hline
 & & & \\
(b)~left-sided & $\rho_{S} \geq 0$ & $\rho_{S}<0$ &
$t_{n}<t_{n-2;\alpha}=-t_{n-2;1-\alpha}$ \\
 & & & \\
\hline
 & & & \\
(c)~right-sided & $\rho_{S} \leq 0$ & $\rho_{S}>0$ &
$t_{n}>t_{n-2;1-\alpha}$ \\
 & & &
\end{tabular}
\end{center}
$p$--values associated with realisations $t_{n}$ of the
\textbf{test statistic}~(\ref{eq:rankcorrelteststat}), which are to
be calculated from the
$\boldsymbol{t}$\textbf{--test distribution}, can be obtained from 
Eqs.~(\ref{eq:pvaluetwo})--(\ref{eq:pvalueright}).

\medskip
\noindent
\underline{\R:} \texttt{cor.test(\textit{variable1},
\textit{variable2}, method = "spearman")}, \\
\texttt{cor.test(\textit{variable1}, \textit{variable2}, 
method = "spearman", alternative = "less")}, \\
\texttt{cor.test(\textit{variable1}, \textit{variable2}, 
method = "spearman", alternative = "greater")} \\
\underline{SPSS:} Analyze $\rightarrow$ Correlate
$\rightarrow$ Bivariate \ldots: Spearman

\medskip
\noindent
\textbf{Effect size:} The practical significance of the phenomenon investigated can be estimated directly from the absolute value of
the scale-invariant sample rank correlation coefficient~$r_{S}$
according to (cf. Cohen (1992)~\ct[Tab.~1]{coh1992})

\medskip
\noindent
\underline{\textbf{Rule of thumb:}}\\
$0.10 \leq |r_{S}| < 0.30$: small effect\\
$0.30 \leq |r_{S}| < 0.50$: medium effect\\
$0.50 \leq |r_{S}|$: large effect.

\section[$\chi^{2}$--test for
independence]{\href{https://www.youtube.com/watch?v=uo2kjAPkYXQ}{$\boldsymbol{\chi}^{2}$--test for independence}}
The non-parametric \textbf{$\boldsymbol{\chi}^{2}$--test for 
independence} constitutes the most generally applicable 
significance test for bivariate statistical associations. Due to 
its formal indifference to the scale level of measurement of the
two-dimensional statistical variable~$(X,Y)$ involved in an
investigation, it may be used for statistical analysis of any kind
of pairwise combinations between nominally, ordinally and
metrically scaled components. The advantage of generality of the
method is paid for at the price of a generally weaker test power.

\medskip
\noindent
Given qualitative and/or quantitative statistical variables~$X$ 
and $Y$ that take values in a spectrum of $k$ mutually exclusive 
categories $a_{1}, \ldots, a_{k}$ resp.~$l$ mutually exclusive 
categories $b_{1}, \ldots, b_{l}$, the intention is to subject 
$H_{0}$ in the pair of alternative

\medskip
\noindent
\textbf{Hypotheses:} \hfill (test for association)
\be
\begin{cases}
H_{0}: \text{There does not exist a statistical association 
between}\ X\ \text{and}\ Y\ \text{in}\ \boldsymbol{\Omega} \\
H_{1}: \text{There does exist a statistical association between}\ 
X\ \text{and}\ Y\ \text{in}\ \boldsymbol{\Omega}
\end{cases}
\ee
to a convenient empirical significance test at level $\alpha$.

\medskip
\noindent
A conceptual issue that requires special attention along the way
is the definition of a reasonable \textbf{zero point} on the
\textbf{scale of statistical dependence} of statistical variables
$X$ and $Y$ (which one aims to establish). This problem is solved
by recognising that a common feature of sample data for statistical 
variables of all scale levels of measurement is the information
residing in the distribution of (relative) frequencies over (all
possible combinations of) categories, and drawing an analogy to the
concept of stochastic independence of two events as expressed in
\textbf{Probability Theory} by Eq.~(\ref{eq:stochindep2}). In this
way, by definition, we refer to variables $X$ and $Y$ as being
mutually \textbf{statistically independent} provided that the
bivariate relative frequencies $h_{ij}$ of \textit{all}
combinations of categories $(a_{i},b_{j})$ are numerically equal to
the products of the univariate marginal relative frequencies
$h_{i+}$ of $a_{i}$ and $h_{+j}$ of $b_{j}$ (cf.
Sec.~\ref{sec:konttaf}), i.e.,
\be
h_{ij} = h_{i+}h_{+j} \ .
\ee
Translated into the language of random sample variables, 
viz.~introducing \textbf{sample observed frequencies}, this 
operational \textbf{independence condition} is re-expressed by 
$O_{ij} = E_{ij}$, where the $O_{ij}$ denote the bivariate
\textbf{observed frequencies} of the category combinations
$(a_{i},b_{j})$ in a \textbf{cross tabulation} underlying a
specific random sample of size $n$, and the quantities $E_{ij}$,
which are defined in terms of (i)~the univariate sum $O_{i+}$ of
observed frequencies in row $i$, see Eq.~(\ref{eq:margfreq1}),
(ii)~the univariate sum $O_{+j}$ of observed frequencies in column
$j$, see Eq.~(\ref{eq:margfreq2}), and (iii)~the sample size $n$ by 
$\displaystyle E_{ij}:=\frac{O_{i+}O_{+j}}{n}$, are interpreted as 
the \textbf{expected frequencies} of $(a_{i},b_{j})$, given that
$X$ and $Y$ are statistically independent. Expressing differences 
between observed and (under independence) expected frequencies via 
the \textbf{residuals} $O_{ij} -  E_{ij}$, the hypotheses may be 
reformulated as

\medskip
\noindent
\textbf{Hypotheses:} \hfill (test for association)
\be
\begin{cases}
H_{0}: O_{ij} - E_{ij} = 0 \qquad\text{for all}
\ i = 1, \ldots, k \ \text{and}
\ j = 1, \ldots, l \\
H_{1}: O_{ij} - E_{ij} \neq 0 \qquad\text{for at least one}
\ i \ \text{and}\ j
\end{cases} \ .
\ee

\medskip
\noindent
For the subsequent test procedure to be reliable, it is
\textit{very important (!)} that the empirical prere\-quisite
\be
E_{ij} \stackrel{!}{\geq} 5
\ee
holds for all values of $i=1\ldots,k$ and $j=1,\ldots,l$,
such that one avoids the possibility of individual rescaled 
squared residuals $\displaystyle 
\frac{(O_{ij}-E_{ij})^{2}}{E_{ij}}$ becoming artificially 
magnified. The latter constitute the core of the

\medskip
\noindent
\textbf{Test statistic:}
\be
\lb{eq:chisqindepteststat}
\fbox{$\displaystyle
T_{n} := \sum_{i=1}^{k}\sum_{j=1}^{l}\frac{(O_{ij}-E_{ij}
)^{2}}{E_{ij}}
\ \stackrel{H_{0}}{\approx}\ \chi^{2}[(k-1)\times(l-1)] \ ,
$}
\ee
which, under $H_{0}$, approximately satisfies a 
$\boldsymbol{\chi^{2}}$\textbf{--test distribution} with $df=(k-1)
\times (l-1)$ degrees of freedom; cf. Sec.~\ref{sec:chi2verteil}.

\medskip
\noindent
\textbf{Test decision:} The rejection region for $H_{0}$ at 
significance level $\alpha$ is given by (right-sided test)
\be
t_{n}>\chi^{2}_{(k-1)\times(l-1);1-\alpha} \ .
\ee
By Eq.~(\ref{eq:pvalueright}), the $p$--value associated with a 
realisation $t_{n}$ of the \textbf{test
statistic}~(\ref{eq:chisqindepteststat}), which is to
be calculated from the $\boldsymbol{\chi^{2}}$\textbf{--test
distribution}, amounts to
\be
p = P(T_{n}>t_{n}|H_{0}) = 1-P(T_{n}\leq t_{n}|H_{0})
= 1-\chi^{2}\texttt{cdf}\left(0,t_{n},(k-1)\times(l-1)\right) \ .
\ee

\medskip
\noindent
\underline{\R:} \texttt{chisq.test(\textit{row variable},
\textit{column variable})} \\
\underline{GDC:} mode \texttt{STAT} $\rightarrow$ \texttt{TESTS}
$\rightarrow$ \texttt{$\chi^{2}$-Test\ldots} \\
\underline{SPSS:} Analyze $\rightarrow$ Descriptive Statistics
$\rightarrow$ Crosstabs \ldots $\rightarrow$ Statistics \ldots:
Chi-square

\medskip
\noindent
The $\chi^{2}$--test for independence can establish the 
\textbf{existence} of a significant association in the joint
distribution of a two-dimensional statistical variable~$(X,Y)$. 
The \textbf{strength} of the association, on the other hand, 
may be measured in terms of \textbf{Cram\'{e}r's} $\boldsymbol{V}$
(Cram\'{e}r (1946)~\ct{cra1946}), which has a normalised range of 
values given by $0 \leq V \leq 1$; cf. Eq.~(\ref{eq:cramv}) and 
Sec.~\ref{sec:2Dnom}. Low values of $V$ in the case of significant 
associations between components $X$ and $Y$ typically indicate the 
statistical influence of additional \textbf{control variables}.

\medskip
\noindent
\underline{\R:} \texttt{assocstats(\textit{contingency table})}
(package: \texttt{vcd}, by Meyer \textit{et al}
(2017)~\ct{meyetal2017}) \\
\underline{SPSS:} Analyze $\rightarrow$ Descriptive Statistics
$\rightarrow$ Crosstabs \ldots $\rightarrow$ Statistics \ldots:
Phi and Cramer's V

\medskip
\noindent
\textbf{Effect size:} The practical significance of the phenomenon investigated can be estimated and interpreted by means of the
effect size measure~$w$ defined in Eq.~(\ref{eq:eschisq});
cf. Cohen (1992)~\ct[Tab.~1]{coh1992}.


\addcontentsline{toc}{chapter}{Outlook}
\chapter*{Outlook}
Our discussion on the foundations of statistical methods of data
analysis and their application to specific quantitative problems
ends here. We have focused on the description of uni- and bivariate
data sets and making inferences from corresponding random samples
within the frequentist approach to Probability Theory. At this
stage, the attentive reader should feel well-equipped for
confronting problems concerning more complex, multivariate data
sets, and adequate methods for tackling them by statistical means.
Many modules at the Master degree level review a broad spectrum of
advanced topics such as multiple linear regression, generalised
linear models, principal component analysis, or cluster analysis,
which in turn relate to computational techniques presently employed
in the context of machine learning. The ambitious reader might even
think of getting involved with proper research and work towards a
Ph.D.~degree in an empirical scientific discipline. To gain
additional data analytical flexibility, and to increase chances on
obtaining transparent and satisfactory research results, it is
strongly recommended to consult the conceptually compelling
inductive Bayes--Laplace approach to statistical inference. In
order to leave behind the methodological shortcomings uncovered by
the recent replication crisis (cf., e.g., Refs.~\ct{eco2013},
\ct{nuz2014}, or \ct{vas2017}), strict adherence to accepted
scientific standards cannot be compromised with.\footnote{With
regard to the replication crisis, the interested reader might be
aware of the international initiative known as the Open Science
Framework. URL (cited on August 17, 2019):
\href{https://osf.io}{https://osf.io}.}

\medskip
\noindent
Beyond activities within the scientific community, the dedicated
reader may feel encouraged to use her/his solid topical
qualification in statistical methods of data analysis for careers
in either field of higher education, public health, renewable
energy supply chains, evaluation of climate change adaptation,
development of plans for sustainable production in agriculture and
global economy, civil service, business management, marketing,
logistics, or the financial services, amongst a multitude of other
inspirational possibilities.

\medskip
\noindent
Not every single matter of human life is amenable to
quantification, or, acknowledging an individual freedom of making
choices, needs to be quantified in the first place. Blind faith in
the powers of quantitative methods is certainly misplaced. Thorough reflection and introspection on the options available for action
and their implied consequences, together with a critical evaluation
of relevant tangible facts, might suggest a viable alternative
approach to a given research or practical problem. Generally, there
is a potential for looking behind curtains, shifting horizons,
or anticipating prospects and opportunities. Finally, more often
than not, there exists a dimension of non-knowledge on the part of
the individual investigator that needs to be taken into account as
an integral part of the boundary conditions of the overall problem
in question.  The adventurous mind will always excel in view of the
intricate challenge of making inferences on the basis of incomplete 
information.

\appendix
\chapter[Simple principal component analysis]{Principal component 
analysis of a $\boldsymbol{(2 \times 2)}$ correlation matrix}
\lb{app1}
Consider a real-valued
$\boldsymbol{(2 \times 2)}$ {\bf correlation matrix} expressed by
\be
\boldsymbol{R} =
\left(\begin{array}{cc}
1 & r \\
r & 1
\end{array}\right) \ , \qquad
-1 \leq r \leq +1 \ ,
\ee
which, by construction, is symmetric.
Its {\bf trace} amounts to $\mathrm{Tr}(\boldsymbol{R})=2$, while
its {\bf determinant} is $\det(\boldsymbol{R}) = 1-r^{2}$. 
Consequently, $\boldsymbol{R}$ is regular
as long as $r \neq \pm 1$. We seek to determine the {\bf 
eigenvalues} and corresponding 
{\bf eigenvectors} (or {\bf principal components}) of 
$\boldsymbol{R}$, i.e., real numbers~$\lambda$ and real-valued 
vectors~$\vec{v}$ such that the condition
\be
\lb{eq:eigenv}
\boldsymbol{R}\,\vec{v} \stackrel{!}{=} \lambda\,\vec{v}
\qquad\Leftrightarrow\qquad
(\boldsymbol{R}-\lambda\boldsymbol{1})\,\vec{v}
\stackrel{!}{=} \boldsymbol{0}
\ee
applies. The determination of non-trivial solutions of this 
algebraic problem leads to the {\bf characteristic equation}
\be
0 \stackrel{!}{=} \det(\boldsymbol{R}-\lambda\boldsymbol{1})
= (1-\lambda)^{2} - r^{2} = (\lambda-1)^{2} - r^{2} \ .
\ee
Hence, by completing squares, it is clear that $\boldsymbol{R}$ 
possesses the two {\bf eigenvalues}
\be
\lambda_{1} = 1+r \qquad\text{and}\qquad
\lambda_{2} = 1-r \ ,
\ee
showing that $\boldsymbol{R}$ is {\bf positive-definite} whenever
$|r| < 1$. The normalised {\bf eigenvectors} associated with 
$\lambda_{1}$ and $\lambda_{2}$, obtained from 
Eq.~(\ref{eq:eigenv}), then are
\be
\vec{v}_{1} = \frac{1}{\sqrt{2}}
\left(\begin{array}{c}
1 \\ 1
\end{array}\right) \qquad\text{and}\qquad
\vec{v}_{2} = \frac{1}{\sqrt{2}}
\left(\begin{array}{r}
- 1\\ 1
\end{array}\right) \ ,
\ee
and constitute a right-handedly oriented basis of the 
two-dimensional {\bf eigenspace} of $\boldsymbol{R}$. 
Note that due to the symmetry of $\boldsymbol{R}$ it holds that 
$\vec{v}_{1}^{T}\cdot\vec{v}_{2}=0$, i.e., the eigenvectors are 
mutually orthogonal.

\medskip
\noindent
The normalised eigenvectors of $\boldsymbol{R}$ define a regular
orthogonal {\bf transformation matrix} $\boldsymbol{M}$, and an
inverse $\boldsymbol{M}^{-1}=\boldsymbol{M}^{T}$, given by resp.
\be
\boldsymbol{M} =
\frac{1}{\sqrt{2}}\left(\begin{array}{cr}
1 & -1 \\
1 &  1
\end{array}\right)
\qquad\text{and}\qquad
\boldsymbol{M}^{-1} =
\frac{1}{\sqrt{2}}\left(\begin{array}{rc}
 1 & 1 \\
-1 & 1
\end{array}\right)
= \boldsymbol{M}^{T} \ ,
\ee
where $\mathrm{Tr}(\boldsymbol{M})=\sqrt{2}$ and 
$\det(\boldsymbol{M})=1$. The correlation matrix $\boldsymbol{R}$ 
can now be {\bf diagonalised} by means of a rotation with 
$\boldsymbol{M}$ according to\footnote{Alternatively one 
can write
\[
\boldsymbol{M} = \left(\begin{array}{cr}
\cos(\pi/4) & -\sin(\pi/4) \\
\sin(\pi/4) & \cos(\pi/4)
\end{array}\right) \ ,
\]
thus emphasising the character of a rotation of $\boldsymbol{R}$ 
by an angle $\varphi=\pi/4$.
}
\bea
\boldsymbol{R}_\mathrm{diag} & = & \boldsymbol{M}^{-1}\boldsymbol{R}
\boldsymbol{M} \nonumber \\
& = & \frac{1}{\sqrt{2}}\left(\begin{array}{rc}
 1 & 1 \\
-1 & 1
\end{array}\right)
\left(\begin{array}{cc}
1 & r \\
r & 1
\end{array}\right)
\frac{1}{\sqrt{2}}\left(\begin{array}{cr}
1 & -1 \\
1 &  1
\end{array}\right)
= \left(\begin{array}{cc}
1+r &  0 \\
0        &  1-r
\end{array}\right) \ .
\eea
Note that $\mathrm{Tr}(\boldsymbol{R}_\mathrm{diag})=2$ and
$\det(\boldsymbol{R}_\mathrm{diag}) = 1-r^{2}$, i.e., the trace 
and determinant of $\boldsymbol{R}$ remain {\bf invariant} under 
the diagonalising transformation.

\medskip
\noindent
The concepts of eigenvalues and 
eigenvectors (principal components), as well as of diagonalisation 
of symmetric matrices, generalise in a straightforward though 
computationally more demanding fashion to arbitrary real-valued 
{\bf correlation matrices} $\boldsymbol{R} \in \mathbb{R}^{m 
\times m}$, with $m \in \mathbb{N}$.

\medskip
\noindent
\underline{\R:} \texttt{prcomp(\textit{data matrix})}


\chapter[Distance measures in Statistics]{Distance measures in 
Statistics}
\lb{app2}
{\bf Statistics} employs a number of different measures of {\bf 
distance} $d_{ij}$ to quantify the separation in an $m$--D space 
of metrically scaled statistical variables $X, Y, \ldots, Z$ of
two statistical units $i$ and $j$ ($i, j = 1, \ldots, n$). Note 
that, by construction, these measures $d_{ij}$ exhibit the 
properties $d_{ij} \geq 0$, $d_{ij} = d_{ji}$ and $d_{ii} = 0$. In 
the following, $X_{ik}$ is the entry of the data matrix 
$\boldsymbol{X} \in \mathbb{R}^{n \times m}$ 
relating to the $i$th statistical unit and the $k$th statistical 
variable, etc. The $d_{ij}$ define the elements of a
$\boldsymbol{(n \times n)}$ {\bf proximity matrix} $\boldsymbol{D} 
\in \mathbb{R}^{n \times n}$.

\subsection*{Euclidian distance \hfill (dimensionful)}
This most straightforward, dimensionful distance measure is named 
after the ancient Greek (?) mathematician 
\href{http://www-history.mcs.st-and.ac.uk/Biographies/Euclid.html}{Euclid of Alexandria (ca.~325BC--ca.~265BC)}. It is defined by
\be
\lb{eq:eucliddist}
\fbox{$\displaystyle
d_{ij}^{E} := 
\sqrt{\sum_{k=1}^{m}\sum_{l=1}^{m}(X_{ik}-X_{jk})\delta_{kl}
(X_{il}-X_{jl})} \ ,
$}
\ee
where $\delta_{kl}$ denotes the elements of the unit matrix 
$\boldsymbol{1} \in \mathbb{R}^{m \times m}$; cf. 
Ref.~\ct[Eq.~(2.2)]{hve2009}.

\subsection*{Mahalanobis distance \hfill (dimensionless)}
A more sophisticated, {\bf scale-invariant} distance measure in 
{\bf Statistics} was devised by the Indian applied statistician 
\href{http://www-history.mcs.st-and.ac.uk/Biographies/Mahalanobis.html}{Prasanta Chandra Mahalanobis (1893--1972)}; cf. Mahalanobis 
(1936)~\ct{mah1936}. It is defined by
\be
\lb{eq:mahaladist}
\fbox{$\displaystyle
d_{ij}^{M} := 
\sqrt{\sum_{k=1}^{m}\sum_{l=1}^{m}(X_{ik}-X_{jk})(S^{2})^{-1}_{kl}
(X_{il}-X_{jl})} \ ,
$}
\ee
where $(S^{2})^{-1}_{kl}$ denotes the elements of the inverse
covariance matrix $(\boldsymbol{S^{2}})^{-1} \in
\mathbb{R}^{m \times m}$ relating to $X, Y, \ldots, Z$; cf.
Sec.~\ref{subsec:covar}. The Mahalanobis distance thus accounts for
inter-variable correlations and so eliminates a potential source of
bias.

\medskip
\noindent
\underline{\R:} \texttt{mahalanobis(\textit{data matrix})}


\chapter[List of online survey tools]{List of online survey tools}
\lb{app3}
A first version of the following list of online survey tools for 
the Social Sciences, the use of some of which is free of charge, 
was compiled and released courtesy of an investigation by Michael 
R\"{u}ger (IMC, year of entry 2010):
\begin{itemize}

\item \href{http://easy-feedback.de/de/startseite}{easy-feedback.de/de/startseite}

\item \href{http://www.evalandgo.de/}{www.evalandgo.de}

\item \href{http://www.limesurvey.org}{www.limesurvey.org}

\item \href{http://www.netigate.de}{www.netigate.de}

\item \href{http://polldaddy.com/}{polldaddy.com}

\item \href{http://q-set.de}{q-set.de}

\item \href{http://www.qualtrics.com}{www.qualtrics.com}

\item \href{https://www.soscisurvey.de/}{www.soscisurvey.de}

\item \href{https://www.surveymonkey.com/}{www.surveymonkey.com}

\item \href{http://www.umfrageonline.com}{www.umfrageonline.com}

\end{itemize}
%


\chapter[Glossary of technical terms (GB -- D)]{Glossary of 
technical terms (GB -- D)}
\lb{app4}

\noindent
{\bf A}\\
additive: additiv, summierbar\\
ANOVA: Varianzanalyse\\
arithmetical mean: arithmetischer Mittelwert\\
association: Zusammenhang, Assoziation\\
attribute: Auspr\"{a}gung, Eigenschaft

\medskip
\noindent
{\bf B}\\
bar chart: Balkendiagramm\\
Bayes' theorem: Satz von Bayes\\
Bayesian probability: Bayesianischer Wahrscheinlichkeitsbegriff\\
best-fit model: Anpassungsmodell\\
bin: Datenintervall\\
binomial coefficient: Binomialkoeffizient\\
bivariate: bivariat, zwei variable Gr\"{o}\ss en betreffend\\
box plot: Kastendiagramm

\medskip
\noindent
{\bf C}\\
category: Kategorie\\
causality: Kausalit\"{a}t\\
causal relationship: Kausalbeziehung\\
census: statistische Vollerhebung\\
central limit theorem: Zentraler Grenzwertsatz\\
centre of gravity: Schwerpunkt\\
centroid: geometrischer Schwerpunkt\\
certain event: sicheres Ereignis\\
class interval: Auspr\"{a}gungsklasse\\
cluster analysis: Klumpenanalyse\\
cluster random sample: Klumpenzufallsstichprobe\\
coefficient of determination: Bestimmtheitsma\ss\\
coefficient of variation: Variationskoeffizient\\
combination: Kombination\\
combinatorics: Kombinatorik\\
compact: geschlossen, kompakt\\
complementation of a set: Bilden der Komplement\"{a}rmenge\\
concentration: Konzentration\\
conditional distribution: bedingte Verteilung\\
conditional probability: bedingte Wahrscheinlichkeit\\
confidence interval: Konfidenzintervall\\
conjunction: Konjunktion, Mengenschnitt\\
contingency table: Kontingenztafel\\
continuous data: stetige Daten\\
control variable: St\"{o}rvariable\\
convenience sample: Gelegenheitsstichprobe\\
convexity: Konvexit\"{a}t\\
correlation matrix: Korrelationsmatrix\\
covariance matrix: Kovarianzmatrix\\
critical value: kritischer Wert\\
cross tabulation: Kreuztabelle\\
cumulative distribution function ({\tt cdf}): theoretische 
Verteilungsfunktion

\medskip
\noindent
{\bf D}\\
data: Daten\\
data matrix: Datenmatrix\\
decision: Entscheidung\\
deductive method: deduktive Methode\\
degree-of-belief: Glaubw\"{u}rdigkeitsgrad, Plausibilit\"{a}t\\
degrees of freedom: Freiheitsgrade\\
dependent variable: abh\"{a}ngige Variable\\
descriptive statistics: Beschreibende Statistik\\
deviation: Abweichung \\
difference: Differenz\\
direction: Richtung\\
discrete data: diskrete Daten\\
disjoint events: disjunkte Ereignisse, einander ausschlie\ss end\\
disjunction: Disjunktion, Mengenvereinigung\\
dispersion: Streuung\\
distance: Abstand\\
distortion: Verzerrung\\
distribution: Verteilung\\
distributional properties: Verteilungseigenschaften

\medskip
\noindent
{\bf E}\\
econometrics: \"{O}konometrie\\
effect size: Effektgr\"{o}\ss e\\
eigenvalue: Eigenwert\\
elementary event: Elementarereignis\\
empirical cumulative distribution function: empirische 
Verteilungsfunktion\\
estimator: Sch\"{a}tzer\\
Euclidian distance: Euklidischer Abstand\\
Euclidian space: Euklidischer (nichtgekr\"{u}mmter) Raum\\
event: Ereignis\\
event space: Ereignisraum\\
evidence: Anzeichen, Hinweis, Anhaltspunkt, Indiz\\
expectation value: Erwartungswert \\
extreme value: extremer Wert

\medskip
\noindent
{\bf F}\\
fact: Tatsache, Faktum\\
factorial: Fakult\"{a}t\\
falsification: Falsifikation\\
five number summary: F\"{u}nfpunktzusammenfassung\\
frequency: H\"{a}ufigkeit\\
frequentist probability: frequentistischer 
Wahrscheinlichkeitsbegriff

\medskip
\noindent
{\bf G}\\
Gini coefficient: Ginikoeffizient\\
goodness-of-the-fit: Anpassungsg\"{u}te

\medskip
\noindent
{\bf H}\\
Hessian matrix: Hesse'sche Matrix\\
histogram: Histogramm\\
homoscedasticity: Homoskedastizit\"{a}t, homogene Varianz\\
hypothesis: Hypothese, Behauptung, Vermutung

\medskip
\noindent
{\bf I}\\
inclusion of a set: Mengeninklusion\\
independent variable: unabh\"{a}ngige Variable\\
inductive method: induktive Methode\\
inferential statistics: Schlie\ss ende Statistik\\
interaction: Wechselwirkung\\
intercept: Achsenabschnitt\\
interquartile range: Quartilsabstand\\
interval scale: Intervallskala\\
impossible event: unm\"{o}gliches Ereignis

\medskip
\noindent
{\bf J}\\
joint distribution: gemeinsame Verteilung

\medskip
\noindent
{\bf K}\\
$k\sigma$--rule: $k\sigma$--Regel\\
kurtosis: W\"{o}lbung

\medskip
\noindent
{\bf L}\\
latent variable: latente Variable, nichtbeobachtbares Konstrukt\\
law of large numbers: Gesetz der gro\ss en Zahlen\\
law of total probability: Satz von der totalen Wahrscheinlichkeit\\
Likert scale: Likertskala, Verfahren zum Messen von 
eindimensionalen latenten Variablen\\
linear regression analysis: lineare Regressionsanalyse\\
location parameter: Lageparameter\\
Lorenz curve: Lorenzkurve

\medskip
\noindent
{\bf M}\\
Mahalanobis distance: Mahalanobis'scher Abstand\\
manifest variable: manifeste Variable, Observable\\
marginal distribution: Randverteilung\\
marginal frequencies: Randh\"{a}ufigkeiten\\
measurement: Messung, Datenaufnahme\\
method of least squares: Methode der kleinsten Quadrate\\
median: Median\\
metrical: metrisch\\
mode: Modalwert

\medskip
\noindent
{\bf N}\\
nominal: nominal

\medskip
\noindent
{\bf O}\\
observable: beobachtbare/messbare Variable, Observable\\
observation: Beobachtung\\
odds: Wettchancen \\
operationalisation: Operationalisieren, latente Variable messbar 
gestalten\\
opinion poll: Meinungsumfrage\\
ordinal: ordinal\\
outlier: Ausrei\ss er

\medskip
\noindent
{\bf P}\\
$p$--value: $p$--Wert\\
partition: Zerlegung, Aufteilung\\
percentile value: Perzentil, $\alpha$--Quantil\\
pie chart: Kreisdiagramm\\
point estimator: Punktsch\"{a}tzer\\
population: Grundgesamtheit\\
power: Testst\"{a}rke\\
power set: Potenzmenge\\
practical significance: praktische Signifikanz, Bedeutung\\
principal component analysis: Hauptkomponentenanalyse\\
probability: Wahrscheinlichkeit\\
probability density function ({\tt pdf}): 
Wahrscheinlichkeitsdichte\\
probability function: Wahrscheinlichkeitsfunktion\\
probability measure: Wahrscheinlichkeitsma\ss\\
probability space: Wahrscheinlichkeitsraum\\
projection: Projektion\\
proportion: Anteil\\
proximity matrix: Distanzmatrix

\medskip
\noindent
{\bf Q}\\
quantile: Quantil\\
quartile: Quartil\\
questionnaire: Fragebogen

\medskip
\noindent
{\bf R}\\
randomness: Zuf\"{a}lligkeit\\
random experiment: Zufallsexperiment\\
random sample: Zufallsstichprobe\\
random variable: Zufallsvariable\\
range: Spannweite\\
rank: Rang\\
rank number: Rangzahl\\
rank order: Rangordnung\\
ratio scale: Verh\"{a}ltnisskala\\
raw data set: Datenurliste\\
realisation: Realisierung, konkreter Messwert f\"{u}r eine 
Zufallsvariable\\
regression analysis: Regressionsanalyse\\
regression coefficient: Regressionskoeffizient\\
regression model: Regressionsmodell\\
regression toward the mean: Regression zur Mitte\\
rejection region: Ablehnungsbereich\\
replication: Nachahmung\\
research: Forschung\\
research question: Forschungsfrage\\
residual: Residuum, Restgr\"{o}\ss e\\
risk: Risiko (berechenbar)

\medskip
\noindent
{\bf S}\\
$\sigma$--algebra: $\sigma$--Algebra\\
$6\sigma$--event: $6\sigma$--Ereignis\\
sample: Stichprobe\\
sample correlation coefficient: 
Stichprobenkorrelationskoeffizient\\
sample covariance: Stichprobenkovarianz\\
sample mean: Stichprobenmittelwert\\
sample size: Stichprobenumfang\\
sample space: Ergebnismenge\\
sample variance: Stichprobenvarianz\\
sampling distribution: Stichprobenkenngr\"{o}\ss enverteilung\\
sampling error: Stichprobenfehler\\
sampling frame: Auswahlgesamtheit\\
sampling unit: Stichprobeneinheit\\
scale-invariant: skaleninvariant\\
scale level: Skalenniveau\\
scale parameter: Skalenparameter\\
scatter plot: Streudiagramm\\
scientific method: Wissenschaftliche Methode\\
shift theorem: Verschiebungssatz\\
significance level: Signifikanzniveau\\
simple random sample: einfache Zufallsstichprobe\\
skewness: Schiefe\\
slope: Steigung\\
spectrum of values: Wertespektrum\\
spurious correlation: Scheinkorrelation\\
standard error: Standardfehler\\
standardisation: Standardisierung\\
statistical (in)dependence: statistische (Un)abh\"{a}ngigkeit\\
statistical unit: Erhebungseinheit\\
statistical significance: statistische Signifikanz\\
statistical variable: Merkmal, Variable\\
stochastic: stochastisch, wahrscheinlichkeitsbedingt\\
stochastic independence: stochastische Unabh\"{a}ngigkeit\\
stratified random sample: geschichtete Zufallsstichprobe\\
strength: St\"{a}rke\\
summary table: Zusammenfassungstabelle\\
survey: statistische Erhebung, Umfrage

\medskip
\noindent
{\bf T}\\
test statistic: Teststatistik, statistische
Effektmessgr\"{o}\ss e\\
type I error: Fehler 1.~Art\\
type II error: Fehler 2.~Art

\medskip
\noindent
{\bf U}\\
unbiased: erwartungstreu, unverf\"{a}lscht, unverzerrt\\
uncertainty: Unsicherheit (nicht berechenbar)\\
univariate: univariat, eine variable Gr\"{o}\ss e betreffend\\
unit: Einheit\\
urn model: Urnenmodell

\medskip
\noindent
{\bf V}\\
value: Wert\\
variance: Varianz\\
variation: Variation\\
Venn diagram: Venn--Diagramm\\
visual analogue scale: visuelle Analogskala

\medskip
\noindent
{\bf W}\\
weighted mean: gewichteter Mittelwert

\pagebreak
\medskip
\noindent
{\bf Z}\\
$Z$ scores: $Z$--Werte\\
zero point: Nullpunkt


\addcontentsline{toc}{chapter}{Bibliography}


\end{document}